\documentclass[preprint,amsmath,amssymb]{revtex4}
\usepackage{bm}
\usepackage{graphicx}

\begin{document}

\newcommand{\uu}[1]{\underline{#1}}
\newcommand{\pp}[1]{\phantom{#1}}
\newcommand{\be}{\begin{eqnarray}}
\newcommand{\ee}{\end{eqnarray}}
\newcommand{\ve}{\varepsilon}
\newcommand{\vs}{\varsigma}
\newcommand{\Tr}{{\,\rm Tr\,}}
\newcommand{\pol}{\textstyle{\frac{1}{2}}}
\newcommand{\lbar}{l_{^{\!\bar{}}}}
\newcommand{\Texp}{\textrm{Texp}}

\title{Regularization just by quantization
--- a new approach to the old problem of infinities in quantum field theory\\
(Draft of lecture notes)}
\author{Marek Czachor}
\affiliation{
Katedra Fizyki Teoretycznej i Informatyki Kwantowej\\
Politechnika Gda\'nska, 80-952 Gda\'nsk, Poland
}
\date{\today}

\maketitle

\eject

\section*{Preface}

In 1994-95 I worked as a postdoc in Dave Pritchard's atom interferometer group at MIT. My task was to understand scattering of a laser beam by an atom whose center-of-mass wave function was a two-peaked Gaussian. The problem was subtle and difficult, actually too difficult, and at certain stage I decided to replace atoms by harmonic oscillators. The presence of the center-of-mass degree of freedom led to correlations between the frequency $\omega$ of the oscillation and momentum of the center of mass (a kind of Doppler effect). Since center-of-mass coordinate was described by a position operator, the resulting $\omega$ ceased to be a parameter and turned into an observable. In effect, the beam of particles in my interferometer consisted of a finite number of oscillators whose frequencies were represented by eigenvalues of some operators. Distances between peaks of center-of-mass wave functions $\psi(\bm R)$ were macroscopic, but the oscillators were formally pointlike (due to the dipole approximation). The field propagating through the interferometer had many properties of a quantum field, but consisted of a finite number of particles. On the other hand, this did not mean that the number of frequencies was finite. All the frequencies were present due to the Doppler formula $\omega=\omega(\bm V)=\omega(\bm P/M)$ where $\bm P$ was the center-of-mass momentum, and $\bm P$ was distributed according to the Fourier transform of $\psi(\bm R)$: The sharper $\psi(\bm R)$, the flatter the probability density of $\omega$s. So even for a beam consisting of a single oscillator I had all the $\omega$s. The result looked completely natural and physical.
I have never managed to complete the project. But I remember that already at that time my intuition was that the oscillators of Born, Heisenberg and Jordan, from their first paper on field quantization (1925), should be {\it at least as quantum\/} as ``my" oscillators in center-of-mass interferometers.

Four years later, in Clausthal, I worked on a Darboux-integrable toy model of a 2-level system interacting with harmonic oscillators. The goal was to have an exactly solvable soliton model that could serve as an intermediate step towards a quantum field theoretic generalization. At certain stage I noticed that the formulas I obtained seemed to have similarity to something I once saw in Feynman's lectures on statistical physics. However, my equations, when confronted with those of Feynman, exhibited unexpected differences. It turned out that I instinctively treated all the oscillators in analogy to those from the center-of-mass interferometer. In my example all the frequencies were represented by eigenvalues of operators and occurred with probabilities determined by a wave function. The resulting averages, even for a single oscillator, looked like those for an ensemble of oscillators but with different $\omega$s appearing with different probabilities.
In particular, an average energy of a {\it single\/} oscillator looked like this:
\be
\sum_n\int d\omega |\psi(\omega,n)|^2 \hbar\omega(n+1/2).\nonumber
\ee
Feynman, on the contrary, treated all these numbers as classical parameters so he had no $\omega$ in $\psi$.
His result was
\be
\sum_n\int d\omega |\psi(n)|^2 \hbar\omega(n+1/2),\nonumber
\ee
which was infinite and had to be amended by adding an ad hoc
cutoff function $\chi(\omega)$,
\be
\sum_n\int d\omega |\psi(n)|^2 \hbar\omega(n+1/2)\to \sum_n\int d\omega \chi(\omega)|\psi(n)|^2 \hbar\omega(n+1/2).
\nonumber
\ee
My $|\psi(\omega,n)|^2$ automatically appeared in the correct place. I remember the thrill I felt when I understood what I did. Would it be a rule even in more complicated examples? Was it possible that I discovered, by accident, a fundamental origin of all the regularizations that were put by hand in quantum field theory?

The more I contemplated the assumption I unwillingly made, the more I was convinced that I was right. Thinking of physical examples of oscillators, a pendulum or a charge in magnetic field, I realized that $\omega$ is always an observable, and not a parameter. The length in $\omega=\sqrt{g/l}$ is a difference of center-of-mass position operators, the magnetic field in $\omega=eB/m$ is an operator... And so on with practically all the examples. So why should we assume that the oscillators behind field quantization are {\it more classical\/}? I checked in Max Jammer's book that when Born, Heisenberg and Jordan were publishing their seminal paper they did not yet understand the role of eigenvalues in quantum mechanics. David Hilbert, who mentioned in conversation with Born and Heisenberg that he saw links of Heisenberg quantization to eigenvalue problems, even later laughed at them that had they treated him more seriously, they would have discovered Sch\"odinger equation before Sch\"odinger.

I plunged into more advanced calculations and quickly posted on arxiv.org the first preprint, where I tested the idea of field quantization with only one oscillator, just to have a copyright for the main idea. Roughly in a year from the first insight I published a paper on nonrelativistic quantum optics where the number of oscillators was arbitrary or indefinite, but anyway independent of the number of $\omega$s, just like in the center-of-mass interferometer (the analysis of blackbody radiation I gave in this paper was premature). In the next step I concentrated on fermions and Poincar\'e covariance. The collaboration with Marcin Wilczewski led to three papers on vacuum Rabi oscillations in lossy cavities in the context of experimental data of S.~Haroche et al. We also understood there the link of my quantization to the quantum weak law of large numbers, thermodynamic limit, and R\'enyi statistics. The next group of papers (with Jan Naudts and Klaudia Wrzask) went yet deeper into relativistic properties of electromagnetic fields interacting with classical sources. ``My" automatic regularization was especially elegant in the context of infrared divergences normally accompanying classical pointlike charges. Preliminary results for EPR correlation of polarizations showed agreement with standard quantum optical predictions. 

This collection of ``lecture notes" does not contain any previously published material. My intention was to present the whole structure of the theory on a relatively simple example, where unnecessary technical details are reduced to a minimum. I also wanted to go much further than what I managed to publish so far. The main goals were loop diagrams and the Casimir force. The loop diagrams involve qualitatively different divergences, basically reducible to squares of Dirac deltas, so this had to be clarified. The Casimir effect is a consequence of nontrivial boundary conditions. Moreover, the Casimir force is sensitive to the structure of vacuum and is often ``cutoff dependent". The dependence on cutoff is, from my perspective, a dependence on the wave function of vacuum so, in principle, measurements of the force might reveal fine details of the vacuum state. However, the calculations I will present in final sections show that at least in parallel-plate configurations modifications of the Casimir pressure are negligible. An interesting by-product is the formula for modified potentials produced by a pointlike charge. I show that agreement with current experimental data does not exclude a potential that changes sign at large distances (of the order of thousands of astronomical units).

\eject

\section*{Units and notation}

Wave functions $\psi(\vec p)$ in momentum space are assumed to be normalizable with respect to relativistically invariant measures,
\be
\int \frac{d^3p}{(2\pi\hbar)^3 2\sqrt{\vec p^2+(m_0c)^2}}|\psi(\vec p)|^2=1
\ee
where $m_0^2\geq 0$. The physical unit of the measure is $(m_0c)^3/(\hbar^3 m_0c)=m_0^2c^2/\hbar^3$ and thus the unit of the probability density is $\hbar^3/(m_0^2c^2)$. However, an essential role in the formalism we will discuss in these notes will be played by a number $Z$ that has, on one hand, the same dimension as $|\psi(\vec p)|^2$, but on the other hand should be dimensionless. The reason is that if $q$ is a bare charge parameter, the ``renormalized" parameter $q_{\rm ph}=q\sqrt{Z}$ also should have a meaning of charge. It follows that we have to work with dimensionless $|\psi(\vec p)|^2$. The simplest way of guaranteeing this property is to work from the outset with dimensionless physical quantities. This can be achieved for the price of introducing a fundamental unit of length, the most natural one being the Planck length $\ell=\sqrt{\hbar G/c^3}$. The remaining units can be derived as follows: momentum $\hbar/\ell$, time $\ell/c$, mass $\hbar/(\ell c)$, energy $\hbar c/\ell$. I propose the following notation for position and momentum vectors:  $\vec x=\bm x \ell$, $\vec p=\bm p \hbar/\ell$. Now $\bm x$ and $\bm p$ are dimensionless. The formula $E^2-c^2\vec p^2=m_0^2c^4=E^2-\bm p^2 c^2 \hbar^2/\ell^2$ transforms into its dimensionless analogue
\be
E^2 \ell^2/(c\hbar)^2-\bm p^2=m_0^2c^2 \ell^2/\hbar^2=m^2
\ee
where the right side is a dimensionless mass $m$ squared. Instead of dimensionless energy it is natural to work with dimensionless $p_0$ satisfying $p_0^2-\bm p^2=m^2$. If $m=0$ the corresponding dimensionless momentum will be denoted by $\bm k$. The normalization is thus
\be
\int \frac{d^3p}{(2\pi)^3 2\sqrt{\bm p^2+m^2}}|\psi(\bm p)|^2 &=&1,\\
\int \frac{d^3k}{(2\pi)^3 2|\bm k|}|\psi(\bm k)|^2 &=&1.
\ee
Let $\Box$ be the d'Alembertian operator in dimensionless coordinates. The classical Klein-Gordon wave equation
\be
(\Box+m^2)\phi(x_0,\bm x)=0,
\ee
expressed in dimensionless variables implies a dimensionless Noether current
\be
t{_\mu}{^r}(x)
&=&
2C\partial_\mu \phi(x)\partial^r \phi(x)
-
C\Big(\partial_\nu\phi(x)\partial^\nu\phi(x)-m^2\phi(x)^2\Big)
g{_\mu}{^r}
\ee
where $C$ is a dimensionless constant. In order to correctly identify observable quantities we have to control the transition to ordinary variables.
Let $\Phi(ct,\vec x)$ be the Klein-Gordon field written in terms of ordinary position and time, whose relation to $\phi(x)$ is
\be
\phi(x_0,\bm x)=D \Phi(x_0\ell,\bm x\ell),
\ee
$D$ being a constant which makes $\phi$ dimensionless.
The field satisfies
\be
\Bigg(\hbar^2\frac{\partial^2}{c^2\partial t^2}-\hbar^2\frac{\partial^2}{\partial \vec x{}^2}+m_0^2c^2\Bigg)\Phi(ct,\vec x)=0
\ee
The 3-momentum reads
\be
\vec P
&=&
2C\int d^3\vec x\, \frac{\partial}{c\partial t}\Phi(ct,\vec x)\frac{\partial}{\partial \vec x} \Phi(ct,\vec x)
\nonumber\\
&=&
2CD^2\ell^3\int d^3\bm x\, \frac{\partial}{\ell\partial x_0}\phi(x)\frac{\partial}{\ell\partial \bm x} \phi(x)
\nonumber\\
&=&
2CD^2\ell\int d^3\bm x\, \frac{\partial}{\partial x_0}\phi(x)\frac{\partial}{\partial \bm x} \phi(x)
\nonumber
\ee
It follows that $D^2\ell$ has the dimension of momentum, i.e. $\hbar/\ell$. Let us take $D=\sqrt{\hbar}/\ell$.
The force associated with $\vec P$ satisfies
\be
\vec F=\frac{d\vec P}{dt}=\frac{c\hbar}{\ell^2}\frac{d\bm P}{dx_0}=\frac{c\hbar}{\ell^2}\bm F
\ee
and thus the pressure reads
\be
\frac{\vec F}{L_0^2}=\frac{c\hbar}{\ell^4}\frac{\bm F}{L^2},
\ee
where $L$ is a dimensionless distance. It follows that the Casimir pressure we will compute in the dimensionless variables at the end of these notes will have to be multiplied by
$c\hbar/\ell^4$ in order to make it directly comparable with experiment.
\eject

\noindent
{\it Sensible mathematics involves neglecting a quantity when it is small --- not neglecting it just because it is infinitely great and you do not want it!\/}

P. A. M. Dirac (1975)\cite{Dirac0}
\bigskip

\noindent
{\it For many the infinities that still haunt physics cry for further and deeper quantization,
but there has been little agreement on exactly what and how far to quantize.\/}

D. R. Finkelstein (2006) \cite{DRF}

\section{Introduction}

When solving problems in quantum mechanics one often notices that discrete energy levels are determined by practically one mathematical condition ---  finite norm of eigenvectors. The reason for the finiteness is simple --- squared norm represents sums of probabilities associated with maximal sets of jointly measurable physical quantities. It should be normalizable to 1 and thus cannot be infinite. Finiteness of another important quantity, $\langle\psi|H^2|\psi\rangle$, guarantees that fluctuations $\Delta H=\sqrt{\langle H^2\rangle-\langle H\rangle^2}$ are finite. From a mathematical point of view $\langle\psi|H^2|\psi\rangle<\infty$ means that $|\psi\rangle$ and $H|\psi\rangle$ belong to the same Hilbert space, a condition that defines the domain of $H$. As we can see, the apparently trivial ``law of finiteness of quantities that should be finite'' is raised in quantum mechanics to the level of important technical principle. It determines values of physical quantities (spectra of operators) and sets of acceptable states of physical systems (domains of operators). If one allowed for discrete spectra and infinite norms then harmonic oscillators would possess infinitely many arbitrarily large negative energy levels.

In quantum field theory we become much more tolerant. Infrared and ultraviolet ``catastrophes" (infinities occurring at various stages of calculations) are regularized and renormalized. Algorithms for extracting physically reasonable numbers from infinite theoretical predictions were invented and resulted in several Nobel Prizes in physics. The procedures generally seem to work, but how to explain to a beginner why we cannot proceed in a similar way in standard quantum mechanics with ``divergent probabilities associated with non-normalizable eigenvectors"? We could take the inverted Gaussian associated with  $E=-\hbar\omega/2$, regularize the divergent norm by a cutoff, divide the eigenvector by such a regularized norm, compute average energy, and finally remove the cutoff. The procedure would certainly imply some kind of ``catastrophe'', but basically what would be wrong with it? A similar level of logical rigor seems acceptable in quantum field theory.

Yet, in spite of all the successes of quantum field theory I'm convinced that it is the paradigm of quantum mechanics that is physically correct. We have to determine mathematical structures by the well-definiteness of physical quantities. Not all mathematically well defined models are physical, but models that are mathematically inconsistent are unphysical for sure. The fact that it is practically impossible to perform a nontrivial quantum field theoretic calculation without encountering an ill defined mathematical expression shows, in my opinion, that the most fundamental theory of physics is formulated in a wrong way.

What I write is, of course, not original. Dirac repeatedly stressed that a deep change is needed in conceptual foundations of field quantization. In his opinion quantum field theory is still at the stage analogous to atomic physics before the advent of Schr\"odinger's equation --- we had agreement between theory and experiment (Bohr's model of atom predicted correct spectra), but theoretical ideas were completely wrong. In his last two papers \cite{Dirac1984a,Dirac1984b}, published in the year of his death, he formulated his desiderata for fundamental physical theories. He believed that quantum dynamics should always be described by Heisenberg equation of motion with appropriate Hamiltonian. Modifications are expected in algebraic properties of dynamical variables since here freedom is immense. He stressed the unexplored potential of reducible representations of fundamental symmetries.

In this context it should be reminded that the Heisenberg equation is completely unrelated to classical Euler-Lagrange equations. Links with the latter can only be established after having decided what are the commutation relations between the dynamical variables, and which representation to choose.

So, which algebra should be satisfied by the dynamical variables?

Finkelstein suggests \cite {DRF,FM} that one should quantize in terms of irreducible representations of those Lie algebras whose physically meaningful representations are finite dimensional. In these notes I want to concentrate on an approach which is, in a sense, complementary to the philosophy of Finkelstein (I prefer reducible but infinitely-dimensional representations), and whose main ideas can be found already in the two preliminary papers of mine \cite{0,I}. However, copyright for the term ``regularization by quantization" is due to David Finkelstein.

Restricting  the discussion to spinless fields I propose to consider the following two options:
Canonical commutation relations (CCR),
\be
{[a(\bm p),a(\bm p')^{\dag}]}
&=&
\delta(\bm p,\bm p'),\label{CCR}
\ee
and harmonic oscillator Lie algebra (HOLA), whose nontrivial commutators read
\be
{[a(\bm p),a(\bm p')^{\dag}]}
&=&
\delta(\bm p,\bm p')I(\bm p),\label{HO1}\\
{[a(\bm p),n(\bm p')]}
&=&
\delta(\bm p,\bm p')a(\bm p),\label{HO2}\\
{[a(\bm p)^{\dag},n(\bm p')]}
&=&
-\delta(\bm p,\bm p')a(\bm p)^{\dag},\label{HO3}
\ee
with $\delta(\bm p,\bm p')$ playing the role of structure constants.

The idea of replacing Poisson brackets by commutators leads to (\ref{CCR}) as a natural candidate. An interpretation of fields as ensembles of harmonic oscillators suggests (\ref{HO1})--(\ref{HO3}). The usual identifications
$n(\bm p)=a(\bm p)^{\dag}a(\bm p)$ and $I(\bm p)=1$ may seem to imply that CCR determines the form of HOLA, but this is not the case, as we shall see in a moment.
Moreover, whatever algebra one selects, one yet has to tell something about $\delta(\bm p,\bm p')$, which is less obvious than one might expect.

\section{CCR does not determine HOLA}

Let us now show that a concrete representation of CCR does not yet fix the form of an associated HOLA. Let $[a_1,a_1^{\dag}]=I_1$ be the representation of CCR in a Hilbert space ${\cal H}_1$ ($I_1$ is the identity map in ${\cal H}_1$). Defining $n_1=a_1^{\dag}a_1$ we obtain a representation of HOLA:
\be
{[a_1,a_1^{\dag}]} &=& I_1,\\
{[a_1,n_1]} &=& a_1,\\
{[a_1^{\dag},n_1]} &=& -a_1^{\dag}.
\ee
Now take
\be
a_2 &=&\frac{1}{\sqrt{2}}\Big(a_1\otimes I_1+I_1\otimes a_1\Big),\label{a_2}\\
a_2^{\dag} &=&\frac{1}{\sqrt{2}}\Big(a_1^{\dag}\otimes I_1+I_1\otimes a_1^{\dag}\Big).
\ee
One checks that
\be
{[a_2,a_2^{\dag}]} &=& I_1\otimes I_1=I_2,
\ee
i.e. this is a representation of CCR in ${\cal H}_2={\cal H}_1\otimes {\cal H}_1$.
In order to extend it to HOLA we have to find an appropriate $n_2$. Obviously, one choice is
\be
n_2 &=& a_2^{\dag}a_2\label{n_2}\\
&=&
\frac{1}{2}\Big(a_1^{\dag}a_1\otimes I_1+I_1\otimes a_1^{\dag}a_1+a_1^{\dag}\otimes a_1+a_1\otimes a_1^{\dag}\Big).\nonumber\\
\ee
The Hamiltonian $H_2=\hbar\omega n_2$ represents a system of two {\it interacting\/} oscillators of frequency $\omega/2$, where
\be
H_2^{(0)}
=
\frac{\hbar\omega}{2}a_1^{\dag}a_1\otimes I_1+I_1\otimes \frac{\hbar\omega}{2}a_1^{\dag}a_1
\ee
is the free part. The interaction term
\be
H_2^{(1)}
=
\frac{\hbar\omega}{2}\Big(a_1^{\dag}\otimes a_1+a_1\otimes a_1^{\dag}\Big)
\ee
is responsible for energy exchange. The algebra is as it should be
\be
{[a_2,a_2^{\dag}]} &=& I_2,\\
{[a_2,n_2]} &=& a_2,\\
{[a_2^{\dag},n_2]} &=& -a_2^{\dag}.
\ee
The problem is that $n_2$ could be replaced by
\be
\tilde n_2
&=&
n_1\otimes I_1+I_1\otimes n_1,\label{tilde n_2}
\ee
still satisfying the same algebra
\be
{[a_2,a_2^{\dag}]} &=& I_2,\\
{[a_2,\tilde n_2]} &=& a_2,\\
{[a_2^{\dag},\tilde n_2]} &=& -a_2^{\dag},
\ee
with the same $a_2$, $a_2^{\dag}$ and $I_2$.
Physically the new representation corresponds to two {\it non-interacting\/} oscillators of frequency $\omega$ if we employ the same correspondence between Hamiltonian and the number-of-excitations operator:
\be
\tilde H_2 &=& \hbar\omega \tilde n_2\\
&=&
\hbar\omega a_1^{\dag}a_1\otimes I_1+I_1\otimes \hbar\omega a_1^{\dag}a_1.
\ee
It is not surprising that still further possibilities exist,
\be
\tilde{\tilde n}{}_2
&=&
a_1^{\dag}\otimes a_1+a_1\otimes a_1^{\dag},
\ee
or an arbitrary convex combination $p_1{\tilde n}{}_2+p_2\tilde{\tilde n}{}_2$, $p_1+p_2=1$. An example of the latter has already been encountered,
\be
n_2 &=& \frac{1}{2}{\tilde n}{}_2+\frac{1}{2}\tilde{\tilde n}{}_2.
\ee
We will later see that the above ambiguities imply certain level of freedom in definitions of quantum-field 4-momenta and number operators.

\section{Number-of-particles representation (bosonic Fock space)}

Before we discuss field quantization in more detail let me introduce here the familiar number-of-particles representation of multi particle systems. The construction is interesting in itself and is often regarded as {\it the\/} representation appropriate for quantum fields. We will later see that the latter statement is not necessarily true, or at least is less obvious than what one is generally led to believe.

One often reads that ``there is no quantum field theory, there is only a theory of multi-particle systems''. In the next chapter
I will give arguments why formal similarities between multi-particle systems and quantum fields may be misleading, a subtlety closely related to the distinction between particles and quasi-particles. But first we have to understand the standard construction, although in the end it will {\it not\/} be employed in my own approach to field quantization.

We begin with a Hilbert space $\cal H$ whose countable basis will be denoted by $|j\rangle$, $j=1,2\dots$. Hilbert spaces that possess a countable basis are called separable (the standard Hilbert space from the first semester of quantum mechanics is separable, an example of the countable basis being provided by Hermite polynomials). Any operator $A$ has matrix elements $\langle k|A|l\rangle=A_{kl}$ and can be written as
\be
A=\sum_{k}|k\rangle\langle k|A\sum_{l}|l\rangle\langle l|=\sum_{kl}|k\rangle\langle k|A|l\rangle\langle l|=\sum_{kl}A_{kl}|k\rangle\langle l|.
\ee
If $|\tilde j\rangle$, $j=1,2\dots$, $j=1,2\dots$ is another basis in $\cal H$, then
\be
V=\sum_j|\tilde j\rangle\langle j|
\ee
is unitary,
\be
VV^\dag &=&\sum_j|\tilde j\rangle\langle j|\sum_k|k\rangle\langle \tilde k|=\sum_{jk}|\tilde j\rangle\langle j|k\rangle\langle \tilde k|=\sum_{jk}|\tilde j\rangle\delta_{jk}\langle \tilde k|
=
\sum_{j}|\tilde j\rangle\langle \tilde j|={\mathbb I},\\
V^\dag V&=&\sum_k|k\rangle\langle \tilde k|\sum_j|\tilde j\rangle\langle j|=\sum_{jk}|k\rangle\langle \tilde k|\tilde j\rangle\langle j|=\sum_{jk}|k\rangle\delta_{jk}\langle j|
=
\sum_{j}|j\rangle\langle j|={\mathbb I},
\ee
relates the two bases,
\be
V|k\rangle &=& \sum_j|\tilde j\rangle\langle j|k\rangle= \sum_j|\tilde j\rangle \delta_{jk}= |\tilde k\rangle,
\ee
and has matrix elements
\be
\langle k|V|l\rangle
&=&
\langle k|\sum_j|\tilde j\rangle\langle j|l\rangle
=
\sum_j\langle k|\tilde j\rangle\delta_{jl}=\langle k|\tilde l\rangle.\label{U kl}
\ee
Moreover,
\be
|\tilde k\rangle
&=&
V|k\rangle
=
\sum_j|j\rangle\langle j|V|k\rangle
=
\sum_j V_{jk}|j\rangle \label{U_kl'}
\ee
shows that numbers $V_{lk}$ allow us to write the new basis as a linear combination of the old one.

Now let us take a countable collection of operators $a_j$ satisfying CCR $[a_k,a_l^\dag]=\delta_{kl}$, and let $|0)$ be their common ``vacuum state'',
\be
a_j|0) &=& 0, \quad j=1,2,\dots\\
(0|0) &=& 1.
\ee
We do not assume that $|0)$ belongs to $\cal H$. It is a completely independent object, belonging to some linear space $\cal F$, say. Let $|j)=a_j^\dag |0)$, $(j|=(0|a_j$. The new vectors are orthonormal
\be
(k|l) &=& (0|a_ka_l^\dag|0)=(0|a_ka_l^\dag-a_l^\dag a_k|0)=\delta_{kl}(0|0)=\delta_{kl},
\ee
and (by definition) also belong to $\cal F$. $|j)$ are orthogonal to $|0)$,
\be
(j|0) &=& (0|a_j|0) =0.
\ee
Taking numbers $V_{kl}=\langle k|V|l\rangle$, defined in (\ref{U kl}), we can define a new set of vectors
\be
\sum_j |j)\langle j|V|k\rangle
&=&
\sum_j V_{jk}|j)=\underbrace{\sum_j V_{jk}a_j^\dag}_{\tilde a_k^\dag} |0)=\tilde a_k^\dag|0)=|\tilde k),\\
(j|\tilde k)
&=&
\langle j|V|k\rangle
=
\langle j|\tilde k\rangle.
\ee
Note that
\be
\tilde a_k &=&
\Big(\sum_j V_{jk}a_j^\dag\Big)^\dag
=
\sum_j \overline{V_{jk}}a_j
=
\sum_j V^\dag_{kj}a_j,
\ee
so that
\be
{[\tilde a_k,\tilde a_m^{\dag}]}
&=&
\big[\sum_j V^\dag_{kj}a_j,\sum_l V_{lm}a_l^\dag\big]
=
\sum_j V^\dag_{kj}\sum_l V_{lm}[a_j,a_l^\dag]
=
\sum_{jl} V^\dag_{kj}V_{lm}\delta_{jl}\nonumber\\
&=&
\sum_{j} V^\dag_{kj}V_{jm}
=
(V^\dag V)_{km}
=
\langle k|V^\dag V|m\rangle
=
\langle \tilde k|\tilde m\rangle=\delta_{km}.
\ee
The fact that
\be
{[\tilde a_k,\tilde a_m^{\dag}]}
&=&
[a_k,a_m^{\dag}]=\delta_{km}
\ee
suggests that there exists a unitary transformation $U$ satisfying
\be
Ua_k^{\dag}U^\dag
&=&
\tilde a_k^{\dag}
=
\sum_j V_{jk}a_j^{\dag}.
\ee
I will construct $U$ explicitly and show that
\be
U|0) &=& U^\dag|0)=|0),\\
U|k)
&=&
Ua_k^{\dag}|0)
=
Ua_k^{\dag}U^\dag|0)
=
\tilde a_k^{\dag}|0)
=
|\tilde k)=
\sum_j V_{jk}a_j^{\dag}|0)=
\sum_j V_{jk}|j).
\ee
As we can see, $U$ will play in $\cal F$  a similar role to that of $V$ in $\cal H$,
\be
|\tilde k) &=& U|k)=\sum_j V_{jk}|j),\\
|\tilde k\rangle &=& V|k\rangle=\sum_j V_{jk}|j\rangle.
\ee
Let us now perform the construction of $U$. To do so we first note that any unitary transformation can be written in an exponential way.
Indeed, eigenvalues $\lambda$ of a unitary operator $V$, say, satisfy
\be
V|\lambda\rangle &=& \lambda|\lambda\rangle,\\
V^{-1}|\lambda\rangle &=& \lambda^{-1}|\lambda\rangle,\\
\langle\lambda|V|\lambda\rangle &=& \lambda\langle\lambda|\lambda\rangle,\\
\overline{\langle\lambda|V|\lambda\rangle} &=& \langle\lambda|V^\dag|\lambda\rangle
=
\overline{\lambda}\langle\lambda|\lambda\rangle
=\langle\lambda|V^{-1}|\lambda\rangle
=\lambda^{-1}\langle\lambda|\lambda\rangle,
\\
\overline{\lambda} &=& \lambda^{-1}.
\ee
Therefore, $\lambda=e^{i\varphi(\lambda)}$ for some real an unique $\varphi(\lambda)\in [0,2\pi)$. Let $\Phi =\sum_\lambda \varphi(\lambda) |\lambda\rangle\langle\lambda|$. Then,
\be
e^{i\Phi} &=&\sum_\lambda e^{i\varphi(\lambda)} |\lambda\rangle\langle\lambda|=\sum_\lambda \lambda |\lambda\rangle\langle\lambda|
\ee
commutes with $V$ and has the same eigenvectors and eigenvalues as $V$. Accordingly, $V=e^{i\Phi}$ with $\Phi^\dag=\Phi$.

In particular,
there exists $X$ such that $U=e^X$, $U^\dag=e^{-X}$. It is known that for any $X$ and $Y$
\be
e^X Y e^{-X}
&=&
Y+[X,Y]+\frac{1}{2!}[X,[X,Y]]+\frac{1}{3!}[X,[X,[X,Y]]]+\dots\label{Y1}
\ee
(Of course, provided products and commutators of operators are meaningful, which is not obvious if the operators are unbounded.)
In order to prove this, let us consider the family of operators
\be
Y(t) &=& e^{tX} Y e^{-tX},\label{Y2}\\
Y(0) &=& Y.
\ee
Differentiating (\ref{Y2}) we find
\be
\frac{d}{dt}Y(t)
&=&
Xe^{tX} Y e^{-tX}
-
e^{tX} Y e^{-tX}X
=
[X,Y(t)].
\ee
On the other hand
\be
{}&{}&\frac{d}{dt}
\Big(
Y+t[X,Y]+\frac{t^2}{2!}[X,[X,Y]]+\frac{t^3}{3!}[X,[X,[X,Y]]]+\dots
\Big)\nonumber\\
&{}&\pp=
=
[X,Y]+t[X,[X,Y]]+\frac{t^2}{2!}[X,[X,[X,Y]]]+\dots
=
[X,Y(t)].
\ee
We conclude that
\be
\frac{d}{dt}Y(t)
&=&
\frac{d}{dt}
\Big(
Y+t[X,Y]+\frac{t^2}{2!}[X,[X,Y]]+\frac{t^3}{3!}[X,[X,[X,Y]]]+\dots
\Big),
\ee
where the differentiated objects satisfy the same initial condition at $t=0$. By uniqueness of solution of a first order linear differential equation
one arrives at
\be
Y(t)
&=&
e^{tX} Y e^{-tX}=
Y+t[X,Y]+\frac{t^2}{2!}[X,[X,Y]]+\frac{t^3}{3!}[X,[X,[X,Y]]]+\dots
\ee
for any $t$. Putting $t=1$ we end the proof.

Now take $X$ of the form
\be
X &=& \sum_{kl}a_k^\dag x_{kl}a_l.\label{X}
\ee
Then
\be
{[X,a_j^\dag]}
&=&
\sum_{k}a_k^\dag x_{kj},\\
{[X,[X,a_j^\dag]]}
&=&
\sum_{k}\sum_{nm}[a_n^\dag x_{nm}a_m,a_k^\dag x_{kj}]
=
\sum_{k}\sum_{nm}a_n^\dag x_{nm}\delta_{mk} x_{kj}
=
\sum_{nm}a_n^\dag x_{nm}x_{mj}
\nonumber\\
&=&
\sum_{k}a_k^\dag (x^2)_{kj}
\ee
Repeating the procedure, we find
\be
Ua_j^\dag U^\dag
&=&
a_j^\dag+
[X,a_j^\dag]+\frac{1}{2!}[X,[X,a_j^\dag]]+\frac{1}{3!}[X,[X,[X,a_j^\dag]]]+\dots\nonumber\\
&=&
\sum_{k}a_k^\dag \Big(\delta_{kj}+x_{kj}+\frac{1}{2!}(x^2)_{kj}+\frac{1}{3!}(x^3)_{kj}+\dots\Big)\nonumber\\
&=&
\sum_{k}a_k^\dag \big(e^x\big)_{kj}.
\ee
It follows that in order to construct $U$ we first have to express $V$ as $V=e^x$ for some antiunitary $x=-x^\dag$. The matrix elements
$x_{kj}=\langle k|x|j\rangle$ are then employed in construction of $X$ satisfying $U=e^X$.
Since $X|0)=0$ we get
\be
U|0) &=&\Big( \mathbb{I}+X +\frac{1}{2!}X^2 +\frac{1}{3!}X^3+\dots \Big)|0)=\mathbb{I}|0)=|0).
\ee
In consequence, once we have $a_j$ corresponding to some basis $|j)$, we know how to construct $\tilde a_j$ such that $|\tilde j)$ is related to $|j)$ in a given way, determined by the link between
$|\tilde j\rangle$ and $|j\rangle$.

The vectors $|j)$ in $\cal F$ so constructed are completely analogous to $|j\rangle$ in $\cal H$. Having any vector
\be
|\psi\rangle &=& \sum_j\psi_j |j\rangle \in {\cal H}
\ee
we can construct
\be
|\psi) &=& \sum_j\psi_j |j)=\underbrace{\sum_j\psi_j a_j^\dag}_{\hat\psi}|0)=\hat\psi|0) \in {\cal F}
\ee
whose properties are ``identical" to those of $|\psi\rangle$. $\hat \psi$ is an example of {\it field operator\/}.

It is interesting to check the action of $X$ given by (\ref{X}) on $\hat\psi|0)$,
\be
X\hat\psi|0)
&=&
\sum_{kl}a_k^\dag x_{kl}a_l\sum_j\psi_j a_j^\dag |0)
=
\sum_{klj}a_k^\dag x_{kl}\psi_j [a_l, a_j^\dag] |0)
=
\sum_{klj}a_k^\dag x_{kl}\psi_j \delta_{lj} |0)
\nonumber\\
&=&
\sum_{kl}a_k^\dag x_{kl}\psi_l |0)
=
\sum_{k}(x\psi)_k a_k^\dag |0)
=
\widehat{x\psi}|0).
\ee
We find $X\hat\psi|0)=\widehat{x\psi}|0)$ but this is true only in action on $|0)$. The general rule for $X$ and $\hat\psi$ of the above form is
\be
[X,\hat\psi] &=&\widehat{x\psi}.
\ee
If $\psi_\lambda$ is an eigenvector of $x$, i.e. $x\psi_\lambda=\lambda \psi_\lambda$, then the $\cal F$-space counterpart of the eigenvalue problem is
\be
[X,\hat\psi_\lambda] &=&\lambda \hat\psi_\lambda,
\ee
implying
\be
[X,\hat\psi_{\lambda_1}\dots \hat\psi_{\lambda_K}] &=&(\lambda_1+\dots+\lambda_K) \hat\psi_{\lambda_1}\dots \hat\psi_{\lambda_K},\\
X \hat\psi_{\lambda_1}\dots \hat\psi_{\lambda_K}|0) &=&(\lambda_1+\dots+\lambda_K) \hat\psi_{\lambda_1}\dots \hat\psi_{\lambda_K}|0).
\ee
The latter formula illustrates the essential difference between $\cal H$ and $\cal F$, namely the fact that in $\cal F$ it makes sense to consider vectors of the form
\be
|jk) &=& a_j^\dag a_k^\dag|0),\\
|jkl) &=& a_j^\dag a_k^\dag a_l^\dag|0),\\
&\vdots&\nonumber
\ee
all of them orthogonal to $|l)$, $l=0,1,2,\dots$. The operator
\be
n &=& \sum_m a_m^\dag a_m,
\ee
acts on these vectors as follows
\be
n|0) &=& 0,\\
n|j) &=& |j),\\
n|jk) &=& 2|jk),\\
n|jkl) &=& 3|jkl),
\ee
so it simply counts how many times one acted with creation operators on $|0)$. For this reason $n$ is termed the number-of-particles operator. The subspace consisting of vectors satisfying $n|\psi)=|\psi)$ is called the one-particle sector of the (bosonic) Fock space $\cal F$. This is the subspace of $\cal F$ that is ``identical'' to the initial Hilbert space $\cal H$ we have started with. $\cal F$ is a countable direct sum of $n$-particle sectors, each of which is in itself a separable Hilbert space. Accordingly, $\cal F$ is also a separable Hilbert space that could become a starting point for yet another Fock space --- and so on and so forth.

In 1925 Heisenberg and his coworkers did not know the notion of a Fock space when they wrote the first paper on field quantization. In fact, they did not even know quantum mechanics. They did not hear of Hilbert spaces or Schr\"odinger's equation. But they formulated cornerstones of modern quantum field theory, with all its advantages and defects.
\medskip

\noindent
{\it Digression\/}: It is instructive to consider also the following construction.
Let us denote by ${\cal H}_n$ the Hilbert space spanned by tensor products
\be
|j_1,\dots,j_n\rangle
&=&
|j_1\rangle\otimes \dots\otimes |j_n\rangle,
\ee
and let $A_n$ and $S_n$ be the projectors on subspaces of, respectively, anti-symmetric and symmetric states in ${\cal H}_n$ (for $n=0$ we take ${\cal H}_0=\mathbb{C}$; $A_0=S_0$ and $A_1=S_1$ are identity operators in ${\cal H}_0$ and ${\cal H}_1$, respectively). Let ${\cal H}=\bigoplus_{n=0}^\infty {\cal H}_n$ and $\alpha_j,\alpha_j^\dag:{\cal H}\to {\cal H}$ be defined by
\be
\alpha_j^\dag |j_1,\dots,j_n\rangle &=& \bar c_{n+1}|j,j_1,\dots,j_n\rangle,\\
\alpha_j |j_0,j_1,\dots,j_n\rangle &=& c_{n+1}\delta_{jj_0}|j_1,\dots,j_n\rangle,\\
\alpha_j |0\rangle &=& 0,\\
\alpha_j^\dag |0\rangle &=&\bar c_1|j\rangle,\\
\alpha_j |j_1\rangle &=&c_1\delta_{jj_1}|0\rangle.
\ee
$c_{n+1}$ and $\bar c_{n+1}$ are constants to be specified later. Let $A=\bigoplus_{n=0}^\infty A_n$,
$S=\bigoplus_{n=0}^\infty S_n$, and
\be
a_j &=& S\alpha_j S,\\
a_j^\dag &=& S\alpha_j^\dag S,\\
b_j &=& A\alpha_j A,\\
b_j^\dag &=& A\alpha_j^\dag A.
\ee
Let us check:
\be
a_ja_j^\dag |0\rangle
&=&
a_j|j\rangle=c_1|0\rangle,\nonumber\\
a_j^\dag a_j |0\rangle
&=&
0,\nonumber\\
{[a_j,a_j^\dag]} |0\rangle
&=&
c_1|0\rangle,\nonumber\\
a_ja_j^\dag |j_1\rangle
&=&
\bar c_2a_j\frac{1}{2}\big(|j,j_1\rangle+|j_1,j\rangle\big)
=
\bar c_2 S\alpha_j\frac{1}{2}\big(|j,j_1\rangle+|j_1,j\rangle\big)
=
\bar c_2 c_2\frac{1}{2}\big(|j_1\rangle+\delta_{jj_1}|j\rangle\big)
\nonumber\\
a_j^\dag a_j |j_1\rangle
&=&
c_1\delta_{jj_1}a_j^\dag |0\rangle
=
c_1\bar c_1\delta_{jj_1}|j\rangle,\nonumber\\
{[a_j,a_j^\dag]} |j_1\rangle
&=&
\frac{\bar c_2 c_2}{2}|j_1\rangle+\frac{\bar c_2 c_2}{2}\delta_{jj_1}|j\rangle-c_1\bar c_1\delta_{jj_1}|j\rangle,\nonumber\\
a_ja_j^\dag  |j_1,j_2\rangle
&=&
a_j \alpha_j^\dag \frac{1}{2}\big(|j_1,j_2\rangle+|j_2,j_1\rangle\Big)
=
\bar c_3 a_j S\frac{1}{2}\big(|j\rangle|j_1,j_2\rangle+|j\rangle|j_2,j_1\rangle\Big)
\nonumber\\
&=&
\bar c_3 a_j \frac{1}{6}\big(|j,j_1,j_2\rangle+|j,j_2,j_1\rangle
+
|j_1,j,j_2\rangle+|j_1,j_2,j\rangle
+
|j_2,j_1,j\rangle+|j_2,j,j_1\rangle
\Big)\nonumber\\
&=&
\bar c_3 c_3S \frac{1}{6}\big(|j_1,j_2\rangle+|j_2,j_1\rangle
+
\delta_{jj_1}|j,j_2\rangle+\delta_{jj_1}|j_2,j\rangle
+
\delta_{jj_2}|j_1,j\rangle+\delta_{jj_2}|j,j_1\rangle
\Big)\nonumber\\
&=&
\bar c_3 c_3\frac{1}{6}\big(|j_1,j_2\rangle+|j_2,j_1\rangle
+
\delta_{jj_1}|j,j_2\rangle+\delta_{jj_1}|j_2,j\rangle
+
\delta_{jj_2}|j_1,j\rangle+\delta_{jj_2}|j,j_1\rangle
\Big)\nonumber\\
a_j^\dag a_j  |j_1,j_2\rangle
&=&
a_j^\dag \alpha_j \frac{1}{2}\big(|j_1,j_2\rangle+|j_2,j_1\rangle\Big)
=
c_2S\alpha_j^\dag  S\frac{1}{2}\big(\delta_{jj_1}|j_2\rangle+\delta_{jj_2}|j_1\rangle\Big)
\nonumber\\
&=&
c_2\bar c_2S\frac{1}{2}\big(\delta_{jj_1}|j,j_2\rangle+\delta_{jj_2}|j,j_1\rangle\Big)
\nonumber\\
&=&
c_2\bar c_2\frac{1}{4}\big(\delta_{jj_1}|j,j_2\rangle+\delta_{jj_1}|j_2,j\rangle+\delta_{jj_2}|j,j_1\rangle+\delta_{jj_2}|j_1,j\rangle\Big)
\nonumber\\
{[a_j,a_j^\dag]}  |j_1,j_2\rangle
&=&
\bar c_3 c_3\frac{1}{6}\big(|j_1,j_2\rangle+|j_2,j_1\rangle
+
\delta_{jj_1}|j,j_2\rangle+\delta_{jj_1}|j_2,j\rangle
+
\delta_{jj_2}|j_1,j\rangle+\delta_{jj_2}|j,j_1\rangle
\Big)\nonumber\\
&\pp=&
-
c_2\bar c_2\frac{1}{4}\big(\delta_{jj_1}|j,j_2\rangle+\delta_{jj_1}|j_2,j\rangle+\delta_{jj_2}|j,j_1\rangle+\delta_{jj_2}|j_1,j\rangle\Big)
\nonumber
\ee
The formulas get incredibly simplified if we take $c_n=\bar c_n=\sqrt{n}$. Then
\be
{[a_j,a_j^\dag]}  |0\rangle &=& |0\rangle,\nonumber\\
{[a_j,a_j^\dag]}  |j_1\rangle &=& |j_1\rangle,\nonumber\\
{[a_j,a_j^\dag]}  |j_1,j_2\rangle &=& \frac{1}{2}\big( |j_1,j_2\rangle+|j_2,j_1\rangle\big)=S|j_1,j_2\rangle\nonumber.
\ee
Note that since $S$ is a projector then ${[a_j,a_j^\dag]}={[a_j,a_j^\dag]}S$, hence
\be
{[a_j,a_j^\dag]}  S|j_1,j_2\rangle &=& S|j_1,j_2\rangle\nonumber.
\ee
Now take any $n>1$ and let $\{\sigma=(\sigma(1),\dots,\sigma(n))\}$ be the set of all the permutations of the sequence $(1,\dots,n)$. Then
\be
{[a_j,a_j^\dag]}|j_1,\dots,j_n\rangle
&=& a_j\frac{1}{n!}\sum_\sigma\alpha_j^\dag|j_{\sigma(1)},\dots,j_{\sigma(n)}\rangle
-a_j^\dag \frac{1}{n!}\sum_\sigma\alpha_j|j_{\sigma(1)},\dots,j_{\sigma(n)}\rangle\nonumber\nonumber\\
&=& a_j\frac{\sqrt{n+1}}{n!}\sum_\sigma|j,j_{\sigma(1)},\dots,j_{\sigma(n)}\rangle
-a_j^\dag \frac{\sqrt{n}}{n!}\sum_\sigma\delta_{jj_{\sigma(1)}}|j_{\sigma(2)},\dots,j_{\sigma(n)}\rangle\nonumber\nonumber\\
&=& S\alpha_j\frac{\sqrt{n+1}}{n!}\sum_\sigma S|j,j_{\sigma(1)},\dots,j_{\sigma(n)}\rangle
-S\alpha_j^\dag \frac{\sqrt{n}}{n!}\sum_\sigma\delta_{jj_{\sigma(1)}}S|j_{\sigma(2)},\dots,j_{\sigma(n)}\rangle\nonumber\nonumber\\
&=& S\alpha_j\frac{\sqrt{n+1}}{n!}\frac{1}{n+1}\sum_\sigma
\Big(
|j,j_{\sigma(1)},\dots,j_{\sigma(n)}\rangle
+
\dots
+
|j_{\sigma(1)},\dots,j_{\sigma(n)},j\rangle
\Big)
\nonumber\\
&\pp=&
-S\alpha_j^\dag \frac{\sqrt{n}}{n!}\sum_\sigma\delta_{jj_{\sigma(1)}}|j_{\sigma(2)},\dots,j_{\sigma(n)}\rangle\nonumber\nonumber
\ee
\be
&=& S\frac{1}{n!}\sum_\sigma
\Big(
|j_{\sigma(1)},\dots,j_{\sigma(n)}\rangle
+
\delta_{jj_{\sigma(1)}}|j,j_{\sigma(2)},\dots,j_{\sigma(n)}\rangle
+
\dots
+
\delta_{jj_{\sigma(1)}}|j_{\sigma(2)},\dots,j_{\sigma(n)},j\rangle
\Big)
\nonumber\\
&\pp=&
-S\frac{n}{n!}\sum_\sigma\delta_{jj_{\sigma(1)}}|j,j_{\sigma(2)},\dots,j_{\sigma(n)}\rangle\nonumber\nonumber\\
&=& S\frac{1}{n!}\sum_\sigma
\Big(
|j_{\sigma(1)},\dots,j_{\sigma(n)}\rangle
+
\delta_{jj_{\sigma(1)}}|j,j_{\sigma(2)},\dots,j_{\sigma(n)}\rangle
+
\dots
+
\delta_{jj_{\sigma(1)}}|j_{\sigma(2)},\dots,j_{\sigma(n)},j\rangle
\Big)
\nonumber\\
&\pp=&
-S\frac{n}{n!}\frac{1}{n}\sum_\sigma\delta_{jj_{\sigma(1)}}\Big(
|j,j_{\sigma(2)},\dots,j_{\sigma(n)}\rangle
+
\dots
+
|j_{\sigma(2)},\dots,j_{\sigma(n)},j\rangle
\Big)
\nonumber\\
&=& S\frac{1}{n!}\sum_\sigma
|j_{\sigma(1)},\dots,j_{\sigma(n)}\rangle
\nonumber\\
&=& S
|j_1,\dots,j_n\rangle\nonumber
\ee
Finally,
\be
{[a_j,a_j^\dag]}=S
\ee
Analogously
\be
b_jb_j^\dag |0\rangle
&=&
b_j|j\rangle=c_1|0\rangle,\nonumber\\
b_j^\dag b_j |0\rangle
&=&
0,\nonumber\\
{\{b_j,b_j^\dag\}]} |0\rangle
&=&
c_1|0\rangle,\nonumber\\
b_jb_j^\dag |j_1\rangle
&=&
\bar c_2b_j\frac{1}{2}\big(|j,j_1\rangle-|j_1,j\rangle\big)
=
\bar c_2 A\alpha_j\frac{1}{2}\big(|j,j_1\rangle-|j_1,j\rangle\big)
=
\bar c_2 c_2\frac{1}{2}\big(|j_1\rangle-\delta_{jj_1}|j\rangle\big)
\nonumber\\
b_j^\dag b_j |j_1\rangle
&=&
c_1\delta_{jj_1}b_j^\dag |0\rangle
=
c_1\bar c_1\delta_{jj_1}|j\rangle,\nonumber\\
\{b_j,b_j^\dag\} |j_1\rangle
&=&
\frac{\bar c_2 c_2}{2}|j_1\rangle-\frac{\bar c_2 c_2}{2}\delta_{jj_1}|j\rangle+c_1\bar c_1\delta_{jj_1}|j\rangle,\nonumber\\
b_jb_j^\dag  |j_1,j_2\rangle
&=&
b_j \alpha_j^\dag \frac{1}{2}\big(|j_1,j_2\rangle-|j_2,j_1\rangle\Big)
=
\bar c_3 b_j A\frac{1}{2}\big(|j,j_1,j_2\rangle-|j,j_2,j_1\rangle\Big)
\nonumber\\
&=&
\bar c_3 A\alpha_j \frac{1}{6}\big(
|j,j_1,j_2\rangle-|j,j_2,j_1\rangle
-
|j_1,j,j_2\rangle+|j_1,j_2,j\rangle
-
|j_2,j_1,j\rangle+|j_2,j,j_1\rangle
\Big)\nonumber\\
&=&
\bar c_3 c_3A \frac{1}{6}\big(
|j_1,j_2\rangle-|j_2,j_1\rangle
-
\delta_{jj_1}|j,j_2\rangle+\delta_{jj_1}|j_2,j\rangle
-
\delta_{jj_2}|j_1,j\rangle+\delta_{jj_2}|j,j_1\rangle
\Big)\nonumber\\
&=&
\bar c_3 c_3\frac{1}{6}\big(
|j_1,j_2\rangle-|j_2,j_1\rangle
-
\delta_{jj_1}|j,j_2\rangle+\delta_{jj_1}|j_2,j\rangle
-
\delta_{jj_2}|j_1,j\rangle+\delta_{jj_2}|j,j_1\rangle
\Big)\nonumber\\
b_j^\dag b_j  |j_1,j_2\rangle
&=&
a_j^\dag \alpha_j \frac{1}{2}\big(|j_1,j_2\rangle-|j_2,j_1\rangle\Big)
=
c_2A\alpha_j^\dag  A\frac{1}{2}\big(\delta_{jj_1}|j_2\rangle-\delta_{jj_2}|j_1\rangle\Big)
\nonumber\\
&=&
c_2\bar c_2A\frac{1}{2}\big(\delta_{jj_1}|j,j_2\rangle-\delta_{jj_2}|j,j_1\rangle\Big)
\nonumber\\
&=&
c_2\bar c_2\frac{1}{4}\big(\delta_{jj_1}|j,j_2\rangle-\delta_{jj_1}|j_2,j\rangle-\delta_{jj_2}|j,j_1\rangle+\delta_{jj_2}|j_1,j\rangle\Big)
\nonumber\\
\{b_j,b_j^\dag\}  |j_1,j_2\rangle
&=&
\bar c_3 c_3\frac{1}{6}\big(
|j_1,j_2\rangle-|j_2,j_1\rangle
-
\delta_{jj_1}|j,j_2\rangle+\delta_{jj_1}|j_2,j\rangle
-
\delta_{jj_2}|j_1,j\rangle+\delta_{jj_2}|j,j_1\rangle
\Big)\nonumber\\
&\pp=&
+
c_2\bar c_2\frac{1}{4}\big(\delta_{jj_1}|j,j_2\rangle-\delta_{jj_1}|j_2,j\rangle-\delta_{jj_2}|j,j_1\rangle+\delta_{jj_2}|j_1,j\rangle\Big)
\nonumber
\ee
As before, choosing $c_n=\bar c_n=\sqrt{n}$, we simplify the formulas. Let $\sum_\sigma'=\sum_\sigma {\rm sign}(\sigma)$.
\be
\{b_j,b_j^\dag\}|j_1,\dots,j_n\rangle
&=& b_j\frac{1}{n!}\sum_\sigma{'}\alpha_j^\dag|j_{\sigma(1)},\dots,j_{\sigma(n)}\rangle
+b_j^\dag \frac{1}{n!}\sum_\sigma{'}\alpha_j|j_{\sigma(1)},\dots,j_{\sigma(n)}\rangle\nonumber\nonumber\\
&=& b_j\frac{\sqrt{n+1}}{n!}\sum_\sigma{'}|j,j_{\sigma(1)},\dots,j_{\sigma(n)}\rangle
+b_j^\dag \frac{\sqrt{n}}{n!}\sum_\sigma{'}\delta_{jj_{\sigma(1)}}|j_{\sigma(2)},\dots,j_{\sigma(n)}\rangle\nonumber\nonumber\\
&=& A\alpha_j\frac{\sqrt{n+1}}{n!}\sum_\sigma{'}A|j,j_{\sigma(1)},\dots,j_{\sigma(n)}\rangle
+A\alpha_j^\dag \frac{\sqrt{n}}{n!}\sum_\sigma{'}\delta_{jj_{\sigma(1)}}A|j_{\sigma(2)},\dots,j_{\sigma(n)}\rangle\nonumber\nonumber\\
&=& A\alpha_j\frac{\sqrt{n+1}}{n!}\frac{1}{n+1}\sum_\sigma{'}
\Big(
|j,j_{\sigma(1)},\dots,j_{\sigma(n)}\rangle
+
\dots
+
(-1)^n|j_{\sigma(1)},\dots,j_{\sigma(n)},j\rangle
\Big)
\nonumber\\
&\pp=&
+A\alpha_j^\dag \frac{\sqrt{n}}{n!}\sum_\sigma{'}\delta_{jj_{\sigma(1)}}|j_{\sigma(2)},\dots,j_{\sigma(n)}\rangle\nonumber\nonumber
\ee
\be
&=& A\frac{1}{n!}\sum_\sigma{'}
\Big(
|j_{\sigma(1)},\dots,j_{\sigma(n)}\rangle
+
(-1)^1\delta_{jj_{\sigma(1)}}|j,j_{\sigma(2)},\dots,j_{\sigma(n)}\rangle
+\dots +
(-1)^n\delta_{jj_{\sigma(1)}}|j_{\sigma(2)},\dots,j_{\sigma(n)},j\rangle
\Big)
\nonumber\\
&\pp=&
+A\frac{n}{n!}\sum_\sigma{'}\delta_{jj_{\sigma(1)}}|j,j_{\sigma(2)},\dots,j_{\sigma(n)}\rangle\nonumber\nonumber\\
&=& A\frac{1}{n!}\sum_\sigma{'}
\Big(
|j_{\sigma(1)},\dots,j_{\sigma(n)}\rangle
+
(-1)^1\delta_{jj_{\sigma(1)}}|j,j_{\sigma(2)},\dots,j_{\sigma(n)}\rangle
+
\dots
+
(-1)^n\delta_{jj_{\sigma(1)}}|j_{\sigma(2)},\dots,j_{\sigma(n)},j\rangle
\Big)
\nonumber\\
&\pp=&
+A\frac{n}{n!}\frac{1}{n}\sum_\sigma{'}\delta_{jj_{\sigma(1)}}\Big(
|j,j_{\sigma(2)},\dots,j_{\sigma(n)}\rangle
+
\dots
+
(-1)^{n-1}|j_{\sigma(2)},\dots,j_{\sigma(n)},j\rangle
\Big)
\nonumber\\
&=& A\frac{1}{n!}\sum_\sigma{'}
|j_{\sigma(1)},\dots,j_{\sigma(n)}\rangle
\nonumber\\
&=& A
|j_1,\dots,j_n\rangle\nonumber
\ee
Finally,
\be
\{b_j,b_j^\dag\}=A.
\ee
In the subspace of symmetric states ${\cal S}\subset \cal H$, $S{\cal S}={\cal S}$, the commutator ${[a_j,a_j^\dag]}=S$ acts as the identity operator,
\be
{[a_j,a_j^\dag]} {\cal S}={\cal S}.
\ee
Similarly, in the subspace of antisymmetric states ${\cal A}\subset \cal H$, $A{\cal A}={\cal A}$, the anti-commutator $\{b_j,b_j^\dag\}=A$ acts as the identity operator,
\be
\{b_j,b_j^\dag\} {\cal A}={\cal A}.
\ee
The two algebras (canonical commutation and anti-commutation relations) in physical applications will correspond to different-spin representations of SU(2) or SL(2,C) groups. Therefore formulas such as $a_j^\dag |0\rangle=|j\rangle$ and $b_j^\dag |0\rangle=|j\rangle$ should not both generate the same $|j\rangle$. The simplest solution is to start with $|0,0\rangle=|0\rangle\otimes |0\rangle$,
\be
a_j^\dag |0\rangle\otimes |0\rangle &=& |j\rangle\otimes |0\rangle \in {\cal S}\otimes {\cal A},\\
b_j^\dag |0\rangle\otimes |0\rangle &=& |0\rangle\otimes |j\rangle \in {\cal S}\otimes {\cal A},
\ee
and so on. The algebras then read
\be
{[a_j,a_j^\dag]} &=& S\otimes I,\\
\{b_j,b_j^\dag\} &=& I\otimes A.
\ee
This type of construction can be found in the Berezin book \cite{Berezin} but, apparently, goes back to the original works of V. Fock.
It is interesting that the right-hand sides of the algebras are not given by the identity operator but by projectors on subspaces of totally symmetric or anti-symmetric states.  $\blacktriangle$

\section{Ensembles of indefinite-frequency oscillators}

Energies of classical free fields look in Fourier space analogously to those of ensembles of oscillators. This observation was at the roots of the first approach to field quantization, formulated already in 1925 by Heisenberg, Born and Jordan \cite{HBJ1925}. One should bear in mind that the first paper of Schr\"odinger, explaining the role of eigenvalues of operators, appeared a few months later \cite{S1926}. The authors of \cite{HBJ1925} apparently did not yet understand the idea of quantum superposition (as suggested by Max Jammer in \cite{Jammer}). Having to consider oscillators with different frequencies they basically had no other option but at least one oscillator per frequency. In this respect nothing essential has changed in mainstream quantum field theory since 1925.

I will now show that their basic assumption is not at all natural. To do so, let us consider a simple 2D pendulum in linear approximation. Classically, its ground state would correspond to no oscillation at all. Quantum mechanically it would imply vanishing momentum and fixed position, a possibility excluded by the uncertainty principle. In consequence one finds the ground state oscillation with energy $\frac{1}{2}\hbar\sqrt{g/l}=\frac{1}{2}\hbar \omega(l)$ ($l$ is the pendulum length). It also practically means that the ``lowest atom" of the pendulum is described by the center-of-mass wave packet $\Psi(X)$ where $|\Psi(X)|^2$ is a Gaussian.

If the pendulum is suspended at the origin $(X,Y)=(0,0)$, the length is given by $l=|Y|=-Y$. For a true pendulum the 2D wave function $\Psi(X,Y)$ would have to be smeared out also in the $Y$ direction and, in fact, a more realistic model should employ a nonseparable potential $U(X,Y)=m\omega(Y)^2X^2/2$, with deep conditional minimum at $Y=-l$. Note that the frequency is no longer a parameter but an eigenvalue  $\omega(Y)$, in position representation, of some operator $\Omega$. Similarly to Schr\"odinger's cat that exists in superposition of being dead and alive, our pendulum exists in a superposition of all its possible lengths, so that many different frequencies can be associated with a single oscillator. The example is generic --- in sufficiently realistic cases the $\omega$s are not just classical parameters but functions of other observables.

I don't see any reason why at levels as fundamental as those related to field quantization the appropriate $\omega$s and wave vectors $\bm k$ should be more classical than $\Omega$ from the preceding example. I would rather expect the wave vectors and frequencies to be eigenvalues of some operators. In order to understand formal implications of the latter postulate we first have to understand the simplest examples based on nonrelativistic oscillators.
\subsection{Single indefinite-frequency oscillator}

The special case $\Omega=\omega {\mathbb I}$, with parameter $\omega$ and identity operator $\mathbb I$, is equivalent to the standard harmonic oscillator.

The simplest nontrivial generalization of $\Omega=\omega {\mathbb I}$ occurs if the operator $\Omega$ has discrete spectrum and commutes with canonical momentum and position. So let $\hat p$ and $\hat x$ be the canonical momentum and position acting in the Hilbert space spanned by the number-of-excitations eigenvectors $|n\rangle$.
Now consider the following representation
\be
\Omega &=& \sum_\omega \omega|\omega\rangle\langle \omega|\otimes I,\label{Omega}\\
P &=& \sum_\omega |\omega\rangle\langle \omega|\otimes \hat p=I_\Omega\otimes \hat p,\\
Q &=& \sum_\omega |\omega\rangle\langle \omega|\otimes \hat x=I_\Omega\otimes \hat x,
\ee
with Hamiltonian of the usual form
\be
H &=& \frac{P^2}{2m}+\frac{m\Omega^2 Q^2}{2}
= \frac{\hbar\Omega}{2}\Big(a_\Omega^{\dag}a_\Omega+a_\Omega a_\Omega^{\dag}\Big)
=
\hbar\Omega\Big(a_\Omega^{\dag}a_\Omega+\frac{1}{2}I_\Omega \otimes I\Big),
\ee
and CCR
\be
a_\Omega
&=&
\frac{1}{\sqrt{2\hbar m \Omega}}
\left( m \Omega Q + i P\right),\\
a^\dagger_\Omega
&=&\frac{1}{\sqrt{2\hbar m \Omega}}
\left( m \Omega Q - i P\right),
\ee
acting in the Hilbert space spanned by $|\omega,n\rangle = |\omega\rangle\otimes|n\rangle$. For any $\omega$
\be
H|\omega,n\rangle &=& \hbar\omega\Big(n+\frac{1}{2}\Big)|\omega,n\rangle,\\
a_\Omega|\omega,0\rangle &=& 0.
\ee
Let
\be
|\psi\rangle=\sum_{\omega,n}\psi_{\omega,n}|\omega,n\rangle
\ee
be an arbitrary state. The average
\be
\langle\psi|H|\psi\rangle=\sum_{\omega,n}\hbar\omega\Big(n+\frac{1}{2}\Big)|\psi_{\omega,n}|^2
\ee
looks as an average energy of an ensemble of harmonic oscillators, different frequencies occurring with probabilities
\be
p_\omega =\sum_{n=0}^\infty |\psi_{\omega,n}|^2.
\ee
We can also write
\be
\langle\psi|H|\psi\rangle
&=&
\sum_{\omega}\hbar\omega\sum_{n=0}^\infty\Big(n+\frac{1}{2}\Big)|\psi_{\omega,n}|^2\nonumber\\
&=&
\sum_{\omega}\frac{\hbar\omega}{2} \langle\psi|\Big(|\omega\rangle\langle \omega|\otimes
\big(\hat a_\omega^{\dag}\hat a_\omega+\hat a_\omega \hat a_\omega^{\dag}\big)\Big)|\psi\rangle.\nonumber\\
&=&
\sum_{\omega}\frac{\hbar\omega}{2} \langle\psi|
\big(a_\omega^{\dag}a_\omega+a_\omega a_\omega^{\dag}\big)|\psi\rangle.
\ee
Here
\be
\hat a_\omega
&=&
\frac{1}{\sqrt{2\hbar m \omega}}
\left( m \omega \hat x + i \hat p\right)=
\sqrt{\frac{ m \omega}{2\hbar}} \hat x + \frac{i}{\sqrt{2\hbar m \omega}}\hat p
,\label{hat a}\\
\hat a^\dagger_\omega
&=&\frac{1}{\sqrt{2\hbar m \omega}}
\left( m \omega \hat x - i \hat p\right)=
\sqrt{\frac{ m \omega}{2\hbar}} \hat x - \frac{i}{\sqrt{2\hbar m \omega}}\hat p,\label{hat a*}\\
\hat x
&=&
\sqrt{\frac{2\hbar}{m\omega}}
\Big(
\hat a_\omega
+
\hat a^\dagger_\omega
\Big),
\ee
are the usual creation and annihilation operators, and
\be
a_\omega
&=&
|\omega\rangle\langle \omega|\otimes \hat a_\omega\label{a omega}
,\\
a^\dagger_\omega
&=&
|\omega\rangle\langle \omega|\otimes \hat a_\omega^\dagger,\label{a omega*}\\
|\omega\rangle\langle \omega|\otimes \hat x
&=&
|\omega\rangle\langle \omega|\otimes \sqrt{\frac{2\hbar}{m\omega}}
\Big(
\hat a_\omega
+
\hat a^\dagger_\omega
\Big)
\\
&=&
\sqrt{\frac{2\hbar}{m\omega}}
\Big(
a_\omega
+
a^\dagger_\omega
\Big)=Q_\omega.
\ee
We have arrived at the decomposition
\be
H &=&\sum_\omega H_\omega,\\
H_\omega
&=&
\frac{\hbar\omega}{2} \big(a_\omega^{\dag}a_\omega+a_\omega a_\omega^{\dag}\big)\\
&=&
\hbar\omega\, n_\omega+\frac{\hbar\omega}{2}I_\omega\\
\langle\psi|H_\omega|\psi\rangle
&=&
\hbar\omega\sum_{n=0}^\infty\Big(n+\frac{1}{2}\Big)|\psi_{\omega,n}|^2.
\ee
As we can see, $H_\omega$ has properties of Hamiltonian of an oscillator whose frequency is $\omega$, and $H$ is a sum of such Hamiltonians taken over all the eigenvalues of $\Omega$. The new feature is the fact that $H$ describes a single harmonic oscillator existing in superposition of different $\omega$s.

In Heisenberg picture we find
\be
a_\omega(t)
&=&
e^{iHt/\hbar}a_\omega e^{-iHt/\hbar}=a_\omega e^{-i\omega t}, \\
a_\omega(t)^\dag
&=&
e^{iHt/\hbar}a_\omega^\dag e^{-iHt/\hbar}=a_\omega^\dag e^{i\omega t},\\
\hat x(t)
&=&
\sqrt{\frac{2\hbar}{m\omega}}
\Big(
\hat a_\omega e^{-i\omega t}
+
\hat a^\dagger_\omega e^{i\omega t}
\Big),\\
Q_\omega(t)
&=&
\sqrt{\frac{2\hbar}{m\omega}}
\Big(
a_\omega e^{-i\omega t}
+
a^\dagger_\omega e^{i\omega t}
\Big),
\ee
which are again the standard formulas. However, in spite of all these similarities to the known textbook results, the representation of HOLA is here a nonstandard one:
\be
n_\omega
&=&
a_\omega^{\dag}a_\omega
=
|\omega\rangle\langle \omega|\otimes \hat a_\omega^\dag\hat a_\omega,\label{red n}\\
I_\omega
&=&
|\omega\rangle\langle \omega|\otimes I,\\
{[a_\omega,a_{\omega'}^{\dag}]} &=& \delta_{\omega,\omega'}I_\omega,\label{red HOLA}\\
{[a_\omega,n_{\omega'}]} &=& \delta_{\omega,\omega'}a_\omega,\\
{[a_\omega^\dag,n_{\omega'}]} &=& -\delta_{\omega,\omega'}a_\omega^\dag.
\ee
$I_\omega$ commutes with all the elements of the Lie algebra, but is not proportional to the identity. By Shur's lemma the representation is thus reducible. Its irreducible representation components are spanned by $|\omega,n\rangle$, with fixed $\omega$ and arbitrary $n$. The projector on this subspace is given by $I_\omega$ itself.

Physically, the above representation describes a single oscillator whose states are wave packets consisting of superpositions of different eigenvalues $\omega$.
\medskip

\noindent
{\it Remark\/}: Let us note that (\ref{hat a}) and (\ref{hat a*}) imply
\be
{[\hat a_\omega,\hat a_{\omega'}^\dag]}
&=&
\Big[\frac{1}{\sqrt{2\hbar m \omega}}
\left( m \omega \hat x + i \hat p\right),\frac{1}{\sqrt{2\hbar m \omega'}}
\left( m \omega' \hat x - i \hat p\right)\Big]
\nonumber\\
&=&
\frac{1}{2\hbar m\sqrt{\omega\omega'}}
\big[m \omega \hat x , - i \hat p\big]
+
\frac{1}{2\hbar m\sqrt{\omega\omega'}}
\big[i \hat p,m \omega' \hat x \big]
\nonumber\\
&=&
\frac{\omega+\omega'}{2\sqrt{\omega\omega'}}=
\iota(\omega,\omega').\label{J}
\ee
The right side of (\ref{J}) is a function satisfying
\be
\iota(\omega,\omega') &=& \iota(\omega',\omega),\\
\iota(\omega,\omega) &=& 1,
\ee
which is neither 1 nor any kind of delta.
The explicit form of $\iota(\omega,\omega')$ does not occur in (\ref{red HOLA}) due to the presence of $|\omega\rangle\langle \omega|$ in (\ref{a omega}) and (\ref{a omega*}). We should keep this observation in mind when we perform field quantization.$\blacktriangle$

\subsection{Several indefinite-frequency oscillators}

Now, what about two such oscillators? One can trivially extend all the operators by
\be
a_\omega &\to& a_\omega \otimes I(1),\\
a_\omega &\to& I(1) \otimes a_\omega,
\ee
and so on, where
\be
I(1)=\sum_\omega I_\omega\label{I(1)}
\ee
is the identity in the one-oscillator Hilbert space spanned by $|\omega,n\rangle$. However, if one additionally requires their bosonic statistics, something one expects for spin-0 fields, say, the operators should preserve symmetry of states. The natural bosonic generalization is then
\be
a_\omega(2)
&=&
c_2\Big(a_\omega \otimes I(1)+I(1) \otimes a_\omega\Big)
\ee
where $c_2$ is a constant, and
\be
{[a_\omega(2),a_{\omega'}(2)^{\dag}]}
&=&
\delta_{\omega,\omega'}|c_2|^2\Big(I_\omega \otimes I(1)+I(1) \otimes I_\omega\Big),\nonumber\\
&=&
\delta_{\omega,\omega'} I_\omega(2).
\ee
Let us note that
\be
\sum_\omega I_\omega(2) &=& 2|c_2|^2 I(1) \otimes I(1)=2|c_2|^2 I(2)
\ee
suggesting the normalization $c_2=1/\sqrt{2}$, analogous to (\ref{a_2}). In order to get the whole HOLA we have to define $n_\omega(2)$. The first guess
\be
n_\omega(2)
&=&
a_\omega(2)^\dag a_\omega(2)
\ee
leads to
\be
{[a_\omega(2),n_\omega(2)]}
&=&
\delta_{\omega,\omega'} I_\omega(2)a_\omega(2)
\ee
which is not even a Lie algebra. (Certain observables do satisfy such generalized Lie algebras --- Hamiltonian, angular momentum, and the Runge-Lentz vector of the Coulomb problem provide an example.) But we remember that in standard representations of CCR we have encountered the ambiguity of (\ref{n_2}) versus (\ref{tilde n_2}). In the present context the second option reads
\be
\tilde n_\omega(2)
&=&
n_\omega \otimes I(1)+I(1) \otimes n_\omega,
\ee
leading to HOLA
\be
{[a_\omega(2),a_{\omega'}(2)^{\dag}]} &=& \delta_{\omega,\omega'}I_\omega(2),\\
{[a_\omega(2),\tilde n_{\omega'}(2)]} &=& \delta_{\omega,\omega'}a_\omega(2),\\
{[a_\omega(2)^\dag,\tilde n_{\omega'}(2)]} &=& -\delta_{\omega,\omega'}a_\omega(2)^\dag.
\ee
This representation is again reducible since
\be
I_\omega(2)
&=&
\frac{1}{2}\Big(I_\omega \otimes I(1)+I(1) \otimes I_\omega\Big)\neq I(2),\\
\sum_\omega I_\omega(2)
&=&
I(2).\label{sum I(2)}
\ee
$\tilde n_\omega(2)$ counts the number of excitations of the two-oscillator system.
Especially interesting is the form of $I_\omega(2)$: This is the number-of-successes (in two trials) operator known from works on probabilistic interpretation of quantum mechanics \cite{LLN,LLN1,LLN2,LLN3,LLN4}. Eigenvalues of $I_\omega(2)$ are $0,1/2,1$ and describe the fraction of positive answers (in two trials) to the question ``is the frequency of the oscillator equal to $\omega$?", if $\omega$ is selected at random.
Due to the resolution of identity (\ref{sum I(2)}) the element $I_\omega(2)$ is a positive operator-valued measure \cite{POVM}.

An extension to arbitrary number $N$ of oscillators is now clear, with $c_N=1/\sqrt{N}$. The fact that $I_\omega(N)$ becomes the $N$-trial number-of-successes operator is essential for the limit $N\to\infty$ which can (and later will) be treated by weak laws of large numbers.

Before I make a digression on scalar fields let me note that in Heisenberg picture, for arbitrary $N\geq 1$, we find
\be
a_\omega(t,N)
&=&
e^{i\tilde H(N)t/\hbar}a_\omega(N) e^{-i\tilde H(N)t/\hbar}=a_\omega(N) e^{-i\omega t},
\ee
if the Hamiltonian is given by $\tilde H(N)=\sum_\omega\hbar\omega\,\tilde n_\omega(N)$. For $H(N)=\sum_\omega\hbar\omega\,a_\omega(N)^\dag a_\omega(N)$ we would get
\be
a_\omega(t,N)
&=&
e^{iH(N)t/\hbar}a_\omega(N) e^{-iH(N)t/\hbar}=a_\omega(N) e^{-i\omega I_\omega(N) t}.
\ee
This is a strong argument in favor of $\tilde H(N)$ in contrast to $H(N)$... (Still, as mentioned already in \cite{I}, I have doubts here --- the choice of $H(N)$ is appealing for various reasons, so we will devote some time also to this option.)

Canonical position operators
\be
Q_\omega(t,N)
&=&
\sqrt{\frac{2\hbar}{m\omega}}
\Big(
a_\omega(N) e^{-i\omega t}
+
a_\omega(N)^\dag e^{i\omega t}
\Big),\label{Q omega}\\
Q(t,N)
&=&
\sum_\omega\sqrt{\frac{2\hbar}{m\omega}}
\Big(
a_\omega(N) e^{-i\omega t}
+
a_\omega(N)^\dag e^{i\omega t}
\Big),\label{Q}
\ee
are formally very similar to scalar-field operators. This observation is behind the basic physical intuition that leads to quantum fields, quantized in ``my" way.

\section{Frequency-of-successes operator and HOLA}

For $N=1$ the operators $\Pi_\omega^{(1)}=I_\omega$ and $\Pi_\omega^{(0)}=I(1)-I_\omega$ are orthogonal projectors : $(\Pi_\omega^{(0)})^2=\Pi_\omega^{(0)}$, $(\Pi_\omega^{(1)})^2=\Pi_\omega^{(1)}$,  $\Pi_\omega^{(0)}\Pi_\omega^{(1)}=0$, $\Pi_\omega^{(0)}+\Pi_\omega^{(1)}=I(1)$.
The central element
\be
I_\omega(N)
&=&
\frac{1}{N}
\Big(
I_\omega\otimes I(1)\otimes\dots\otimes I(1)
+
\dots
+
I(1)\otimes\dots\otimes I(1)\otimes  I_\omega
\Big)
\ee
has eigenvalues $0/N$, $1/N$, $2/N$,..., $(N-1)/N$, $N/N$. Its spectral decomposition can be deduced from
\be
I(1) &=&\Pi_\omega^{(0)}+\Pi_\omega^{(1)} =\sum_{s=0}^1 \Pi_\omega^{(s)},\\
I_\omega(N)
&=&
\frac{1}{N}
\Big(
\Pi_\omega^{(1)}\otimes I(1)\otimes\dots\otimes I(1)
+
\dots
+
I(1)\otimes\dots\otimes I(1)\otimes  \Pi_\omega^{(1)}
\Big)\nonumber\\
&=&
\frac{0}{N}\Pi_\omega^{(0)}\otimes \dots\otimes \Pi_\omega^{(0)}
\nonumber\\
&\pp=&+
\frac{1}{N}
\Big(
\Pi_\omega^{(1)}\otimes I(1)\otimes\dots\otimes I(1)
+
\dots
+
I(1)\otimes\dots\otimes I(1)\otimes  \Pi_\omega^{(1)}
\Big)
\ee
and
\be
I_\omega(N)\Pi_\omega^{(s_1)}\otimes \dots\otimes \Pi_\omega^{(s_N)}
&=&
\frac{s_1+\dots +s_N}{N}\Pi_\omega^{(s_1)}\otimes \dots\otimes \Pi_\omega^{(s_N)}.
\ee
Since
\be
\sum_{s_1,\dots,s_N=0}^1
\Pi_\omega^{(s_1)}\otimes \dots\otimes \Pi_\omega^{(s_N)}
=
I(N)
\ee
we arrive at
\be
I_\omega(N)
&=&
\sum_{s_1,\dots,s_N=0}^1
I_\omega(N)\Pi_\omega^{(s_1)}\otimes \dots\otimes \Pi_\omega^{(s_N)}\nonumber\\
&=&
\sum_{s_1,\dots,s_N=0}^1
\frac{s_1+\dots +s_N}{N}\Pi_\omega^{(s_1)}\otimes \dots\otimes \Pi_\omega^{(s_N)}=
\sum_{s=0}^N
\frac{s}{N}\Pi_\omega\Big(\frac{s}{N}\Big),\\
\Pi_\omega\Big(\frac{s}{N}\Big)
&=&
\sum_{s_1+\dots+s_N=s}
\Pi_\omega^{(s_1)}\otimes \dots\otimes \Pi_\omega^{(s_N)}.
\ee
Now let $|\psi,1\rangle=\sum_{\omega,n} \psi_{\omega,n} |\omega,n\rangle$, $|\psi,N\rangle=|\psi,1\rangle\otimes\dots\otimes |\psi,1\rangle$.
The average
\be
\langle \psi,1|I_\omega|\psi,1\rangle
&=&
\langle \psi,1|\Pi_\omega^{(1)}|\psi,1\rangle
=
\sum_{n}|\psi_{\omega,n}|^2=p_\omega,
\ee
is the probability of finding $\omega$. The average
\be
\langle \psi,N|\Pi_\omega\Big(\frac{s}{N}\Big)|\psi,N\rangle
&=&
\sum_{s_1+\dots+s_N=s}
\langle \psi,N|\Pi_\omega^{(s_1)}\otimes \dots\otimes \Pi_\omega^{(s_N)}|\psi,N\rangle\nonumber\\
&=&
\sum_{s_1+\dots+s_N=s}
\langle \psi,1|\Pi_\omega^{(s_1)}|\psi,1\rangle\dots \langle \psi,1|\Pi_\omega^{(s_N)}|\psi,1\rangle\nonumber\\
&=&
\binom{N}{s}p_\omega^s(1-p_\omega)^{N-s}
\ee
is the probability of finding $\omega$ exactly $s$ times in $N$ measurements, performed on each of the $N$ oscillators once, if each oscillator is in state $|\psi,1\rangle$.

If $F:[0,1]\to \mathbb{R}$ is continuous, then by the weak law of large numbers for binomial distribution (or, more generally, Feller's theorem \cite{Feller})
\be
\lim_{N\to\infty}\langle \psi,N|F\big(I_\omega(N)\big)|\psi,N\rangle
&=&
\lim_{N\to\infty}\langle \psi,N|\sum_{s=0}^NF\Big(\frac{s}{N}\Big)\Pi_\omega\Big(\frac{s}{N}\Big)|\psi,N\rangle\nonumber\\
&=&
\lim_{N\to\infty}\sum_{s=0}^NF\Big(\frac{s}{N}\Big)\binom{N}{s}p_\omega^s(1-p_\omega)^{N-s}\nonumber\\
&=&
F(p_\omega).
\ee
So, in practice, predictions concerning the gas consisting of $N$ indefinite-frequency oscillators are in the limit $N\to\infty$ close to those of the standard oscillators with the modified CCR $[a_\omega,a_{\omega'}^\dag]=p_\omega \delta_{\omega\omega'}$.
This result will be the basis of the correspondence principle relating standard regularized quantum field theory with the one I propose in these notes. In fact, as discussed in detail in \cite{MCMW, MWMC1,MW PhD}, there exist two physically meaningful limits $N\to\infty$. One is just the weak law of large numbers while the second one is a kind of thermodynamic limit.

\section{States of indefinite-frequency oscillators}

States of indefinite-frequency oscillators have properties that will be later used in construction of  vacuum, $n$-particle and coherent states of quantum fields.

\subsection{$N$-oscillator analogues of $n$-photon states}

The ``vacuum subspace'' consists of vectors that are annihilated by all annihilation operators.
The $N$-oscillator vacuum  subspace is spanned by $|\omega_1,0\rangle\otimes \dots \otimes|\omega_N,0\rangle$. I will assume that vacua are given by pure product states
\begin{eqnarray}
|O,N\rangle
&=&
\sum_{\omega_1,\ldots,\omega_N}
O_{\omega_1}\ldots O_{\omega_N}
|\omega_1,0\rangle \otimes \dots \otimes |\omega_N,0\rangle,
\end{eqnarray}
where the $O_\omega$s play a role of a single-oscillator wave function, normalized by
\begin{eqnarray}
\sum_\omega
|O_\omega|^2
&=&
1.
\end{eqnarray}
$p_\omega=|O_\omega|^2$ is the probability that  a given oscillator has frequency $\omega$.

Let us begin with the analog of an ordinary $1$-photon state, that is $a_\omega(N)^\dagger |O,N\rangle$. Its squared norm is
\begin{eqnarray}
\langle O,N|a_\omega(N)a_\omega(N)^\dagger |O,N\rangle
&=&
\langle O,N|I_\omega(N) |O,N\rangle\nonumber
\\
&=&
\langle O,1|I_\omega |O,1\rangle\nonumber
\\
&=&
|O_\omega|^2=p_\omega.
\label{eq-global-11}
\end{eqnarray}
Whatever representation of $n_\omega(N)$ satisfying HOLA $[n_\omega(N),a_{\omega'}(N)^\dag]=\delta_{\omega\omega'}a_{\omega'}(N)^\dag$ and annihilating $|O,N\rangle$ we take, we find that $a_\omega(N)^\dagger |O,N\rangle$ is a ``single-excitation" state, i.e.
\be
n_\omega(N)a_\omega(N)^\dagger |O,N\rangle
&=&
[n_\omega(N),a_\omega(N)^\dagger] |O,N\rangle
=
a_\omega(N)^\dagger |O,N\rangle.
\ee
The peculiarity of this representation of HOLA is that there exist ``single-excitation" states that are not spanned by vectors
$a_\omega(N)^\dagger |O,N\rangle$. A simple (and generic) example is provided by any state of the form
\be
F\big(I_\omega(N)\big)a_\omega(N)^\dagger |O,N\rangle.
\ee
In general, $n$-excitation states are any states spanned by
\be
|\omega_1,n_1\rangle\otimes \dots \otimes |\omega_N,n_N\rangle, \quad n_1+\dots+n_N=n.\label{n exc}
\ee
Their particular sub-class is given by
\be
a_{\omega_1}(N)^\dagger \dots a_{\omega_n}(N)^\dagger|O,N\rangle.\label{nexc}
\ee
Indeed, each action of a creation operator adds one excitation to the state, and $|O,N\rangle$ has zero excitations.

\subsection{$N$-oscillator analogues of coherent states}

Quantum mechanics textbooks give (at least) two definitions of coherent states of a harmonic oscillator: States generated by displacement operators acting on the oscillator ground state and eigenstates of annihilation operators. In reducible representations of HOLA the two definitions are not equivalent.

The displacement operator
\begin{eqnarray}
{\cal D}(\alpha,N)
&=&
\exp\Big({\sum_\omega \alpha_\omega  a_\omega(N)^\dagger -\sum_\omega \overline{\alpha_\omega}  a_\omega(N)}\Big)
\nonumber\\
&=&
\exp\Big(-\frac{1}{2} \sum_\omega |\alpha_\omega |^2 I_\omega(N)\Big)
\exp\Big(\sum_\omega \alpha_\omega a_\omega(N)^\dagger  \Big) \exp\Big(-\sum_\omega \overline{\alpha_\omega}a_\omega(N)\Big).
\nonumber\\
\label{eq-global-150}
\end{eqnarray}
is unitary
\be
{\cal D}(\alpha,N)^\dag
&=&
{\cal D}(\alpha,N)^{-1}
=
{\cal D}(-\alpha,N).
\ee
The operator can be also represented in the following useful way
\be
{\cal D}(\alpha,N)
&=&
e^{\frac{1}{\sqrt{N}}\sum_\omega \big(\alpha_\omega  a_\omega(1)^\dagger -\overline{\alpha_\omega}  a_\omega(1)\big)}
\otimes
\dots
\otimes
e^{\frac{1}{\sqrt{N}}\sum_\omega \big(\alpha_\omega  a_\omega(1)^\dagger -\overline{\alpha_\omega}  a_\omega(1)\big)}
\nonumber\\
&=&
{\cal D}(\alpha/\sqrt{N},1)
\otimes
\dots
\otimes
{\cal D}(\alpha/\sqrt{N},1)\nonumber\\
&=&{\cal D}(\alpha/\sqrt{N},1)^{\otimes N},\\
{\cal D}(\alpha,1)
&=&
\exp\Big(\sum_\omega |\omega\rangle\langle\omega|\otimes\big(\alpha_\omega  \hat a^\dagger -\overline{\alpha_\omega}  \hat a\big)\Big)
\nonumber\\
&=&
\sum_\omega |\omega\rangle\langle\omega|\otimes \exp\big(\alpha_\omega  \hat a^\dagger -\overline{\alpha_\omega}  \hat a\big)
\nonumber\\
&=&
\sum_\omega |\omega\rangle\langle\omega|\otimes \exp\big(-|\alpha_\omega|^2/2\big)
\exp\big(\alpha_\omega  \hat a^\dagger\big)
\exp\big(-\overline{\alpha_\omega}  \hat a\big)
\nonumber\\
&=&
\sum_\omega |\omega\rangle\langle\omega|\otimes \hat{\cal D}(\alpha_\omega).
\ee
$\hat{\cal D}(\alpha_\omega)$ is the ordinary displacement operator known from quantum mechanics textbooks.

The name of the operator comes from the following property
\be
{\cal D}(\alpha,N)^\dag a_\omega(N) {\cal D}(\alpha,N)
&=&
a_\omega(N)+\alpha_\omega I_\omega(N),\\
{\cal D}(\alpha,N)^\dag a_\omega(N)^\dag {\cal D}(\alpha,N)
&=&
a_\omega(N)^\dag+\overline{\alpha_\omega} I_\omega(N),\\
{\cal D}(\alpha,N)^\dag I_\omega(N) {\cal D}(\alpha,N)
&=&
I_\omega(N).
\ee
Combining these formulas we obtain a generalized eigenvalue problem
\be
a_\omega(N) {\cal D}(\alpha,N)|O,N\rangle
&=&
{\cal D}(\alpha,N) {\cal D}(\alpha,N)^\dag a_\omega(N) {\cal D}(\alpha,N)|O,N\rangle
\nonumber\\
&=&
{\cal D}(\alpha,N)
\Big(
a_\omega(N)+\alpha_\omega I_\omega(N)
\Big)|O,N\rangle
\nonumber\\
&=&
\alpha_\omega {\cal D}(\alpha,N)I_\omega(N)|O,N\rangle
\nonumber\\
&=&
\alpha_\omega I_\omega(N){\cal D}(\alpha,N)|O,N\rangle
.
\ee
The coherent state
\be
|\alpha,N\rangle
&=&
{\cal D}(\alpha,N)|O,N\rangle\\
&=&
\exp\Big(-\frac{1}{2} \sum_\omega |\alpha_\omega |^2 I_\omega(N)\Big)
\exp\Big(\sum_\omega \alpha_\omega a_\omega(N)^\dagger  \Big)|O,N\rangle
\\
&=&
{\cal D}(\alpha/\sqrt{N},1)|O,1\rangle
\otimes
\dots
\otimes
{\cal D}(\alpha/\sqrt{N},1)|O,1\rangle
\\
&=&
|\alpha/\sqrt{N},1\rangle
\otimes
\dots
\otimes
|\alpha/\sqrt{N},1\rangle
\ee
satisfies
\be
a_\omega(N) |\alpha,N\rangle
&=&
\alpha_\omega I_\omega(N)|\alpha,N\rangle.\label{eigenvalue}
\ee
The latter is a generalized eigenvalue equation in the sense that it combines several ordinary eigenvalue problems that can be extracted from (\ref{eigenvalue}) by means of $\Pi_\omega(s/N)$:
\be
a_\omega(N) \Pi_\omega(s/N)|\alpha,N\rangle
&=&
\Pi_\omega(s/N)a_\omega(N)|\alpha,N\rangle
\nonumber\\
&=&
\Pi_\omega(s/N)\alpha_\omega I_\omega(N)|\alpha,N\rangle\nonumber\\
&=&
\frac{s}{N}\alpha_\omega \Pi_\omega(s/N)|\alpha,N\rangle.
\ee
Vector $\Pi_\omega(s/N)|\alpha,N\rangle$ is the eigenvector of $a_\omega(N)$ with the eigenvalue $s\alpha_\omega/N$.

The difference with respect to the ``ordinary'' harmonic oscillator known from textbooks is that the latter involves $I_\omega=1$ so that the coherent state is just a single eigenvector with eigenvalue $\alpha_\omega$. We will later see that the presence of $s/N$ may make various quantum field theoretical predictions more physical (if quantum fields are defined by means of an analogous representation of HOLA).

\subsection{Statistics of excitations}

The $n$-excitation subspace is spanned by vectors (\ref{n exc}). The projector on this subspace is given by
\be
\Pi(n,N)
&=&
\sum_{n_1+\dots+n_N=n}\sum_{\omega_1\dots\omega_N}
|\omega_1,n_1\rangle\langle \omega_1,n_1|\otimes \dots\otimes |\omega_N,n_N\rangle\langle \omega_N,n_N|
\nonumber\\
&=&
\sum_{n_1+\dots+n_N=n}
\Big(I_\Omega\otimes|n_1\rangle\langle n_1|\Big)\otimes \dots\otimes \Big(I_\Omega\otimes|n_N\rangle\langle n_N|\Big).
\ee
Probability of finding $n$ excitations in a 1-oscillator coherent state is
\be
p(n,1)
&=&
\langle \alpha,1|\Pi(n,1)|\alpha,1\rangle
=\sum_\omega |O_\omega|^2 \frac{(|\alpha_\omega|^2)^n}{n!}e^{-\sum_\omega |\alpha_\omega|^2}.\label{p(n,1)}
\ee
(\ref{p(n,1)}) is the Poisson distribution typical of standard coherent states weighted by probability that frequency of the oscillator equals $\omega$. This is precisely the result one expected for a single oscillator wave packet.

Probability of finding $n$ excitations in an $N$-oscillator coherent state can be computed if one recalls that (\ref{nexc}) belongs to the $n$-excitation subspace. Therefore,
\be
p(n,N)
&=&
\langle \alpha,N|\Pi(n,N)|\alpha,N\rangle\nonumber\\
&=&
\frac{1}{(n!)^2}\langle O,N|\exp\Big(-\sum_\omega |\alpha_\omega |^2 I_\omega(N)\Big)
\Big(\sum_{\omega_1} \overline{\alpha_{\omega_1}} a_{\omega_1}(N)\Big)^n
\Big(\sum_{\omega_2} \alpha_{\omega_2} a_{\omega_2}(N)^\dagger\Big)^n|O,N\rangle
\nonumber
\ee
Taking into account
\be
{\Big[\sum_{\omega_1} \overline{\alpha_{\omega_1}} a_{\omega_1}(N),\Big(\sum_{\omega_2} \alpha_{\omega_2} a_{\omega_2}(N)^\dagger\Big)^n\Big]}
&=&
\sum_{\omega_1} \overline{\alpha_{\omega_1}} \Big[a_{\omega_1}(N),\Big(\sum_{\omega_2} \alpha_{\omega_2} a_{\omega_2}(N)^\dagger\Big)^n\Big]
\nonumber\\
&=&
n\sum_{\omega_1} |\alpha_{\omega_1}|^2 I_{\omega_1}(N)\Big(\sum_{\omega_2} \alpha_{\omega_2} a_{\omega_2}(N)^\dagger\Big)^{n-1}
\nonumber
\ee
and acting with
\be
{}&{}&
\Big[\Big(\sum_{\omega_1} \overline{\alpha_{\omega_1}} a_{\omega_1}(N)\Big)^n,\Big(\sum_{\omega_2}\alpha_{\omega_2}a_{\omega_2}(N)^\dagger\Big)^n\Big]
\nonumber\\
&{}&\pp=
=
\Big(\sum_{\omega_1} \overline{\alpha_{\omega_1}} a_{\omega_1}(N)\Big)^{n-1}\Big[\sum_{\omega_1} \overline{\alpha_{\omega_1}} a_{\omega_1}(N),\Big(\sum_{\omega_2} \alpha_{\omega_2} a_{\omega_2}(N)^\dagger\Big)^n\Big]
\nonumber\\
&{}&\pp{==}
+
\Big[\Big(\sum_{\omega_1} \overline{\alpha_{\omega_1}} a_{\omega_1}(N)\Big)^{n-1},\Big(\sum_{\omega_2} \alpha_{\omega_2} a_{\omega_2}(N)^\dagger\Big)^n\Big]
\sum_{\omega_1} \overline{\alpha_{\omega_1}} a_{\omega_1}(N)
\nonumber\\
&{}&\pp=
=
n\sum_{\omega_1} |\alpha_{\omega_1}|^2 I_{\omega_1}(N)
\Big(\sum_{\omega_1} \overline{\alpha_{\omega_1}} a_{\omega_1}(N)\Big)^{n-1}
\Big(\sum_{\omega_2} \alpha_{\omega_2} a_{\omega_2}(N)^\dagger\Big)^{n-1}
\nonumber\\
&{}&\pp{==}
+
\Big[\Big(\sum_{\omega_1} \overline{\alpha_{\omega_1}} a_{\omega_1}(N)\Big)^{n-1},\Big(\sum_{\omega_2} \alpha_{\omega_2} a_{\omega_2}(N)^\dagger\Big)^n\Big]
\sum_{\omega_1} \overline{\alpha_{\omega_1}} a_{\omega_1}(N)
\ee
on $|O,N\rangle$, one finds
\be
{}&{}&
\Big(\sum_{\omega_1} \overline{\alpha_{\omega_1}} a_{\omega_1}(N)\Big)^n\Big(\sum_{\omega_2}\alpha_{\omega_2}a_{\omega_2}(N)^\dagger\Big)^n |O,N\rangle
\nonumber\\
{}&{}&\pp=
=
\Big[\Big(\sum_{\omega_1} \overline{\alpha_{\omega_1}} a_{\omega_1}(N)\Big)^n,\Big(\sum_{\omega_2}\alpha_{\omega_2}a_{\omega_2}(N)^\dagger\Big)^n\Big]|O,N\rangle
\nonumber\\
&{}&\pp=
=
\Big(\sum_{\omega_1} \overline{\alpha_{\omega_1}} a_{\omega_1}(N)\Big)^{n-1}\Big[\sum_{\omega_1} \overline{\alpha_{\omega_1}} a_{\omega_1}(N),\Big(\sum_{\omega_2} \alpha_{\omega_2} a_{\omega_2}(N)^\dagger\Big)^n\Big]|O,N\rangle
\nonumber\\
&{}&\pp=
=
n\sum_{\omega_1} |\alpha_{\omega_1}|^2 I_{\omega_1}(N)
\Big(\sum_{\omega_1} \overline{\alpha_{\omega_1}} a_{\omega_1}(N)\Big)^{n-1}
\Big(\sum_{\omega_2} \alpha_{\omega_2} a_{\omega_2}(N)^\dagger\Big)^{n-1}|O,N\rangle
\nonumber
\\
&{}&\pp=
=
n!\Big(\sum_{\omega_1} |\alpha_{\omega_1}|^2 I_{\omega_1}(N)\Big)^n
|O,N\rangle.
\nonumber
\ee
The end result
\be
p(n,N)
&=&
\langle O,N|\exp\Big(-\sum_\omega |\alpha_\omega |^2 I_\omega(N)\Big)
\frac{1}{n!}\Big(\sum_{\omega_1} |\alpha_{\omega_1}|^2 I_{\omega_1}(N)\Big)^n|O,N\rangle
\nonumber\\
&=&
\frac{1}{n!}\frac{d^n}{d\lambda^n}
\langle O,N|\exp\Big(\lambda\sum_\omega |\alpha_\omega |^2 I_\omega(N)\Big)|O,N\rangle\Big|_{\lambda=-1}.
\label{p(n,N)}
\ee
is a generalized Poisson distribution. In order to better understand the generalization we have found
let us take a closer look at the generating function
\be
\langle O,N|\exp\Big(\lambda\sum_\omega |\alpha_\omega |^2 I_\omega(N)\Big)|O,N\rangle
&=&
\langle O,1|\exp\Big(\lambda\sum_\omega \frac{1}{N}|\alpha_\omega |^2 I_\omega\Big)|O,1\rangle^N
\nonumber\\
&=&
\langle O,1|\sum_\omega I_\omega\exp\Big(\frac{1}{N}\lambda|\alpha_\omega |^2 \Big)|O,1\rangle^N
\nonumber\\
&=&
\Big(\sum_\omega |O_\omega|^2e^{\lambda\frac{1}{N}|\alpha_\omega |^2} \Big)^N.
\nonumber
\ee
The possibility of taking the sum in front of the exponent comes from the fact that $I_\omega=|\omega\rangle\langle\omega|\otimes 1$ is a projector.
Now, let us introduce a new parameter $q$ satisfying $1-q=1/N$,
\be
\Big(\sum_\omega |O_\omega|^2e^{(1-q)\lambda|\alpha_\omega |^2} \Big)^\frac{1}{1-q}
&=&
\exp\ln \Big(\sum_\omega |O_\omega|^2e^{(1-q)\lambda|\alpha_\omega |^2} \Big)^\frac{1}{1-q}
\nonumber\\
&=&
\exp\Bigg(\frac{1}{1-q}\ln \Big(\sum_\omega |O_\omega|^2e^{(1-q)\lambda|\alpha_\omega |^2} \Big)\Bigg).
\ee
Expression under the exponent,
\be
\frac{1}{1-q}\ln \Big(\sum_\omega |O_\omega|^2e^{(1-q)\lambda|\alpha_\omega |^2} \Big)
\ee
is well known in probability and information theory: This is the Kolmogorov-Nagumo average of the random variable $\lambda|\alpha_\omega |^2$ \cite{Kolmogorov-Nagumo}, introduced by Alfr\'ed R\'enyi in his derivation of generalized entropies \cite{Renyi}.  Since
\be
\lim_{q\to 1}
\ln \Big(\sum_\omega |O_\omega|^2e^{(1-q)\lambda|\alpha_\omega |^2} \Big)
=
\ln \Big(\underbrace{\sum_\omega |O_\omega|^2}_{\langle O,1|O,1\rangle=1}\Big)=0,
\ee
we can use the de l'Hospital rule to compute the limit
\be
\lim_{q\to 1}
\frac{1}{1-q}\ln \Big(\sum_\omega |O_\omega|^2e^{(1-q)\lambda|\alpha_\omega |^2} \Big)
=
\lambda\sum_\omega |O_\omega|^2|\alpha_\omega |^2.
\ee
The limiting generating function
\be
\lim_{N\to \infty}
\Big(\sum_\omega |O_\omega|^2e^{\lambda\frac{1}{N}|\alpha_\omega |^2} \Big)^N
&=&
\lim_{q\to 1}
\exp
\Bigg(\frac{1}{1-q}\ln \Big(\sum_\omega |O_\omega|^2e^{(1-q)\lambda|\alpha_\omega |^2} \Big)\Bigg)
\nonumber
\\
&=&
\exp\Big(\lambda\sum_\omega |O_\omega|^2|\alpha_\omega |^2\Big)
\ee
generates the Poisson distribution
\be
p(n,\infty)
&=&
\frac{1}{n!}\frac{d^n}{d\lambda^n}
\exp\Big(\lambda\sum_\omega |O_\omega|^2|\alpha_\omega |^2\Big)
\Big|_{\lambda=-1}.
\ee
Asymptotically, for large $N$, the gas of coherent-state indefinite-frequency oscillators possesses Poissonian statistics of excitations. What is interesting the parameter of the Poisson distribution is not just $|\alpha_\omega|^2$ but $|O_\omega|^2|\alpha_\omega|^2$, a result of great importance for my formulation of quantum field theory. For finite $N$ the distribution is a R\'enyi-deformed Poissonian.

Let me summarize this section. For finite $N$ we get
\be
\langle O,N|\exp\Big(\lambda\sum_\omega |\alpha_\omega |^2 I_\omega(N)\Big)|O,N\rangle.
\ee
The $N\to\infty$ limiting case is
\be
\exp\Big(\lambda\sum_\omega p_\omega|\alpha_\omega |^2\Big), \quad p_\omega=|O_\omega|^2.
\ee
The standard formalism based on infinitely many standard oscillators would imply
\be
\exp\Big(\lambda\sum_\omega |\alpha_\omega |^2\Big)
\ee
which, in practice, is often replaced with
\be
\exp\Big(\lambda\sum_\omega \chi_\omega|\alpha_\omega |^2\Big),
\ee
where $0\leq \chi_\omega\leq 1$, $\lim_{\omega\to\infty}\chi_\omega=0$, is a cut-off function introduced by hand if $\sum_\omega |\alpha_\omega |^2=\infty$. So our $p_\omega$ has {\it automatically\/} appeared in the place where standard formalism is artificially amended by adding $\chi_\omega$.
\medskip

\noindent
{\it Remark\/}:
The limit $q\to 1$ is known in information theory as the Shannon limit because the Kolmogorov-Nagumo average of random variable $\ln (1/|O_\omega|^2)$ (the amount of information obtained by observation of an event whose probability is $|O_\omega|^2$),
\be
\frac{1}{1-q}\ln \Big(\sum_\omega |O_\omega|^2e^{(1-q)\ln (1/|O_\omega|^2)} \Big)
&=&
\frac{1}{1-q}\ln \Big(\sum_\omega |O_\omega|^2e^{\ln |O_\omega|^{2(q-1)}} \Big)
\nonumber\\
&=&
\frac{1}{1-q}\ln \Big(\sum_\omega (|O_\omega|^2)^q \Big)
\nonumber
\ee
tends to Shannon's entropy
\be
\lim_{q\to 1}
\frac{1}{1-q}\ln \Big(\sum_\omega |O_\omega|^2e^{(1-q)\ln (1/|O_\omega|^2)} \Big)
&=&
\sum_\omega |O_\omega|^2 \ln (1/|O_\omega|^2)
=
-\sum_\omega |O_\omega|^2 \ln |O_\omega|^2.\nonumber
\ee
Generalized entropy $\frac{1}{1-q}\ln \Big(\sum_\omega p_\omega^q \Big)$ is termed the R\'enyi $q$-entropy of probability distribution $p_\omega$.
Similarly to Shannon's entropy it is additive for independent events
\be
\frac{1}{1-q}\ln \Big(\sum_{\omega_1,\omega_2} (p_{\omega_1}\tilde p_{\omega_2})^q \Big)
&=&
\frac{1}{1-q}\ln \Big(\sum_{\omega_1} p_{\omega_1}^q\sum_{\omega_1} \tilde p_{\omega_1}^q \Big)\nonumber\\
&=&
\frac{1}{1-q}\ln \Big(\sum_{\omega_1} p_{\omega_1}^q \Big)
+
\frac{1}{1-q}\ln \Big(\sum_{\omega_1} \tilde p_{\omega_1}^q \Big)\nonumber.
\ee
A less obvious property of the Kolmogorov-Nagumo-R\'enyi average is the following analog of
\be
\langle A+C\rangle=\langle A\rangle+C,
\ee
for a constant $C$ and random variable $A$:
\be
\langle A+C\rangle_q
&=&
\frac{1}{1-q}\ln \Big(\sum_\omega p_\omega e^{(1-q)(A_\omega+C)} \Big)\nonumber\\
&=&
\frac{1}{1-q}\ln \Big(e^{(1-q)C}\sum_\omega p_\omega e^{(1-q)A_\omega} \Big)\nonumber\\
&=&
\frac{1}{1-q}\ln \Big(\sum_\omega p_\omega e^{(1-q)A_\omega} \Big)
+
\frac{1}{1-q}\ln \Big(e^{(1-q)C}\Big)
\nonumber\\
&=&
\langle A\rangle_q+C.
\ee
General Kolmogorov-Nagumo averages corresponding to a monotonic function $\phi$ are defined as
\be
\langle A\rangle_\phi
&=&
\phi^{-1}\Big(\sum_\omega p_\omega \phi(A_\omega)\Big).
\ee
It is interesting that only for exponential or linear $\phi$ one finds $\langle A+C\rangle_\phi=\langle A\rangle_\phi+C$.$\blacktriangle$

\section{Digression on free scalar fields}\label{sec dig on sf}

Consider a free scalar field of mass $m$. The latter means that in momentum (i.e. Fourier) space the field is defined on the manifold of 4-momenta $p$ satisfying the constraint $p^2=p_0^2-\bm p^2=m^2$. Due to the constraint the four components of $p$ are not independent: $p_0=\pm\sqrt{\bm p^2+m^2}$. The manifold is a 3-dimensional hyperboloid in $\mathbb{R}^4$, consisting of two sheets corresponding to the two signs of $p_0$. Depending on the sign we speak of hyperboloids of, respectively, future-pointing ($p_0>0$) and past-pointing ($p_0<0$) 4-momenta. If $m=0$ the hyperboloid is termed the light cone.

Lorentz transformations are linear transformations $p'=Lp$ that do not change $p^2\neq 0$. If $p^2=0$, with $p\neq 0$, the (conformal) group of transformations is larger than the Lorentz group. It contains, in particular, rescalings of the form $p'=5p$ and the like. If $p=0$ then all linear transformations preserve $p^2$. For any Lorentz transformation $\det L=\pm 1$, a condition implying that
$dp'_0d^3p'=d^4p'=\pm d^4p=\pm dp_0d^3p$.

Now assume that $p_0=\sqrt{\bm p^2+m^2}>0$. Changing variables from $(p_0,\bm p)$ to $(m,\bm p)$ we find that
\be
d^4p &=& dm\,d^3p \frac{m}{\sqrt{\bm p^2+m^2}}=d(m^2) \frac{d^3p}{2\sqrt{\bm p^2+m^2}}.
\ee
Since $(p')^2=(Lp)^2=p^2=m^2$ it follows that
\be
d^4p &=& d(m^2) \frac{d^3p}{2\sqrt{\bm p^2+m^2}}=\pm d^4p'\nonumber\\
&=&
\pm d(m^2) \frac{d^3p'}{2\sqrt{\bm p'^2+m^2}}
\ee
and thus (for any $m$, even $m=0$)
\be
\frac{d^3p}{2\sqrt{\bm p^2+m^2}}
&=&
\pm \frac{d^3p'}{2\sqrt{\bm p'^2+m^2}},
\ee
if the {\it four\/} components of $p$ and $p'$ are related by a Lorentz transformation.
The same argument can be applied on the past-pointing part of the mass-$m$ hyperboloid --- just change variables from $(p_0,\bm p)$ to $(-m,\bm p)$.

In consequence, in any integral we can change variables according to the recipe
\be
\int_{\mathbb{R}^3}\frac{d^3p}{2\sqrt{\bm p^2+m^2}}
F(\bm p)
&=&
\int_{\mathbb{R}^3}\frac{d^3p'}{2\sqrt{\bm p'^2+m^2}}
F(\bm p')
=
\int_{\mathbb{R}^3}\frac{d^3p}{2\sqrt{\bm p^2+m^2}}
F(\bm{Lp}).
\ee
The Lorentz invariant measure on mass-$m$ hyperboloid (divided by $(2\pi)^3$ for certain reasons related to Fourier transforms) will be denoted by $dp$, i.e.
\be
dp=\frac{d^3p}{(2\pi)^3 2\sqrt{\bm p^2+m^2}}.
\ee
In many applications it is convenient to work from the outset with 3D integrals involving $dp$ and not just $d^3p$. Now let $px=p_0x_0-\bm p\cdot\bm x$ (I will sometimes denote $px$ by $p\cdot x$ or $p_\mu x^\mu$). Two arbitrary solutions
\be
\phi_1(x) &=& \int dp \Big(a_1(\bm p)e^{-ipx}+b_1(\bm p)^\dag e^{ipx}\Big),\\
\phi_2(x) &=& \int dp \Big(a_2(\bm p)e^{-ipx}+b_2(\bm p)^\dag e^{ipx}\Big),
\ee
of the Klein-Gordon wave equation
\be
(\Box+m^2)\phi(x)=0,
\ee
lead to a family of conserved Noether currents, $\partial^\mu T{_\mu}{^r}=0$,
\be
T{_\mu}{^r}(x)
&=&
C\Big(
\partial^r \phi_1(x)\partial_\mu \phi_2(x)
+
\partial_\mu \phi_1(x)\partial^r \phi_2(x)
\Big)
\\
&\pp=&
-
C\Big(\partial_\nu\phi_1(x)\partial^\nu\phi_2(x)-m^2\phi_1(x)\phi_2(x)\Big)
g{_\mu}{^r}.\label{C const in T}
\ee
We interpret $a(\bm p)$ and $b(\bm p)$ as amplitudes describing particles and antiparticles, respectively.
$C$ is a normalization constant [typically one takes $C=1$ for charged fields, and $C=1/2$ for neutral fields i.e. those whose particles equal to their antiparticles: $a(\bm p)=b(\bm p)$; we will later see that the correct choices are, respectively, $C=1/Z$ and $C=1/(2Z)$, where $Z$ is a renormalization constant related to the choice of vacuum in reducible representations of HOLA].
The 4-vector
\be
P_\mu[\phi_1,\phi_2]
&=&
\int d^3x\, T_{\mu 0}(x_0,\bm x)\label{xP}\\
&=&
\int dp \,p_\mu \Big(b_1(\bm p)^\dag a_2(\bm p)+ a_1(\bm p)b_2(\bm p)^\dag\Big)
\label{pP}
\ee
is time-independent.
Another conserved current is
\be
j_\mu(x)=iq\phi_1(x)\partial_\mu\phi_2(x)-iq\partial_\mu\phi_1(x)\phi_2(x)
\ee
with the time-independent scalar
\be
\hat q[\phi_1,\phi_2]
&=&
\int d^3x\, J_{0}(x_0,\bm x)\label{xQ}\\
&=&
q\int dp \Big(b_1(\bm p)^\dag a_2(\bm p)- a_1(\bm p)b_2(\bm p)^\dag\Big).\label{pQ}
\ee
The transition from (\ref{xP}) to (\ref{pP}), and from (\ref{xQ}) to (\ref{pQ}), involves the assumption that it is justified to use the Fourier transform representation of Dirac's delta,
\be
\delta^{(3)}(\bm p) &=& \frac{1}{(2\pi)^3}\int d^3x\, e^{i\bm p\cdot\bm x},\label{delta F}
\ee
which I will comment on later. (Of course, one could start with (\ref{pP}), (\ref{pQ}), and not with, (\ref{xP}), (\ref{xQ}); then the comment on applicability of (\ref{delta F}) would be irrelevant.)

Inserting
\be
\phi_2(x) &=& \phi(x)=\int dp \Big(a(\bm p)e^{-ipx}+b(\bm p)^\dag e^{ipx}\Big),\label{phi_2}\\
\phi_1(x) &=& \phi(x)^\dag =\int dp \Big(b(\bm p)e^{-ipx}+a(\bm p)^\dag e^{ipx}\Big),\label{phi_1}
\ee
we find the 4-momentum
\be
P_\mu[\phi^\dag,\phi]
&=&
\int dp \,p_\mu \Big(a(\bm p)^\dag a(\bm p)+ b(\bm p)b(\bm p)^\dag\Big)\label{P}
\ee
and charge
\be
\hat q[\phi^\dag,\phi]
&=&
q\int dp \Big(a(\bm p)^\dag a(\bm p)- b(\bm p)b(\bm p)^\dag\Big).\label{hat q}
\ee
What one usually finds in the literature are the particular cases (\ref{phi_2}) and (\ref{phi_1}), while the general ones are not mentioned.
The general expressions are complex (or non-Hermitian, if quantized) and this is probably why they may seem ``unphysical", although their real and imaginary (Hermitian and anti-Hermitian, respectively) parts are separately conserved.

However, when one comes to field quantization it  may pay to maintain their general forms. Let me give several applications.

\subsection{Bosonic Fock space}

Let us begin with the orthodox formulation based on bosonic Fock space representation: $a(\bm p)=\sum_j a_j\psi_j(\bm p)$, $b(\bm p)=\sum_j b_j\psi_j(\bm p)$.
$a_j$ and $b_j$ are bosonic annihilation operators and $\psi_j(\bm p)$ are orthogonal functions
\be
\int dp\,\overline{\psi_j(\bm p)}\psi_k(\bm p)=\delta_{jk}.
\ee
Then
\be
P_\mu[\phi^\dag,\phi]
&=&
\int d^3x\, T_{\mu 0}(x_0,\bm x)\\
&=&
\int dp \,p_\mu \Big(a(\bm p)^\dag a(\bm p)+ b(\bm p)b(\bm p)^\dag\Big)\\
&=&
\sum_{jk}\int dp \,p_\mu \,
\Big(
\overline{\psi_j(\bm p)}\psi_k(\bm p)a_j^\dag a_k
+
\psi_k(\bm p)\overline{\psi_j(\bm p)}b_k b_j^\dag
\Big)\\
&=&
\sum_{jk}\int dp \,p_\mu \,
\overline{\psi_j(\bm p)}\psi_k(\bm p)
\frac{1}{2}\Big(a_j^\dag a_k+b_k b_j^\dag\Big)
\\
&=&
\sum_{jk}(P_{\mu})_{jk}
\Big(a_j^\dag a_k+b_kb_j^\dag\Big)\\
&=&
\sum_{jk}
\Big(a_j^\dag (P_{\mu})_{jk}a_k+b_j^\dag (P_{\mu})_{jk}b_k\Big)
+
\sum_{j}(P_{\mu})_{jj}.\\
\hat q[\phi^\dag,\phi]
&=&
\int d^3x\, J_{0}(x_0,\bm x)\\
&=&
q\int dp \Big(a(\bm p)^\dag a(\bm p)- b(\bm p)b(\bm p)^\dag\Big)\\
&=&
q\sum_{j}
\Big(a_j^\dag a_j-b_j^\dag b_j\Big)
-
q\sum_{j}1.
\ee
The ``vacuum terms'' $\sum_{j}(P_{\mu})_{jj}$ and $-q\sum_{j}1$ are divergent. The infinite negative charge of vacuum may be intuitively interpreted as the charge of the ``Dirac sea" of antiparticles. The remaining charge term
\be
:\hat q:&=&\hat q-\hat q_{\textrm{vacuum}}
=
q\sum_{j}
\Big(a_j^\dag a_j-b_j^\dag b_j\Big)= q(n_+-n_-)
\ee
looks reasonable.

A similar analysis can be performed for the case of 4-momentum.
For example, the Hamiltonian $H=P_0$ leads to vacuum energy
\be
H_{\textrm{vacuum}}
&=&
\sum_{j}\int dp \sqrt{\bm p^2+m^2}
|\psi_j(\bm p)|^2
>
m\sum_{j}\int dp \,
|\psi_j(\bm p)|^2=m\sum_{j}1.
\ee
On the other hand
\be
:H: &=&H-H_{\textrm{vacuum}}
=
\sum_{jk}
\Big(a_j^\dag H_{jk}a_k+b_j^\dag H_{jk}b_k\Big)
\ee
has the form we know from general considerations on the Fock space $\cal F$.
In spite of these partly acceptable predictions the presence of infinite terms makes both 4-momentum and charge ill defined. ``Sensible mathematics" does not allow us to ignore these infinities ``because we do not want them". And even if we ignore the infinities at this stage, they will be reappearing again and again in various (practically all) calculations. Quantization based on the Fock space simply does not correctly work, although in some sense it must be close to a correct theory.

\subsection{Reducible representation of HOLA}

Now assume that
\be
a_1(\bm p) &=& |\bm p\rangle\langle\bm p|\otimes \hat b(\bm p),\\
b_1(\bm p) &=& |\bm p\rangle\langle\bm p|\otimes \hat a(\bm p),\\
a_2(\bm p) &=& 1\otimes \hat a(\bm p),\\
b_2(\bm p) &=& 1\otimes \hat b(\bm p).
\ee
I do not specify at this moment the exact mathematical meanings of $|\bm p\rangle\langle\bm p|$ and the hatted operators --- we will return to it. We obtain
\be
P_\mu[\phi_1,\phi_2]
&=&
\int dp \,p_\mu \Big(b_1(\bm p)^\dag a_2(\bm p)+ a_1(\bm p)b_2(\bm p)^\dag\Big)\nonumber\\
&=&
\int dp \,p_\mu |\bm p\rangle\langle\bm p|\otimes \Big(\hat a(\bm p)^\dag \hat a(\bm p)+ \hat b(\bm p)\hat b(\bm p)^\dag\Big)
\nonumber\\
&=&
\int dp \,p_\mu |\bm p\rangle\langle\bm p|\otimes \Big(\hat a(\bm p)^\dag \hat a(\bm p)+ \hat b(\bm p)^\dag\hat b(\bm p)\Big)
+\int dp \,p_\mu |\bm p\rangle\langle\bm p|\otimes \iota(\bm p,\bm p),\nonumber\\
\label{P12}
\\
\hat q[\phi_1,\phi_2]
&=&
q\int dp \Big(b_1(\bm p)^\dag a_2(\bm p)- a_1(\bm p)b_2(\bm p)^\dag\Big)\nonumber\\
&=&
q\int dp \,|\bm p\rangle\langle\bm p|\otimes \Big(\hat a(\bm p)^\dag \hat a(\bm p)-\hat b(\bm p)\hat b(\bm p)^\dag\Big)\nonumber\\
&=&
q\int dp \,|\bm p\rangle\langle\bm p|\otimes \Big(\hat a(\bm p)^\dag \hat a(\bm p)-\hat b(\bm p)^\dag\hat b(\bm p)\Big)
-
q\int dp \,|\bm p\rangle\langle\bm p|\otimes
\iota(\bm p,\bm p),\label{Q12}
\ee
where $\iota(\bm p,\bm p)=[\hat b(\bm p),\hat b(\bm p)^\dag]$ is the analogue of (\ref{J}).
The resulting operators are, of course, Hermitian.

As we can see, the 4-momentum and charge involve number operators of the form (\ref{red n}).
Several approaches to relativistic field theoretic formalism, based on $\iota(\bm p,\bm p)=1$, were described in detail in \cite{II,III,IV,V}. Later on in these notes we will reconsider all the steps involved in such a choice of $\iota(\bm p,\bm p)$.

\subsection{Sequential approach}

The next application is the following. Assume we have two sequences of operators, $a(\bm p,n)$, $b(\bm p,n)$, convergent in some sense to $a(\bm p)$ and $b(\bm p)$. Now take
\be
\phi_2(x) &=& \int dp \Big(a(\bm p,n_2)e^{-ipx}+b(\bm p,n_2)^\dag e^{ipx}\Big),\nonumber\\
\phi_1(x) &=& \int dp \Big(b(\bm p,n_1)e^{-ipx}+a(\bm p,n_1)^\dag e^{ipx}\Big).\nonumber
\ee
Then
\be
P_\mu[\phi_1,\phi_2]
&=&
\int d^3x\, T_{\mu 0}(x_0,\bm x;n_1,n_2)\label{xPnn}\\
&=&
\int dp \,p_\mu \Big(a(\bm p,n_1)^\dag a(\bm p,n_2)+ b(\bm p,n_1)b(\bm p,n_2)^\dag\Big),\label{xPnn'}\\
\hat q[\phi_1,\phi_2]
&=&
\int d^3x\, J_{0}(x_0,\bm x;n_1,n_2)\label{xQnn}\\
&=&
q\int dp \Big(a(\bm p,n_1)^\dag a(\bm p,n_2)- b(\bm p,n_1)b(\bm p,n_2)^\dag\Big).\label{xQnn'}
\ee
are time independent for all $n_1$ and $n_2$. We will later consider situations where the Fourier form of Dirac delta (\ref{delta F}) is rigorously applicable in (\ref{xPnn}) and (\ref{xQnn}) if $n_1$ and $n_2$ are finite. Integration  in (\ref{xPnn}) and (\ref{xQnn}) must be then performed before the limits $n_1\to\infty$, $n_2\to\infty$ are evaluated. Alternatively, the limits $n_1\to\infty$, $n_2\to\infty$ may be performed under momentum integrals in (\ref{xPnn'}) and (\ref{xQnn'}), but not under position integrals in (\ref{xPnn}) and (\ref{xQnn}).

\section{Sequential approach to Dirac deltas}

Dirac described his $\delta(x)$ as follows:
``To get a picture of $\delta(x)$, take a function of the real variable $x$ which vanishes
everywhere except inside a small domain, of length $\ve$ say, surrounding the origin $x = 0$,
and which is so large inside this domain that its integral over this domain is unity. The
exact shape of the function inside this domain does not matter, provided there are no
unnecessarily wild variations (for example provided the function is always of order $\ve^{-1}$).
Then in the limit $\ve\to 0$ this function will go over into  $\delta(x)$'' \cite{D-PoQM}.

One possible meaning of Dirac's delta was formalized by Laurent Schwartz \cite{LS}, who interpreted it as an ``evaluation-at-zero map", i.e. a linear functional that, given a function $f$, returns its value at $x=0$. In the standard functional-analysis notation we would write  $\langle\delta|f\rangle=f(0)$ or, more generally, $\langle\delta_x|f\rangle=f(x)$; in the bra-ket notation of Dirac one would expresses the same as $\langle x|f\rangle=f(x)$.

An alternative to the Schwartz formalism was developed by the Polish mathematician Jan Mikusi\'nski \cite{JM} and his coworkers, a program that culminated in the textbook \cite{AMS}.
The authors wanted to make precise the intuition of Dirac that
\be
\int_{-\infty}^{\infty}dx\,\delta(x)f(x)
=
\lim_{n\to\infty}
\int_{-\infty}^{\infty}dx\,\delta_n(x)f(x)\label{int Dirac},
\ee
and, on this basis, formulate the whole of theory of distributions.
In introduction to \cite{AMS} they wrote:
``We shall not avail ourselves to the methods of functional analysis and we shall not define distributions as functionals. In applied mathematics distributions are regarded as ordinary functions, e.g. the function $\delta(x)$ of Dirac. Essentially, however, distributions are not functions but in an intuitive sense, they may be approximated by functions. Approximation, strictly defined, is our starting point for the definition of distributions."

The idea of ``filtering integrals", satisfying $\lim_{n\to\infty}
\int_{-\infty}^{\infty}dx\,\delta_n(x)f(x)=[f(0_-)+f(0_+)]/2$, had been known to Cauchy, Hermite, Poisson, Kirchhoff, and Heaviside many decades before Dirac rediscovered the notion for the purposes of quantum mechanics (for a review see the fifth chapter of \cite{Bremmer}).
The Mikusi\'nski sequential approach goes deeper, generalizing the idea to general distributions. Distributions are in this formalism  equivalence classes of fundamental sequences of ordinary functions, much the same way as in Cantor's theory real numbers are defined as equivalence classes of fundamental sequences of rational numbers.
It follows that one first has to define fundamental sequences of functions, and then introduce an equivalence relation that leads to equivalence classes.

Let me begin with recalling some basic notions. We say that $f_n$ converges uniformly to $f$, if for any $\ve>0$  one can find $n_0$ such that $|f_n(x)-f(x)|<\ve$ for any $n>n_0$. The sequence $f_n(x)=x/n$ is not uniformly convergent to $f(x)=0$, since no matter what $\ve>0$ and $n<\infty$ we take, we will always find $x$ for which $|x/n-0|>\ve$ (just take any $x>n\ve$). We say that $f_n$ converges in the interval $A<x<B$ almost uniformly to $f$, if it converges to $f$ uniformly on each finite closed interval contained in the interval $A<x<B$ (this definition of almost uniform convergence is employed in \cite{AMS}). The sequence $f_n(x)=x/n$ is almost uniformly convergent to $f(x)=0$ on the whole real axis since $x/n$ converges uniformly to 0 on each interval $-\infty<a\leq x\leq b<\infty$ (now $|x|$ cannot be arbitrarily large and $|x/n-0|<\ve$, for each $x$ in the interval, if $n$ is sufficiently large).

In part I of \cite{AMS} a delta-sequence is defined as any sequence $\delta_n$ of continuous functions, satisfying
\be
\int_{-\infty}^{\infty}dx\,\delta_n(x) &=& 1,\\
\delta_n(x) &=& F_n''(x),
\ee
where $F_n$ are twice differentiable functions almost uniformly convergent to
\be
F(x) &=& \left\{
\begin{array}{rcl}
0 & {\rm for} & x<0\\
x & {\rm for} & x\geq 0
\end{array}
\right.
,\label{F}
\ee
and the prime denotes the derivative.

The intuition behind the construction is the following. First,
\be
F'(x) &=& \theta(x)=\left\{
\begin{array}{rcl}
0 & {\rm for} & x<0\\
1 & {\rm for} & x\geq 0
\end{array}
\right.
.
\ee
The derivatives at 0 are defined by right and left limits.
$\theta'(x)$ is zero everywhere, with the exception of $x=0$, where $\theta(x)$ ``jumps from 0 to 1 infinitely fast" --- this is, roughly, how Dirac imagined his delta ``function".

A nontrivial and useful example of a delta-sequence is obtained by means of Cauchy's principal value,
\be
\int_{-\infty}^\infty dx\,f(x)\frac{e^{ixn}}{x}
=
\int_{-\infty}^\infty dx\,f(x/n)\frac{e^{ix}}{x}
=
\lim_{\epsilon\to 0_+}
\Big(\int_{-1/\epsilon}^{-\epsilon}+\int_{\epsilon}^{1/\epsilon}\Big) dx\,f(x/n)\frac{e^{ix}}{x}.
\ee
If $f(x)$ does not grow too fast with $x\to\pm \infty$ (i.e. when for large $n$ the function $f(x/n)$ is sufficiently slowly changing if compared to $1/x$) then
\be
\lim_{n\to\infty}
\int_{-\infty}^\infty dx\,f(x/n)\frac{e^{ix}}{x}
&=&
f(0)
\lim_{\epsilon\to 0_+}
\Big(\int_{-1/\epsilon}^{-\epsilon}+\int_{\epsilon}^{1/\epsilon}\Big) dx\,\frac{e^{ix}}{x}
\nonumber\\
&=&
i\pi f(0)
=
i\pi \int_{-\infty}^\infty dx\,f(x)\delta(x).
\ee
In this sense,
\be
\frac{e^{ix n}}{i\pi x}=\delta_n(x)\label{complex delta s}.
\ee
As we can see, delta-sequences can be given also by complex-valued functions.

In parts II and III of the book the authors further restrict delta-sequences to the narrower class of those $\delta_n$ that are smooth, and $\delta_n(x)=0$ for $|x|>\alpha_n$, where $\alpha_n$ is a sequence of real numbers convergent to 0 (now $\delta_n(x)$ is exactly vanishing outside a given integral --- the shorter, the greater $n$). The authors explain in the introduction that there are certain differences between different parts of the book since they correspond to different versions of the theory, developed by the Mikusi\'nski group over different time periods. In particular, the requirement of smoothness and exact vanishing of $\delta_n(x)$ outside of an interval is {\it not needed\/} for (\ref{int Dirac}) to hold, but turns out convenient for some applications.

I stress the latter property of the Mikusi\'nski-Antosik-Sikorski approach since in what follows I want to perform a similar step and restrict admissible delta-sequences to functions with some specified properties.
\begin{figure}
\includegraphics[width=8cm]{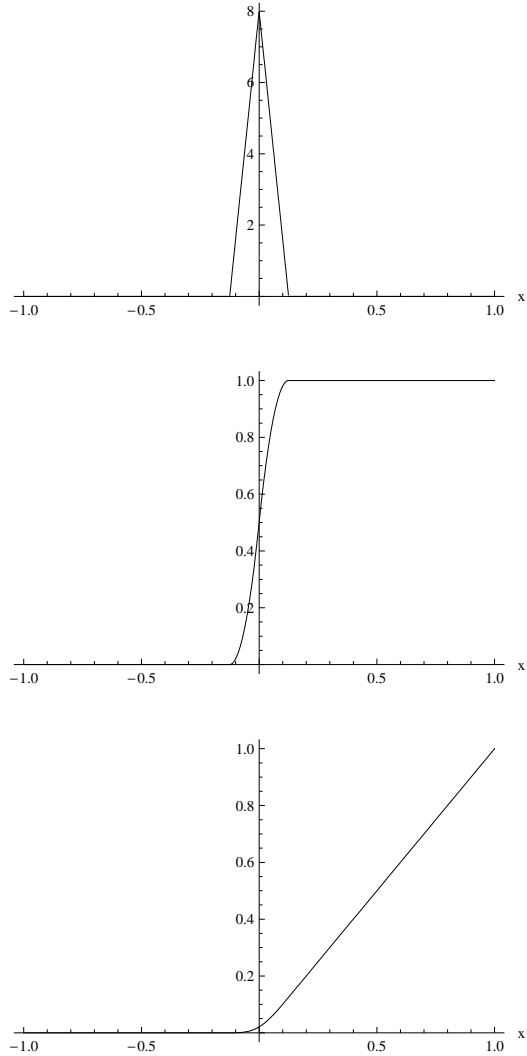}
\caption{$\Lambda$-shaped $\delta_n(x)$ (upper), $\theta_n(x)=\int_{-\infty}^x dx_1\,\delta_n(x_1)$ (middle), and $F_n(x)=
\int_{-\infty}^{x} dx_2\int_{-\infty}^{x_2} dx_1\,\delta_n(x_1)$ (lower). $\delta_n(x)=F_n''(x)$. The support of $\delta_n(x)$ is here given by the interval $[-\frac{1}{n},\frac{1}{n}]$ with $n=8$. $F_n$ uniformly converges to $F$ given by (\ref{F}).}
\end{figure}

Now, as we have developed some intuitions for the sequential approach, let us describe in more detail the equivalence relation that defines Dirac's delta. We say that the sequence of continuous functions $f_n$ defined for $A<x<B$ is fundamental if there exists an integer $k\geq 0$ and an almost uniformly convergent sequence $F_n$ of functions, satisfying $d^k F_n(x)/dx^k=f_n(x)$. $k=2$ for $\delta_n$. The sequences $f_n$ and $g_n$ are equivalent (we write $f_n\sim g_n$), if $d^k F_n(x)/dx^k=f_n(x)$, $d^k G_n(x)/dx^k=g_n(x)$, for the same $k$, and both $F_n$ and $G_n$ converge almost uniformly to the same function.

One can prove (cf. p. 11 in \cite{AMS}) that $\sim$ is an equivalence relation. The equivalence class $[f_n]=:f$ is the distribution associated with equivalent fundamental sequences. Any $f_n$ from the equivalence class is a representative of the distribution $f$.

Let us explicitly show an appropriate construction of $\delta$. Let $\delta_n$ be the function of the type shown in Fig.~1. It is clear that plots similar to the middle and lower ones would be obtained if one replaced this concrete form of $\delta_n$ by any function with the same support (and integrable to unity), or by a sufficiently narrow Gaussian supported on the whole of $\mathbb R$. All such delta-sequences thus belong to the same equivalence class $\delta=[\delta_n]$. Delta sequences that tend to infinity at $x=0$ will be termed $\Lambda$-shaped.

\section{Further splitting of equivalence classes:
Discontinuities and delta-sequences regular at zero}

In Eq. (\ref{int Dirac}) I have purposefully avoided one more natural identification, namely
\be
\int_{-\infty}^{\infty}dx\,\delta(x)f(x)
=
\lim_{n\to\infty}
\int_{-\infty}^{\infty}dx\,\delta_n(x)f(x)=f(0)=\langle\delta|f\rangle\label{int Dirac'}.
\ee
The reason for this omittance is that (\ref{int Dirac'}) is true only for functions continuous at $x=0$.
In case of discontinuity the delta-sequence shown at Fig.~1 would imply
\be
\lim_{n\to\infty}
\int_{-\infty}^{\infty}dx\,\delta_n(x)f(x) &=& \frac{f(0_-)+f(0_+)}{2}\label{int Dirac''},
\ee
where $f(0_\pm)$ denotes the left and right limits of $f(x)$ at $x=0$.

But what would have happened had we replaced this concrete delta-sequence by a new (fundamental) sequence
$\tilde\delta_n(x)=\delta_n(x-\frac{1}{n})$? All the plots from Fig.~1 would be simply shifted to the right by $1/n$, so this is again a delta-sequence, but
\be
\lim_{n\to\infty}
\int_{-\infty}^{\infty}dx\,\tilde\delta_n(x)f(x) &=& f(0_+)\label{int Dirac'''}.
\ee
Of particular interest are also the following two examples (M-shaped delta sequence, Fig.~2)
\be
\tilde{\tilde\delta}{}_n(x) &=& \frac{1}{2}\delta_n\Big(x-\frac{1}{n}\Big)+\frac{1}{2}\delta_n\Big(x+\frac{1}{n}\Big)\label{Mdelta}
\ee
and ($\Lambda\Lambda$-shaped delta sequence, Fig.~3)
\be
\tilde{\tilde\delta}{}_n(x) &=& \frac{1}{2}\delta_n\Big(x-\frac{2}{n}\Big)+\frac{1}{2}\delta_n\Big(x+\frac{2}{n}\Big)\label{MMdelta},
\ee
both yielding
\be
\lim_{n\to\infty}
\int_{-\infty}^{\infty}dx\,\tilde{\tilde\delta}{}_n(x)f(x) &=& \frac{f(0_-)+f(0_+)}{2}\label{int Dirac"}.
\ee
\begin{figure}
\includegraphics[width=8cm]{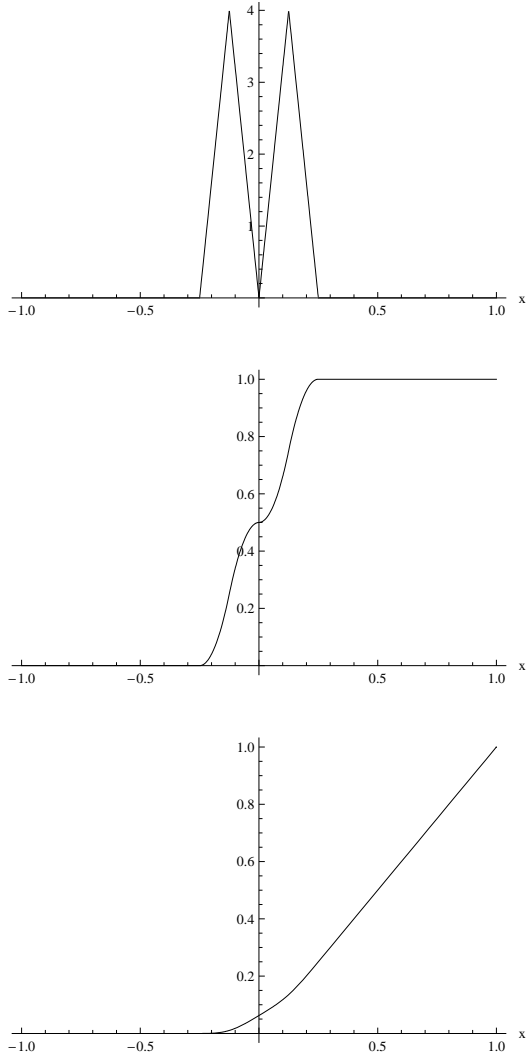}
\caption{Analogues of the plots from Fig.~1 but for M-shaped delta-sequence (\ref{Mdelta}). $F_n''(0)=0$ but the third derivative is not continuous at $x=0$. $F_n$ uniformly converges to $F$ given by (\ref{F}).}
\end{figure}
\begin{figure}
\includegraphics[width=8cm]{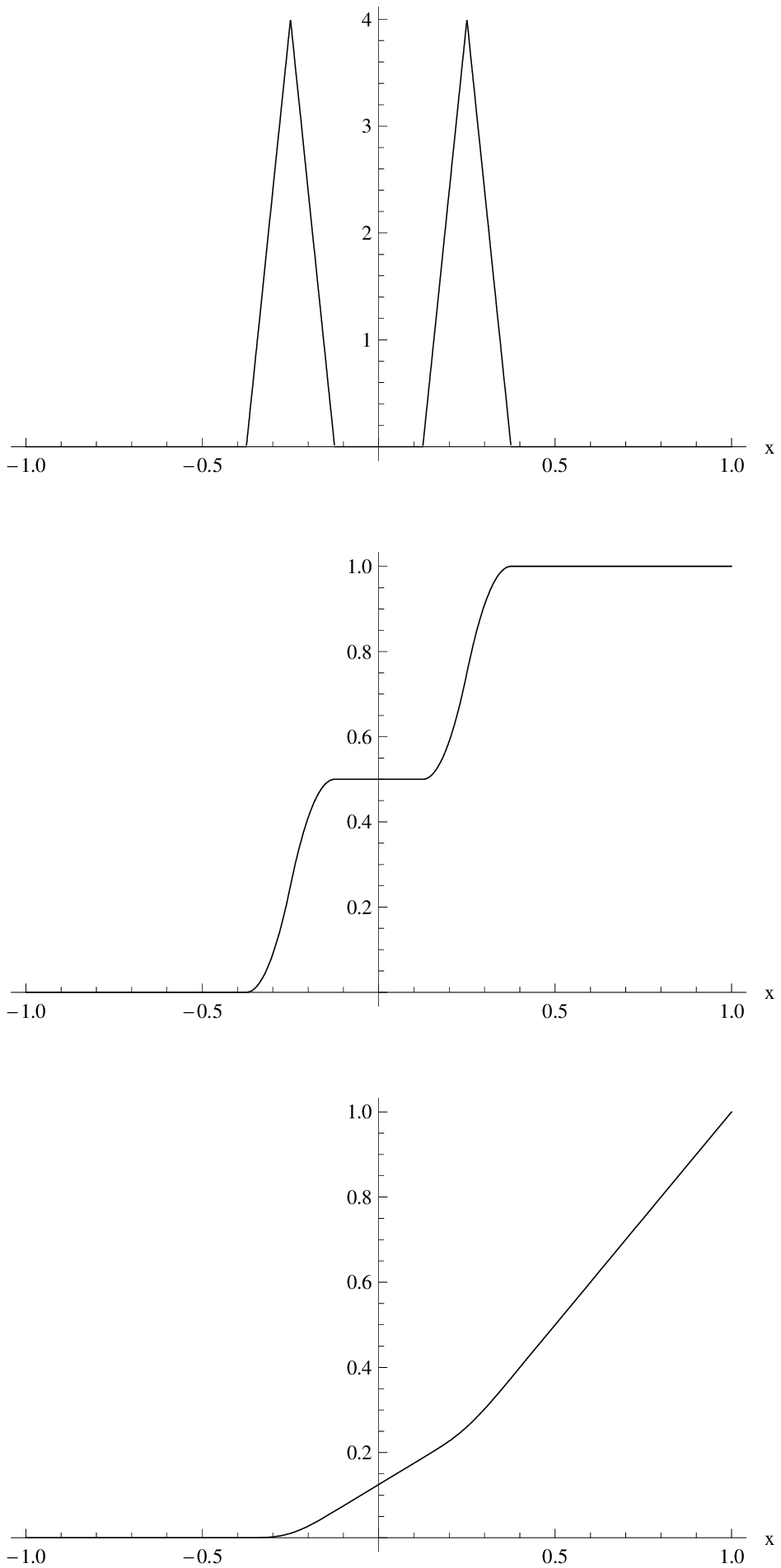}
\caption{Analogues of the plots from Fig.~1 but for $\Lambda\Lambda$-shaped  delta-sequence (\ref{MMdelta}).  $F_n$ uniformly converges to $F$ given by (\ref{F}), and all derivatives of $F'_n$ vanish at $x=0$.}
\end{figure}

It is clear that the equivalence relation we have discussed in the previous section can only be true if Dirac deltas ``act" on continuous $f$. Schwartz and the other authors go typically even further and assume that one deals with functions that are smooth.

However, in quantum field theory one encounters expressions of the form
\be
\int_{-\infty}^{\infty}dx\,\delta(x)\delta(x)f(x)\label{delta^2}
\ee
(or even worse), involving ``functions" $\delta(x)f(x)$ that are quite far from any form of continuity. This is one of the sources of the infinities that plague ``the so-called quantum electrodynamics" (the phrase of Dirac \cite{Dirac0}), especially in loop-amplitude calculations. Intuitively one expects that (\ref{delta^2}) equals $\delta(0)f(0)$, whatever this means. It is interesting that such a rule can be indeed derived from dimensional regularization techniques for path integrals, but $\delta(0)$ is then treated as the ``infinite quantity $\int dk/(2\pi)$" \cite{Kleinert}.

From a sequential point of view an appropriate calculation could read
\be
\int_{-\infty}^{\infty}dx\,\delta(x)\delta(x)f(x)
&=&
\lim_{n\to\infty}\lim_{m\to\infty}\int_{-\infty}^{\infty}dx\,\delta_n(x)\tilde\delta_m(x)f(x)\\
&=&
\lim_{m\to\infty}\lim_{n\to\infty}\int_{-\infty}^{\infty}dx\,\delta_n(x)\tilde\delta_m(x)f(x)\\
&=&
\lim_{m\to\infty}\tilde\delta_m(0)f(0)\\
&=&
\lim_{n\to\infty}\delta_n(0)f(0),
\ee
where $\delta_n(x)$ and $\tilde\delta_n(x)$ are, in principle, different representatives of $\delta=[\delta_n]=[\tilde\delta_n]$.
It is obvious that in order that the expression be well defined one has to require
\be
\lim_{n\to\infty}\tilde\delta_n(0)f(0)=\lim_{n\to\infty}\delta_n(0)f(0).
\ee
The examples of delta-sequences we have discussed above would imply
\be
\lim_{n\to\infty}\delta_n(0) f(0)&=& \infty\times \frac{f(0_-)+f(0_+)}{2},\\
\lim_{n\to\infty}\tilde\delta_n(0) f(0_+)&=& 0\times f(0_+),\\
\lim_{n\to\infty}\tilde{\tilde\delta}_n(0)\frac{f(0_-)+f(0_+)}{2}
&=&0\times \frac{f(0_-)+f(0_+)}{2},
\ee
so that
\be
\int_{-\infty}^{\infty}dx\,\delta(x)\delta(x)f(x)
\ee
becomes ambiguous, unless one restricts delta sequences to a subset of $[\delta_n]$. In other words, we have to modify the equivalence relation.

{\it Remark\/}: The proof that $\delta^2$ does not exist, given in \cite{AMS}, is based on the assumption that the sequential definition
of (\ref{delta^2}) should read
\be
\int_{-\infty}^{\infty}dx\,\delta(x)\delta(x)f(x)
&=&
\lim_{n\to\infty}\int_{-\infty}^{\infty}dx\,\delta_n(x)^2f(x),
\ee
which indeed does not exist for any delta-sequence. It seems that the reason for such a choice of definition lies in ambiguities implied by the form of $\sim$. However, we are free to change the equivalence relation.$\blacktriangle$

One option is the following: Two delta-sequences $\delta_n$ and $\tilde\delta_n$ are equivalent if $\delta_n\sim\tilde\delta_n$ with respect to the Mikusi\'nski-Antosik-Sikorski relation discussed above and, in addition,
\be
\lim_{n\to\infty}\delta_n(0) &=&\lim_{n\to\infty}\tilde\delta_n(0)=:\delta(0),\\
\lim_{n\to\infty}
\int_{-\infty}^{\infty}dx\,{\tilde\delta}{}_n(x)f(x) &=& \lim_{n\to\infty}
\int_{-\infty}^{\infty}dx\,\delta{}_n(x)f(x)\nonumber\\
&=&
\frac{1}{2}f(0_-)+\frac{1}{2}f(0_+),
\ee
for any $f$. Equivalence classes with respect to this new equivalence relation define different Dirac deltas.
Deltas belonging to the same equivalence class can be uniquely multiplied,
\be
\int_{-\infty}^{\infty}dx\,\delta(x)^j f(x)
&=&
\lim_{n_1\to\infty}\dots\lim_{n_j\to\infty}\int_{-\infty}^{\infty}dx\,\delta^{(1)}_{n_1}(x)\dots \delta^{(j)}_{n_j}(x)f(x)
\\
&=& \frac{1}{2}\delta(0_-)^{j-1}f(0_-)+\frac{1}{2}\delta(0_+)^{j-1}f(0_+)\\
&=& \delta(0)^{j-1}\frac{1}{2}\Big(f(0_-)+f(0_+)\Big).
\ee
Here $\delta^{(1)}_{n_1}(x)\dots \delta^{(j)}_{n_j}(x)$ means that we are free to take any (continuous) representatives $\delta^{(i)}_{n_i}(x)$ for any of the delta-sequences. The examples show that there is no difficulty with assuming even $\delta(0)=0$, an option suggested by relativistic invariance, as we shall see in a moment.

Delta-sequences vanishing at 0 are not a new concept (cf. Chapter V, Eq.~(38) in \cite{Bremmer}). An intriguing example of such a delta-sequence was recently found in the context of time-of-arrival operator \cite{Galapon1,Galapon2}. In spite of this, many authors who generalize the concept of Dirac's delta stick to the ``obvious" requirement of either divergence or indefiniteness of $\delta(0)$ (cf. \cite{Cortizo,Ortiz,Tsallis}).

\section{{\tt M}-shaped and M-shaped Dirac deltas}

In what follows we will mostly work in ``momentum space" so that delta-sequences will depend on arguments $p$, $k$, etc., whereas $x$ will be reserved for their Fourier-transform arguments. This is perhaps different from what one is accustomed to, but is more convenient from the point of view of applications we have in mind.

\subsection{{\tt M}-shaped delta-sequences}

Let us consider the function shown in the upper part of Fig.~4. It is a particular example, for $a=1$ and $\epsilon=1/2$, of
\be
\delta(k,a,\epsilon)
&=&
\left\{
\begin{array}{crc}
0 & \textrm{for} & k < -\frac{\epsilon}{2} \\
\big(\frac{4k}{\epsilon} +2\big)\big(\frac{2}{\epsilon} - \frac{a}{2}\big) & \textrm{for} & -\frac{\epsilon}{2} \leq k < -\frac{\epsilon}{4}\\
-\frac{4k}{\epsilon}\big(\frac{2}{\epsilon} - \frac{3a}{2}\big) + a & \textrm{for} & -\frac{\epsilon}{4} \leq k < 0\\
\frac{4k}{\epsilon}\big(\frac{2}{\epsilon} - \frac{3a}{2}\big) + a & \textrm{for} & 0\leq k<\frac{\epsilon}{4}\\
\big(-\frac{4k}{\epsilon} +2\big)\big(\frac{2}{\epsilon} - \frac{a}{2}\big) & \textrm{for} & \frac{\epsilon}{4} \leq k < \frac{\epsilon}{2}\\
0 & \textrm{for} & \frac{\epsilon}{2}\leq k,
\end{array}\label{1}
\right.\nonumber
\\
\ee
($a\geq 0$, $\epsilon>0$).
\begin{figure}
\includegraphics[width=8cm]{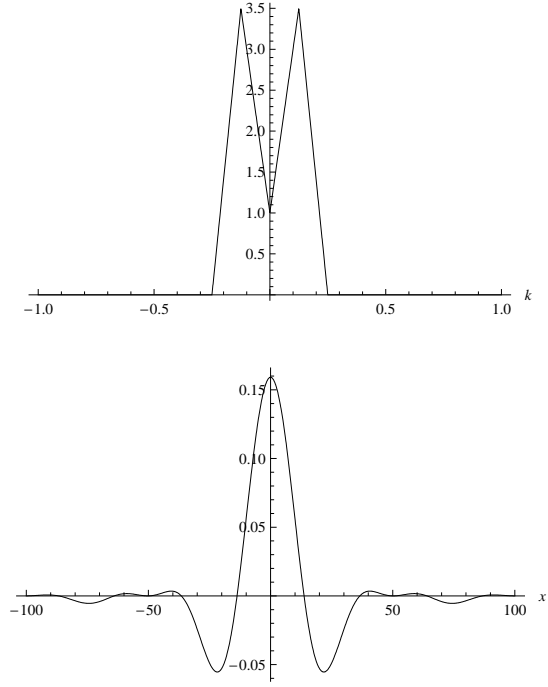}
\caption{The {\tt M}-shaped function (\ref{1}) with $a=1$, $\epsilon=1/2$ (upper), and its Fourier transform (lower).}
\end{figure}
Its Fourier transform,
\be
\hat\delta(x,a,\epsilon)
&=&
\frac{1}{2\pi}
\int_{-\infty}^\infty \delta(k,a,\epsilon) e^{ikx}dk\nonumber\\
&=&
\frac{8}{\pi}
\frac{\epsilon a + (4 - \epsilon a) \cos\frac{\epsilon x}{4}}{\epsilon^2 x^2}\sin^2\frac{\epsilon x}{8},
\\
\lim_{\epsilon\to 0}\hat\delta(x,a,\epsilon) &=& \frac{1}{2\pi},
\ee
is the real function shown in the lower part of Fig.~4 . The sequence $\delta_n(k,a)=\delta(k,a,\textstyle{\frac{1}{n}})$, with natural $n$ (i.e. $\epsilon=1/n$), is an example of what I call an {\tt M}-shaped delta-sequence, which is again a particular example of the delta-sequence in the sense of \cite{AMS}.
Indeed,
\be
\int_{-\infty}^\infty \delta(k,a,\textstyle{\frac{1}{n}})dk &=& 1,
\ee
and
\be
\lim_{n\to\infty}\int_{-\infty}^\infty f(k)\delta(k,a,\textstyle{\frac{1}{n}})dk &=&\frac{f(0_-)+f(0_+)}{2}.
\ee
{\tt M}-shaped delta-sequences do not have to vanish at 0,
\be
\delta(0,a,\textstyle{\frac{1}{n}}) &=& a,
\ee
for all $n$, so that
\be
\lim_{n\to\infty}\delta(0,a,\textstyle{\frac{1}{n}}) &=& a.
\ee
For each $a$ we deal with a sequence belonging to a different equivalence class. We will now show that $a=0$ we have encountered before is an important special case.

\subsection{{\tt M}-shaped deltas with respect to more general measures}

Let us assume that instead of $dp$ we have to use a measure $d\mu(p)=\rho(p)dp$, and an appropriate delta is needed,
\be
\int d\mu(p')\delta(p,p')f(p')=f(p),
\ee
with $\delta(p,p)=a$, say, where $a$ is a constant (that is, not a function of $p$).
The standard solution
\be
\delta(p,p')=\rho(p')^{-1}\delta(p-p'),\label{mu}
\ee
if generalized to {\tt M}-shaped deltas implies $\delta(p,p)=\rho(p)^{-1}\delta(0)$ and will not lead to $a$ independent of $p$ (unless $\delta(0)=a=0$).

So let us try a different option.
Let $\delta(p,a,\frac{1}{n})$, $\delta(0,a,\frac{1}{n})=a$, be an arbitrary {\tt M}-shaped delta-sequence. The sequence
\be
\delta_n(p,p')
&=&
\rho(p')^{-1}\delta\big(p-p',a\rho(p),\textstyle{\frac{1}{n}}\big),
\ee
$\rho(p)=d\mu(p)/dp$, has the required properties
\be
\lim_{n\to \infty}
\int d\mu(p')\delta_n(p,p')f(p')
&=&
\frac{f(p_-)+f(p_+)}{2},\\
\delta_n(p,p)
&=&
a
\ee
As an exercise, compute in two different ways
\be
{}&{}&\int d\mu(p')\delta(p,p')\delta(p+k,p'+k)f(p')
\nonumber\\
&{}&\pp=
=
\lim_{n\to\infty}\lim_{m\to\infty}
\int dp'\rho(p')\rho(p')^{-1}\delta\big(p-p',a\rho(p),\textstyle{\frac{1}{n}}\big)
\rho(p'+k)^{-1}\delta\big(p-p',a\rho(p+k),\textstyle{\frac{1}{m}}\big)f(p')\nonumber\\
&{}&\pp=
=
\lim_{n\to\infty}
\delta\big(p-p,a\rho(p),\textstyle{\frac{1}{n}}\big)
\rho(p+k)^{-1}f(p)\nonumber\\
&{}&\pp=
=
a\rho(p)\rho(p+k)^{-1}f(p).
\ee
Reversing the order of limits we would get
\be
{}&{}&\int d\mu(p')\delta(p,p')\delta(p+k,p'+k)f(p')
\nonumber\\
&{}&\pp=
=
\lim_{m\to\infty}\lim_{n\to\infty}
\int dp'\rho(p')\rho(p')^{-1}\delta\big(p-p',a\rho(p),\textstyle{\frac{1}{n}}\big)
\rho(p'+k)^{-1}\delta\big(p-p',a\rho(p+k),\textstyle{\frac{1}{m}}\big)f(p')\nonumber\\
&{}&\pp=
=
\lim_{m\to\infty}
\rho(p+k)^{-1}\delta\big(p-p,a\rho(p+k),\textstyle{\frac{1}{m}}\big)f(p)\nonumber\\
&{}&\pp=
=
\rho(p+k)^{-1}a\rho(p+k)f(p)\nonumber\\
&{}&\pp=
=af(p).
\ee
Consistency of the two calculations requires that
\be
a\rho(p)\rho(p+k)^{-1}=a\label{a rho}
\ee
for all $p$ and $k$, so that either $a\neq 0$ and then $\rho(p)=$~const, or $\rho(p)\neq$~const and $a=0$.

In non relativistic quantum mechanics $\rho(p)=$~const so that $a$ can be arbitrary. However, relativistic measures involve a nontrivial $\rho$ and thus we have to assume $a=0$. From now on, if I write of M-shaped delta-sequences, as opposed to the
{\tt M}-shaped
ones, I mean those corresponding to $a=0$. A particular class of M-shaped
delta sequences are the $\Lambda\Lambda$-shaped sequences whose derivatives of all orders vanish at 0.

\subsection{M-shaped versus $\Lambda$-shaped delta-sequences}

Let us discuss in more detail the properties of delta-sequences of the types shown in Figs.~1--3.
All of them are derived from a single $\delta_n(k)$, and can be regarded as particular cases of the formula
\be
\delta_n(k,j) &=& \frac{1}{2}\delta_n\Big(k-\frac{j}{n}\Big)+\frac{1}{2}\delta_n\Big(-k-\frac{j}{n}\Big)=\delta_n(-k,j),
\quad j=0,1,2,\dots
\ee
The Fourier transform
\be
\hat\delta_n(x,j) &=& \frac{1}{2\pi}\int_{-\infty}^\infty dk\, \delta_n(k,j) e^{ikx}
\\
&=&
\frac{1}{2}\frac{1}{2\pi}\int_{-\infty}^\infty dk\, \delta_n\Big(k-\frac{j}{n}\Big) e^{ikx}
+
\frac{1}{2}\frac{1}{2\pi}\int_{-\infty}^\infty dk\, \delta_n\Big(-k-\frac{j}{n}\Big) e^{ikx}
\\
&=&
\frac{1}{2}\frac{1}{2\pi}\int_{-\infty}^\infty dk\, \delta_n(k) e^{ikx}e^{ijx/n}
+
\frac{1}{2}\frac{1}{2\pi}\int_{-\infty}^\infty dk\, \delta_n(k) e^{-ikx}e^{-ijx/n}
\\
&=&
\frac{1}{2}\hat\delta_n(x)e^{ijx/n}+\frac{1}{2}\overline{\hat\delta_n(x)}e^{-ijx/n}
\ee
is real.

It is instructive to pause here for a moment and perform all these calculations for the explicit choice of $\Lambda$-shaped delta-sequence from Fig.~1 (now the argument is $k$ and not $x$),
\be
\delta_n(k)
&=&
\left\{
\begin{array}{cll}
0 & \textrm{for} & k<-\frac{1}{n},\\
n^2k+n & \textrm{for} & -\frac{1}{n}\leq k<0,\\
-n^2k+n & \textrm{for} & 0 \leq k < \frac{1}{n},\\
0 & \textrm{for} & k\geq \frac{1}{n}.
\end{array}
\right.\label{explicit}
\ee
Its Fourier transform
\be
\hat \delta_n(x)
&=&
\frac{1}{2\pi}\int_{-\infty}^\infty dk\, \delta_n(k) e^{ikx}
\nonumber\\
&=&
\frac{1}{2\pi}
\Big(\frac{x}{2n}\Big)^{-2}\sin^2\frac{x}{2n}
\ee
is real, symmetric, bounded, $|\hat \delta_n(x)|\leq 1/(2\pi)$, and almost uniformly (but not uniformly) convergent to $1/(2\pi)$.

For all delta-sequences generated from a symmetric $\delta_n(k)=\delta_n(k,0)$, $\overline{\hat\delta_n(x)}=\hat\delta_n(x)$, we can further simplify
\be
\hat\delta_n(x,j)
&=&
\hat\delta_n(x)\cos \frac{jx}{n},\\
\lim_{n\to\infty}\hat\delta_n(x,j)
&=&
\frac{1}{2\pi}\lim_{n\to\infty}\cos \frac{jx}{n}\lim_{n\to\infty}\int_{-\infty}^\infty dk\, \delta_n(k) e^{ikx}=
\frac{1}{2\pi}.
\ee
The latter formulas indicate very clearly that there cannot be much difference between $\Lambda$-shaped delta sequences we are accustomed to, and the regular-at-zero M-shaped and $\Lambda\Lambda$-shaped ones. In fact, they all define the same distribution in the sense of Mikusi\'nski, but
$\delta_n(k,0)$ and $\delta_n(k,j)$, $j>0$, belong to different equivalence classes with respect to our modified relation. In particular, products of Dirac deltas $\delta(k,j)=[\delta_n(k,j)]$ can be taken for all integer $j>0$, but $\delta(k,0)\delta(k,j)$ is ill defined.

If both $f(0_-)$ and $f(0_+)$ are finite, we get $\int dk\,\delta(k,j)^Nf(k)=0$ for all $N=2,3,\dots$ and $j=1,2,\dots$. If any of $f(0_\pm)$ is infinite, the expression is still not well defined. This is why a modification of Dirac's delta cannot solve all the problems of field quantization. However, the situation is much better if we additionally take a nontrivially chosen $I(\bm p)$ occurring in HOLA. I will discuss the issue later on in these notes.

\subsection{$\delta[f(k)]$ for $M$-shaped Dirac deltas}

Let $f(k)$ be a function that vanishes at some $k_l$, i.e. $f(k_l)=0$, and let $\delta_n(k)$ be a delta-sequence such that supports of $\delta_n(k-k_1)$ and $\delta_n(k-k_2)$ do not overlap, no matter which $k_1$ and $k_2$, $f(k_1)=f(k_2)=0$, one takes. I will now show how to derive in the context of $M$-shaped deltas the analogue of the standard formula
\be
\delta[f(k)]
=
\sum_l \frac{\delta(k-k_l)}{|f'(k_l)|}.\label{delta f(k)}
\ee
Let us first assume that $f'(k_l)$ exists.
The first trick is to replace, in a neighborhood of a $k_l$, the function $f(k)$ by $g_l(k)=f'(k_l)(k-k_l)$ i.e. its tangent line at $k_l$. Then
\be
\lim_{n\to\infty}\int_{-\infty}^\infty dk\, \delta_n[f(k)]F(k)
&=&
\sum_l\lim_{n\to\infty}\int_{-\infty}^\infty dk\, \delta_n[g_l(k)]F(k)
\nonumber\\
&=&
\sum_l\lim_{n\to\infty}\frac{1}{|f'(k_l)|}\int_{-\infty}^\infty dk\, \delta_n(k)F\left(\frac{k+f'(k_l)k_l}{f'(k_l)}\right)
\nonumber\\
&=&
\sum_l\frac{1}{|f'(k_l)|}\frac{F(k_{l+})+F(k_{l-})}{2}
\ee
which coincides with (\ref{delta f(k)}). As an example consider the important case of
\be
f(p_0) &=& p_0^2-\bm p^2-m^2,\\
f'(p_0) &=& 2p_0,\\
p_l &=& \pm\sqrt{\bm p^2+m^2}.
\ee
Let $p_1=-\sqrt{\bm p^2+m^2}$. Then
\be
g_1(p_0) &=& -2\sqrt{\bm p^2+m^2}(p_0+\sqrt{\bm p^2+m^2}),\\
g_2(p_0) &=& 2\sqrt{\bm p^2+m^2}(p_0-\sqrt{\bm p^2+m^2}).
\ee
For $m=0$ the point $\bm k=0$ is singular since $f'(k_1)=f'(k_2)=0$. In expressions such as
\be
\phi(x)
&=&
\int d^4k\,\delta(k_0^2-\bm k^2)\tilde \phi(k_0,\bm k)e^{-ikx}
\ee
the function $\tilde \phi(k_0,\bm k)$ must satisfy
\be
\lim_{k_0\to 0_\pm}\frac{\tilde \phi(k_0,\bm 0)}{k_0}=0.
\ee
The fact that $\delta$ is M-shaped is thus irrelevant in this context.

\section{Plane waves and M-shaped Dirac deltas}

In Dirac's bra-ket notation one encounters formulas such as $\langle k|k'\rangle=2\pi\delta(k-k')$ or
$\frac{1}{2\pi} \int_{-\infty}^\infty dk\,|k\rangle\langle k|=1$. In what follows I want to show in what sense they can be interpreted in the language of M-shaped delta sequences.

Let $\delta_n(k)=\delta_n(k,j)$ for some positive integer $j$.
We start with convolution of two M-shaped delta-sequences,
\be
\delta_{nm}^*(k)
&=&
\delta_{n}*\delta_{m}(k)
=
\int_{-\infty}^\infty dk'\,\delta_n(k-k')\delta_m(k')
=
\delta_{mn}^*(k),\\
\lim_{m\to\infty}
\delta_{nm}^*(k)
&=&
\lim_{m\to\infty}
\int_{-\infty}^\infty dk'\,\delta_n(k-k')\delta_m(k')
=
\delta_n(k),\\
\lim_{n\to\infty}
\delta_{nm}^*(k)
&=&
\lim_{n\to\infty}
\int_{-\infty}^\infty dk'\,\delta_n(k-k')\delta_m(k')
=
\delta_m(k).
\ee
The new sequence is again a delta-sequence,
\be
\int_{-\infty}^\infty dk\,\delta_{nm}^*(k)
&=&
\int_{-\infty}^\infty dk\int_{-\infty}^\infty dk'\,\delta_n(k-k')\delta_m(k')
\nonumber\\
&=&
\int_{-\infty}^\infty dk'\,\delta_m(k')\int_{-\infty}^\infty dk\,\delta_n(k-k')
\nonumber\\
&=&
\int_{-\infty}^\infty dk'\,\delta_m(k')
=1,
\ee
\be
\lim_{n\to\infty}\lim_{m\to\infty}\int_{-\infty}^\infty dk\,
 f(k)\delta_{nm}^*(k)
&=&
\lim_{n\to\infty}\lim_{m\to\infty}\int_{-\infty}^\infty dk\int_{-\infty}^\infty dk'\,f(k)\delta_n(k-k')\delta_m(k')
\nonumber\\
&=&
\lim_{n\to\infty}\int_{-\infty}^\infty dk\,f(k)\delta_n(k)
=
\frac{f(0_-)+f(0_+)}{2},
\ee
but
\be
\delta_{nm}^*(0)
&=&
\int_{-\infty}^\infty\delta_n(0-k')\delta_m(k')dk'
\nonumber\\
&=&
\int_{-\infty}^\infty\delta_n(k')\delta_m(k')dk'
\ee
in general depends on $n$ and $m$. The other properties are nevertheless analogous to M-shaped delta-sequences,
\be
\lim_{m\to\infty}\delta_{nm}^*(0)
&=&
\lim_{m\to\infty}\int_{-\infty}^\infty\delta_n(k')\delta_m(k')dk'=
\delta_n(0)=0,
\ee
and
\be
\lim_{n\to\infty}\delta_{nn}^*(0)
&=&
\lim_{n\to\infty}\int_{-\infty}^\infty\delta_n(k')\delta_n(k')dk'=\infty
\nonumber.
\ee
Moreover, there exists a sequence $\alpha_{nm}$,
\be
\lim_{m\to\infty}\lim_{n\to\infty}\alpha_{nm}=\lim_{n\to\infty}\lim_{m\to\infty}\alpha_{nm}=0,
\ee
such that
$\delta_{nm}^*(k)=0$ for $|k|\geq \alpha_{nm}$.
Employing the Fourier transform
\be
\hat\delta_{nm}^*(x)
&=&
\frac{1}{2\pi}
\int_{-\infty}^\infty dk\,\delta_{nm}^*(k)e^{ikx}
=
2\pi \hat\delta_n(x)\hat\delta_m(x)\label{FF}
\ee
we can write
\be
\delta_{nm}^*(k-k')
&=&
\int_{-\infty}^\infty dx\,\hat\delta_{nm}^*(x) e^{-i(k-k')x}
\nonumber\\
&=&
2\pi\int_{-\infty}^\infty dx\,
\overline{\hat\delta_n(x)e^{ikx}}\hat\delta_m(x)e^{ik'x}\label{2 pi delta delta}
\ee
since $\hat\delta_n(x)$ is real. The evaluation map $\langle x|\psi\rangle=\hat\psi(x)$, returning the value of Fourier transform of a given $\psi$, can be used to denote
\be
\langle x|k,n\rangle
&=&
2\pi\hat\delta_n(x)e^{ikx}=2\pi\hat\delta_n(x)\langle x|k\rangle,\\
\lim_{n\to\infty}\langle x|k,n\rangle
&=&
e^{ikx}=:\langle x|k\rangle,\\
\langle k,n|x\rangle
&=&
\overline{\langle x|k,n\rangle}
=
2\pi\hat\delta_n(x)e^{-ikx}=2\pi\hat\delta_n(x)\langle k|x\rangle,\\
\lim_{n\to\infty}\langle k,n|x\rangle
&=&
e^{-ikx}=:\langle k|x\rangle.
\ee
Accordingly
\be
\delta_{nm}^*(k-k')
&=&
\frac{1}{2\pi}\int_{-\infty}^\infty dx\,
\langle k,n|x\rangle
\langle x|k',m\rangle\\
&=&
\frac{1}{2\pi}\langle k,n|k',m\rangle.
\ee
\begin{figure}
\includegraphics[width=8cm]{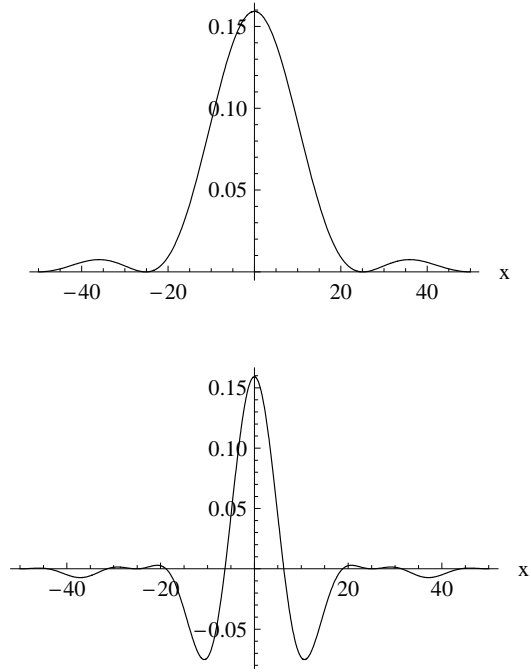}
\caption{Plots of (\ref{ogolny j}) for $n=4$ and different values of $j$:  $j=0$ (upper) and $j=1$ (lower). For $j=1$ the integral
(\ref{int ogolny j}) vanishes due to the oscillating term, whereas the integrand in (\ref{int ogolny j'}) is non-negative.}
\end{figure}
The fact that $\hat\delta_n(x)$ is a Fourier transform of a square-integrable $\delta_n(k)$ implies,
by Plancherel's theorem \cite{Kaiser}, that $\langle x|k,n\rangle$ is square integrable for finite $n$,
\be
\int_{-\infty}^\infty dx \,|\langle x|k,n\rangle|^2
&=&
\int_{-\infty}^\infty dx \,|2\pi\hat\delta_n(x)e^{ikx}|^2
=
(2\pi)^2
\int_{-\infty}^\infty dx \,|\hat\delta_n(x)|^2
<\infty.
\ee
A similar result holds for
\be
\big|\delta_{nm}^*(k-k')\big|^2
&=&
\frac{1}{(2\pi)^2}\big|\langle k,n|k',m\rangle\big|^2
\\
&\leq& \frac{1}{(2\pi)^2} \langle k,n|k,n\rangle \langle k',m|k',m\rangle<\infty.
\ee
Let us finally return to (\ref{2 pi delta delta}) and take the (well defined) limit,
\be
\lim_{m\to\infty}\delta_{nm}^*(k-k')
&=&
2\pi\lim_{m\to\infty}\int_{-\infty}^\infty dx\,
\overline{\hat\delta_n(x)e^{ikx}}\hat\delta_m(x)e^{ik'x},\nonumber\\
&=&
\int_{-\infty}^\infty dx\,
\hat\delta_n(x)e^{-i(k-k')x}=\delta_n(k-k'),
\ee
which is just the inverse Fourier transform of a square integrable function.
The next limit would be, however, purely formal
\be
\lim_{n\to\infty}\delta_n(k-k')
&\stackrel{?}{=}&
\frac{1}{2\pi}\int_{-\infty}^\infty dx\,
e^{-i(k-k')x}\stackrel{?}{=}\delta(k-k')\stackrel{?}{=}\delta^*(k-k'),
\ee
still showing that the familiar expression for the Fourier form of Dirac's delta is true also for deltas constructed by means M-shaped delta sequences vanishing at 0. But, then, what about the celebrated divergency
\be
\delta(0)
&\stackrel{?}{=}&
\frac{1}{2\pi}\int_{-\infty}^\infty dx\stackrel{?}{=}\infty?
\ee
Let us do the calculation more precisely,
\be
\lim_{m\to\infty}\delta_{nm}^*(0)
&=&
\int_{-\infty}^\infty dx\,
\hat\delta_n(x)=\delta_n(0).
\ee
Inserting our explicit example,
\be
\hat\delta_n(x)
&=&
\frac{1}{2\pi}\Big(\frac{x}{2n}\Big)^{-2}\sin^2\frac{x}{2n}\cos\frac{jx}{n},\label{ogolny j}
\ee
we find that the integral vanishes for M-shaped delta sequences ($j=1,2,\dots$)
\be
\frac{1}{2\pi}\int_{-\infty}^\infty dx\,\Big(\frac{x}{2n}\Big)^{-2}\sin^2\frac{x}{2n}\cos\frac{jx}{n}=0, \label{int ogolny j}
\ee
and for the $\Lambda$-shaped one ($j=0$),
\be
\frac{1}{2\pi}\int_{-\infty}^\infty dx\,\Big(\frac{x}{2n}\Big)^{-2}\sin^2\frac{x}{2n}=n,\label{int ogolny j'}
\ee
diverges in the limit $n\to\infty$, as expected (compare Fig.~5).

\section{M-shaped Dirac deltas and spectral theorem}

Following Dirac, we will denote scalar products of square-integrable functions by the same bra-ket symbol as the evaluation map. Let $\hat\psi(x)=\langle x|\psi\rangle$ be square-integrable. Then
\be
\langle k,n|\psi\rangle
&=&
\int_{-\infty}^\infty dx \,\langle k,n|x\rangle\langle x|\psi\rangle
\nonumber\\
&=&
\int_{-\infty}^\infty dx \,2\pi\hat\delta_n(x)e^{-ikx}\hat\psi(x).\label{wFT}
\ee
(\ref{wFT}) is a kind of windowed Fourier transform \cite{Kaiser} of $\psi(x)$, with the window function $2\pi\hat\delta_n(x)$.

Let us again be as explicit as possible and employ the concrete $\Lambda$-shaped sequence (\ref{explicit}),
\be
\langle k,n|\psi\rangle
&=&
\int_{-\infty}^\infty dx \,2\pi\frac{1}{2\pi}
\Big(\frac{x}{2n}\Big)^{-2}\sin^2\frac{x}{2n}\cos\frac{jx}{n}e^{-ikx}\hat\psi(x)
\nonumber\\
&=&
\int_{-\infty}^\infty dx \,
e^{-ikx}\underbrace{\Big(\frac{x}{2n}\Big)^{-2}\sin^2\frac{x}{2n}\cos\frac{jx}{n}\hat\psi(x)}_{\hat\psi_n(x,j)}
\nonumber\\
&=&
\int_{-\infty}^\infty dx \,
e^{-ikx}\hat\psi_n(x,j).
\ee
The Parseval theorem,
\be
\frac{1}{2\pi} \int_{-\infty}^\infty dk\,\langle \psi|k,m\rangle\langle k,n|\psi\rangle
&=&
\int_{-\infty}^\infty dx\,\overline{\hat\psi_m(x,j)}\hat\psi_n(x,j),
\ee
with
\be
\overline{|\hat\psi_m(x,j)}\hat\psi_n(x,j)|=|\hat\psi(x)|^2
\underbrace{\Big|\Big(\frac{x}{2m}\Big)^{-2}\sin^2\frac{x}{2m}\cos\frac{jx}{m}\Big(\frac{x}{2n}\Big)^{-2}\sin^2\frac{x}{2n}\cos\frac{jx}{n}\Big|}_{\leq 1}
\leq |\hat\psi(x)|^2,\nonumber
\ee
allows us to use the dominated convergence theorem (Theorem~6 in \cite{lim pod int}) and take the limits, in any order, under the integral sign,
\be
{}&{}&
\lim_{n\to\infty}\lim_{m\to\infty}
\frac{1}{2\pi}
\int_{-\infty}^\infty dk\,\langle \psi|k,m\rangle\langle k,n|\psi\rangle
\nonumber\\
&{}&\pp=
=
\lim_{n\to\infty}\lim_{m\to\infty}
\int_{-\infty}^\infty dx\,
|\hat\psi(x)|^2
\Big(\frac{x}{2m}\Big)^{-2}\Big(\frac{x}{2n}\Big)^{-2}\sin^2\frac{x}{2m}\cos\frac{jx}{m}\sin^2\frac{x}{2n}\cos\frac{jx}{n}
\nonumber\\
&{}&\pp=
=
\int_{-\infty}^\infty dx\,
|\hat\psi(x)|^2
\lim_{n\to\infty}\lim_{m\to\infty}
\Big(\frac{x}{2m}\Big)^{-2}\Big(\frac{x}{2n}\Big)^{-2}\sin^2\frac{x}{2m}\cos\frac{jx}{m}\sin^2\frac{x}{2n}\cos\frac{jx}{n}
\nonumber\\
&{}&\pp=
=
\int_{-\infty}^\infty dx\,
|\hat\psi(x)|^2.
\ee
The same argument can be applied to the diagonal limit ($n=m$),
\be
\lim_{n\to\infty}\frac{1}{2\pi} \int_{-\infty}^\infty dk\,\langle \psi|k,n\rangle\langle k,n|\psi\rangle
&=&
\int_{-\infty}^\infty dx\,|\hat\psi(x)|^2
\lim_{n\to\infty}\Big|\Big(\frac{x}{2n}\Big)^{-2}\sin^2\frac{x}{2n}\cos\frac{jx}{n}\Big|^2
\nonumber\\
&=&
\int_{-\infty}^\infty dx\,|\hat\psi(x)|^2
=\langle \psi|\psi\rangle
\nonumber\\
&=&
\frac{1}{2\pi} \int_{-\infty}^\infty dk\,\lim_{n\to\infty}\langle \psi|k,n\rangle\langle k,n|\psi\rangle
\nonumber.
\ee
Expressing scalar products in terms of norms (the polarization identity \cite{Kaiser}),
\be
\langle f|g\rangle
&=&
\frac{1}{4}\Big(
|| f+g ||^2 - || f-g ||^2+i|| f+ig ||^2-i|| f-ig ||^2
\Big),
\ee
we can extend the result to pairs of arbitrary square-integrable functions $\psi$ and $\phi$,
\be
\lim_{n\to\infty}\frac{1}{2\pi} \int_{-\infty}^\infty dk\,\langle \phi|k,n\rangle\langle k,n|\psi\rangle
&=&
\frac{1}{2\pi} \int_{-\infty}^\infty dk\,\lim_{n\to\infty}\langle \phi|k,n\rangle\langle k,n|\psi\rangle
\nonumber\\
&=&
\lim_{m\to\infty}\lim_{n\to\infty}\frac{1}{2\pi} \int_{-\infty}^\infty dk\,\langle \phi|k,m\rangle\langle k,n|\psi\rangle
\nonumber\\
&=&
\frac{1}{2\pi} \int_{-\infty}^\infty dk\,\lim_{m\to\infty}\lim_{n\to\infty}\langle \phi|k,m\rangle\langle k,n|\psi\rangle
\nonumber\\
&=&
\langle \phi|\psi\rangle.
\ee
In this sense
\be
{\mathbb I}&=&
\lim_{n\to\infty}\frac{1}{2\pi} \int_{-\infty}^\infty dk\,|k,n\rangle\langle k,n|
\nonumber\\
&=&
\frac{1}{2\pi} \int_{-\infty}^\infty dk\,\lim_{n\to\infty}|k,n\rangle\langle k,n|
\nonumber\\
&=&
\lim_{m\to\infty}\lim_{n\to\infty}\frac{1}{2\pi} \int_{-\infty}^\infty dk\,|k,m\rangle\langle k,n|
\nonumber\\
&=&
\frac{1}{2\pi} \int_{-\infty}^\infty dk\,\lim_{m\to\infty}\lim_{n\to\infty}|k,m\rangle\langle k,n|
\nonumber\\
&=&
\frac{1}{2\pi} \int_{-\infty}^\infty dk\,|k\rangle\langle k|.
\ee
Now consider the matrix element
\be
\langle\phi|A|\psi\rangle
&=&
\lim_{m\to\infty}
\lim_{n\to\infty}\frac{1}{2\pi} \int_{-\infty}^\infty dk\,A(k)\langle\phi|k,m\rangle\langle k,n|\psi\rangle
\nonumber\\
&=&
\lim_{m\to\infty}
\lim_{n\to\infty}\frac{1}{2\pi}
\int_{-\infty}^\infty dk\,A(k)
\int_{-\infty}^\infty dx\,\langle\hat\phi|x\rangle\langle x|k,m\rangle
\int_{-\infty}^\infty dy\,\langle k,n|y\rangle\langle y|\hat\psi\rangle
\nonumber\\
&=&
\lim_{m\to\infty}
\lim_{n\to\infty}\frac{1}{2\pi} \int_{-\infty}^\infty dk\,A(k)
\int_{-\infty}^\infty dx\,\overline{\hat\phi(x)}2\pi\hat\delta_m(x)e^{ikx}
\int_{-\infty}^\infty dy\,2\pi\hat\delta_n(y)\hat\psi(y)e^{-iky}
\nonumber\\
&=&
\lim_{m\to\infty}
\lim_{n\to\infty}\frac{1}{2\pi} \int_{-\infty}^\infty dk\,A(k)
\overline{\int_{-\infty}^\infty dx\,2\pi\hat\delta_m(x)\hat\phi(x)e^{-ikx}}
\int_{-\infty}^\infty dy\,2\pi\hat\delta_n(y)\hat\psi(y)e^{-iky}.\nonumber
\ee
With our definitions of Fourier transforms,
\be
\hat f(x) &=& \frac{1}{2\pi}\int_{-\infty}^\infty dk\,f(k)e^{ikx},\\
f(k) &=& \int_{-\infty}^\infty dx\,\hat f(x)e^{-ikx},\\
\int_{-\infty}^\infty dx\,\hat f(x)\hat g(x)e^{-ikx}
&=&
\frac{1}{2\pi}\int_{-\infty}^\infty dk'\,f(k-k')g(k'),
\ee
we obtain
\be
\langle\phi|A|\psi\rangle
&=&
\lim_{m\to\infty}\lim_{n\to\infty}\frac{1}{2\pi}
\int_{-\infty}^\infty dk\, A(k)
\underbrace{\int_{-\infty}^\infty dk'\,
\delta_m(k-k')\overline{\phi(k')}}_{\delta_m *\overline{\phi(k)}}
\underbrace{\int_{-\infty}^\infty dk''\,
\delta_n(k-k'')\psi(k'')}_{\delta_n *\psi(k)}\nonumber.
\ee
As a by-product of the above calculation we have shown that $\langle k,n|\psi\rangle=\delta_n *\psi(k)$.

Of great importance is the following
\medskip

\noindent
{\it Lemma\/}: (a variant of Theorem II.3.1.1 in \cite{AMS})
If $\psi(k)$ is continuous then $\langle k,n|\psi\rangle=\delta_n *\psi(k)$ converges to $\psi(k)$ almost uniformly.
\medskip

\noindent
{\it Proof\/}: Assume for simplicity that the delta-sequence is non-negative, $\delta_n(k)\geq 0$, and vanishes for $|k|>\alpha_n$, $\lim_{n\to\infty}\alpha_n = 0$, like in all our previous examples. Assume $k$ belongs to a finite interval $[a,b]$. Then
\be
\big|\delta_n *\psi(k)-\psi(k)\big|
&=&
\Big|
\int_{-\infty}^\infty dk'\,\delta_n(k-k')\psi(k')-\psi(k)\int_{-\infty}^\infty dk'\,\delta_n(k-k')\Big|
\nonumber\\
&=&
\Big|
\int_{-\infty}^\infty dk'\,\delta_n(k-k')\big(\psi(k')-\psi(k)\big)\Big|
\nonumber\\
&\leq&
\int_{-\infty}^\infty dk'\,
\delta_n(k-k')\big|\psi(k')-\psi(k)\big|
\nonumber\\
&=&
\int_{-\alpha_n}^{\alpha_n} dk'\,
\delta_n(k')\big|\psi(k-k')-\psi(k)\big|
\nonumber
\ee
By definition of continuity of $\psi(k)$ at $k$, for any $\ve>0$ there exists $n_0$ such that $\big|\psi(k-k')-\psi(k)\big|<\ve$ for all $n>n_0$ and $-\alpha_n\leq k'\leq \alpha_n$. So
\be
\big|\delta_n *\psi(k)-\psi(k)\big|
&\leq&
\int_{-\alpha_n}^{\alpha_n} dk'\,
\delta_n(k')\big|\psi(k-k')-\psi(k)\big|
\nonumber\\
&\leq&
\ve
\int_{-\alpha_n}^{\alpha_n} dk'\,
\delta_n(k')
=\ve
\nonumber
\ee
for all $n>n_0$ and $a\leq k\leq b$.$\blacksquare$

We can take the limits under the integral sign in
\be
\langle\phi|A|\psi\rangle
&=&
\lim_{m\to\infty}\lim_{n\to\infty}\frac{1}{2\pi}
\int_{-\infty}^\infty dk\, A(k)
\delta_m *\overline{\phi(k)}
\delta_n *\psi(k)\nonumber
\ee
in several cases. The simplest is the one of $A(k)\delta_m *\overline{\phi(k)}\delta_n *\psi(k)$ converging uniformly to
$A(k)\delta_m *\overline{\phi(k)}\psi(k)$ and $A(k)\overline{\phi(k)}\delta_n *\psi(k)$. For, example let $A(k)$ be continuous for $a<k<b$, and $A(k)=0$ for $k\geq b$, $k\leq a$. Almost uniform convergence implies uniform convergence on any closed interval, and
\be
\langle\phi|A|\psi\rangle
&=&
\lim_{m\to\infty}\lim_{n\to\infty}\frac{1}{2\pi}
\int_a^b dk\, A(k)
\delta_m *\overline{\phi(k)}
\delta_n *\psi(k)\nonumber\\
&=&
\frac{1}{2\pi}
\int_a^b dk\, A(k)
\lim_{m\to\infty}\delta_m *\overline{\phi(k)}
\lim_{n\to\infty}\delta_n *\psi(k)\nonumber\\
&=&
\frac{1}{2\pi}
\int_a^b dk\, A(k)
\overline{\phi(k)}
\psi(k)\nonumber\\
&=&
\frac{1}{2\pi}
\int_a^b dk\, A(k)
\langle\phi|k\rangle\langle k|\psi\rangle\nonumber.
\ee
A generalization going beyond finite intervals $a\leq k\leq b$ can be based on the observation that dominated convergence theorem guarantees
\be
\lim_{m\to\infty}\lim_{n\to\infty}\frac{1}{2\pi}
\int_{-\infty}^\infty dk\, A(k)
\delta_m *\overline{\phi(k)}
\delta_n *\psi(k)
&=&
\frac{1}{2\pi}
\int_{-\infty}^\infty dk\, A(k)
\overline{\phi(k)}
\psi(k) \label{A<infty}
\ee
if the right-hand-side of (\ref{A<infty}) is finite and
\be
|A(k)
\delta_m *\overline{\phi(k)}
\delta_n *\psi(k)|
&\leq&
\textrm{const}_1\times |A(k)
\delta_m *\overline{\phi(k)}
\psi(k)|
\,\leq\,
\textrm{const}_2\times |A(k)
\overline{\phi(k)}\psi(k)|.\nonumber
\ee

Let us now turn to the issue of eigenvalues of $A$.
Let $\phi(k)$ be continuous. The eigenvalue problem for $A$ should be understood in this formalism in the following sense
\be
\langle\phi|A|p\rangle
&=&
\lim_{l\to \infty}
\langle\phi|A|p,l\rangle\nonumber\\
&=&
\lim_{l\to \infty}\lim_{m\to \infty}\lim_{n\to \infty}
\frac{1}{2\pi}
\int_{-\infty}^\infty dk\, A(k)
\langle\phi|k,m\rangle\langle k,n|p,l\rangle\nonumber\\
&=&
\lim_{l\to \infty}\lim_{m\to \infty}\lim_{n\to \infty}
\int_{-\infty}^\infty dk\, A(k)
\langle\phi|k,m\rangle\delta_{nl}(k-p)\label{order lmn}\\
&=&
\lim_{m\to \infty}
\frac{1}{2}\Big(A(p_-)+A(p_+)\Big)
\langle\phi|p,m\rangle\nonumber\\
&=&
\frac{1}{2}\Big(A(p_-)+A(p_+)\Big)
\langle\phi|p\rangle.\label{A(p_pm)}
\ee
Let us recall that $\langle\phi|k,m\rangle$ converges almost uniformly to a continuous $\langle\phi|k\rangle=\overline{\phi(k)}$, while the support of $\delta_{nl}(k-p)$ is compact. Therefore, the sequence
$A(k)\langle\phi|k,m\rangle\delta_{nl}(k-p)$ converges uniformly to $A(k)\langle\phi|k\rangle\delta_{nl}(k-p)$ if $A(k)$ is bounded on any compact subset of $\mathbb R$.
In consequence, one can alternatively compute the eigenvalue problem as follows [compare (\ref{order lmn})]
\be
\langle\phi|A|p\rangle
&=&
\lim_{l\to \infty}\lim_{n\to \infty}
\int_{-\infty}^\infty dk\, A(k)
\langle\phi|k\rangle\delta_{nl}(k-p)\\
&=&
\frac{1}{2}\Big(A(p_-)+A(p_+)\Big)\langle\phi|p\rangle.
\ee
If continuous $\phi(k)$ is itself of compact support then $\langle\phi|k,m\rangle$ converges to $\langle\phi|k\rangle$ uniformly.

Although (\ref{A(p_pm)}) shows that the sequential approach automatically defines eigenvalues also at points of discontinuity, in what follows I simplify discussion and assume that $A(k)$ is continuous. In such a case the formalism based on M-shaped deltas does not seem to essentially differ from other mathematical approaches to the Dirac bra-ket formalism for continuous spectra (cf. \cite{Bohm1,Bohm2,de Graaf}). Anyway, what we modify is the value of $\langle k|k'\rangle$ on the diagonal $k=k'$. Looking at the rigged Hilbert space approach, say, one does not find any point where the formalism employs a concrete value (infinite or not) of $\langle k|k\rangle$.

The main difference with respect to more standard formalisms thus lies in definition of diagonal elements of operators whose spectrum is continuous. Indeed,
\be
\langle p|A|p\rangle
&=&
\lim_{r\to \infty}\lim_{s\to \infty}
\langle p,r|A|p,s\rangle\nonumber\\
&=&
\lim_{r\to \infty}\lim_{s\to \infty}\lim_{m\to \infty}\lim_{n\to \infty}
\frac{1}{2\pi}
\int_{-\infty}^\infty dk\, A(k)
\langle p,r|k,m\rangle\langle k,n|p,s\rangle\nonumber\\
&=&
\lim_{r\to \infty}\lim_{s\to \infty}\lim_{m\to \infty}\lim_{n\to \infty}
2\pi
\int_{-\infty}^\infty dk\, A(k)
\delta_{rm}(p-k)\delta_{ns}(k-p)\nonumber\\
&=&
\lim_{s\to \infty}\lim_{n\to \infty}
2\pi A(p) \delta_{ns}(0)=0.
\ee
In particular, $\langle k|k\rangle=0$. The latter, of course, does not mean that $|k\rangle=\lim_{n\to\infty}|k,n\rangle$ is vanishing. Simply, as has been stressed here many times,
neither $\langle k|k'\rangle$ nor $\langle k|\psi\rangle$ should be regarded as scalar products.

It follows that $|k\rangle\langle k|$ is not a projector,
\be
|k\rangle\langle k|k\rangle\langle k|=\delta(0)|k\rangle\langle k|=0.
\ee
However, let $\chi_X(k)$ be the characteristic function of $X\subset \mathbb R$, i.e. $\chi_X(k)=1$ if $k\in X$, and $\chi_X(k)=0$ if $k\notin X$. The operators
\be
E(X)
&=&
\frac{1}{2\pi}
\int_{-\infty}^\infty dk\,\chi_X(k)\, |k\rangle\langle k|,\\
E([a,b])
&=&
\frac{1}{2\pi}\int_{a}^b dk\,|k\rangle\langle k|,
\ee
are projectors ($\emptyset$ is the empty set):
\be
E(X)E(Y) &=& E(X\cap Y),\\
E(\emptyset) &=& 0,\\
E({\mathbb R}) &=& {\mathbb I},\\
E([a,b])\frac{1}{2\pi}\int_{-\infty}^\infty dk\,A(k)\, |k\rangle\langle k|
&=&
\frac{1}{2\pi}\int_{a}^b dk\,A(k)\, |k\rangle\langle k|.
\ee
The operator  $|k\rangle\langle k|$ is not a projector but a POVM \cite{POVM}, since
\be
\langle\psi|\Big(|k\rangle\langle k|\Big)|\psi\rangle
&=&
\lim_{n_1\to\infty}\dots \lim_{n_4\to\infty}
\frac{1}{2\pi}\int_{-\infty}^\infty dp\,\overline{\psi(p)}\langle p,n_1|k,n_2\rangle\langle k,n_3|\frac{1}{2\pi}\int_{-\infty}^\infty dp'\,\psi(p')|p',n_4\rangle
\nonumber\\
&=&
\lim_{n_1\to\infty}\dots \lim_{n_4\to\infty}
\frac{1}{2\pi}\int_{-\infty}^\infty dp\frac{1}{2\pi}\int_{-\infty}^\infty dp'\,\overline{\psi(p)}\psi(p')\langle p,n_1|k,n_2\rangle\langle k,n_3|p',n_4\rangle
\nonumber\\
&=&
\lim_{n_1\to\infty}\dots \lim_{n_4\to\infty}
\int_{-\infty}^\infty dp\int_{-\infty}^\infty dp'\,\overline{\psi(p)}\psi(p')\delta_{n_1n_2}(p-k)\delta_{n_3n_4}(k-p')
\nonumber\\
&=& \langle\psi|k\rangle\langle k|\psi\rangle=|\psi(k)|^2\geq 0,
\ee
and ${\mathbb I}=\int dk/(2\pi)|k\rangle\langle k|$.
\medskip

\noindent
{\it Remark\/}: The above calculations show what should be meant by products of operators. For example, let $A$, $B$, $C$ commute. Then
\be
ABC
&=&
\int_{-\infty}^\infty \frac{dk}{2\pi}A(k)\, |k\rangle\langle k|
\int_{-\infty}^\infty \frac{dk'}{2\pi}B(k')\, |k'\rangle\langle k'|
\int_{-\infty}^\infty \frac{dk''}{2\pi}C(k'')\, |k''\rangle\langle k''|\nonumber\\
&=&
\lim_{n_1\to\infty}\dots \lim_{n_2''\to\infty}
\int_{-\infty}^\infty \frac{dk}{2\pi}A(k)\, |k,n_1\rangle\langle k,n_2|
\int_{-\infty}^\infty \frac{dk'}{2\pi}B(k')\, |k',n_1'\rangle\langle k',n_2'|\nonumber\\
&\pp=&\times
\int_{-\infty}^\infty \frac{dk''}{2\pi}C(k'')\, |k'',n_1''\rangle\langle k'',n_2''|\label{ABC}\\
&=&
\lim_{n_1\to\infty}\dots \lim_{n_2''\to\infty}
\int_{\mathbb{R}^3}\frac{d^3k}{(2\pi)^3}A(k)B(k')C(k'')
|k,n_1\rangle\langle k,n_2|k',n_1'\rangle\langle k',n_2'|k'',n_1''\rangle\langle k'',n_2''|\nonumber\\
&=&
\int_{-\infty}^\infty \frac{dk}{2\pi}A(k)B(k)C(k)
|k\rangle\langle k|\nonumber
\ee
independently of the order of limits, the latter property being equivalent to associativity of the product.$\blacktriangle$
\medskip

As we can see, one can work with M-shaped deltas with no difficulty. Restricting the class of Mikusi\'nski deltas to a subset of $[\delta_n]$ we obtain the usual Dirac deltas but equipped with new and useful properties. The fact that we work with kets of ``zero length" is no more paradoxical than what we face when we deal with those of ``infinite length".

Paraphrasing physics textbook presentations of $\delta(x)$  we could say that M-shaped delta is a ``function which is zero everywhere, but whose integral over any interval containing 0 equals unity". Such a ``definition" is neither worse nor better than the usual one with infinity at 0 --- the difference between them is on the set of zero measure!

Let me end this section with remarks on relativistic M-shaped Dirac deltas.
In what follows two types of expressions involving Dirac deltas will be encountered. First,
\be
\delta^{*(3)}(\bm k) &=& \delta^*(k_1)\delta^*(k_2)\delta^*(k_3)=[\delta_{m_1n_1}^*(k_1)][\delta_{m_2n_2}^*(k_2)][\delta_{m_3n_3}^*(k_3)],\\
1 &=& \int_{{\mathbb R}^3} d^3k\, \delta^{*(3)}(\bm k),
\ee
is M-shaped in 1D and 3D: $\delta^*(0)=0$, $\delta^{*(3)}(\bm 0)=0$.
Sometimes it may be useful to employ the integral formula (\ref{delta F}) for $\delta^{(3)}(\bm p)$ (under appropriate integrals) but then $\delta^{(3)}(\bm 0)$ is meaningless. From now on I will stick to the convention that $\bm p$ denotes the spacelike part of the 4-vector $p=(\sqrt{\bm p^2+m^2},\bm p)=(p_0,\bm p)$, whereas $\bm k$ stands for an analogous part of the null 4-vector $k=(|\bm k|,\bm k)=(k_0,\bm k)$. Relativistically covariant kets will be normalized as follows,
\be
\langle \bm p|\bm p'\rangle &=&(2\pi)^3 2\sqrt{\bm p^2+m^2}\,\delta^{*(3)}(\bm p-\bm p')=:\delta_m(\bm p,\bm p'),\\
\langle \bm k|\bm k'\rangle &=&(2\pi)^3 2|\bm k|\,\delta^{*(3)}(\bm k-\bm k')=:\delta_0(\bm k,\bm k'),\\
\delta_m(\bm p,\bm p) &=& \delta_0(\bm k,\bm k) =0.
\ee
The resolutions of unity are
\be
\int_{{\mathbb R}^3} \frac{d^3k}{(2\pi)^3 2|\bm k|}|\bm k\rangle\langle \bm k|
&=&
\int_{{\mathbb R}^3} dk\,|\bm k\rangle\langle \bm k|={\mathbb I},\\
\int_{{\mathbb R}^3} \frac{d^3p}{(2\pi)^3 2\sqrt{\bm p^2+m^2}}|\bm p\rangle\langle \bm p|
&=&
\int_{{\mathbb R}^3} dp\,|\bm p\rangle\langle \bm p|=
{\mathbb I}.
\ee
Here $dp$ and $dk$ denote relativistic measures on the hyperboloid $p^2=m^2$ and the light cone $k^2=0$, respectively.

\section{Algebra of free field operators revisited}

The ultimate goal of these lecture notes is to discuss a theory of interacting quantum relativistic scalar fields: A charged massive boson $\psi(x)$ and a neutral massless boson $\phi(x)$.
For free fields one finds
\be
\psi(x) &=& \int dp \Big(a(\bm p)e^{-ipx}+b(\bm p)^\dag e^{ipx}\Big),\\
\phi(x) &=& \int dk \Big(c(\bm k)e^{-ikx}+c(\bm k)^\dag e^{ikx}\Big).
\ee
The amplitude particle and antiparticle operators are assumed to satisfy bosonic HOLAs:
\be
{[a(\bm p),a(\bm p')^{\dag}]}
&=&
\delta_m(\bm p,\bm p')I_m(\bm p),\\
{[a(\bm p),n_+(\bm p')]}
&=&
\delta_m(\bm p,\bm p')a(\bm p),\\
{[a(\bm p)^{\dag},n_+(\bm p')]}
&=&
-\delta_m(\bm p,\bm p')a(\bm p)^{\dag},\\
{[b(\bm p),b(\bm p')^{\dag}]}
&=&
\delta_m(\bm p,\bm p')I_m(\bm p),\\
{[b(\bm p),n_-(\bm p')]}
&=&
\delta_m(\bm p,\bm p')b(\bm p),\\
{[b(\bm p)^{\dag},n_-(\bm p')]}
&=&
-\delta_m(\bm p,\bm p')b(\bm p)^{\dag}.
\ee
The massless neutral field satisfies a similar algebra but with the light cone deltas,
\be
{[c(\bm k),c(\bm k')^{\dag}]}
&=&
\delta_0(\bm k,\bm k')I_0(\bm k),\\
{[c(\bm k),n_0(\bm k')]}
&=&
\delta_0(\bm k,\bm k')c(\bm k),\\
{[c(\bm k)^{\dag},n_0(\bm k')]}
&=&
-\delta_0(\bm k,\bm k')c(\bm k)^{\dag}.
\ee
The remaining commutators vanish.
\medskip

\noindent
{\it Remark\/}:
Let us immediately note the peculiarity of M-shaped deltas:
\be
{[a(\bm p),a(\bm p)^{\dag}]}
&=&
[b(\bm p),b(\bm p)^{\dag}]
\,=\,
[c(\bm k),c(\bm k)^{\dag}]\,=\,0.\label{peculiarity}
\ee
This type of commutation relations is, in fact, implicitly employed in standard quantum field theory whenever one applies the normal ordering operation ($:\hat q:$ instead of $\hat q$, etc.). I think this is yet another argument for quantization in terms of M-shaped deltas.$\blacktriangle$
\medskip

\noindent
{\it Remark\/}:
Let us divide $\mathbb{R}^3$ into disjoined sets $X_n$, $\bigcup_{n=0}^\infty X_n=\mathbb{R}^3$,  and let $\chi_{n}(\bm p)$ be their characteristic functions satisfying $\sum_{n=0}^\infty \chi_{n}(\bm p)=1$ and $\chi_{n}(\bm p)\chi_{j}(\bm p)=\chi_{n}(\bm p)\delta_{nj}$. Here $\delta_{nj}$ is the Kronecker delta. Defining
\be
a_n &=& \int dp\, \chi_{n}(\bm p)a(\bm p)
\ee
we find that
\be
{[a_n,a_n^\dag]}
&=&
\Big[\int dp\, \chi_{n}(\bm p)a(\bm p),\int dp'\, \chi_{n}(\bm p')a(\bm p')^\dag\Big]
\nonumber\\
&=&
\int dp\int dp'\, \chi_{n}(\bm p)\chi_{n}(\bm p')[a(\bm p),a(\bm p')^\dag]
\nonumber\\
&=&
\int dp\int dp'\, \chi_{n}(\bm p)\chi_{n}(\bm p')\delta_m(\bm p,\bm p')I_m(\bm p)\nonumber\\
&=&
\int dp\, \chi_{n}(\bm p)^2I_m(\bm p)\nonumber\\
&=&
\int dp\, \chi_{n}(\bm p)I_m(\bm p)=I_{m,n}\neq 0.
\ee
The example shows that commutability (\ref{peculiarity}) typical of M-shaped deltas does {\it not\/} imply that a discrete version of the representation should be automatically taken in the form $[a_n,a_n^\dag]=0$. The discrete counterpart of HOLA will thus read
\be
{[a_n,a_j^\dag]}
&=&
I_n\delta_{nj}
\ee
with appropriately chosen $I_n$.
$\blacktriangle$
\medskip

The above fields will be used by me in two nontrivial toy models. The first model involves interaction term (in interaction picture)
\be
H_1(t) &=& \int_{\mathbb{R}^3} d^3x j_0(t,\bm x)\phi(t,\bm x)\\
&=& iq\int_{\mathbb{R}^3} d^3x
\Big(
\psi(t,\bm x)\dot\psi(t,\bm x)-\dot\psi(t,\bm x)\psi(t,\bm x)
\Big)
\phi(t,\bm x)\\
&=& \int_{\mathbb{R}^3} d^3x\,
\hat\rho(t,\bm x)
\phi(t,\bm x),
\ee
where
\be
j_\mu(x)=iq\psi(x)\partial_\mu\psi(x)-iq\partial_\mu\psi(x)\psi(x)
\ee
is the charge-density current of the massive field, and $\hat\rho(t,\bm x)$ is the charge-density operator,
\be
\int_{\mathbb{R}^3} d^3x\,\hat\rho(t,\bm x)
&=&
\int_{\mathbb{R}^3} d^3x\,\hat\rho(0,\bm x)
=\hat q[\psi^\dag,\psi].
\ee
The fields that define the Hamiltonian are {\it free\/}, a fact typical of the interaction picture.
Note that I do not assume normal ordering of $\hat\rho(t,\bm x)$.

The Hamiltonian has the essential properties of quantum-electrodynamical interaction, but is not relativistically covariant.
For this reason I will also consider a more complicated interaction, with the same charge current,
\be
H_1(t) &=& \int_{\mathbb{R}^3} d^3x j_\mu(t,\bm x)A^\mu(t,\bm x),
\ee
where $A^\mu(t,\bm x)=\phi(t,\bm x)\otimes R^\mu $. $R^\mu$ will be any constant 4-vector operator, commuting with all the other operators occurring in the model. Analysis of (at least) loop diagrams for the latter interaction is the principal goal of these lecture notes.

\subsection{Reducible $N\geq 1$ representation of HOLA --- an indefinite-frequency harmonic oscillator representation of $\psi(x)$}

Let $[\hat a(\bm p),\hat a(\bm p')^\dag]=\iota_+(\bm p,\bm p')$, $[\hat b(\bm p),\hat b(\bm p')^\dag]=\iota_-(\bm p,\bm p')$, where $\iota_\pm$ are any functions (or distributions) satisfying $\iota_\pm(\bm p,\bm p)=1$. We assume that all commutators between the $a$s and the $b$s vanish.
Then $N=1$ reducible representation is constructed analogously to the indefinite-frequency harmonic oscillator discussed earlier in these notes. The particle part,
\be
a(\bm p,1) &=& |\bm p\rangle\langle \bm p|\otimes \hat a(\bm p),\\
a(\bm p,1)^\dag &=& |\bm p\rangle\langle \bm p|\otimes \hat a(\bm p)^\dag,\\
n_+(\bm p,1) &=& |\bm p\rangle\langle \bm p|\otimes \hat a(\bm p)^\dag\hat a(\bm p),\\
I_m(\bm p,1) &=& |\bm p\rangle\langle \bm p|\otimes 1,\\
\int dp\, I_m(\bm p,1) &=& I_m(1),\quad \textrm{(resolution of identity)}
\ee
and the antiparticle part,
\be
b(\bm p,1) &=& |\bm p\rangle\langle \bm p|\otimes \hat b(\bm p),\\
b(\bm p,1)^\dag &=& |\bm p\rangle\langle \bm p|\otimes \hat b(\bm p)^\dag,\\
n_-(\bm p,1) &=& |\bm p\rangle\langle \bm p|\otimes \hat b(\bm p)^\dag\hat b(\bm p),\\
I_m(\bm p,1) &=& |\bm p\rangle\langle \bm p|\otimes 1,
\ee
can be explicitly constructed in various ways.
The simplest case occurs if $\hat a(\bm p)$, $\hat b(\bm p)$ are $\bm p$-independent, i.e.
\be
a(\bm p,1) &=& |\bm p\rangle\langle \bm p|\otimes (\hat a\otimes 1),\label{hola 1 s}\\
a(\bm p,1)^\dag &=& |\bm p\rangle\langle \bm p|\otimes (\hat a^\dag\otimes 1),\\
I_m(\bm p,1) &=& |\bm p\rangle\langle \bm p|\otimes (1\otimes 1),\\
n_+(\bm p,1) &=& |\bm p\rangle\langle \bm p|\otimes (\hat a^\dag\hat a\otimes 1),
\ee
\be
b(\bm p,1) &=& |\bm p\rangle\langle \bm p|\otimes (1\otimes \hat a),\\
b(\bm p,1)^\dag &=& |\bm p\rangle\langle \bm p|\otimes (1\otimes \hat a^\dag),\\
I_m(\bm p,1) &=& |\bm p\rangle\langle \bm p|\otimes (1\otimes 1),\\
n_-(\bm p,1) &=& |\bm p\rangle\langle \bm p|\otimes (1\otimes \hat a^\dag\hat a),\label{hola 2 s}
\ee
with $[\hat a,\hat a^\dag]=1$. At the other extreme is $\iota_\pm(\bm p,\bm p')$ that equals an $\tt M$-shaped Dirac delta normalized to 1 at $\bm 0$.
Both cases are worthy of consideration.

For $N>1$
\be
a(\bm p,N) &=&
\frac{1}{\sqrt{N}}
\Big(
a(\bm p,1)\otimes I_m(1)\otimes \dots \otimes I_m(1)+\dots +I_m(1)\otimes \dots \otimes I_m(1)\otimes  a(\bm p,1)
\Big)
,\nonumber\\
\\
a(\bm p,N)^\dag &=& \frac{1}{\sqrt{N}}
\Big(
a(\bm p,1)^\dag \otimes I_m(1)\otimes \dots \otimes I_m(1)+\dots +I_m(1)\otimes \dots \otimes I_m(1)\otimes  a(\bm p,1)^\dag
\Big)
,\nonumber\\
\\
n_+(\bm p,N) &=& n_+(\bm p,1)\otimes I_m(1)\otimes \dots \otimes I_m(1)+\dots +I_m(1)\otimes \dots \otimes I_m(1)\otimes n_+(\bm p,1)
,\nonumber\\
\\
I_m(\bm p,N) &=& \frac{1}{N}
\Big(
I_m(\bm p,1)\otimes I_m(1)\otimes \dots \otimes I_m(1)+\dots +I_m(1)\otimes \dots \otimes I_m(1)\otimes  I_m(\bm p,1)
\Big),\nonumber\\
\label{I_m(bm p,N)}
\\
I_m(N) &=& \int dp\, I_m(\bm p,N),\quad \textrm{(resolution of identity)}
\ee
\be
b(\bm p,N) &=&
\frac{1}{\sqrt{N}}
\Big(
b(\bm p,1)\otimes I_m(1)\otimes \dots \otimes I_m(1)+\dots +I_m(1)\otimes \dots \otimes I_m(1)\otimes  b(\bm p,1)
\Big)
,\nonumber\\
\\
b(\bm p,N)^\dag &=& \frac{1}{\sqrt{N}}
\Big(
b(\bm p,1)^\dag \otimes I_m(1)\otimes \dots \otimes I_m(1)+\dots +I_m(1)\otimes \dots \otimes I_m(1)\otimes  b(\bm p,1)^\dag
\Big)
,\nonumber\\
\\
n_-(\bm p,N) &=& n_-(\bm p,1)\otimes I_m(1)\otimes \dots \otimes I_m(1)+\dots +I_m(1)\otimes \dots \otimes I_m(1)\otimes n_-(\bm p,1)
.\nonumber\\
\ee
Note that
\be
a(\bm p,1)^2 &=& |\bm p\rangle\langle \underbrace{\!\bm p|\bm p}_0\!\rangle\langle \bm p|\otimes \hat a(\bm p)^2=0
\ee
and similarly with products of all the other elements of HOLA provided they are taken at $\bm p'=\bm p$. In particular,
\be
n_+(\bm p,1) &=& \Big(|\bm p\rangle\langle \bm p|\otimes \hat a(\bm p)^\dag\Big)\Big({\mathbb I}\otimes \hat a(\bm p)\Big)
=
a(\bm p,1)^\dag\Big({\mathbb I}\otimes \hat a(\bm p)\Big)\neq a(\bm p,1)^\dag a(\bm p,1)=0.\nonumber
\ee
This does not mean that the algebra is trivial or Abelian. Indeed,
the 4-momentum (\ref{P12})
\be
P_\mu(1)
&=&
\int dp \,p_\mu |\bm p\rangle\langle\bm p|\otimes \Big(\hat a(\bm p)^\dag \hat a(\bm p)+ \hat b(\bm p)\hat b(\bm p)^\dag\Big)
\nonumber\\
&=&
\int dp \,p_\mu |\bm p\rangle\langle\bm p|\otimes \Big(\hat a(\bm p)^\dag \hat a(\bm p)+ \hat b(\bm p)^\dag\hat b(\bm p)\Big)
+\int dp \,p_\mu\, |\bm p\rangle\langle\bm p|\otimes \iota_-(\bm p,\bm p),\nonumber\\
&=&
\int dp \,p_\mu \Big(n_+(\bm p,1)+n_-(\bm p,1)\Big)
+\int dp \,p_\mu\, I_m(\bm p,1),\label{P(1)}\\
P_\mu(N) &=& P_\mu(1)\otimes I_m(1)\otimes \dots \otimes I_m(1)+\dots +I_m(1)\otimes \dots \otimes I_m(1)\otimes P_\mu(1),
\ee
possesses the required properties of generator of 4-translations: For any $N\geq 1$
\be
e^{iP(N) x}a(\bm p,N)e^{-iP(N) x} &=& a(\bm p,N) e^{-ixp},\\
e^{iP(N) x}b(\bm p,N)e^{-iP(N) x} &=& b(\bm p,N) e^{-ixp},\\
e^{iP(N) x}a(\bm p,N)^\dag e^{-iP(N) x} &=& a(\bm p,N)^\dag  e^{ixp},\\
e^{iP(N) x}b(\bm p,N)^\dag e^{-iP(N) x} &=& b(\bm p,N)^\dag  e^{ixp},\\
e^{iP(N)) x}\psi(0,N)e^{-iP(N) x} &=& \psi(x,N),\\
e^{-iP(N) y}\psi(x,N)e^{iP(N) y} &=& \psi(x-y,N).
\ee
Here $\psi(x,N)$ denotes $\psi(x)$ but taken in the reducible $N\geq 1$ representation of HOLA.
The vacuum term
\be
P_\mu(N)_{\textrm{vacuum}}=N\int dp \,p_\mu\, I_m(\bm p,N)
\ee
is well defined and commutes with all elements of HOLA. In analogy to the standard Fock-space formalism we can work with
\be
:P_\mu(N):
&=&
P_\mu(N)-P_\mu(N)_{\textrm{vacuum}}
=
\int dp \,p_\mu \Big(n_+(\bm p,N)+n_-(\bm p,N)\Big)
\ee
but now, as opposed to the standard approach, the subtraction is well defined.
\medskip

\noindent
{\it Remark\/}: We know that one is not allowed to divide by 0. Why? Because then arithmetics would become ambiguous. Think of $1\cdot 0=2\cdot 0$. If division by 0 would be acceptable then we would conclude that $1=2$. This is the actual reason why $1/0$ is not allowed.
The same situation occurs with $1+\infty=2+\infty$. If we could subtract infinities then $1=2$. So subtraction of infinities is as forbidden in arithmetics as division by 0.$\blacktriangle$

\noindent
{\it Remark\/}: Due to the property of M-shaped deltas,
\be
\int dp \,p_\mu \, \Big(a(\bm p,1)^\dag a(\bm p,1)+ b(\bm p,1)b(\bm p,1)^\dag\Big)
=0\nonumber
\ee
so this would not be a correct definition of free-field 4-momentum. For higher $N$
\be
{}&{}&
\int dp \,p_\mu \, \Big(a(\bm p,N)^\dag a(\bm p,N)+ b(\bm p,N)b(\bm p,N)^\dag\Big)
\nonumber\\
&{}&\pp=
=
\int dp \,p_\mu \, \Big(a(\bm p,N)^\dag a(\bm p,N)+ b(\bm p,N)^\dag b(\bm p,N)\Big)
\neq 0\nonumber
\ee
but for reasons mentioned earlier I do not think this is the correct free-field Hamiltonian. Although, who knows...$\blacktriangle$

\subsection{Reducible $N\geq 1$ representation of HOLA --- an indefinite-frequency harmonic oscillator representation of $\phi(x)$}

The constant in the Noether current is $C=1/2$. Take $\hat c(\bm k)=a$. For $N=1$
\be
c(\bm k,1) &=& |\bm k\rangle\langle \bm k|\otimes \hat a,\label{hola 3 s}\\
c(\bm k,1)^\dag &=& |\bm k\rangle\langle \bm k|\otimes \hat a^\dag,\\
I_0(\bm k,1) &=& |\bm k\rangle\langle \bm k|\otimes 1,\\
n_0(\bm k,1) &=& |\bm k\rangle\langle \bm k|\otimes \hat a^\dag\hat a.\label{hola 4 s}
\ee
For $N>1$,
\be
c(\bm k,N) &=&
\frac{1}{\sqrt{N}}
\Big(
c(\bm k,1)\otimes I_0(1)\otimes \dots \otimes I_0(1)+\dots +I_0(1)\otimes \dots \otimes I_0(1)\otimes  c(\bm k,1)
\Big)
,\nonumber\\
\\
c(\bm k,N)^\dag &=& \frac{1}{\sqrt{N}}
\Big(
c(\bm p,1)^\dag \otimes I_0(1)\otimes \dots \otimes I_0(1)+\dots +I_0(1)\otimes \dots \otimes I_0(1)\otimes  c(\bm k,1)^\dag
\Big)
,\nonumber\\
\\
n_0(\bm k,N) &=& n_0(\bm k,1)\otimes I_0(1)\otimes \dots \otimes I_0(1)+\dots +I_0(1)\otimes \dots \otimes I_0(1)\otimes n_0(\bm k,1)
,\nonumber\\
\\
I_0(\bm k,N) &=& \frac{1}{N}
\Big(
I_0(\bm k,1)\otimes I_0(1)\otimes \dots \otimes I_0(1)+\dots +I_0(1)\otimes \dots \otimes I_0(1)\otimes  I_0(\bm k,1)
\Big)
,\nonumber\\
\\
I_0(N) &=& \int dk\, I_0(\bm k,N),\quad \textrm{(resolution of identity)}
\ee
The indefinite-frequency-type 4-momentum
\be
P_\mu(1)
&=&
\int dk \,k_\mu |\bm k\rangle\langle\bm k|\otimes \hat a^\dag \hat a
+\frac{1}{2}\int dk \,k_\mu\, |\bm k\rangle\langle\bm k|\otimes 1\nonumber\\
&=&
\int dk \,k_\mu\, n_0(\bm k,1)
+\frac{1}{2}\int dk \,k_\mu\, I_0(\bm k,1)
\ee
acts as a generator of 4-translations
\be
e^{iP(1) x}c(\bm k,1)e^{-iP(1) x} &=& c(\bm k,1) e^{-ixk},\\
e^{iP(1) x}c(\bm k,1)^\dag e^{-iP(1) x} &=& c(\bm k,1)^\dag  e^{ixk},\\
e^{iP(1) x}\phi(0,1)e^{-iP(1) x} &=& \phi(x,1),\\
e^{-iP(1) y}\phi(x,1)e^{iP(1) y} &=& \phi(x-y,1).
\ee

\subsection{Physical interpretation of field operators}

Returning to canonical position operator (\ref{Q})
\be
Q(t,N)
&=&
\sum_\omega\sqrt{\frac{2\hbar}{m\omega}}
\Big(
a_\omega(N) e^{-i\omega t}
+
a_\omega(N)^\dag e^{i\omega t}
\Big),
\ee
and comparing it with $\phi(x,N)$ evaluated at the origin $\bm x=\bm 0$,
\be
\phi(t,\bm 0,N) &=& \int_{\mathbb{R}^3}\frac{d^3k}{(2\pi)^3 2 \omega(\bm k)} \Big(c(\bm k,N)e^{-i\omega(\bm k)t}
+
c(\bm k,N)^\dag e^{i\omega(\bm k)t}\Big),
\ee
one understands that for $N=1$ the field at the origin is a canonical position of some indefinite-frequency operator. The field taken at an arbitrary point $\bm x$ is obtained from the one at the origin by translation. An elementary quantum field $\phi(x,1)$, corresponding to $N=1$, consists of a single oscillator that exists ``everywhere" in space, i.e. is in superposition of different localizations. Representations characterized by $N>1$ have physical interpretation of a gas consisting of $N$ indefinite-frequency noninteracting bosonic oscillators.

\subsection{$(N,N')$-oscillator representation of HOLA:
$N$ massive charged and $N'$ massless neutral oscillators}

Let ${\cal H}_m(N)$ and ${\cal H}_0(N)$ be the Hilbert spaces of the above two types of $N\geq 1$ scalar field representations. The full Hilbert space appropriate for the model of interacting fields is
\be
{\cal H}(N,N')
&=&
{\cal H}_m(N)\otimes {\cal H}_0(N')
\ee
with two in principle independent parameters $N$ and $N'$. The representations are constructed by the embeddings
\be
a(\bm p,N,N')
&=&
a(\bm p,N)\otimes I_0(N'),\\
c(\bm k,N,N')
&=&
I_m(N)\otimes c(\bm k,N'),
\ee
and so on.

\section{States  and their boundary conditions for $(N,N')$-representation}

We have not yet needed explicit forms of analogues of vacuum, $n$-particle or coherent states of quantum fields. Their construction is analogous to what we have done with states of $N$ indefinite-frequency oscillators.

Vacuum appropriate for representations (\ref{hola 1 s})--(\ref{hola 2 s}), (\ref{hola 3 s})--(\ref{hola 4 s}), is constructed as follows:
\be
|O_m,1\rangle
&=&
\int dp\, O_m(\bm p)|\bm p, 0,0\rangle,\\
|O_0,1\rangle
&=&
\int dk\, O_0(\bm k)|\bm k, 0\rangle,\\
|O,N,N'\rangle
&=&
|O_m,1\rangle^{\otimes N}\otimes
|O_0,1\rangle ^{\otimes N'}.
\ee
Vacuum states are normalizable
\be
\int dp\, |O_m(\bm p)|^2=\int dk\, |O_0(\bm k)|^2=1
\ee
Square integrability implies, roughly speaking, vanishing of $O_m(\bm p)$ and $O_0(\bm k)$ in the limits $|\bm p|\to\infty$, $|\bm k|\to\infty$. This is one of those features of field quantization in $(N,N')$-representations that will in the end make mathematics of field quantization ``sensible" (in Dirac's sense). For massless fields, such as $O_0(\bm k)$, a boundary condition has to be imposed also at $k=(k_0,\bm k)=(|\bm k|,\bm k)=0$. For massive fields, with $p^2=m^2>0$, the point $p=0$ does not belong to the mass-$m$ hyperboloid: $(\pm\sqrt{m^2+\bm p^2},\bm p)\to (\pm m,\bm 0)\neq 0$ with $|\bm p|\to 0$.

For $m=0$ the point $\bm k=\bm 0$ is Lorentz invariant. It corresponds to $k=0$ and $k'=Lk=L0=0$ for any $L$ so that $\bm k'=\bm 0$ as well. The boundary condition $O_0(\bm 0)=0$ is also Lorentz invariant (in fact, it is invariant under the whole Poincar\'e group). Well-definiteness of field theory that involves a massless field will require a yet stronger boundary condition, namely
\be
\lim_{\bm k\to\bm 0}\frac{O_0(\bm k)}{|\bm k|^n} &=& 0,\label{boundary k=0}\\
\int dk\,\frac{|O_0(\bm k)|^2}{|\bm k|^n} &<& \infty,\label{boundary k=0'}
\ee
for all natural $n$. The fact that the tip $k=0$ of the light cone is Lorentz invariant has deep consequences for the structure of unitary representations of the Poincar\'e group and geometry of the light cone. Hilbert spaces of massless $k=0$ and $k\neq 0$ representations are completely different (the representations are induced from different little groups \cite{Wigner}) and there is no Lorentz transformation that could map $k=0$ into $k'\neq 0$, even if $k'^2=0$. The light cone splits in a Lorentz invariant way into the tip $k_0=0$ and the physical cone ($k_0\neq 0$). The requirement that the probability density  $|O_0(\bm k)|^2$ tends to 0 as $\bm k\to\bm 0$ is thus mathematically very natural. From a physical point of view it eliminates massless fields of infinite wavelength and zero frequency.

\medskip

\noindent
{\it Remark\/}:
Consider a future-pointing time-like 4-vector $y^\mu$. If $k^\mu$ is nonzero null and future-pointing then $y\cdot k>0$. A function that satisfies all the imposed requirements is
\be
O_0(\bm k) &=& C e^{-\frac{\lambda^2}{2y\cdot k}-\frac{y\cdot k}{2}},\label{O_0-1/k}
\ee
where $\lambda$ is some real parameter.
In order to show explicitly that the normalization factor is Lorentz invariant compute
\be
1
&=&
\int dk|O_0(\bm k)|^2
\nonumber\\
&=&
|C|^2\int \frac{d^3k}{(2\pi)^3 2|\bm k|}e^{-\frac{\lambda^2}{y\cdot k}- y\cdot k}
\nonumber\\
&=&
|C|^2\int \frac{d^3k}{(2\pi)^3 2|\bm k|}e^{-\frac{\lambda^2}{y'_0|\bm k|}- y'_0|\bm k|}
\nonumber\\
&=&
|C|^2\frac{4\pi}{2(2\pi)^3}\int_0^\infty \frac{d\kappa\,\kappa^2}{\kappa}e^{-\frac{\lambda^2}{y'_0\kappa}-y'_0\kappa}
\nonumber\\
&=&
|C|^2\frac{1}{(2\pi)^2}\int_0^\infty d\kappa\,\kappa\, e^{-\frac{\lambda^2}{y'_0\kappa}- y'_0\kappa}
\nonumber\\
&=&
|C|^2\frac{\lambda^2K_2(2\lambda)}{2\pi^2(y'_0)^2}.
\ee
$K_2$ is the modified Bessel function of the second kind.
We have employed the fact that the measure $dk$ is Lorentz invariant and changed reference frame $y'=\Lambda y$ in such a way that $y'=(y'_0,\bm 0)$. This, on the other hand, implies that $(y'_0)^2=y'\cdot y'=y\cdot y=y^2$ is Lorentz invariant. The normalization is thus indeed Lorentz invariant
\be
|C|^2=\frac{2\pi^2y^2}{\lambda^2K_2(2\lambda)}
\ee
Since we work with dimensionless $k$, the same must be true of $y$ and $\lambda$. $\blacktriangle$
\medskip

The rest of the construction is completely analogous to what I did for $N$-oscillator states. (However, we have to remember that
$\langle \bm p|\bm p\rangle=\langle \bm k|\bm k\rangle=0$.) In particular, the displacement operator is defined as
\be
{\cal D}(\alpha,\beta,\gamma,N,N')
&=&
\exp\int dp\Big(\alpha(\bm p)a(\bm p,N,N')^\dag+\beta(\bm p)b(\bm p,N,N')^\dag-\textrm{H.c.}\Big)
\nonumber\\
&\pp=&\times \exp\int dk\Big(\gamma(\bm k)c(\bm k,N,N')^\dag-\textrm{H.c.}\Big)\\
&=&
\exp\int dp\Big(\alpha(\bm p)a(\bm p,N)^\dag+\beta(\bm p)b(\bm p,N)^\dag-\textrm{H.c.}\Big)
\nonumber\\
&\pp=&\otimes\, \exp\int dk\Big(\gamma(\bm k)c(\bm k,N')^\dag-\textrm{H.c.}\Big)
\ee
Corresponding statistics of excitations is R\'enyi-deformed Poissonian.

\section{Relativistic covariance}

Let $x$ be a point in dimensionless Minkowski space.
The Poincar\'e transformation
\be
x'=Lx+y,
\ee
where $L$ is a Lorentz transformation and $y$ a 4-vector, is equivalent to
\be
x'^\alpha &=& L{^\alpha}{_\beta}x^\beta+y^\alpha=x^\beta L^{-1}{_\beta}{^\alpha}+y^\alpha,\\
x'_\alpha &=& L{_\alpha}{^\beta}x_\beta+y_\alpha=x_\beta L^{-1}{^\beta}{_\alpha}+y_\alpha.
\ee
Poincar\'e transformations are represented unitarily by operators $U(L,y,N,N')$ constructed below.
To begin with, the unitary representation of 4-translations is generated by
\be
P_\mu(N,N')
&=&
\int dp\, p_\mu \Big( n_+(\bm p,N)+n_-(\bm p,N)\Big)\otimes I_0(N')
+
I_m(N)\otimes \int dk\, k_\mu\, n_0(\bm k,N')
\nonumber\\
&\pp=&
+
N\int dp \,p_\mu\, I_m(\bm p,N)\otimes I_0(N')
+
I_m(N)\otimes \frac{N'}{2}\int dk \,k_\mu\, I_0(\bm k,N')
\label{P(N,N')}
\ee

\subsection{Action of the Poincar\'e group on field operators}

A pure 4-translation $x'=x+y$ is represented by
\be
U(1,y,N,N')
&=&
e^{i y^\mu P_\mu(N,N')}=e^{i y P(N,N')}.
\ee
It multiplies annihilation operators by phase factors
\be
U(1,y,N,N')^\dag
a(\bm p,N,N')
U(1,y,N,N')
&=&
a(\bm p,N,N')e^{ipy},\\
U(1,y,N,N')^\dag
b(\bm p,N,N')
U(1,y,N,N')
&=&
b(\bm p,N,N')e^{ipy},\\
U(1,y,N,N')^\dag
c(\bm k,N,N')
U(1,y,N,N')
&=&
c(\bm k,N,N')e^{iky}.
\ee
Recall that $py=\sqrt{\bm p^2+m^2}y^0-\bm p\cdot\bm y$, $ky=|\bm k|y^0-\bm k\cdot\bm y$. Now let $L$ be a Lorentz transformation and $\bm{Lp}$ be the spacelike part of $Lp$. Unitary spin-0 representation of the Lorentz group is given by
\be
U_m(L,0,1)
&=&
\int dp |\bm p\rangle\langle \bm{L^{-1}p}|\otimes (1\otimes 1),\\
U_0(L,0,1)
&=&
\int dk |\bm k\rangle\langle \bm{L^{-1}k}|\otimes 1,\\
U(L,0,N,N')
&=&
U_m(L,0,1)^{\otimes N}
\otimes
U_0(L,0,1)^{\otimes N'}.
\ee
$U(L,0,N,N')$ changes momenta of the amplitude operators (the Doppler effect),
\be
U(L,0,N,N')^\dag
a(\bm p,N,N')
U(L,0,N,N')
&=&
a(\bm{L^{-1}p},N,N'),\\
U(L,0,N,N')^\dag
b(\bm p,N,N')
U(L,0,N,N')
&=&
b(\bm{L^{-1}p},N,N'),\\
U(L,0,N,N')^\dag
c(\bm k,N,N')
U(L,0,N,N')
&=&
c(\bm{L^{-1}k},N,N').
\ee
Four-translations and Lorentz transformations are combined into full Poincar\'e transformations $(L,y)=(1,y)(L,0)$:
\be
(L,y)x=(1,y)(L,0)x=(1,y)(Lx+0)=1(Lx+0)+y=Lx+y.
\ee
Therefore
\be
U(L,y,N,N')=U(1,y,N,N')U(L,0,N,N')
\ee
is a unitary representation of a general Poincar\'e transformation. As an example of explicit Poincar\'e transformation consider the massive charged field
\be
\psi(x,N,N')
&=&
\int dp \Big(a(\bm p,N,N')e^{-ipx}+b(\bm p,N,N')^\dag e^{ipx}\Big).
\ee
Then
\be
{}&{}&
U(L,y,N,N')^\dag \psi(x,N,N')U(L,y,N,N')
\nonumber\\
&{}&\pp=
= \int dp \Big(U(L,0,N,N')^\dag U(1,y,N,N')^\dag a(\bm p,N,N')U(1,y,N,N')U(L,0,N,N')e^{-ipx}
\nonumber\\
&{}&\pp{= \int dp \Big(+}+U(L,0,N,N')^\dag U(1,y,N,N')^\dag b(\bm p,N,N')^\dag U(1,y,N,N')U(L,0,N,N') e^{ipx}\Big)\nonumber\\
&{}&\pp=
= \int dp \Big(U(L,0,N,N')^\dag a(\bm p,N,N')U(L,0,N,N')e^{-ip(x-y)}
\nonumber\\
&{}&\pp{= \int dp \Big(+}+U(L,0,N,N')^\dag b(\bm p,N,N')^\dag U(L,0,N,N') e^{ip(x-y)}\Big)\nonumber\\
&{}&\pp=
= \int dp \Big(a(\bm{L^{-1}p},N,N')e^{-ip(x-y)}+b(\bm{L^{-1}p},N,N')^\dag  e^{ip(x-y)}\Big)\nonumber\\
&{}&\pp=
= \int dp \Big(a(\bm{L^{-1}p},N,N')e^{-iLL^{-1}p(x-y)}+b(\bm{L^{-1}p},N,N')^\dag  e^{iLL^{-1}p(x-y)}\Big)\nonumber\\
&{}&\pp=
= \int dp \Big(a(\bm{p},N,N')e^{-iLp(x-y)}+b(\bm{p},N,N')^\dag  e^{iLp(x-y)}\Big)\nonumber\\
&{}&\pp=
= \int dp \Big(a(\bm{p},N,N')e^{-ipL^{-1}(x-y)}+b(\bm{p},N,N')^\dag  e^{ipL^{-1}(x-y)}\Big)\nonumber\\
&{}&\pp=
=
\psi\big(L^{-1}(x-y),N,N'\big).
\ee
I have used here the Lorentz invariance : $d(L^{-1}p)=dp$ and $Lp\cdot Lx=p\cdot x=p_\mu x^\mu$, implying
$$
Lp\cdot  (x-y)=Lp\cdot LL^{-1} (x-y)=p\cdot L^{-1} (x-y).
$$
$L^{-1}(x-y)$ is the inverse of $Lx+y$. Indeed,
\be
L^{-1} \big((Lx+y)-y\big)
&=&
x,\\
L\big(L^{-1} (x-y)\big)+y
&=&
x.
\ee
The same transformation rule holds for the massless neutral field
\be
U(L,y,N,N')^\dag \phi(x,N,N')U(L,y,N,N')
=
\phi\big(L^{-1}(x-y),N,N'\big).
\ee
Number operators transform as translation invariant scalar fields in momentum space
\be
U(L,y,N,N')^\dag
n_\pm(\bm p,N,N')
U(L,y,N,N')
&=&
n_\pm(\bm{L^{-1}p},N,N'),\\
U(L,y,N,N')^\dag
n_0(\bm k,N,N')
U(L,y,N,N')
&=&
n_0(\bm{L^{-1}k},N,N').
\ee
In consequence, the 4-momentum is a 4-vector:
\be
\int dp\,p_\mu n_\pm(\bm{L^{-1}p},N,N')
&=&
\int dp\,(LL^{-1}p)_\mu n_\pm(\bm{L^{-1}p},N,N')
\nonumber\\
&=&
\int dp\,(Lp)_\mu n_\pm(\bm{p},N,N')
\nonumber\\
&=&
L{_\mu}{^\nu}\int dp\,p_\nu n_\pm(\bm{p},N,N')
\nonumber\\
\int dk\,k_\mu n_0(\bm{L^{-1}k},N,N')
&=&
\int dk\,(LL^{-1}k)_\mu n_0(\bm{L^{-1}k},N,N')
\nonumber\\
&=&
L{_\mu}{^\nu}\int dk\,k_\nu n_0(\bm{k},N,N'),
\nonumber
\ee
so that
\be
U(L,y,N,N')^\dag
P_\mu(N,N')
U(L,y,N,N')
&=&
L{_\mu}{^\nu}P_\nu(N,N').
\ee
\subsection{Action of the Poincar\'e group on vacuum}

Vacuum states are neither exactly translation invariant (in consequence of the ``zero energy" 4-momentum, which we do not neglect)
\be
{}&{}&
U(1,y,N,N')|O,N,N'\rangle
\nonumber\\
&{}&\pp=
=
e^{i y^\mu P_\mu(N,N')}|O,N,N'\rangle
\nonumber\\
&{}&\pp=
=
\Big(
\int dp
\,e^{i yp}O_m(\bm{p})|\bm p\rangle\otimes|0,0\rangle
\Big)^{\otimes N}
\otimes
\Big(
\int dk
\,e^{i y k/2}O_0(\bm{k})|\bm k\rangle\otimes|0\rangle
\Big)^{\otimes N'}\nonumber
\ee
nor Lorentz invariant
\be
U(L,0,N,N')|O,N,N'\rangle
&=&
\Big(
U_m(L,0,1)^{\otimes N}
\otimes
U_0(L,0,1)^{\otimes N'}
\Big)
\Big(
|O_m,1\rangle^{\otimes N}\otimes
|O_0,1\rangle ^{\otimes N'}
\Big)
\nonumber\\
&=&
\Big(U_m(L,0,1)|O_m,1\rangle\Big)^{\otimes N}\otimes
\Big(U_0(L,0,1)|O_0,1\rangle\Big)^{\otimes N'}
\nonumber\\
U_m(L,0,1)|O_m,1\rangle
&=&
\int dp |\bm p\rangle\langle \bm{L^{-1}p}|\otimes (1\otimes 1)
\int dp'\,O_m(\bm p')|\bm p'\rangle\otimes|0,0\rangle\nonumber\\
&=&
\int dp
\int dp'\,O_m(\bm p')|\bm p\rangle\langle\bm{L^{-1}p}|\bm p'\rangle\otimes|0,0\rangle\nonumber\\
&=&
\int dp
\,O_m(\bm{L^{-1}p})|\bm p\rangle\otimes|0,0\rangle,\nonumber\\
U_0(L,0,1)|O_0,1\rangle
&=&
\int dk |\bm k\rangle\langle \bm{L^{-1}k}|\otimes 1
\int dk'\,O_0(\bm k')|\bm k'\rangle\otimes|0\rangle\nonumber\\
&=&
\int dk
\int dk'\,O_0(\bm k')|\bm p\rangle\langle\bm{L^{-1}k}|\bm k'\rangle\otimes|0\rangle\nonumber\\
&=&
\int dk
\,O_0(\bm{L^{-1}k})|\bm k\rangle\otimes|0\rangle\nonumber.
\ee
We can say that vacuum wave functions behave under Poincar\'e transformations as classical scalar fields on mass-$m$ hyperboloid and light cone, respectively,
\be
O_m(\bm{p}) &\to& e^{i yp}O_m(\bm{p}),\\
O_m(\bm{p}) &\to& O_m(\bm{L^{-1}p}),\\
O_0(\bm{k}) &\to& O_0(\bm{L^{-1}k}),\\
O_0(\bm{k}) &\to& e^{i yk/2}O_0(\bm{k}).
\ee
The ``zero-energy" parts of 4-momentum can be removed by a well defined unitary transformation. Such a new vacuum will become 4-translation invariant, but one cannot do the same with Lorentz transformations.
Lorentz {\it invariance\/} would require constant $O_m(\bm{p})$ and $O_0(\bm{k})$, a condition inconsistent with square-integrability
\be
\int dp\,|O_m(\bm{p})|^2=\int dk\,|O_0(\bm{k})|^2=1
\ee
which we assume.

The entire subspace of vacuum states is nevertheless Poincar\'e invariant.
It follows that the projector on the vacuum subspace
\be
\Pi_0(N,N')
&=&
\Big(
\int dp \,|\bm p\rangle\langle \bm p|\otimes |0,0\rangle\langle 0,0|
\Big)^{\otimes N}
\otimes
\Big(
\int dk \,|\bm k\rangle\langle \bm k|\otimes |0\rangle\langle 0|
\Big)^{\otimes N'}
\ee
is Poincar\'e invariant
\be
U(L,y,N,N')^\dag\Pi_0(N,N')U(L,y,N,N') &=& \Pi_0(N,N'),
\ee
and commutes with $U(L,y,N,N')$,
\be
\Pi_0(N,N')U(L,y,N,N') &=& U(L,y,N,N')\Pi_0(N,N').
\ee
\subsection{Vacuum in 4-position space and nontrivial boundary conditions}

The fact that vacuum wave functions behave under Lorentz transformations as classical scalar fields on mass-$m$ hyperboloid and light cone, respectively, suggests their interpretation as Fourier-space amplitudes of classical scalar fields
\be
O_m(x)
&=&
\int dp\, \Big(O_m(\bm{p})e^{-ipx}+\overline{O_m(\bm{p})}e^{ipx}\Big),\\
O_0(x)
&=&
\int dk\, \Big(O_0(\bm{k})e^{-ikx/2}+\overline{O_0(\bm{k})}e^{ikx/2}\Big).
\ee
The fields $O_0(x)$ and $O_m(x)$ provide a means of imposing nontrivial boundary conditions associated with concrete geometry.

This is an interesting point that will be discussed later in the context of the Casimir effect and fields in cavities.
At this place it is just good to know that boundary conditions should be in my approach imposed on both vacuum states and field operators.

\medskip

\noindent
{\it Remark\/}:
The essential mathematical difference between $|O,N,N'\rangle$ and the vacuum $|0)$ of the bosonic Fock space is related to Poincar\'e invariance. Uniqueness and Poincar\'e invariance of the vacuum state is one of those Wightman axioms of standard Fock-space based quantum field theory \cite{CPT,Haag} that are not satisfied by my formalism. (Note, however, that vacuum understood as the subspace of vacuum states is both unique and Poincar\'e invariant even in my formalism; in Wightmanian formalism this space is 1-dimensional so both types of invariance and uniqueness are then equivalent.) In the context of indefinite-frequency oscillators the state $|O,N,N'\rangle$ represents a gas of $N$ massive and $N'$ massless bosonic oscillator wave packets at zero temperature. As such it should be neither unique nor Poincar\'e invariant, similarly to states of actual Bose-Einstein condensates occurring in atomic physics (clouds of trapped cold atoms are not even translation invariant). In relativistic theories all states should be at least Poincar\'e {\it covariant\/}.$\blacktriangle$
\medskip

\noindent
{\it Remark\/}: In solid-state physics one distinguishes between ``particles", such as atoms forming a given medium, and ``quasi-particles" (phonons, plasmons, etc.)  --- quantized oscillations of the medium. The physical difference between the Bose-Einstein condensate $|O,N,N'\rangle$ and the Fock vacuum $|0)$ can be better understood if one thinks of Lewenstein-You-Cooper-Burnett Fock-space formulation of atomic Bose-Einstein condensates \cite{You}. The vacuum there corresponds to the Fockian $|0)$, i.e. the state of no atoms. This is a purely formal object, a ``ground state" meaning ``no atoms in the trap". A ``Bose-Einstein condensate at zero temperature" is a state analogous to $|O,N,N'\rangle$. It is not necessarily the lowest energy state of the total Hamiltonian since electrons in atoms may exist in higher excited states, but rather a ground state of a part of the Hamiltonian (typically representing the center-of-mass kinetic energy). In my formalism a kind of $|0)$ could correspond to $N=0$, $N'=0$ (in the first version of the formalism published in \cite{I} I indeed considered this type of dual particle/quasi-particle Fock-type structure).
$n$-quasi-particle states, that is those containing $n$ excitations of the oscillators, are regarded in these notes as representing $n$ particles. The notion of particle statistics refers only to the quasi-particle level. Numbers of quasi-particles may change in time as opposed to $N$ and $N'$ that remain fixed. The ``medium" composed of our $N$ and $N'$ particles is static and does not explicitly take part in dynamical processes.$\blacktriangle$
\medskip

\noindent
{\it Remark\/}: The so called loop diagrams I will describe in detail later correspond to ``vacuum to vacuum" processes, such as annihilation of particles that were spontaneously created from vacuum. In my formalism the appropriate formulas will be obtained by sandwiching evolution operators between {\it projectors\/} on vacuum and not between vacuum {\it state vectors\/}, as would be the case in more standard approaches.$\blacktriangle$

\section{Interaction-picture dynamics}

Let $\phi$ be any operator and $H=H_0+H_1$ a dimensionless Hamiltonian. Its dynamics in Heisenberg picture, with respect to dimensionless time $t$,  is given by
\be
\hat\phi(t)
&=&
e^{iH t}\phi e^{-iHt}=
U_1(t)^\dag e^{iH_0 t}\phi e^{-iH_0t}U_1(t),\quad \hat\phi(0) = \phi,\\
U_1(t)
&=&
e^{iH_0t}e^{-iHt},\label{U_1(t)}\\
i\dot U_1(t)
&=&
e^{iH_0t}(-H_0+H)e^{-iHt}=e^{iH_0t}H_1e^{-iH_0t}U_1(t).
\ee
Denoting
\be
\phi(t)
&=&
e^{iH_0 t}\phi e^{-iH_0t},\\
H_1(t)
&=&
e^{iH_0 t}H_1 e^{-iH_0t},
\ee
we can write
\be
\hat\phi(t)
&=&
U_1(t)^\dag \phi(t)U_1(t),\\
i\dot U_1(t)
&=&
H_1(t)U_1(t).
\ee
Let us note that $\hat\phi(t)$ and $\phi(t)$ are equal at $t=0$ --- this is an example of initial condition for Heisenberg dynamics.
It is often convenient to take the initial condition $\hat\phi(t_0)=\phi(t_0)$ at an arbitrary $t_0$,
\be
\hat\phi(t)
&=&
e^{iH (t-t_0)}\phi (t_0)e^{-iH (t-t_0)}\\
&=&
e^{iH (t-t_0)}e^{iH_0 t_0}\phi e^{-iH_0 t_0}e^{-iH (t-t_0)}\\
&=&
e^{iH (t-t_0)}e^{-iH_0 (t-t_0)}\phi(t) e^{iH_0 (t-t_0)}e^{-iH (t-t_0)}
\\
&=&
V_1(t,t_0)^\dag\phi(t)V_1(t,t_0).
\ee
The evolution operator
\be
V_1(t,t_0)
&=&
e^{iH_0 (t-t_0)}e^{-iH (t-t_0)},\label{V_1(t)}
\ee
satisfies Schr\"odinger's equation
\be
i\frac{d}{dt} V_1(t,t_0)
&=&
-e^{iH_0 (t-t_0)}H_0e^{-iH (t-t_0)}
+
e^{iH_0 (t-t_0)}He^{-iH (t-t_0)}
\nonumber\\
&=&
e^{iH_0 (t-t_0)}H_1e^{-iH (t-t_0)}
\nonumber\\
&=&
e^{iH_0 (t-t_0)}H_1e^{-iH_0 (t-t_0)}V_1(t,t_0)\nonumber\\
&=&
H_1(t-t_0)V_1(t,t_0),
\ee
and
\be
V_1(t_0,t_0)
&=&
\mathbb{I},\\
V_1(t,t_0)
&=&
V_1(t-t_0,0),
\ee
but in general does not fulfill the composition property, i.e.
\be
V_1(t_1,t_2)V_1(t_2,t_3)
&\neq&
V_1(t_1,t_3).
\ee
Another option is to take
\be
U_1(t,t_0)
&=&
e^{iH_0 t}e^{-iH (t-t_0)}e^{-iH_0t_0}\label{U_1(t,t0)}
\ee
which, as opposed to $V_1(t,t_0)$, satisfies the composition property
\be
U_1(t_1,t_2)U_1(t_2,t_3)
&=&
U_1(t_1,t_3).
\ee
The dynamics is now different:
\be
i\frac{d}{dt} U_1(t,t_0)
&=&
-
e^{iH_0 t}H_0e^{-iH (t-t_0)}e^{-iH_0t_0}
+
e^{iH_0 t}He^{-iH (t-t_0)}e^{-iH_0t_0}
\nonumber\\
&=&
e^{iH_0 t}H_1e^{-iH (t-t_0)}e^{-iH_0t_0}
\nonumber\\
&=&
e^{iH_0 t}H_1e^{-iH_0 t}U_1(t,t_0)
\nonumber\\
&=&
H_1(t)U_1(t,t_0).\label{Dirac picture}
\ee
The initial condition is unchanged,
\be
U_1(t_0,t_0)
&=&\mathbb{I},\label{Dirac picture'}
\ee
but $U_1(t,t_0)\neq U_1(t-t_0,0)$. Assuming $\hat\phi(t_0)=\phi$ we find
\be
\hat\phi(t)
&=&
e^{iH (t-t_0)}\phi e^{-iH (t-t_0)}
\nonumber\\
&=&
e^{-iH_0t_0}
U_1(t,t_0)^\dag
e^{iH_0t}
\phi
e^{-iH_0t}
U_1(t,t_0)
e^{iH_0t_0}
\nonumber\\
&=&
e^{-iH_0t_0}
U_1(t,t_0)^\dag
\phi(t)
U_1(t,t_0)
e^{iH_0t_0}
\ee
Let us note that $\hat\phi(t_0)=\phi=\phi(0)$ whereas the case of $V_1(t,t_0)$ corresponded to $\hat\phi(t_0)=\phi(t_0)$.
For $t_0=0$ the three evolution operators, (\ref{U_1(t)}), (\ref{V_1(t)}), (\ref{U_1(t,t0)}), coincide:
\be
U_1(t)
=
U_1(t,0)=V_1(t,0).
\ee
Equation (\ref{Dirac picture}) is known as the interaction or Dirac picture. (\ref{Dirac picture}) and (\ref{Dirac picture'}) are equivalent to
\be
U_1(t,t_0)
&=&
\mathbb{I}+(-i)\int_{t_0}^{t}d\tau_1 H_1(\tau_1)U_1(\tau_1,t_0)
\nonumber\\
&=&
\mathbb{I}+(-i)\int_{t_0}^{t}d\tau_1 H_1(\tau_1)\Big(\mathbb{I}+(-i)\int_{t_0}^{\tau_1}d\tau_2 H_1(\tau_2)U_1(\tau_2,t_0)\Big).\nonumber
\ee
By iteration
\be
U_1(t,t_0)
&=&
\mathbb{I}+(-i)\int_{t_0}^{t}d\tau_1\, H_1(\tau_1)+(-i)^2\int_{t_0}^{t}d\tau_1 \int_{t_0}^{\tau_1}d\tau_2\,\!\!\!\!\!\!\!
\underbrace{H_1(\tau_1)H_1(\tau_2)}_{\textrm{loop terms originate here}}\!\!\!\!\!\!+\dots
\label{Texp}\\
&=&
\Texp\Big(
-i\int_{t_0}^{t}d\tau_1\, H_1(\tau_1)
\Big).
\ee
Texp$(\dots)$ is the so-called time-ordered exponential.

Another popular (and equivalent) form of Texp$(\dots)$ involves the so called time-ordered products,
\be
U_1(t,t_0)
&=&
\sum_{n=0}^\infty
(-i)^n\int_{t_0}^{t}d\tau_1 \int_{t_0}^{\tau_1}d\tau_2\dots \int_{t_0}^{\tau_{n-1}}d\tau_n\,
H_1(\tau_1)H_1(\tau_2)\dots H_1(\tau_n)\nonumber\\
&=&
\sum_{n=0}^\infty
(-i)^n\int_{t_0}^{t}d\tau_1 \dots \int_{t_0}^{t}d\tau_n\,
\theta(\tau_1-\tau_2)\dots \theta(\tau_{n-1}-\tau_n)
H_1(\tau_1)\dots H_1(\tau_n).
\ee
Now we have $n$ integrals from $t_0$ to $t$. Taking into account all the possible permutations $(\tau_{i_1},\dots,\tau_{i_n})$ of the $n$-tuple
$(\tau_1,\dots,\tau_n)$ and assuming that orders of integration can be interchanged (which is not always obvious, especially if $t_0$ or $t$ are taken at infinity) we obtain
\be
U_1(t,t_0)
&=&
\sum_{n=0}^\infty
\frac{(-i)^n}{n!}\int_{t_0}^{t}d\tau_1 \dots \int_{t_0}^{t}d\tau_n\,
T\Big(H_1(\tau_1)\dots H_1(\tau_n)\Big),
\ee
where the time-ordered product is defined by
\be
T\Big(H_1(\tau_1)\dots H_1(\tau_n)\Big)
&=&
\sum_{(\tau_{i_1},\dots,\tau_{i_n})}
\theta(\tau_{i_1}-\tau_{i_2})\dots \theta(\tau_{i_{n-1}}-\tau_{i_n})
H_1(\tau_{i_1})\dots H_1(\tau_{i_n}).
\ee

The above forms of dynamics can be further generalized as follows. Assume ${\cal O}\subset \mathbb{R}^4$ is a subset of Minkowski space contained between two spacelike hyper-surfaces $\Sigma_{\tau_1}$ and $\Sigma_{\tau_2}$. Let us also assume that there exists a 1-parameter family of hypersurfaces $\Sigma_\tau$, ${\cal O}=\bigcup_{\tau}\Sigma_\tau$, such that for each $x\in {\cal O}$ there exists one and only one $\Sigma_\tau$ containing $x$. The family $\Sigma_\tau$ is termed the foliation of $\cal O$.

Interaction-picture Hamiltonian associated with a given foliation is defined by
\be
V(\tau)
&=&
\int_{\Sigma_\tau}d\Sigma_\tau(x){\cal V}(x).
\ee
${\cal V}(x)$ is a Poincar\'e covariant interaction-Hamiltonian density,
\be
U(L,y)^{\dag}{\cal V}(x)U(L,y) &=& {\cal V}\big(L^{-1}(x-y)\big).
\ee
$U(L,y)$ is a representation of the Poincar\'e group --- at this moment we do not have to specify exactly which representation we have in mind. $d\Sigma_\tau$ is a measure on $\Sigma_\tau$ satisfying
\be
d^4x=d\tau\,d\Sigma_\tau.\label{d^4x}
\ee
Condition (\ref{d^4x}) implicitly enforces relativistic covariance of the formalism.
Let $x'=Lx$ where $L$ is a Lorentz transformation. Then
\be
d^4x &=& dx_0\, d^3x\\
&=& dx'_0\,d^3x'\\
&=&
d\tau \frac{d^3x}{\sqrt{1+\bm x^2/\tau^2}}=d\tau \frac{d^3x'}{\sqrt{1+\bm x'^2/\tau^2}}
\ee
define two hyperplane foliations $\Sigma_{\tau}=\{x;\,x_0=\tau\}$ and $\Sigma_{\tau}=\{x;\,x'_0=\tau\}$ of the Minkowski space, and the hyperbolic foliation $\Sigma_{\tau}=\{x;\, (x-y)^2=\tau^2,\,x_0\geq y_0\}$ of the future causal cone of some event $y$ (the so-called Milne universe \cite{Milne}). The respective measures are
\be
d\Sigma_\tau(x) &=& d^3x,\\
d\Sigma_\tau(x') &=& d^3x',\\
d\Sigma_\tau(x) &=& \frac{d^3x}{\sqrt{1+\bm x^2/\tau^2}}=\frac{d^3x'}{\sqrt{1+\bm x'^2/\tau^2}}=d\Sigma_\tau(x').
\ee
In Dirac's terminology \cite{Dirac-point} the first two foliations define {\it instant-form\/} dynamics. The hyperbolic foliation defines the {\it point-form\/} dynamics in the Milne universe. Point-form quantum optics of classical sources formulated in ``my" approach to quantum field theory was described in detail in \cite{V}. Milne universe is the exceptional case of dynamics where free-field initial condition at $\tau=0$ is physical: At $\tau=0$ the fields are at the boundary of the universe.

Interaction-picture evolution operator with respect to any foliation is a solution of
\be
i\frac{d}{d\tau}U_1(\tau,\tau_1) &=& V(\tau)U_1(\tau,\tau_1),\\
U_1(\tau_1,\tau_1) &=& \mathbb{I}.
\ee
An important property of interaction-picture dynamics is the fact that Hamiltonians $H_1(t)$ are constructed from operators that depend on time through the free-Hamiltonian dynamics generated by $H_0$.

\section{Concrete choice of $\cal V$}

A scalar-scalar interaction that seems formally closest to quantum electrodynamics is
\be
V(\tau)
&=&
\Big(\int_{\mathbb{R}^3} d^3x\,j_\mu(x,N,N')\phi(x,N,N')\Big)\otimes R^\mu, \quad x_0=\tau,\label{scalar V}\\
j_\mu(x,N,N') &=& iq\psi(x,N,N')^\dag\partial_\mu\psi(x,N,N')-iq\partial_\mu\psi(x,N,N')^\dag\psi(x,N,N').\label{VJ}
\ee
The fields $\psi(x,N,N')$, $\phi(x,N,N')$ are free. $R^\mu$ is any 4-translation invariant 4-vector operator, i.e. there exists a unitary representation ${W}(L,y)$ satisfying
\be
{W}(L,y)^\dag
R_\mu
{W}(L,y)
&=&
L{_\mu}{^\nu}R_\nu.
\ee
I will later give an explicit example of $R^\mu$ and ${W}(L,y)$, but for the moment it is irrelevant.

Another interesting case is given by
\be
V(\tau)
&=&
\int_{\mathbb{R}^3} \frac{d^3x}{\sqrt{1+\bm x^2/\tau^2}} \frac{x^\mu}{\tau}j_\mu(x,N,N')\phi(x,N,N')\nonumber\\
&=&
\int_{\mathbb{R}^3} \frac{d^3x}{\sqrt{\tau^2+\bm x^2}} x^\mu j_\mu(x,N,N')\phi(x,N,N')
\label{scalar V tau}
\ee
with the current given by (\ref{VJ}). The point $x=(\sqrt{\tau^2+\bm x^2},\bm x)$ belongs to $\Sigma_{\tau}=\{x;\, x^2=\tau^2,\,x_0\geq 0\}$, so
\be
x/\tau=(\sqrt{1+\bm x^2/\tau^2},\bm x/\tau)\to (1,\bm 0)\quad \textrm{with }\tau\to\infty.
\ee
As our first nontrivial exercise, leaving aside for the moment the issues of relativistic covariance but concentrating entirely on the problem of divergences in loop diagrams, we begin with a simplified version of (\ref{scalar V}),
\be
V(\tau)
&=&
\int_{\mathbb{R}^3} d^3x\,j_0(\tau,\bm x,N,N')\phi(\tau,\bm x,N,N').\label{scalar V2}
\ee
It corresponds to replacing the operator $R^\mu$ by the classical 4-velocity $u^\mu$, $u_\mu u^\mu=1$; the tensor product $\otimes$ reduces to ordinary multiplication by a real number. Alternatively, (\ref{scalar V2}) is the asymptotic form of (\ref{scalar V tau}) for $\tau\to\infty$.

One should not confuse the models with the so-called scalar electrodynamics, a gauge invariant theory describing interaction of charged spin-0 particles with electromagnetic fields. The corresponding interaction Hamiltonian is much more complicated but contains a term analogous to
(\ref{scalar V}) (see example 8.6 on p. 251 in \cite{Greiner-QFT}).

\section{Simplified hamiltonian (\ref{scalar V2}) in momentum space}

We begin with expressing (\ref{scalar V2}) in momentum space.
It is this place where we have to carefully define products and integrals of operators [see the discussion accompanying (\ref{xPnn}) and (\ref{ABC})].
The integrals involve $dp=d^3p/[(2\pi)^32\sqrt{m^2+\bm p^2}]$, $dr=d^3r/[(2\pi)^32\sqrt{m^2+\bm r^2}]$, $dk=d^3k/[(2\pi)^32|\bm k|]$, $p_0=\sqrt{m^2+\bm p^2}$, $r_0=\sqrt{m^2+\bm r^2}$, $k_0=|\bm k|$.
\be
V(\tau)
&=&
q\int d^3x
\int dp \Big(a(p)e^{-ipx}+b(p)^{\dag}e^{ipx}\Big)
i\partial_0\int dr \Big(b(r)e^{-irx}+a(r)^{\dag}e^{irx}\Big)
\phi(x)
\nonumber\\
&\pp=&-
q\int d^3x
i\partial_0\int dp \Big(a(p)e^{-ipx}+b(p)^{\dag}e^{ipx}\Big)
\int dr \Big(b(r)e^{-irx}+a(r)^{\dag}e^{irx}\Big)
\phi(x)
\nonumber\\
&=&
q\int d^3x
\int dp \Big(a(p)e^{-ipx}+b(p)^{\dag}e^{ipx}\Big)
\int dr\,r_0 \Big(b(r)e^{-irx}-a(r)^{\dag}e^{irx}\Big)
\phi(x)
\nonumber\\
&\pp=&-
q\int d^3x
\int dp \,p_0\Big(a(p)e^{-ipx}-b(p)^{\dag}e^{ipx}\Big)
\int dr \Big(b(r)e^{-irx}+a(r)^{\dag}e^{irx}\Big)
\phi(x)
\nonumber\\
&=&
q\int d^3x
\int dpdrdk \,r_0
\nonumber\\
&\pp=&\times
\Big(a(p)e^{-ipx}+b(p)^{\dag}e^{ipx}\Big)
\Big(b(r)e^{-irx}-a(r)^{\dag}e^{irx}\Big)
\Big(c(k)e^{-ikx}+c(k)^{\dag}e^{ikx}\Big)
\nonumber\\
&\pp=&-
q\int d^3x
\int dpdrdk \,p_0
\nonumber\\
&\pp=&\pp=
\times
\Big(a(p)e^{-ipx}-b(p)^{\dag}e^{ipx}\Big)
\Big(b(r)e^{-irx}+a(r)^{\dag}e^{irx}\Big)
\Big(c(k)e^{-ikx}+c(k)^{\dag}e^{ikx}\Big).
\nonumber
\ee
The above expressions should be understood in the following sense
\be
V(\tau)
&=&
q\int d^3x
\int dpdrdk \,r_0
a(p)b(r)c(k)e^{-i(p+r+k)x}+\dots\nonumber\\
&=&
\lim_{n_1,n_2,n_3\to\infty}q\int d^3x
\int dpdrdk \,r_0
\nonumber\\
&\pp=&\times
a(p,n_1,N,N')b(r,n_2,N,N')c(k,n_3,N,N')e^{-i(p+r+k)x}+\dots\nonumber
\ee
Here $a(p,n_1,N,N')$ is the $(N,N')$-oscillator extension of
\be
a(p,n_1,1) &=& |\bm p,n_1\rangle \langle \bm p,n_1|\otimes (\hat a\otimes 1)
\ee
or
\be
a(p,n_1,n_1',1) &=& |\bm p,n_1\rangle \langle \bm p,n_1'|\otimes (\hat a\otimes 1)
\ee
(compare the discussion of spectral theorem for M-shaped Dirac deltas).
Similar definitions apply to the other operators. For finite $n_1$, $n_2$,... we can apply (\ref{delta F})
so that
\be
V(\tau)
&=&
\frac{q}{2}\int dpd^3rdk \,
a(\bm p)b(\bm r)c(\bm k)e^{-i(\sqrt{m^2+\bm p^2}+\sqrt{m^2+\bm r^2}+|\bm k|)\tau }\delta^{(3)}(-\bm r-\bm p-\bm k)
\nonumber\\
&-&
\frac{q}{2}\int d^3pdrdk \,
a(\bm p)b(\bm r)c(\bm k)e^{-i(\sqrt{m^2+\bm p^2}+\sqrt{m^2+\bm r^2}+|\bm k|)\tau }\delta^{(3)}(-\bm p-\bm r-\bm k)
\nonumber\\
&+&
\frac{q}{2}\int dpd^3rdk \,
a(\bm p)b(\bm r)c(\bm k)^{\dag}e^{-i(\sqrt{m^2+\bm p^2}+\sqrt{m^2+\bm r^2}-|\bm k|)\tau }\delta^{(3)}(-\bm r-\bm p+\bm k)
\nonumber\\
&-&
\frac{q}{2}\int d^3pdrdk \,
a(\bm p)b(\bm r)c(\bm k)^{\dag}e^{-i(\sqrt{m^2+\bm p^2}+\sqrt{m^2+\bm r^2}-|\bm k|)\tau }\delta^{(3)}(-\bm p-\bm r+\bm k)
\nonumber\\
&-&
\frac{q}{2}\int dpd^3rdk \,
a(\bm p)a(\bm r)^{\dag}c(\bm k)e^{-i(\sqrt{m^2+\bm p^2}-\sqrt{m^2+\bm r^2}+|\bm k|)\tau }\delta^{(3)}(-\bm r+\bm p+\bm k)
\nonumber\\
&-&
\frac{q}{2}\int d^3pdrdk \,
a(\bm p)a(\bm r)^{\dag}c(\bm k)e^{-i(\sqrt{m^2+\bm p^2}-\sqrt{m^2+\bm r^2}+|\bm k|)\tau }\delta^{(3)}(-\bm p+\bm r-\bm k)
\nonumber\\
&-&
\frac{q}{2}\int dpd^3rdk \,
a(\bm p)a(\bm r)^{\dag}c(\bm k)^{\dag}e^{-i(\sqrt{m^2+\bm p^2}-\sqrt{m^2+\bm r^2}-|\bm k|)\tau }\delta^{(3)}(-\bm r+\bm p-\bm k)
\nonumber\\
&-&
\frac{q}{2}\int d^3pdrdk \,
a(\bm p)a(\bm r)^{\dag}c(\bm k)^{\dag}e^{-i(\sqrt{m^2+\bm p^2}-\sqrt{m^2+\bm r^2}-|\bm k|)\tau }\delta^{(3)}(-\bm p+\bm r+\bm k)
\nonumber\\
&+&
\frac{q}{2}\int dpd^3rdk \,
b(\bm p)^{\dag}b(\bm r)c(\bm k)e^{-i(-\sqrt{m^2+\bm p^2}+\sqrt{m^2+\bm r^2}+|\bm k|)\tau }\delta^{(3)}(-\bm r+\bm p-\bm k)
\nonumber\\
&+&
\frac{q}{2}\int d^3pdrdk \,
b(\bm p)^{\dag}b(\bm r)c(\bm k)e^{-i(-\sqrt{m^2+\bm p^2}+\sqrt{m^2+\bm r^2}+|\bm k|)\tau }\delta^{(3)}(-\bm p+\bm r+\bm k)
\nonumber\\
&+&
\frac{q}{2}\int dpd^3rdk \,
b(\bm p)^{\dag}b(\bm r)c(\bm k)^{\dag}e^{-i(-\sqrt{m^2+\bm p^2}+\sqrt{m^2+\bm r^2}-|\bm k|)\tau }\delta^{(3)}(-\bm r+\bm p+\bm k)
\nonumber\\
&+&
\frac{q}{2}\int d^3pdrdk \,
b(\bm p)^{\dag}b(\bm r)c(\bm k)^{\dag}e^{-i(-\sqrt{m^2+\bm p^2}+\sqrt{m^2+\bm r^2}-|\bm k|)\tau }\delta^{(3)}(-\bm p+\bm r-\bm k)
\nonumber\\
&-&
\frac{q}{2}\int dpd^3rdk \,
b(\bm p)^{\dag}a(\bm r)^{\dag}c(\bm k)e^{-i(-\sqrt{m^2+\bm p^2}-\sqrt{m^2+\bm r^2}+|\bm k|)\tau }\delta^{(3)}(-\bm r-\bm p+\bm k)
\nonumber\\
&+&
\frac{q}{2}\int d^3pdrdk \,
b(\bm p)^{\dag}a(\bm r)^{\dag}c(\bm k)e^{-i(-\sqrt{m^2+\bm p^2}-\sqrt{m^2+\bm r^2}+|\bm k|)\tau }\delta^{(3)}(-\bm p-\bm r+\bm k)
\nonumber\\
&-&
\frac{q}{2}\int dpd^3rdk \,
b(\bm p)^{\dag}a(\bm r)^{\dag}c(\bm k)^{\dag}e^{-i(-\sqrt{m^2+\bm p^2}-\sqrt{m^2+\bm r^2}-|\bm k|)\tau }\delta^{(3)}(-\bm r-\bm p-\bm k)
\nonumber\\
&+&
\frac{q}{2}\int d^3pdrdk \,
b(\bm p)^{\dag}a(\bm r)^{\dag}c(\bm k)^{\dag}e^{-i(-\sqrt{m^2+\bm p^2}-\sqrt{m^2+\bm r^2}-|\bm k|)\tau }\delta^{(3)}(-\bm p-\bm r-\bm k)
\nonumber
\ee
\be
V(\tau)
&=&
\frac{q}{2}\int dpdk \,
\Big(
a(\bm p)b(-\bm p-\bm k)-a(-\bm p-\bm k)b(\bm p)
\Big)
c(\bm k)
e^{-i(\sqrt{m^2+\bm p^2}+\sqrt{m^2+(\bm p+\bm k)^2}+|\bm k|)\tau }
\nonumber\\
&+&
\frac{q}{2}\int dpdk \,
\Big(
b(-\bm p-\bm k)^{\dag}a(\bm p)^{\dag}-b(\bm p)^{\dag}a(-\bm p-\bm k)^{\dag}
\Big)
c(\bm k)^{\dag}
e^{i(\sqrt{m^2+\bm p^2}+\sqrt{m^2+(\bm p+\bm k)^2}+|\bm k|)\tau }
\nonumber\\
&+&
\frac{q}{2}\int dpdk \,
\Big(
a(\bm p)b(-\bm p+\bm k)-a(-\bm p+\bm k)b(\bm p)
\Big)
c(\bm k)^{\dag}
e^{-i(\sqrt{m^2+\bm p^2}+\sqrt{m^2+(\bm p-\bm k)^2}-|\bm k|)\tau }
\nonumber\\
&+&
\frac{q}{2}\int dpdk \,
\Big(
b(-\bm p+\bm k)^{\dag}a(\bm p)^{\dag}-b(\bm p)^{\dag}a(-\bm p+\bm k)^{\dag}
\Big)
c(\bm k)
e^{i(\sqrt{m^2+\bm p^2}+\sqrt{m^2+(\bm p-\bm k)^2}-|\bm k|)\tau }
\nonumber\\
&+&
\frac{q}{2}\int dpdk \,
\Big(
b(\bm p+\bm k)^{\dag}b(\bm p)-\underline{a(\bm p)a(\bm p+\bm k)^{\dag}}
\Big)
c(\bm k)
e^{-i(\sqrt{m^2+\bm p^2}-\sqrt{m^2+(\bm p+\bm k)^2}+|\bm k|)\tau }
\nonumber\\
&+&
\frac{q}{2}\int dpdk \,
\Big(
b(\bm p)^{\dag}b(\bm p+\bm k)-\underline{a(\bm p+\bm k)a(\bm p)^{\dag}}
\Big)
c(\bm k)^{\dag}
e^{i(\sqrt{m^2+\bm p^2}-\sqrt{m^2+(\bm p+\bm k)^2}+|\bm k|)\tau }
\nonumber\\
&+&
\frac{q}{2}\int dpdk \,
\Big(
b(\bm p)^{\dag}b(\bm p-\bm k)-\underline{a(\bm p-\bm k)a(\bm p)^{\dag}}
\Big)
c(\bm k)
e^{-i(-\sqrt{m^2+\bm p^2}+\sqrt{m^2+(\bm p-\bm k)^2}+|\bm k|)\tau }
\nonumber\\
&+&
\frac{q}{2}\int dpdk \,
\Big(
b(\bm p-\bm k)^{\dag}b(\bm p)-\underline{a(\bm p)a(\bm p-\bm k)^{\dag}}
\Big)
c(\bm k)^{\dag}
e^{i(-\sqrt{m^2+\bm p^2}+\sqrt{m^2+(\bm p-\bm k)^2}+|\bm k|)\tau }.
\nonumber
\ee
I have underlined the four terms that are not normally ordered. Defining the normally ordered part of $V(\tau)$ by
\be
:V(\tau):
&=&
\frac{q}{2}\int dpdk \,
\Big(
a(\bm p)b(-\bm p-\bm k)-a(-\bm p-\bm k)b(\bm p)
\Big)
c(\bm k)
e^{-i(\sqrt{m^2+\bm p^2}+\sqrt{m^2+(\bm p+\bm k)^2}+|\bm k|)\tau }
\nonumber\\
&+&
\frac{q}{2}\int dpdk \,
\Big(
b(-\bm p-\bm k)^{\dag}a(\bm p)^{\dag}-b(\bm p)^{\dag}a(-\bm p-\bm k)^{\dag}
\Big)
c(\bm k)^{\dag}
e^{i(\sqrt{m^2+\bm p^2}+\sqrt{m^2+(\bm p+\bm k)^2}+|\bm k|)\tau }
\nonumber\\
&+&
\frac{q}{2}\int dpdk \,
\Big(
a(\bm p)b(-\bm p+\bm k)-a(-\bm p+\bm k)b(\bm p)
\Big)
c(\bm k)^{\dag}
e^{-i(\sqrt{m^2+\bm p^2}+\sqrt{m^2+(\bm p-\bm k)^2}-|\bm k|)\tau }
\nonumber\\
&+&
\frac{q}{2}\int dpdk \,
\Big(
b(-\bm p+\bm k)^{\dag}a(\bm p)^{\dag}-b(\bm p)^{\dag}a(-\bm p+\bm k)^{\dag}
\Big)
c(\bm k)
e^{i(\sqrt{m^2+\bm p^2}+\sqrt{m^2+(\bm p-\bm k)^2}-|\bm k|)\tau }
\nonumber\\
&+&
\frac{q}{2}\int dpdk \,
\Big(
b(\bm p+\bm k)^{\dag}b(\bm p)-\underline{a(\bm p+\bm k)^{\dag}a(\bm p)}
\Big)
c(\bm k)
e^{-i(\sqrt{m^2+\bm p^2}-\sqrt{m^2+(\bm p+\bm k)^2}+|\bm k|)\tau }
\nonumber\\
&+&
\frac{q}{2}\int dpdk \,
\Big(
b(\bm p)^{\dag}b(\bm p+\bm k)-\underline{a(\bm p)^{\dag}a(\bm p+\bm k)}
\Big)
c(\bm k)^{\dag}
e^{i(\sqrt{m^2+\bm p^2}-\sqrt{m^2+(\bm p+\bm k)^2}+|\bm k|)\tau }
\nonumber\\
&+&
\frac{q}{2}\int dpdk \,
\Big(
b(\bm p)^{\dag}b(\bm p-\bm k)-\underline{a(\bm p)^{\dag}a(\bm p-\bm k)}
\Big)
c(\bm k)
e^{-i(-\sqrt{m^2+\bm p^2}+\sqrt{m^2+(\bm p-\bm k)^2}+|\bm k|)\tau }
\nonumber\\
&+&
\frac{q}{2}\int dpdk \,
\Big(
b(\bm p-\bm k)^{\dag}b(\bm p)-\underline{a(\bm p-\bm k)^{\dag}a(\bm p)}
\Big)
c(\bm k)^{\dag}
e^{i(-\sqrt{m^2+\bm p^2}+\sqrt{m^2+(\bm p-\bm k)^2}+|\bm k|)\tau },
\nonumber
\ee
where the underlined terms are reordered so that annihilation operators are to the right of creation operators,
one finds
\be
V(\tau)
&=& :V(\tau):\nonumber\\
&\pp=&-
\frac{q}{2}\int dpdk \,
I_m(\bm p)2(2\pi)^3 \sqrt{m^2+\bm p^2}\delta^{*(3)}(\bm k)
c(\bm k)
e^{-i(\sqrt{m^2+\bm p^2}-\sqrt{m^2+(\bm p+\bm k)^2}+|\bm k|)\tau }
\nonumber\\
&\pp=&-
\frac{q}{2}\int dpdk \,
I_m(\bm p)2(2\pi)^3 \sqrt{m^2+\bm p^2}\delta^{*(3)}(\bm k)
c(\bm k)^{\dag}
e^{i(\sqrt{m^2+\bm p^2}-\sqrt{m^2+(\bm p+\bm k)^2}+|\bm k|)\tau }
\nonumber\\
&\pp=&-
\frac{q}{2}\int dpdk \,
I_m(\bm p)2(2\pi)^3 \sqrt{m^2+\bm p^2}\delta^{*(3)}(\bm k)
c(\bm k)
e^{-i(-\sqrt{m^2+\bm p^2}+\sqrt{m^2+(\bm p-\bm k)^2}+|\bm k|)\tau }
\nonumber\\
&\pp=&-
\frac{q}{2}\int dpdk \,
I_m(\bm p)2(2\pi)^3 \sqrt{m^2+\bm p^2}\delta^{*(3)}(\bm k)
c(\bm k)^{\dag}
e^{i(-\sqrt{m^2+\bm p^2}+\sqrt{m^2+(\bm p-\bm k)^2}+|\bm k|)\tau }
\nonumber\\
&=& :V(\tau):\nonumber\\
&\pp=&-
q\int \frac{d^3p}{(2\pi)^3 2\sqrt{m^2+\bm p^2}}I_m(\bm p)(2\pi)^3 2\sqrt{m^2+\bm p^2}
\int\frac{d^3k}{(2\pi)^3 2|\bm k|}\delta^{*(3)}(\bm k)\Big(c(\bm k)+c(\bm k)^{\dag}\Big)\nonumber\\
\ee
Let us note that in the standard approach, where $I_m(\bm p)=1$, the vacuum term
\be
V(\tau)-:V(\tau):
\ee
is badly divergent due to the first integral
\be
\int \frac{d^3p}{(2\pi)^3 2\sqrt{m^2+\bm p^2}}I_m(\bm p)(2\pi)^3 2\sqrt{m^2+\bm p^2}
=
\int_{\mathbb{R}^3} d^3p.
\ee
This is precisely an example of ``ultraviolet catastrophe" since the integral is divergent due to the behavior of its integrand (equal to 1) for large momenta.

The other term
\be
\int\frac{d^3k}{(2\pi)^3 2|\bm k|}\delta^{(3)}(\bm k)\Big(c(\bm k)+c(\bm k)^{\dag}\Big)
=
\lim_{\bm k\to\bm 0}\frac{c(\bm k)+c(\bm k)^{\dag}}{(2\pi)^3 2|\bm k|}\label{k->0 w c}
\ee
would be in standard approaches an example of ``infrared catastrophe". Eq.~(\ref{k->0 w c}) shows that
$
c(\bm k)+c(\bm k)^{\dag}
$
must tend to zero at least as fast as $|\bm k|$ but in the standard formalism it is completely unclear how to achieve it in a mathematically precise way. The ``infrared catastrophe" is typical of massless quantum fields --- $|\bm k|$ in denominators would be replaced for $m>0$ by $\sqrt{m^2+\bm k^2}$.

The most typical way of dealing with the divergences occurring in $V(\tau)-:V(\tau):$ is just to ignore them! We ignore them not because they are small, but ``because we do not want them". One replaces $V(\tau)$ by $:V(\tau):$ and starts with
\be
i\frac{d}{d\tau}U_1(\tau,\tau_1) &=& :V(\tau):U_1(\tau,\tau_1).
\ee
It is not hard to guess that this type of ``solution" will sooner or later lead to another ``catastrophe" that will have to be remedied in one way or another.

The elegance typical of quantum mechanics has been completely lost. What one does is not ``sensible mathematics". This is not even a physical theory but a ``collection of working rules" --- at least in Dirac's opinion.

Let us now see what happens in $(N,N')$ representations of HOLA. First of all, by (\ref{I_m(bm p,N)})
\be
I_m(\bm p,N,N')
&=&
I_m(\bm p,N)\otimes I_0(N')\nonumber\\
&=&
\frac{1}{N}
\Big(
\big(|\bm p\rangle\langle \bm p|\otimes (1\otimes 1)\big)\otimes I_m(1)\otimes\dots\otimes I_m(1)+\dots
\Big)
\otimes I_0(N')\nonumber,
\ee
The expression
\be
\int \frac{d^3p}{(2\pi)^3 2\sqrt{m^2+\bm p^2}}I_m(\bm p,N)(2\pi)^3 2\sqrt{m^2+\bm p^2}
\ee
is $1/N$ times a sum of $N$ operators of the form
\be
\int \frac{d^3p}{(2\pi)^3 2\sqrt{m^2+\bm p^2}}I_m(\bm p,1)(2\pi)^3 2\sqrt{m^2+\bm p^2}
&=&
(2\pi)^3 2
\int dp \,|\bm p\rangle\langle\bm p|\sqrt{m^2+\bm p^2}
\otimes (1\otimes 1)
\nonumber
\\\label{spectral}
\ee
each of them acting in the Hilbert space of a single indefinite-frequency oscillator. (\ref{spectral}) is the spectral representation of a well behaved operator
\be
(2\pi)^3 2\sqrt{m^2+\hat{\bm p}{^2}}\otimes (1\otimes 1),\label{hat p 1}
\ee
where
\be
\hat{\bm p}
&=&
\int dp \,|\bm p\rangle\langle\bm p|\, \bm p.\label{hat p}
\ee
Recall that $\langle\bm p|\bm p\rangle=0$, a fact that will become important later.

The example shows ``regularization by quantization" in action: There is no ``ultraviolet catastrophe" because the ordinary integral
\be
\int dp\, (\dots)
\ee
has been replaced by the {\it spectral integral\/}
\be
\int dp \,|\bm p\rangle\langle\bm p|\,(\dots)
\ee
Operators (\ref{hat p 1}) and (\ref{hat p}) are the analogues of, respectively,
\be
\Omega &=& \sum_\omega\omega|\omega\rangle\langle \omega|\otimes I,\quad \big(\textrm{compare (\ref{Omega})}\big)
\nonumber
\ee
and $\hat\omega = \sum_\omega\omega|\omega\rangle\langle \omega|$  we have started with in our discussion of indefinite-frequency
oscillators.

Now let us discuss the issue of infrared divergence of
\be
{}&{}&
\int\frac{d^3k}{(2\pi)^3 2|\bm k|}\delta^{*(3)}(\bm k)\Big(c(\bm k,N,N')+c(\bm k,N,N')^{\dag}\Big)
\nonumber\\
&{}&\pp=
=
I_m(N)\otimes \int\frac{d^3k}{(2\pi)^3 2|\bm k|}\delta^{*(3)}(\bm k)\Big(c(\bm k,N')+c(\bm k,N')^{\dag}\Big).
\ee
This is essentially a sum of $N'$ single-oscillator operators
\be
\int\frac{d^3k}{(2\pi)^3 2|\bm k|}\delta^{*(3)}(\bm k)\Big(c(\bm k,1)+c(\bm k,1)^{\dag}\Big)
&=&
\int\frac{d^3k}{(2\pi)^3 2|\bm k|}\delta^{*(3)}(\bm k)
|\bm k\rangle\langle\bm k|\otimes(a+a^{\dag})\label{delta c}
\ee
multiplied by $1/\sqrt{N'}$. Now let us act with it on a single-oscillator ($N'=1$) vacuum state
\be
\int\frac{d^3k}{(2\pi)^3 2|\bm k|}\delta^{*(3)}(\bm k)
|\bm k\rangle\langle\bm k|\otimes(a+a^{\dag})
|O_0,1\rangle
&=&
\int dk\,\delta^{*(3)}(\bm k)
|\bm k\rangle\langle\bm k|\otimes a^{\dag}
\int dk'\,O_0(\bm k')|\bm k',0\rangle\nonumber\\
&=&
\int \frac{d^3k}{(2\pi)^3 2|\bm k|}\delta^{*(3)}(\bm k)O_0(\bm k)|\bm k,1\rangle\nonumber\\
&=&
\lim_{\bm k\to\bm 0}\frac{1}{(2\pi)^3 2|\bm k|}O_0(\bm k)|\bm k,1\rangle\nonumber=0.
\ee
The latter follows from the Poincar\'e invariant boundary condition (\ref{boundary k=0})
\be
\lim_{\bm k\to\bm 0}\frac{O_0(\bm k)}{|\bm k|^n}=0.\nonumber
\ee
Now check the action of (\ref{delta c}) on a more general ``one-photon" one-oscillator state
\be
|\psi\rangle
&=&
\int dk' \psi(\bm k')c(\bm k',1)^\dag|O_0,1\rangle,
\ee
\be
{}&{}&\int\frac{d^3k}{(2\pi)^3 2|\bm k|}\delta^{*(3)}(\bm k)\Big(c(\bm k,1)+c(\bm k,1)^{\dag}\Big)
|\psi\rangle
\nonumber\\
&{}&
\pp=
=
\int\frac{d^3k}{(2\pi)^3 2|\bm k|}\delta^{*(3)}(\bm k)c(\bm k,1)
\int dk' \psi(\bm k')c(\bm k',1)^\dag|O_0,1\rangle
\nonumber\\
&{}&
\pp{==}
+
\int\frac{d^3k}{(2\pi)^3 2|\bm k|}\delta^{*(3)}(\bm k)c(\bm k,1)^{\dag}
\int dk' \psi(\bm k')c(\bm k',1)^\dag|O_0,1\rangle
\nonumber\\
&{}&
\pp=
=
\int\frac{d^3k}{(2\pi)^3 2|\bm k|}\delta^{*(3)}(\bm k)\int dk' \psi(\bm k')\delta_0(\bm k,\bm k')I_0(\bm k,1)
\int dk'' O_0(\bm k'')|\bm k''\rangle\otimes|0\rangle
\nonumber\\
&{}&
\pp{==}
+
\int\frac{d^3k}{(2\pi)^3 2|\bm k|}\delta^{*(3)}(\bm k)c(\bm k,1)^{\dag}
\int dk' \psi(\bm k')c(\bm k',1)^\dag \int dk'' O_0(\bm k'')|\bm k''\rangle\otimes|0\rangle
\nonumber\\
&{}&
\pp=
=
\int\frac{d^3k}{(2\pi)^3 2|\bm k|}\delta^{*(3)}(\bm k)\int dk' \psi(\bm k')\delta_0(\bm k,\bm k')
\int dk'' O_0(\bm k'')|\bm k\rangle\langle \bm k|\bm k''\rangle\otimes|0\rangle
\nonumber\\
&{}&
\pp{==}
+
\int\frac{d^3k}{(2\pi)^3 2|\bm k|}\delta^{*(3)}(\bm k)
\int dk' \psi(\bm k')\int dk'' O_0(\bm k'')|\bm k\rangle\langle \bm k|\bm k'\rangle\langle \bm k'|\bm k''\rangle\otimes|2\rangle
\nonumber\\
&{}&
\pp=
=
\int\frac{d^3k}{(2\pi)^3 2|\bm k|}\delta^{*(3)}(\bm k)\int dk' \psi(\bm k')\delta_0(\bm k,\bm k')
\int dk'' O_0(\bm k'') \delta_0(\bm k,\bm k'')|\bm k\rangle\otimes|0\rangle
\nonumber\\
&{}&
\pp{==}
+
\int\frac{d^3k}{(2\pi)^3 2|\bm k|}\delta^{*(3)}(\bm k)
\int dk' \psi(\bm k')\int dk'' O_0(\bm k'')\delta_0(\bm k,\bm k')\delta_0(\bm k',\bm k'')|\bm k\rangle\otimes|2\rangle
\nonumber\\
&{}&
\pp=
=
\int\frac{d^3k}{(2\pi)^3 2|\bm k|}\delta^{*(3)}(\bm k)\psi(\bm k)O_0(\bm k) |\bm k\rangle\otimes|0\rangle
+
\int\frac{d^3k}{(2\pi)^3 2|\bm k|}\delta^{*(3)}(\bm k)
\psi(\bm k)O_0(\bm k)|\bm k\rangle\otimes|2\rangle
\nonumber\\
&{}&
\pp=
=
\lim_{\bm k\to \bm 0}\frac{1}{(2\pi)^3 2|\bm k|}\psi(\bm k)O_0(\bm k) |\bm k\rangle\otimes\big(|0\rangle+|2\rangle\big).
\nonumber
\ee
The term again vanishes unless $\psi(\bm k)$ blows up at the origin faster than any power of $1/|\bm k|$. The argument can be generalized by induction to states generated from vacuum by any number of creation oscillators.

All of this is possible since in the formula
\be
O_0(\bm k)
&=&
\langle O_0,N'|I_0(\bm k,N')|O_0,N'\rangle
\ee
the operator $I_0(\bm k,N')$ is not proportional to the identity. In representations occurring in the standard formalisms one would find similar expressions as we have obtained (since algebraic manipulations with operators would be identical), but with
\be
O_0(\bm k)
&=&
\langle O|I(\bm k)|O\rangle=\langle O|I|O\rangle=1
\ee
and then infrared divergences would be inevitable.

\section{Vacuum-to-vacuum loop diagram in 2nd order perturbation theory}

Vacuum-to-vacuum loop diagram is the 2nd order perturbative correction to $\langle O,N,N'|U_1(\tau_2,\tau_1) |O,N,N'\rangle$, where
$|O,N,N'\rangle$ is {\it a\/} vacuum space. What is very important we want to have a result that does not depend on concrete explicit forms of vacuum wave functions $O_m(\bm p)$ and $O_0(\bm k)$ but only on their general properties such as boundary conditions.

So let $\Pi_0=\Pi_0(N,N')$ be the projector on the vacuum subspace. It is characterized by
\be
a(\bm p)\Pi_0 &=& b(\bm p)\Pi_0 =c(\bm k)\Pi_0 =\Pi_0a(\bm p)^\dag = \Pi_0b(\bm p)^\dag =\Pi_0c(\bm k)^\dag =0,\\
\int dk\,\delta^{*(3)}(\bm k)c(\bm k)^\dag\Pi_0
&=&
\int dk\,\delta^{*(3)}(\bm k)\Pi_0c(\bm k)=
\int dk\,\delta^{*(3)}(\bm k)I_0(\bm k)\Pi_0
=
0.
\ee
\begin{figure}
\includegraphics[width=4cm]{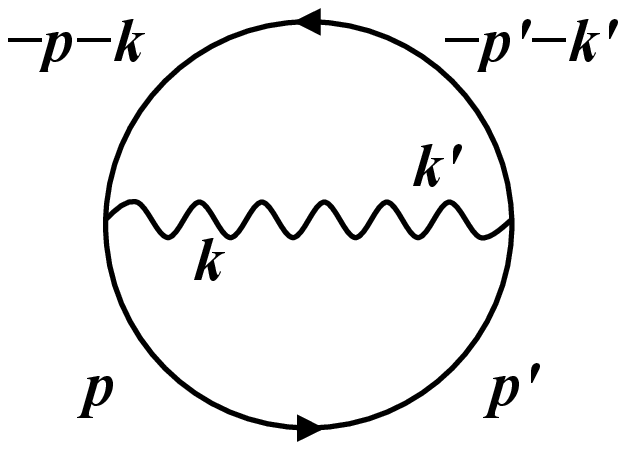}
\caption{The simplest loop vacuum-to-vacuum Feynman diagram represents a process where creation operators $a(\bm p)^\dag$, $b(-\bm p-\bm k)^\dag$, $c(\bm k)^\dag$ act on vacuum and the resulting state is then acted upon by the annihilation operators $a(\bm p')$, $b(-\bm p'-\bm k')$, $c(\bm k')$. So one starts from vacuum and ends up with vacuum.}
\end{figure}
Our goal is to compute in 2nd order perturbation theory the operator $\Pi_0 U_1(\tau_2,\tau_1) \Pi_0$. The appropriate term is
\be
{}&{}&
\Pi_0
{V}(\tau)
{V}(\tau')
\Pi_0\nonumber\\
&{}&=
\Pi_0
:{V}(\tau):
:{V}(\tau'):
\Pi_0
\nonumber\\
&{}&=
\frac{q^2}{4}\int dpdk  dp'dk' \,
e^{-i(\sqrt{m^2+\bm p^2}+\sqrt{m^2+(\bm p+\bm k)^2}+|\bm k|)\tau }
e^{i(\sqrt{m^2+\bm p'^2}+\sqrt{m^2+(\bm p'+\bm k')^2}+|\bm k'|)\tau' }
\nonumber\\
&{}&
\pp=
\times
\Pi_0
\Big(
a(\bm p)b(-\bm p-\bm k)-a(-\bm p-\bm k)b(\bm p)
\Big)
c(\bm k)
\nonumber\\
&{}&
\pp{=\times\Pi_0\Big(}
\times\Big(
b(-\bm p'-\bm k')^{\dag}a(\bm p')^{\dag}-b(\bm p')^{\dag}a(-\bm p'-\bm k')^{\dag}
\Big)
c(\bm k')^{\dag}
\Pi_0
\nonumber\\
&{}&=
\frac{q^2}{4}\int dpdk  dp' \,
e^{-i(\sqrt{m^2+\bm p^2}+\sqrt{m^2+(\bm p+\bm k)^2}+|\bm k|)\tau }
e^{i(\sqrt{m^2+\bm p'^2}+\sqrt{m^2+(\bm p'+\bm k)^2}+|\bm k|)\tau' }
\nonumber\\
&{}&\pp=
\times
\Pi_0
\Big(
a(\bm p)a(\bm p')^{\dag}b(-\bm p-\bm k)b(-\bm p'-\bm k)^{\dag}
+
a(-\bm p-\bm k)a(-\bm p'-\bm k)^{\dag}b(\bm p)b(\bm p')^{\dag}
\nonumber\\
&{}&
\pp{=\times\Pi_0\Big(}
-
a(\bm p)a(-\bm p'-\bm k)^{\dag}b(-\bm p-\bm k)b(\bm p')^{\dag}
-
a(-\bm p-\bm k)a(\bm p')^{\dag}b(\bm p)b(-\bm p'-\bm k)^{\dag}
\Big)
I_0(\bm k)
\Pi_0
\nonumber\\
&{}&=
\frac{q^2}{4}\int dpdk  dp' \,
e^{-i(\sqrt{m^2+\bm p^2}+\sqrt{m^2+(\bm p+\bm k)^2}+|\bm k|)\tau }
e^{i(\sqrt{m^2+\bm p'^2}+\sqrt{m^2+(\bm p'+\bm k)^2}+|\bm k|)\tau' }
\nonumber\\
&{}&\pp=
\times
\Pi_0
\Big(
I_m(\bm p)\delta_m(\bm p,\bm p')I_m(-\bm p-\bm k)\delta_m(-\bm p-\bm k,-\bm p'-\bm k)
\nonumber\\
&{}&
\pp{=\times\Pi_0}
+
I_m(-\bm p-\bm k)\delta_m(-\bm p-\bm k,-\bm p'-\bm k)I_m(\bm p)\delta_m(\bm p,\bm p')
\nonumber\\
&{}&
\pp{=\times\Pi_0}
-
I_m(\bm p)\delta_m(\bm p,-\bm p'-\bm k)I_m(-\bm p-\bm k)\delta_m(-\bm p-\bm k,\bm p')
\nonumber\\
&{}&
\pp{=\times\Pi_0}
-
I_m(-\bm p-\bm k)\delta_m(-\bm p-\bm k,\bm p')I_m(\bm p)\delta_m(\bm p,-\bm p'-\bm k)
\Big)
I_0(\bm k)
\Pi_0
\label{2nd}
\ee
The latter expression is ill defined in standard approaches due to occurrences of squared Dirac deltas. Additional divergences would be generated if one assumed the standard condition $I_0(\bm k)=I_m(\bm p)=1$. This is why the loop diagram is normally an example of a ``difficult infinity" (see for example an analogous calculation made in full electrodynamics in chapter 4.1 of \cite{Scharf}).
In my formalism the calculation is rather trivial.

We have to keep in mind that $dk=\rho_0(\bm k)d^3k$, $dp=\rho_m(\bm p)d^3p$ so that internal consistency conditions analogous to (\ref{a rho}) can be derived. The only solution is thus to work with M-shaped Dirac deltas satisfying $\delta_m(\bm p,\bm p)=0$. The result is
\be
(\ref{2nd})
&{}&=
\frac{q^2}{4}\int dpdk   \,
e^{-i(\sqrt{m^2+\bm p^2}+\sqrt{m^2+(\bm p+\bm k)^2}+|\bm k|)\tau }
e^{i(\sqrt{m^2+\bm p^2}+\sqrt{m^2+(\bm p+\bm k)^2}+|\bm k|)\tau' }
\nonumber\\
&{}&\pp=
\times
\Big(
\delta_m(-\bm p-\bm k,-\bm p-\bm k)
+
\delta_m(-\bm p-\bm k,-\bm p-\bm k)
-
\delta_m(\bm p,\bm p)
-
\delta_m(\bm p,\bm p)
\Big)
\nonumber\\
&{}&\pp=
\times
\Pi_0
I_m(\bm p)I_m(-\bm p-\bm k)I_0(\bm k)
\Pi_0\nonumber\\
&{}&=
0.
\ee
Notice that we have not needed the explicit forms of $O_m(\bm p)$ and $O_0(\bm k)$. In standard quantum field theoretic parlance the result thus ``does not depend on a cutoff".

\section{Massless quantum scalar field produced by a classical pointlike source}

In order to compute radiative corrections to atomic energy levels one first has to decide what kind of potential is associated with atomic nucleus. In simplest approaches the nucleus is modeled by the classical static pointlike charge density
\be
\rho(\bm x)&=& q\delta^{(3)}(\bm x).
\ee
Classically the field produced by the source would be a solution of an inhomogeneous (Maxwell, d'Alembert...) field equation. In the quantum context one should solve the Heisenberg equation with appropriate interaction term.

Let us employ Heisenberg and Dirac picture forms of evolution equations to the dynamics of $\hat\phi(t,\bm x)$ for
${V}(t)=\int_{\mathbb{R}^3}d^3x\, \rho(\bm x)\phi(t,\bm x)$,
\be
\phi(t,\bm x) &=& \int dk \Big(c(\bm k)e^{-i|\bm k|t+i\bm k\cdot \bm x}+c(\bm k)^\dag e^{i|\bm k|t-i\bm k\cdot \bm x}\Big),\\
{V}(t)
&=&
q\int_{\mathbb{R}^3}d^3x\, \delta^{(3)}(\bm x)\phi(t,\bm x)
\nonumber\\
&=&
q \phi(t,\bm 0)\\
&=&
\underbrace{q\int dk\, c(\bm k)e^{-i|\bm k|t}}_{{V}_-(t)}
+
\underbrace{q\int dk\, c(\bm k)^\dag e^{i|\bm k|t}}_{{V}_+(t)={V}_-(t)^\dag}.
\ee
The example is interesting since in standard field quantization $U_1(t,t_0)$ does not exist due to ultraviolet divergences.
We will need
\be
{[{V}_-(t),{V}_+(t')]}
&=&
q^2\Big[\int dk \,c(\bm k)e^{-i|\bm k|t},
\int dk' \,c(\bm k')^\dag e^{i|\bm k'|t'}\Big]\nonumber\\
&=&
q^2\int dk \,I_0(\bm k)e^{i|\bm k|(t'-t)}.
\ee
Of particular interest is the equal-time commutator
\be
{[{V}_-(t),{V}_+(t)]}
&=&
q^2\int dk \,I_0(\bm k).\nonumber
\ee
In reducible $N$-representation of HOLA
\be
{[{V}_-(t),{V}_+(t)]}
&=&
q^2\int dk \,I_0(\bm k,N)\label{ttN}\\
&=&
q^2 I_0(N)
\ee
is a well defined operator (\ref{ttN}). This should be contrasted with the divergent integral that would have occurred here for $I_0(\bm k)=1$.
Since for all times
\be
{[{V}_-(t),{V}_-(t')]}=[{V}_+(t),{V}_+(t')]=0
\ee
and
\be
{\big[[{V}_-(t),{V}_+(t')],{V}_\pm(t'')\big]}=0,\label{double[]}
\ee
we can use the continuous Baker-Campbell-Hausdorff (BCH) formula (cf. Appendix H, Eq.~(13) in \cite{IZBB}),
\be
U_1(t,t_0)
&=&
\Texp\Big(-i\int_{t_0}^{t}d\tau\big({V}_+(\tau)+{V}_-(\tau)\big)\Big)\\
&=&
\exp\Big(-i\int_{t_0}^{t}d\tau\,{V}_+(\tau)\Big)
\exp\Big(-i\int_{t_0}^{t}d\tau\,{V}_-(\tau)\Big)
\nonumber\\
&\pp=&\times\exp\Big(\int_{t_0}^{t}d\tau\int_{t_0}^{\tau}d\tau'\,[{V}_+(\tau'),{V}_-(\tau)]\Big).\label{cBCH}
\ee
Another useful form of continuous BCH identity is based on the ordinary BCH identity $e^Xe^Y=e^{[X,Y]/2}e^{X+Y}$, which can be applied to the right-hand-side of (\ref{cBCH}),
\be
U_1(t,t_0)
&=&
\exp\Big(-i\int_{t_0}^{t}d\tau\,{V}(\tau)\Big)
\exp\Big(-\frac{1}{2}\int_{t_0}^{t}d\tau\int_{t_0}^{t}d\tau'\,[{V}_+(\tau'),{V}_-(\tau)]\Big)
\nonumber\\
&\pp=&\times\exp\Big(\int_{t_0}^{t}d\tau\int_{t_0}^{t}d\tau'\,\theta(\tau-\tau')[{V}_+(\tau'),{V}_-(\tau)]\Big)
\\
&=&
\exp\Big(-i\int_{t_0}^{t}d\tau\,{V}(\tau)\Big)
\exp\Big(\frac{1}{2}\int_{t_0}^{t}d\tau\int_{t_0}^{t}d\tau'\,{\rm sgn\,}(\tau-\tau')[{V}_+(\tau'),{V}_-(\tau)]\Big).
\nonumber\\\label{cBCH'}
\ee
sgn$(x)$ and $\theta(x)$ are, respectively, the sign-of-$x$ and step functions. For $\tau=\tau'$ the commutators in exponents are finite in ``my" representation of HOLA, so exact values at zero of sgn and $\theta$ are for the moment irrelevant. Both exponents at the right side of (\ref{cBCH'}) are unitary. The second of them commutes with all the elements of HOLA, so this is an operator that in practice behaves as numerical phase factor.

Continuous BCH formula can be verified in a straightforward manner. First of all, at $t=t_0$ the left and right sides of (\ref{cBCH}) equal $\mathbb{I}$. It remains to check if they satisfy the same differential equation,
\be
i\frac{dU_1(t,t_0)}{dt}
&=&
{V}_+(t)\exp\Big(-i\int_{t_0}^{t}d\tau\,{V}_+(\tau)\Big)
\exp\Big(-i\int_{t_0}^{t}d\tau\,{V}_-(\tau)\Big)
\nonumber\\
&\pp=&\times\exp\Big(\int_{t_0}^{t}d\tau\int_{t_0}^{\tau}d\tau'\,[{V}_+(\tau'),{V}_-(\tau)]\Big)
\nonumber\\
&\pp=&
+
\exp\Big(-i\int_{t_0}^{t}d\tau\,{V}_+(\tau)\Big)
{V}_-(t)
\exp\Big(-i\int_{t_0}^{t}d\tau\,{V}_-(\tau)\Big)
\nonumber\\
&\pp=&\times\exp\Big(\int_{t_0}^{t}d\tau\int_{t_0}^{\tau}d\tau'\,[{V}_+(\tau'),{V}_-(\tau)]\Big)
\nonumber\\
&\pp=&
+
i\int_{t_0}^{t}d\tau'\,[{V}_+(\tau'),{V}_-(t)]
\exp\Big(-i\int_{t_0}^{t}d\tau\,{V}_+(\tau)\Big)
\exp\Big(-i\int_{t_0}^{t}d\tau\,{V}_-(\tau)\Big)
\nonumber\\
&\pp=&\times\exp\Big(\int_{t_0}^{t}d\tau\int_{t_0}^{\tau}d\tau'\,[{V}_+(\tau'),{V}_-(\tau)]\Big)
\nonumber\\
&=&
\Bigg(
{V}_+(t)
+
e^{-i\int_{t_0}^{t}d\tau\,{V}_+(\tau)}
{V}_-(t)
e^{i\int_{t_0}^{t}d\tau\,{V}_+(\tau)}
+
i\int_{t_0}^{t}d\tau'\,[{V}_+(\tau'),{V}_-(t)]
\Bigg)
U_1(t,t_0)
\nonumber
\ee
Employing (\ref{Y1}) and (\ref{double[]}) we rewrite the second term
\be
e^{-i\int_{t_0}^{t}d\tau\,{V}_+(\tau)}
{V}_-(t)
e^{i\int_{t_0}^{t}d\tau\,{V}_+(\tau)}
&=&
{V}_-(t)
+
\Big[-i\int_{t_0}^{t}d\tau\,{V}_+(\tau),{V}_-(t)\Big],
\nonumber
\ee
which ends the proof:
\be
i\dot U_1(t,t_0)
&=&
\big({V}_+(t)+{V}_-(t)\big)U_1(t,t_0).
\ee
We can finally compute
\be
\hat\phi(t,\bm x)
&=&
\exp\Big(i\int_{t_0}^{t}d\tau\,{V}(\tau)\Big)
\phi(t,\bm x)
\exp\Big(-i\int_{t_0}^{t}d\tau\,{V}(\tau)\Big)\nonumber\\
&=&
\exp\Big(iq\int_{t_0}^{t}d\tau\,\phi(\tau,\bm 0)\Big)
\phi(t,\bm x)
\exp\Big(-iq\int_{t_0}^{t}d\tau\,\phi(\tau,\bm 0)\Big)\nonumber\\
&=&
\phi(t,\bm x)
+
iq\int_{t_0}^{t}d\tau\,[\phi(\tau,\bm 0),\phi(t,\bm x)]
\nonumber\\
&=&
\phi(t,\bm x)
\nonumber\\
&\pp=&
+
iq\int_{t_0}^{t}d\tau \int dk dk'\,
\nonumber\\
&\pp=&
\times
\big[c(\bm k)e^{-i|\bm k|\tau}+c(\bm k)^\dag e^{i|\bm k|\tau},
c(\bm k')e^{-i|\bm k'|t+i\bm k'\cdot \bm x}+c(\bm k')^\dag e^{i|\bm k'|t-i\bm k'\cdot \bm x}\big]
\nonumber\\
&=&
\phi(t,\bm x)
+
iq\int dk I_0(\bm k)
\Big(
e^{i|\bm k|t-i\bm k\cdot \bm x}\int_{t_0}^{t}d\tau e^{-i|\bm k|\tau}
-
e^{-i|\bm k|t+i\bm k\cdot \bm x}\int_{t_0}^{t}d\tau e^{i|\bm k|\tau}
\Big)
\nonumber\\
&=&
\phi(t,\bm x)
\nonumber\\
&\pp=&
-
q\int \frac{d^3k}{(2\pi)^3 2|\bm k|^2} I_0(\bm k)
\Big(
e^{i|\bm k|t-i\bm k\cdot \bm x}\big(e^{-i|\bm k|t}-e^{-i|\bm k|t_0}\big)
+
e^{-i|\bm k|t+i\bm k\cdot \bm x}\big(e^{i|\bm k|t}-e^{i|\bm k|t_0}\big)
\Big)\nonumber\\
&=&
\phi(t,\bm x)
-
q\int \frac{d^3k}{(2\pi)^3 2|\bm k|^2} I_0(\bm k)
\Big(
e^{-i\bm k\cdot \bm x}
+
e^{i\bm k\cdot \bm x}
-
e^{ikx}e^{-i|\bm k|t_0}
-
e^{-ikx}e^{i|\bm k|t_0}
\Big).\nonumber
\ee
$\hat\phi(t,\bm x)$ splits into free-field and source parts. The source term can be further split into the ``generalized Coulomb field"
\be
\phi_{\rm gC}(\bm x)
&=&
-
q\int \frac{d^3k}{(2\pi)^3 |\bm k|^2} I_0(\bm k)
\cos \bm k\cdot \bm x
\label{gen Coulomb}
\ee
and the ``compensating field"
\be
\phi_{\rm c}(t-t_0,\bm x)
&=&
q\int \frac{d^3k}{(2\pi)^3 |\bm k|^2} I_0(\bm k)
\cos(kx-|\bm k|t_0).
\label{compensating-field}
\ee
Compensating and Coulomb fields cancel each another at $t=t_0$. Let us discuss the two terms separately.

\subsection{Compensating field}

In order to better understand the role of the compensating field let us consider an arbitrary coherent state
\be
|\alpha,N\rangle
&=&
{\cal D}_0(\alpha,N)|O_0,N\rangle,\\
{\cal D}_0(\alpha,N)
&=&
\exp\int dk\Big(\alpha(\bm k)c(\bm k,N)^\dag-\overline{\alpha(\bm k)}c(\bm k,N)\Big).
\ee
Since
\be
{\cal D}_0(\alpha,N)^\dag
I_0(\bm k,N)
{\cal D}_0(\alpha,N)
&=&
I_0(\bm k,N)
\ee
one finds
\be
\langle\alpha,N|\phi_{\rm c}(t-t_0,\bm x,N)|\alpha,N\rangle
&=&
q\int \frac{d^3k}{(2\pi)^3 |\bm k|^2} \langle O_0,N|I_0(\bm k,N)|O_0,N\rangle
\cos(kx-|\bm k|t_0)\nonumber\\
&=&
q\int \frac{d^3k}{(2\pi)^3 |\bm k|^2} |O_0(\bm k)|^2
\cos(kx-|\bm k|t_0)\nonumber.
\ee
The same result will be obtained if we replace the displacement operator by any unitary operator constructed from elements of HOLA.
Normalization and boundary conditions imposed on vacuum imply
\be
\int \frac{d^3k}{(2\pi)^3 2|\bm k|} |O_0(\bm k)|^2 &=& 1,\\
\lim_{\bm k\to\bm 0}\frac{|O_0(\bm k)|^2}{|\bm k|^2} &=&0.
\ee
If additionally
\be
\int \frac{d^3k}{(2\pi)^3 2|\bm k|} \frac{|O_0(\bm k)|^2}{|\bm k|}<\infty \label{O<infty}
\ee
then Riemann-Lebesgue lemma (Chapter 2.5, Theorem 2.2 in \cite{RL}) implies
\be
\lim_{t_0\to\pm\infty}
\int \frac{d^3k}{(2\pi)^3 |\bm k|^2} |O_0(\bm k)|^2
\cos(kx)\cos(|\bm k|t_0) &=& 0\nonumber\\
\lim_{t_0\to\pm\infty}
\int \frac{d^3k}{(2\pi)^3 |\bm k|^2} |O_0(\bm k)|^2
\sin(kx)\sin(|\bm k|t_0) &=& 0\nonumber.
\ee
Hence
\be
\lim_{t_0\to\pm\infty}
\langle\psi,N|\phi_{\rm c}(t-t_0,\bm x,N)|\psi,N\rangle
&=& 0
\ee
for all states $|\psi,N\rangle$ generated from vacuum by unitary transformations constructed from the $N$-representation of HOLA.
In this weak sense one can write
\be
\lim_{t_0\to\pm\infty}
\phi_{\rm c}(t-t_0,\bm x,N)
&=& 0.
\ee
Let us note that asymptotic vanishing of compensating fields leads to the additional condition (\ref{O<infty}) that should be satisfied by vacuum wave functions.

For comparison let us check what happens in the usual representation, $I_0(\bm k)=1$,
\be
\phi_{\rm c}(t-t_0,\bm x)
&=&
q\int \frac{d^3k}{(2\pi)^3 |\bm k|^2}
\cos(kx-|\bm k|t_0)\nonumber\\
&=&
q\int \frac{d^3k}{(2\pi)^3 |\bm k|^2}
\cos\big(|\bm k|(t-t_0)-\bm k\cdot\bm x\big)\nonumber\\
&=&
\frac{q}{(2\pi)^3}\int_0^\infty d\kappa\int_0^{2\pi}d\varphi\int_0^\pi d\theta\sin\theta
\cos\big(\kappa(t-t_0)-|\bm x|\kappa\cos\theta\big)\nonumber\\
&=&
-\frac{q}{(2\pi)^2}\int_0^\infty d\kappa \int_1^{-1} du
\cos\big(\kappa(t-t_0)-|\bm x|\kappa u\big)\nonumber\\
&=&
\frac{q}{(2\pi)^2}\int_0^\infty d\kappa \frac{1}{|\bm x|\kappa}
\sin\big(\kappa(t-t_0)-|\bm x|\kappa u\big)\Big|_{1}^{-1}\nonumber\\
&=&
\frac{q}{(2\pi)^2}\int_0^\infty d\kappa \frac{\sin\big(\kappa(t-t_0+|\bm x|)\big)+\sin\big(\kappa(t_0-t+|\bm x|)\big)}{|\bm x|\kappa}.
\nonumber
\ee
Note that I denote $|\bm k|=\kappa$ and $d\kappa$, integration over the radial coordinate, which should not be confused with the measure $dk$ on the light cone.
For $t=t_0$
\be
\phi_{\rm c}(0,\bm x)
&=&
\frac{q}{2\pi^2}\int_0^\infty d\kappa \frac{\sin \kappa|\bm x|}{\kappa|\bm x|}=\frac{q}{4\pi |\bm x|} .
\nonumber
\ee
For $\Delta t=t-t_0\neq 0$ the field is symmetric in time, $\phi_{\rm c}(\Delta t,\bm x)=\phi_{\rm c}(-\Delta t,\bm x)$. Assume $\Delta t>0$,
\be
\phi_{\rm c}(\Delta t,\bm x)
&=&
\frac{q}{(2\pi)^2}\int_0^\infty d\kappa \frac{\sin\big(\kappa(|\bm x|+\Delta t)\big)}{|\bm x|\kappa}
+
\frac{q}{(2\pi)^2}\int_0^\infty d\kappa \frac{\sin\big(\kappa(|\bm x|-\Delta t)\big)}{|\bm x|\kappa}
\nonumber\\
&=&
\frac{q}{(2\pi)^2}\frac{1}{|\bm x|}\frac{\pi}{2}
+
\frac{q}{(2\pi)^2}\frac{|\bm x|-\Delta t}{|\bm x|}\frac{\big||\bm x|-\Delta t\big|}{\big||\bm x|-\Delta t\big|}
\int_0^\infty d\kappa \frac{\sin\big(\kappa(|\bm x|-\Delta t)\big)}{\kappa(|\bm x|-\Delta t)}
\nonumber\\
&=&
\frac{q}{4\pi |\bm x|}\frac{1}{2}\Big(1+{\rm sgn}(|\bm x|-\Delta t)\Big)
=
\frac{q}{4\pi |\bm x|}\theta(|\bm x|-\Delta t).
\ee
For $|\bm x|=\Delta t>0$
\be
\phi_{\rm c}(\Delta t,\bm x)
&=&
\frac{q}{(2\pi)^2}\int_0^\infty d\kappa \frac{\sin 2\kappa|\bm x|}{\kappa|\bm x|}
=
\frac{q}{4\pi^2 |\bm x|}\frac{\pi}{2}=\frac{q}{4\pi |\bm x|}\frac{1}{2}\ee
For any $t$, $t_0$,
\be
\phi_{\rm c}(t-t_0,\bm x)
&=&
\frac{q}{4\pi |\bm x|}\theta\big(|\bm x|-|t-t_0|\big).
\ee
For fixed $t$ and $|\bm x|\neq 0$
\be
\lim_{t_0\to\pm\infty} \phi_{\rm c}(t-t_0,\bm x)
&=&
0.
\ee
As we can see, the step function is defined by
\be
\theta(x)
&=&
\left\{
\begin{array}{ll}
0 & \textrm{for }x<0\\
\frac{1}{2} & \textrm{for }x=0\\
1 & \textrm{for }x>0
\end{array}
\right.
.
\ee
The role of compensating field is similar to the one of compensating currents occurring in some approaches to quantum electrodynamics (cf. \cite{IZBB}).
A difference is that compensating currents are introduced in an ad hoc manner, whereas our compensating field is a consequence of quantum dynamics.

\subsection{Generalized Coulomb field}

In order to understand why I call $\phi_{\rm gC}(\bm x)$ a generalized Coulomb field let us note that setting $I_0(\bm k)=1$ one indeed arrives at the ordinary Coulomb field
\be
-
q\int \frac{d^3k}{(2\pi)^3 |\bm k|^2}
\cos \bm k\cdot \bm x=-\frac{q}{4\pi|\bm x|}=\phi_{\rm C}(\bm x)=-\phi_{\rm c}(0,\bm x).
\ee
However, we know that $I_0(\bm k)=1$ implies $|O_0(\bm k)|^2=1$ (since $|O_0(\bm k)|^2$ is the vacuum average of $I_0(\bm k)$) which would lead to infrared divergences and all types of internal inconsistencies of the formalism. So, take
\be
I_0(\bm k,N)
&=&
\frac{1}{N}
\Big(
I_0(\bm k,1)\otimes I_0(1)\otimes\dots\otimes I_0(1)
+
\dots
+
I_0(1)\otimes\dots\otimes I_0(1)\otimes I_0(\bm k,1)
\Big),
\nonumber\\
\phi_{\rm gC}(\bm x,N)
&=&
\frac{1}{N}
\Big(
\phi_{\rm gC}(\bm x,1)\otimes I_0(1)\otimes\dots\otimes I_0(1)
+
\dots
+
I_0(1)\otimes\dots\otimes I_0(1)\otimes \phi_{\rm gC}(\bm x,1)
\Big),
\nonumber\\
\phi_{\rm gC}(\bm x,1)
&=&
-
q\int \frac{d^3k}{(2\pi)^3 |\bm k|^2}\cos (\bm k\cdot \bm x) \,|\bm k\rangle\langle\bm k|\otimes 1
=
-
2q\int dk
\frac{\cos \bm k\cdot \bm x}{|\bm k|} |\bm k\rangle\langle\bm k|\otimes 1\nonumber\\
&=&
-
2q \frac{\cos \hat{\bm k}\cdot \bm x}{|\hat{\bm k}|} \otimes 1,\\
\hat{\bm k}
&=&
\int dk\,\bm k|\bm k\rangle\langle\bm k|.
\ee
Of course, $\phi_{\rm gC}(\bm x,N)$ is a field {\it operator\/} commuting with all elements of HOLA (the same modification of Coulomb's law was found in full electrodynamics in \cite{V}). Vacuum averages of $\phi_{\rm gC}(\bm x,N)$
are ordinary functions that can be compared with $\phi_{\rm C}(\bm x,N)$.

Let us thus consider the average evaluated in a vacuum state $|O_0,N\rangle$,
\be
\langle O_0,N|\phi_{\rm gC}(\bm x,N)|O_0,N\rangle
&=&
-
q\int \frac{d^3k}{(2\pi)^3 |\bm k|^2} |O_0(\bm k)|^2
\cos \bm k\cdot \bm x.
\ee
Probability $|O_0(\bm k)|^2$ occurs in place where in the usual approach one might, if needed, put a cutoff function $\chi(\bm k)$. Cutoff functions are assumed to vanish for large and small $\bm k$ but somewhere in between infrared and ultraviolet regimes achieve their maximal value 1. It is obvious, however, that the maximal value of $|O_0(\bm k)|^2$ is greater from 0, but may be different from unity. It follows that $|O_0(\bm k)|^2$ itself cannot be regarded as the usual cutoff function. The latter is obtained if one divides $|O_0(\bm k)|^2$ by its maximal value $Z$, say.

We know that Poincar\'e transformations transform $|O_0(\bm k)|$ into $|O_0(\bm{L^{-1}k})|$. Since the maximum
\be
Z=\max_{\bm k}\{ |O_0(\bm k)|^2\}=\max_{\bm k}\{ |O_0(\bm{L^{-1}k})|^2\}
\ee
is Poincar\'e invariant we can define  in Poincar\'e covariant way the normalized ``cutoff function"
\be
\chi_0(\bm k)&=& |O_0(\bm k)|^2/Z,\quad
0\leq \chi_0(\bm k)\leq 1.
\ee
Accordingly,
\be
\langle O_0,N|\phi_{\rm gC}(\bm x,N)|O_0,N\rangle
&=&
-
qZ\int \frac{d^3k}{(2\pi)^3 |\bm k|^2} \chi(\bm k)
\cos \bm k\cdot \bm x,
\ee
suggesting that $qZ$ is the physical charge. This statement is {\it almost\/} true but all the examples studied in the literature so far (cf. \cite{MWMC1,V}) show that the correct physical parameter that should be compared with experiment is $q^2Z=(q\sqrt{Z})^2=q_{\rm ph}^2$. At this stage I propose the readers just to take this statement for granted, without proof. In other words, if we want to test the average with respect to the classical Coulomb field one should compare $-q_{\rm ph}^2/(4\pi |\bm x|)$ with
\be
q\langle O_0,N|\phi_{\rm gC}(\bm x,N)|O_0,N\rangle
&=&
-q_{\rm ph}^2
\int \frac{d^3k}{(2\pi)^3 |\bm k|^2} \chi(\bm k)
\cos \bm k\cdot \bm x,
\ee
where $q_{\rm ph}=q\sqrt{Z}$ is the {\it renormalized\/} (i.e. physical) charge and $Z$ is the renormalization constant (analogous to $Z_3$ known from textbooks \cite{IZBB}). Let us also note here that condition (\ref{O<infty}) makes
\be
\langle O_0,N|\phi_{\rm gC}(\bm 0,N)|O_0,N\rangle
&=&
-
q\int \frac{d^3k}{(2\pi)^3 |\bm k|^2} |O_0(\bm k)|^2
\ee
finite.

To have a feel of general properties of $q\langle O_0,N|\phi_{\rm gC}(\bm x,N)|O_0,N\rangle$ let us take a rotationally invariant
$|O_0(\bm k)|^2$ which vanishes for $0\leq|\bm k|\leq k_1$ and $|\bm k|\geq k_2$ but is otherwise constant (equal to some $Z$).
Then
$Z$ can be then computed from
\be
1&=&
\int dk |O_0(\bm k)|^2
=
4\pi Z\int_{k_1}^{k_2} \frac{d\kappa\,\kappa}{2(2\pi)^3}=
\frac{Z(k_2^2-k_1^2)}{8\pi^2},\\
Z &=&
\frac{8\pi^2}{k_2^2-k_1^2},\\
q_{\rm ph} &=& q\sqrt{Z}=q\frac{2\sqrt{2}\pi }{\sqrt{k_2^2-k_1^2}}.
\ee
Now
\be
q\langle O_0,N|\phi_{\rm gC}(\bm x,N)|O_0,N\rangle
&=&
-\frac{2\pi q_{\rm ph}^2}{(2\pi)^3}
\int_{k_1}^{k_2} d\kappa \int_0^\pi d\theta\sin\theta
\cos (\kappa\cos\theta |\bm x|)\nonumber\\
&=&
\frac{q_{\rm ph}^2}{4\pi^2}
\int_{k_1}^{k_2} d\kappa \int_1^{-1} du
\cos (\kappa|\bm x|u)\nonumber\\
&=&
-\frac{q_{\rm ph}^2}{2\pi^2}
\int_{k_1}^{k_2} d\kappa \frac{\sin \kappa|\bm x|}{\kappa|\bm x|}\nonumber\\
&=&
-\frac{q_{\rm ph}^2}{4\pi |\bm x|}
\frac{{\rm Si}(k_2|\bm x|)-{\rm Si}(k_1|\bm x|)}{\pi/2}.
\ee
\begin{figure}
\includegraphics[width=8cm]{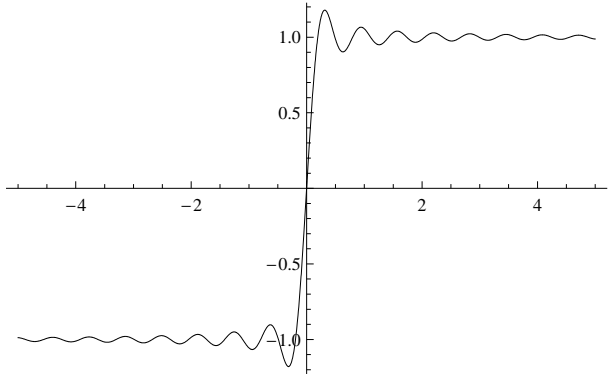}
\caption{Normalized sine integral $\frac{2}{\pi}{\rm Si}(k x)$ for small $x$ behaves as $2kx/\pi$ but for large $|x|$ approaches 1 independently of $k$ (on the plot $k=10$). This is why $\langle O_0,N|\phi_{\rm gC}(\bm x,N)|O_0,N\rangle$ is finite at the origin and for large $|\bm x|$ decays faster than $1/|\bm x|$. If $k_1$ were exactly 0 (no infrared cutoff) then $\langle O_0,N|\phi_{\rm gC}(\bm x,N)|O_0,N\rangle$ would asymptotically approach the Coulomb field. However, $k_1=0$ would be in contradiction with $\lim_{\bm k\to\bm 0}O_0(\bm k)=0$ which we assume.}
\end{figure}
The sine integral function (see Fig.~7)
\be
{\rm Si}(x)
&=&
\int_0^x d\kappa\frac{\sin \kappa}{\kappa}
\ee
has series expansion
\be
{\rm Si}(x)
&=&
x - x^3/18 + x^5/600 - x^7/35280 + x^9/3265920+\dots
\ee
so
\be
\lim_{x\to 0}
\frac{{\rm Si}(k_2x)-{\rm Si}(k_1x)}{x}
&=&
k_2-k_1.
\ee
Another important property of ${\rm Si}(x)$ is the smallest solution of
\be
{\rm Si}(\infty)-{\rm Si}(x)=0
\ee
which is numerically found to be $x\approx 1.92645$. What this physically means is that for a very large $k_2$ (a property one expects on physical grounds) the potential
\be
-\frac{q_{\rm ph}^2}{4\pi |\bm x|}
\frac{{\rm Si}(k_2|\bm x|)-{\rm Si}(k_1|\bm x|)}{\pi/2} \label{phi +->-}
\ee
changes sign if $|\bm x|\approx 1.92645/k_1$. Oscillating changes of sign are implied by oscillations of Si$(x)$ depicted at Fig.~7. It is interesting that suggestions of analogous changes of sign have been seriously considered in cosmology and astrophysics in the context of gravitational Newton law (see \cite{dark} for a systematized guide over the literature of the subject).

Current experimental data show that exact Coulomb law  is indistinguishable, up to $|\vec x|$ as large as the Earth-Sun distance, from the Yukawa law
\be
-\frac{q_{\rm ph}^2}{4\pi |\vec x|}e^{-|\vec x|/\lambda}\label{Yukawa}
\ee
where $\lambda$ is of the order of the Earth-Moon distance (more than 300000 km, corresponding to the Compton wavelength of a particle whose mass is $10^{-48}{\rm g}<m<10^{-47}{\rm g}$, cf. the review \cite{Coulomb}). The experiments are supposed to test the value of photon's rest mass --- the tacit assumption being that for massless photons and pointlike sources one necessarily gets the Coulomb law. However, we have just seen that the result depends also on quantization procedures --- in my formalism the field is massless and the source is pointlike, and yet the Coulomb law is derived in a generalized form.

Anyway, returning to the data, the Yukawa potential (\ref{Yukawa}) effectively deviates from the Coulomb law for large $|\vec x|$, a behavior controlled in my formula by the infrared cutoff $k_1$. The data from \cite{Coulomb} show that varying $\lambda$ between, roughly, $\lambda_{\rm min}=300000$~km  and infinity we remain within the experimental uncertainty even for $|\vec x|$ as large as some $10^9$~km. The error of determining the exact Coulomb law thus can be defined as
\be
\Delta(\bm x)
=
\frac{1}{|\bm x|}
-
\frac{e^{-|\bm x|\ell/\lambda_{\rm min}}}{|\bm x|}.
\ee
The scale of infrared cutoff is determined by $k_1$, which can be estimated on the basis of
\be
\frac{e^{-|\bm x|\ell/\lambda_{\rm min}}}{|\bm x|}-\Delta(\bm x)
=
2\frac{e^{-|\bm x|\ell/\lambda_{\rm min}}}{|\bm x|}-\frac{1}{|\bm x|}
\leq
\frac{1}{|\bm x|}
\frac{{\rm Si}(\infty)-{\rm Si}(k_1|\bm x|)}{\pi/2}
=
\frac{1}{|\bm x|}
-
\frac{2}{\pi|\bm x|}
{\rm Si}(k_1|\bm x|)
\leq
\frac{1}{|\bm x|}
\nonumber
\ee
at least for $0\ll|\bm x|<10^9$~km. So,
\be
2\frac{e^{-|\bm x|\ell/\lambda_{\rm min}}}{|\bm x|}-2\frac{1}{|\bm x|}
\leq
-
\frac{2}{\pi|\bm x|}
{\rm Si}(k_1|\bm x|)
\leq
0\nonumber,
\ee
and finally
\be
0
\leq
\pi(1-e^{-|\bm x|\ell/\lambda_{\rm min}})-{\rm Si}(k_1|\bm x|)\label{ineq}.
\ee
Fig.~9 shows that for $k_1=2\ell/\lambda_{\rm min}$ (or smaller) inequality (\ref{ineq}) is satisfied for {\it all\/} values of $|\bm x|$.

The conclusion is that even the most precise tests of the Coulomb law available so far do not contradict the possibility that $I_0(\bm k)\neq 1$.
$O_0(\bm k)$ may be a nontrivial function that tends to 0 with $\bm k\to\bm 0$. The analysis I have presented is based on a concrete form of $O_0(\bm k)$ but one should not expect drastically different predictions if $O_0(\bm k)$ decayed in the infrared regime in a smoother way.

Another important test of the Coulomb law would be the energy spectrum of hydrogen-like atoms with the {\it operator\/} potential $\phi_{\rm gC}(\bm x)$. I hope to return to this problem later on in these notes.

\begin{figure}
\includegraphics[width=16cm]{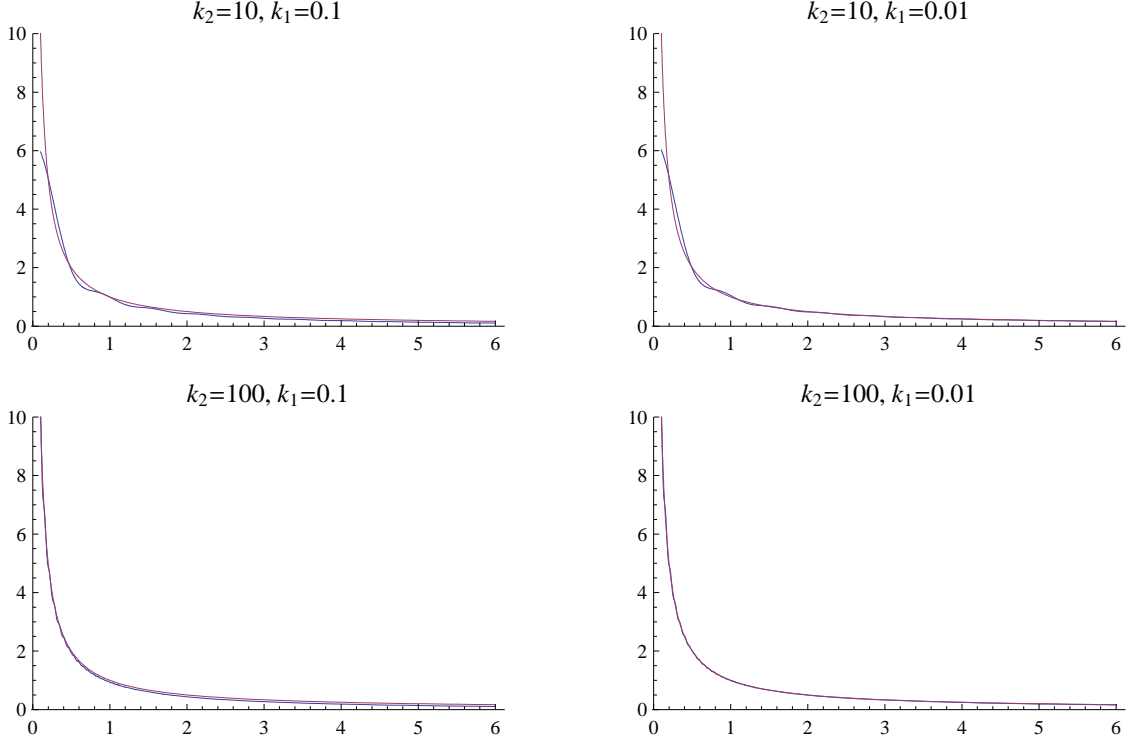}
\caption{Comparison of $1/x$ with $2[{\rm Si}(k_2x)-{\rm Si}(k_1x)]/(\pi x)$ for arbitrarily chosen $k_1$ and $k_2$.}
\end{figure}
\begin{figure}
\includegraphics[width=10cm]{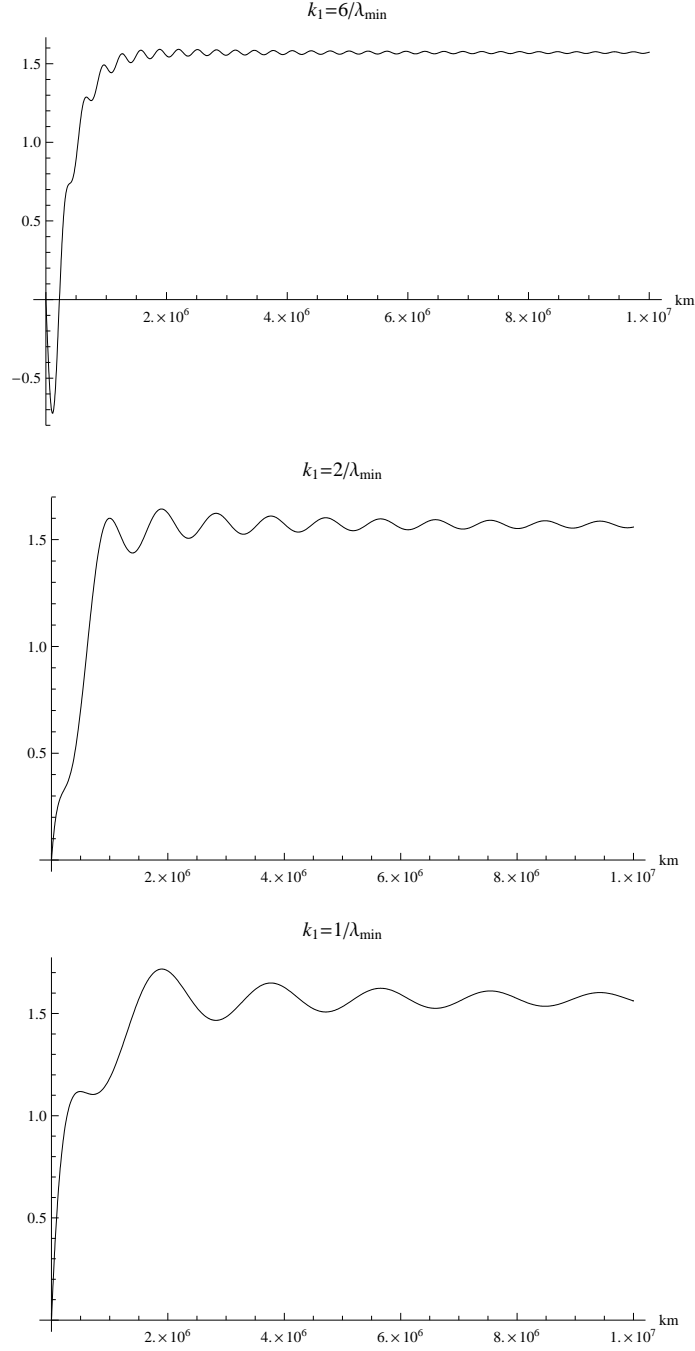}
\caption{Plots of $\pi(1-e^{-|\bm x|\ell/\lambda_{\rm min}})-{\rm Si}(k_1|\bm x|)$ for $\lambda_{\rm min}=300000$~km and different values of $k_1$. For $k_1=2\ell/\lambda_{\rm min}$, or smaller, inequality (\ref{ineq}) is satisfied for all $|\bm x|$.}
\end{figure}

\subsection{Generalized Coulomb law for $O_0(\bm k)$ given by (\ref{O_0-1/k})}

Let us now employ the more realistic wave function (\ref{O_0-1/k}),
\be
|O_0(\bm k)|^2 &=& \frac{2\pi^2y^2}{\lambda^2K_2(2\lambda)} e^{-\frac{\lambda^2}{y\cdot k}-y\cdot k}.\label{O_0-1/k'}
\ee
Probability density (\ref{O_0-1/k'}) achieves its maximum at $y\cdot k=\lambda$, i.e.
\be
Z=\max_{\bm k}\{|O_0(\bm k)|^2\}=\frac{2\pi^2y^2}{\lambda^2K_2(2\lambda)} e^{-2\lambda},
\ee
so the cutoff function is
\be
\chi(\bm k) &=& |O_0(\bm k)|^2/Z=e^{-\frac{\lambda^2}{y\cdot k}-y\cdot k+2\lambda}.
\ee
Accordingly,
\be
q\langle O_0,N|\phi_{\rm gC}(\bm x,N)|O_0,N\rangle
&=&
-q_{\rm ph}^2
\int \frac{d^3k}{(2\pi)^3 |\bm k|^2} \chi(\bm k)
\cos \bm k\cdot \bm x
\nonumber\\
&=&
-q_{\rm ph}^2
\int \frac{d^3k}{(2\pi)^3 |\bm k|^2} e^{-\frac{\lambda^2}{y\cdot k}-y\cdot k+2\lambda}
\cos \bm k\cdot \bm x
\nonumber
\ee
For simplicity consider $y=(y_0,\bm 0)$. Then,
\be
q\langle O_0,N|\phi_{\rm gC}(\bm x,N)|O_0,N\rangle
&=&
-q_{\rm ph}^2
\int \frac{d^3k}{(2\pi)^3 |\bm k|^2} \chi(\bm k)
\cos \bm k\cdot \bm x
\nonumber\\
&=&
-q_{\rm ph}^2
\int \frac{d^3k}{(2\pi)^3 |\bm k|^2} e^{-\frac{\lambda^2}{y_0|\bm k|}-y_0|\bm k|+2\lambda}
\cos \bm k\cdot \bm x
\nonumber\\
&=&
-q_{\rm ph}^2
\int_0^\infty \frac{d\kappa}{(2\pi)^2} \int_0^\pi d\theta \sin\theta e^{-\frac{\lambda^2}{y_0\kappa}-y_0\kappa+2\lambda}
\cos (\kappa|\bm x|\cos\theta)
\nonumber\\
&=&
q_{\rm ph}^2
\int_0^\infty \frac{d\kappa}{(2\pi)^2} \int_1^{-1} du\, e^{-\frac{\lambda^2}{y_0\kappa}-y_0\kappa+2\lambda}
\cos (\kappa|\bm x|u)
\nonumber\\
&=&
q_{\rm ph}^2
\int_0^\infty \frac{d\kappa}{(2\pi)^2}  e^{-\frac{\lambda^2}{y_0\kappa}-y_0\kappa+2\lambda}
\frac{\sin (\kappa|\bm x|u)}{\kappa|\bm x|}\Big|_{u=1}^{-1}
\nonumber\\
&=&
-2q_{\rm ph}^2
\int_0^\infty \frac{d\kappa}{(2\pi)^2}  e^{-\frac{\lambda^2}{y_0\kappa}-y_0\kappa+2\lambda}
\frac{\sin (\kappa|\bm x|)}{\kappa|\bm x|}
\nonumber\\
&=&
-\frac{q_{\rm ph}^2}{2\pi^2|\bm x|}e^{2\lambda}
\int_0^\infty \frac{d\kappa}{\kappa}  e^{-\frac{\lambda^2}{y_0\kappa}-y_0\kappa}
\frac{1}{2}\Big(-ie^{i\kappa|\bm x|}+i e^{-i\kappa|\bm x|}\Big)
\nonumber\\
&=&
i\frac{q_{\rm ph}^2}{4\pi^2|\bm x|}e^{2\lambda}
\int_0^\infty \frac{d\kappa}{\kappa}  e^{-\frac{\lambda^2}{y_0\kappa}-y_0\kappa+i|\bm x|\kappa}+\textrm{c.c.}
\nonumber
\ee
Taking into account the integral representation of the Bessel function
\be
K_0(z) &=& \frac{1}{2}\int_0^\infty \frac{d\kappa}{\kappa}e^{-\kappa-\frac{z^2}{4\kappa}}
=\frac{1}{2}\int_0^\infty \frac{d\kappa}{\kappa}e^{-\frac{1}{\kappa}-\frac{z^2}{4}\kappa}
,
\ee
valid for $\Re(z^2)>0$ (\cite{Watson}, p. 183; $\kappa\leftrightarrow 1/\kappa$ switches between the two forms of the integral)
we find
\be
q\langle O_0,N|\phi_{\rm gC}(\bm x,N)|O_0,N\rangle
&=&
i\frac{q_{\rm ph}^2}{4\pi^2|\bm x|}e^{2\lambda}
\int_0^\infty \frac{d\kappa}{\kappa}  e^{-\frac{\lambda^2}{y_0\kappa}-\lambda^2\frac{y_0\kappa}{\lambda^2}+i|\bm x|\frac{\lambda^2}{y_0}\frac{y_0\kappa}{\lambda^2}}+\textrm{c.c.}
\nonumber\\
&=&
i\frac{q_{\rm ph}^2}{4\pi^2|\bm x|}e^{2\lambda}
\int_0^\infty \frac{d\kappa}{\kappa}  e^{-\frac{1}{\kappa}-\lambda^2\kappa+i|\bm x|\frac{\lambda^2}{y_0}\kappa}+\textrm{c.c.}
\nonumber\\
&=&
i\frac{q_{\rm ph}^2}{4\pi^2|\bm x|}e^{2\lambda}
\int_0^\infty \frac{d\kappa}{\kappa}  e^{-\frac{1}{\kappa}-\big(\lambda^2-i|\bm x|\frac{\lambda^2}{y_0}\big)\kappa}+\textrm{c.c.}
\nonumber
\ee
So here $z^2/4=\lambda^2-i|\bm x|\frac{\lambda^2}{y_0}$, $\Re(z^2)=4\lambda^2>0$, and
\be
q\langle O_0,N|\phi_{\rm gC}(\bm x,N)|O_0,N\rangle
&=&
i\frac{q_{\rm ph}^2}{2\pi^2|\bm x|}e^{2\lambda}
K_0(z)+\textrm{c.c.}
\nonumber\\
&=&
i\frac{q_{\rm ph}^2}{2\pi^2}e^{2\lambda}
\frac{K_0\left(2 \lambda\sqrt{1-\frac{i |\bm x|}{y_0}}  \right)-K_0\left(2\lambda \sqrt{1+\frac{i |\bm x|}{y_0}}  \right)}{|\bm x|}
\ee
The square root symbol denotes the principal branch of $z^{1/2}$, but what about the other branch? The next figure shows that it is the principal branch that gives a function that continuously deforms into the Coulomb field if one replaces $|O_0(\bm k)|^2$ by 1.
\begin{figure}
\includegraphics[width=8cm]{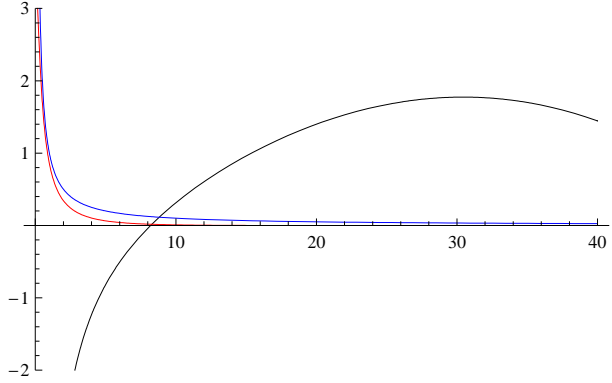}
\caption{The blue line represents $1/x$. The red and black ones correspond to $\lambda=y_0=1/4$, but the red plot involves the principal branch of the square root function.}
\end{figure}
The long-range behavior of $q\langle O_0,N|\phi_{\rm gC}(\bm x,N)|O_0,N\rangle$ is controlled by the term with $-\lambda^2/y_0$. A numerical analysis shows that $q\langle O_0,N|\phi_{\rm gC}(\bm x,N)|O_0,N\rangle$ is experimentally indistinguishable from the Yukawa law if $y_0\ell/\lambda^2\sim 10^{11}$ km (or greater), so is not smaller than the size of the Solar System, where $y_0$ is the length that controls the UV regime. Taking, for example, $y_0\ell\sim 10^{-38}$ km (the Planck length) one finds $\lambda^2\sim 10^{-49}$, or smaller.

Now comes an important observation. We know that the modified potential (\ref{phi +->-})
changes sign if $|\bm x|\approx 1.92645/k_1$. The new potential also changes its sign at the distance determined by the first zero
of
\be
K_0\left(2 \lambda\sqrt{1-\frac{i |\bm x|}{y_0}}  \right)-K_0\left(2\lambda \sqrt{1+\frac{i |\bm x|}{y_0}}  \right)\label{KK->0}.
\ee
Inserting $y_0\ell= 10^{-38}$ km and $\lambda^2= 10^{-49}$ into (\ref{KK->0}) we find that the smallest (non-zero) distance where (\ref{KK->0}) vanishes is
$|\bm x|\ell\approx 3,83\times 10^{11}$~km~$\approx 2560.2$ AU (astronomical units). Aphelion of Pluto is of the order of 50 AU, solar heliopause ends at 230 AU from the Sun...

\subsection{``Photon statistics"}

Averages of generalized Coulomb field did not depend on the parameter $N$ of the representation of HOLA. $N$ becomes essential for the  $n$-``scalar photon" statistics.

Let us begin with the unitary operator
\be
e^{-i\int_{t_0}^{t}d\tau\,{V}(\tau)}
&=&
\exp\Big(-iq\int_{t_0}^{t}d\tau\,\phi(\tau,\bm 0,N)\Big)
\nonumber\\
&=&
\exp\Bigg(
-iq\int_{t_0}^{t}d\tau\int dk
\Big(
c(\bm k,N)e^{-i|\bm k|\tau}+c(\bm k,N)^\dag e^{i|\bm k|\tau}
\Big)
\Bigg)
\nonumber\\
&=&
\exp\Bigg(
-iq\int dk
\Big(
c(\bm k,N)\frac{e^{-i|\bm k|t}-e^{-i|\bm k|t_0}}{-i|\bm k|}+c(\bm k,N)^\dag \frac{e^{i|\bm k|t}-e^{i|\bm k|t_0}}{i|\bm k|}
\Big)
\Bigg)
\nonumber\\
&=&
\exp
\int dk
\Big(
\alpha_{t,t_0}(\bm k)
c(\bm k,N)^\dag-\overline{\alpha_{t,t_0}(\bm k)}c(\bm k,N)
\Big)
\nonumber\\
&=&
{\cal D}_0(\alpha_{t,t_0},N).
\ee
This is a coherent-state displacement operator with
\be
\alpha_{t,t_0}(\bm k)
&=&
q\frac{e^{i|\bm k|t_0}-e^{i|\bm k|t}}{|\bm k|}.
\ee
Repeating calculations analogous to those that led to (\ref{p(n,N)}) we obtain statistics of ``scalar photons" produced by a pointlike charge, with vacuum initial condition at $t=t_0$,
\be
p(n,N)
&=&
\langle O_0,N|\exp\Big(-\int dk |\alpha_{t,t_0}(\bm k)|^2 I_0(\bm k,N)\Big)
\frac{1}{n!}\Big(\int dk_1 |\alpha_{t,t_0}(\bm k_1)|^2 I_0(\bm k_1,N)\Big)^n|O_0,N\rangle
\nonumber\\
&=&
\frac{1}{n!}\frac{d^n}{d\lambda^n}
\langle O_0,N|\exp\Big(\lambda\int dk |\alpha_{t,t_0}(\bm k)|^2 I_0(\bm k,N)\Big)|O_0,N\rangle
\Big|_{\lambda=-1}.
\label{p(n,N)1}
\ee
Generating function occurring in (\ref{p(n,N)1}) can be further transformed, leading to Kolmogorov-Nagumo average
\be
\langle O_0,N|\exp\Big(\lambda\int dk |\alpha_{t,t_0}(\bm k)|^2 I_0(\bm k,N)\Big)|O_0,N\rangle
&=&
\Big(\int dk|O_0(\bm k)|^2e^{\lambda\frac{1}{N}|\alpha_{t,t_0}(\bm k)|^2} \Big)^N.
\ee
The limiting case
\be
\lim_{N\to\infty}
\Big(\int dk|O_0(\bm k)|^2e^{\lambda\frac{1}{N}|\alpha_{t,t_0}(\bm k)|^2} \Big)^N
&=&
\exp\Big(\lambda\int dk|O_0(\bm k)|^2|\alpha_{t,t_0}(\bm k)|^2\Big)
\ee
is a regularized Poisson distribution. Explicitly,
\be
|\alpha_{t,t_0}(\bm k)|^2
&=&
q^2\frac{2-e^{i|\bm k|(t_0-t)}-e^{-i|\bm k|(t_0-t)}}{|\bm k|^2}=
q^2\frac{2-2\cos\big(|\bm k|(t-t_0)\big)}{|\bm k|^2}\nonumber\\
&=&
q^2\frac{\sin^2\big(|\bm k|(t-t_0)/2\big)}{\big(|\bm k|/2\big)^2}
\ee
and
\be
\exp\Big(\lambda\int dk|O_0(\bm k)|^2|\alpha_{t,t_0}(\bm k)|^2\Big)
&=&
\exp\Bigg(\lambda q^2\int dk|O_0(\bm k)|^2\frac{\sin^2\big(|\bm k|(t-t_0)/2\big)}{\big(|\bm k|/2\big)^2}\Bigg)
\nonumber\\
&=&
\exp\Bigg(\lambda q_{\rm ph}^2\int dk\chi(\bm k)\frac{\sin^2\big(|\bm k|(t-t_0)/2\big)}{\big(|\bm k|/2\big)^2}\Bigg)
\label{gen-f}
\ee
Let us take a closer look at the exponent in (\ref{gen-f}). Since,
\be
0\leq
\frac{\sin^2\big(|\bm k|\Delta t/2\big)}{(|\bm k|/2)^2}
=
\Delta t^2\frac{\sin^2\big(|\bm k|\Delta t/2\big)}{(|\bm k|\Delta t/2)^2}
\leq \Delta t^2
\ee
the integral is finite at least for finite $\Delta t=t-t_0$,
\be
\int dk|O_0(\bm k)|^2\frac{\sin^2\big(|\bm k|(t-t_0)/2\big)}{(|\bm k|/2)^2}
\leq
(t-t_0)^2\int dk|O_0(\bm k)|^2=(t-t_0)^2.
\ee
In the ordinary approach we would obtain the same formula (\ref{gen-f}) but with $|O_0(\bm k)|^2=1$, i.e.
\be
\int \frac{d^3k}{(2\pi)^3 2|\bm k|}\frac{\sin^2(|\bm k|\Delta t/2)}{(|\bm k|/2)^2}
&=&
\frac{16\pi}{2(2\pi)^3}\int_0^\infty d\kappa\frac{\sin^2(\kappa\Delta t/2)}{\kappa}.
\ee
The latter integral vanishes for $\Delta t=0$, but becomes divergent for $\Delta t>0$. More precisely, for $0<k_1<k_2$,
\be
\int_{k_1}^{k_2} d\kappa\,\frac{\sin^2 \kappa}{\kappa}=\frac{{\rm Ci}(2k_1)-\ln k_1-{\rm Ci}(2k_2)+\ln k_2}{2}.
\ee
The infrared limit is finite,
\be
\lim_{k_1\to 0}\big({\rm Ci}(2k_1)-\ln k_1\big)
&=&
\gamma+\ln 2
\ee
(Ci and $\gamma\approx 0.577216$ are, respectively, the cosine integral function and the Euler constant). The infinity comes from the ultraviolet divergency of $\ln k_2$. An infrared divergency would, however, also appear if the charge was accelerating (for details of standard calculations cf. \cite{IZBB} and Sec. 4-1-2 in \cite{IZ}; analogous results for $N$-representations of HOLA and true electromagnetic fields are described in detail in \cite{V}). One concludes that the standard approach to pointlike classical sources leads to mathematically ill defined objects such as $e^{\lambda \infty}$ and thus is mathematically inconsistent.

In ``my" formalism one finds that probability of finding $n$ ``scalar photons" is given by the vacuum average of POVM
\be
\Pi(n,N)
&=&
\exp\Big(-\int dk |\alpha_{t,t_0}(\bm k)|^2 I_0(\bm k,N)\Big)
\frac{1}{n!}\Big(\int dk_1 |\alpha_{t,t_0}(\bm k_1)|^2 I_0(\bm k_1,N)\Big)^n
\nonumber\\
&=&
\frac{1}{n!}\frac{d^n}{d\lambda^n}
\exp\Big(\lambda\int dk |\alpha_{t,t_0}(\bm k)|^2 I_0(\bm k,N)\Big)
\Big|_{\lambda=-1},\\
\sum_{n=0}^\infty\Pi(n,N)
&=&I_0(N).
\ee
This POVM is uniquely and well defined, but its average depends on the choice of vacuum, of course.

Now let us discuss the issue of the limit $t_0\to\pm\infty$. Let us return to
\be
\int dk|O_0(\bm k)|^2|\alpha_{t,t_0}(\bm k)|^2
&=&
q^2\int dk|O_0(\bm k)|^2
\frac{2-2\cos\big(|\bm k|(t-t_0)\big)}{|\bm k|^2}
\nonumber\\
&=&
2q^2\int \frac{d^3k}{(2\pi)^3 2|\bm k|}\frac{|O_0(\bm k)|^2}{|\bm k|^2}
\nonumber\\
&\pp=&
-
2q^2\int \frac{d^3k}{(2\pi)^3 2|\bm k|}|O_0(\bm k)|^2
\frac{\cos(|\bm k|t)\cos(|\bm k|t_0)}{|\bm k|^2}
\nonumber\\
&\pp=&
-
2q^2\int \frac{d^3k}{(2\pi)^3 2|\bm k|}|O_0(\bm k)|^2
\frac{\sin(|\bm k|t)\sin(|\bm k|t_0)}{|\bm k|^2}.
\ee
If the first integral
\be
\int \frac{d^3k}{(2\pi)^3 2|\bm k|}\frac{|O_0(\bm k)|^2}{|\bm k|^2}\nonumber
\ee
is finite (compare (\ref{boundary k=0'})) then
$|O_0(\bm k)|^2/|\bm k|^3$
satisfies assumptions of the Riemann-Lebesgue lemma, and
\be
\lim_{t_0\to\pm\infty}\int dk|O_0(\bm k)|^2|\alpha_{t,t_0}(\bm k)|^2
&=&
2q^2\int \frac{d^3k}{(2\pi)^3 2|\bm k|}\frac{|O_0(\bm k)|^2}{|\bm k|^2}
\nonumber.
\ee
\section{Preliminaries on radiative corrections to energy levels}

Let us consider a Hamiltonian $H=H_0+qV$. $q$ is a parameter introduced for later convenience, and $V$ is an operator.
Assuming that we know a solution of
\be
H_0|E_0\rangle
&=&
E_0|E_0\rangle,
\ee
we would like to solve
\be
H|E\rangle
&=&
E|E\rangle,
\ee
in such a way that $\lim_{q\to 0}|E\rangle =|E_0\rangle$. The trick is to begin with the analogue of interaction picture dynamics \cite{GL}
\be
i\frac{d}{dt} U_1(t,t_0;\epsilon)
&=&
qV(t)e^{-\epsilon |t|}U_1(t,t_0;\epsilon),\label{U epsilon}
\ee
where in addition to the usual time dependence $V(t)=e^{iH_0 t}V e^{-iH_0 t}$ we add an explicit time dependent factor $e^{-\epsilon |t|}$.
Solutions of (\ref{U epsilon}) can be expressed in the usual way as
\be
U_1(t,t_0;\epsilon)
&=&
\sum_{n=0}^\infty
\frac{(-iq)^n}{n!}\int_{t_0}^{t}d\tau_1 \dots \int_{t_0}^{t}d\tau_n\,e^{-\epsilon(|\tau_1|+\dots+|\tau_n|)}
T\Big(V(\tau_1)\dots V(\tau_n)\Big).
\ee
Now let $t_0<0$,
\be
U_1(0,t_0;\epsilon)
&=&
\sum_{n=0}^\infty
\frac{(-iq)^n}{n!}\int_{t_0}^{0}d\tau_1 \dots \int_{t_0}^{0}d\tau_n\,e^{\epsilon(\tau_1+\dots+\tau_n)}
T\Big(V(\tau_1)\dots V(\tau_n)\Big),
\ee
and consider
\be
(H_0-E_0)U_1(0,t_0;\epsilon)|E_0\rangle
&=&
[H_0,U_1(0,t_0;\epsilon)]|E_0\rangle
\nonumber\\
&=&
\sum_{n=0}^\infty
\frac{(-iq)^n}{n!}\int_{t_0}^{0}d\tau_1 \dots d\tau_n\,e^{\epsilon\sum_{j=1}^n\tau_j}
[H_0,T\big(V(\tau_1)\dots V(\tau_n)\big)]|E_0\rangle
\nonumber\\
&=&
\sum_{n=1}^\infty
\frac{(-iq)^n}{n!}\int_{t_0}^{0}d\tau_1 \dots d\tau_n\,e^{\epsilon\sum_{j=1}^n\tau_j}
\underbrace{[H_0,T\big(V(\tau_1)\dots V(\tau_n)\big)]}_{(*)}|E_0\rangle
\nonumber
\ee
The commutator
\be
(*)
&=&
\sum_{(\tau_{i_1},\dots,\tau_{i_n})}
\theta(\tau_{i_1}-\tau_{i_2})\dots \theta(\tau_{i_{n-1}}-\tau_{i_n})
[H_0,V(\tau_{i_1})\dots V(\tau_{i_n})]\nonumber\\
&=&
\sum
\theta(\tau_{i_1}-\tau_{i_2})\dots \theta(\tau_{i_{n-1}}-\tau_{i_n})
\Big(
[H_0,V(\tau_{i_1})]\dots V(\tau_{i_n})
+
\dots
+
V(\tau_{i_1})\dots [H_0,V(\tau_{i_n})]
\Big)
\nonumber\\
&=&
-i\sum_{(\tau_{i_1},\dots,\tau_{i_n})}
\theta(\tau_{i_1}-\tau_{i_2})\dots \theta(\tau_{i_{n-1}}-\tau_{i_n})
\sum_{j=i_1}^{i_n}\frac{\partial}{\partial \tau_j}V(\tau_{i_1})\dots V(\tau_{i_n})
\nonumber\\
&=&
-i\frac{\partial}{\partial \tau}\sum_{(\tau_{i_1},\dots,\tau_{i_n})}
\theta(\tau+\tau_{i_1}-\tau-\tau_{i_2})\dots \theta(\tau+\tau_{i_{n-1}}-\tau-\tau_{i_n})
V(\tau+\tau_{i_1})\dots V(\tau+\tau_{i_n})\Big|_{\tau=0}
\nonumber\\
&=&
-i\frac{\partial}{\partial \tau}T\Big(V(\tau+\tau_1)\dots V(\tau+\tau_n)\Big)\Big|_{\tau=0},
\ee
so,
\be
&{}&
(H_0-E_0)U_1(0,t_0;\epsilon)|E_0\rangle
\nonumber\\
&{}&=
\sum_{n=1}^\infty
\frac{(-iq)^n}{n!}
\frac{1}{i}\frac{\partial}{\partial \tau}
\int_{t_0}^{0}d\tau_1 \dots d\tau_n\,e^{\epsilon\sum_{j=1}^n\tau_j}
T\Big(V(\tau+\tau_1)\dots V(\tau+\tau_n)\Big)\Big|_{\tau=0}|E_0\rangle
\nonumber\\
&{}&=
\sum_{n=1}^\infty
\frac{(-iq)^n}{n!}
\frac{1}{i}\frac{\partial}{\partial \tau}e^{-n\epsilon\tau}
\int_{t_0+\tau}^{\tau}d\tau_1 \dots d\tau_n\,e^{\epsilon\sum_{j=1}^n\tau_j}
T\Big(V(\tau_1)\dots V(\tau_n)\Big)\Big|_{\tau=0}|E_0\rangle
\nonumber
\ee
\be
&{}&=
\sum_{n=1}^\infty
\frac{(-iq)^n}{n!}
\frac{-n\epsilon}{i}
\int_{t_0}^{0}d\tau_1 \dots d\tau_n\,e^{\epsilon\sum_{j=1}^n\tau_j}
T\Big(V(\tau_1)\dots V(\tau_n)\Big)|E_0\rangle
\nonumber\\
&{}&\pp=
+
\sum_{n=1}^\infty
\frac{(-iq)^n}{n!}
\frac{1}{i}\frac{\partial}{\partial \tau}
\int_{t_0+\tau}^{\tau}d\tau_1 \dots d\tau_n\,e^{\epsilon\sum_{j=1}^n\tau_j}
T\Big(V(\tau_1)\dots V(\tau_n)\Big)\Big|_{\tau=0}|E_0\rangle
\nonumber\\
&{}&=
q\epsilon\sum_{n=1}^\infty
\frac{(-iq)^{n-1}}{(n-1)!}
\int_{t_0}^{0}d\tau_1 \dots d\tau_n\,e^{\epsilon\sum_{j=1}^n\tau_j}
T\Big(V(\tau_1)\dots V(\tau_n)\Big)|E_0\rangle
\nonumber\\
&{}&\pp=
+
\frac{1}{i}\frac{\partial}{\partial \tau}
U_1(\tau,\tau+t_0;\epsilon)\Big|_{\tau=0}|E_0\rangle
\nonumber\\
&{}&=
q\epsilon i\frac{\partial}{\partial q}\sum_{n=0}^\infty
\frac{(-iq)^{n}}{n!}
\int_{t_0}^{0}d\tau_1 \dots d\tau_n\,e^{\epsilon\sum_{j=1}^n\tau_j}
T\Big(V(\tau_1)\dots V(\tau_n)\Big)|E_0\rangle
\nonumber\\
&{}&\pp=
-
qV(\tau)e^{-\epsilon|\tau|}U_1(\tau,\tau+t_0;\epsilon)\Big|_{\tau=0}|E_0\rangle
+
qU_1(\tau,\tau+t_0;\epsilon)V(\tau+t_0)e^{-\epsilon|\tau+t_0|}\Big|_{\tau=0}|E_0\rangle
\nonumber\\
&{}&=
q\epsilon i\frac{\partial}{\partial q}U_1(0,t_0;\epsilon)|E_0\rangle
\nonumber\\
&{}&\pp=
-
qV(0)U_1(0,t_0;\epsilon)|E_0\rangle
+
qU_1(0,t_0;\epsilon)V(t_0)e^{-\epsilon|t_0|}|E_0\rangle
\nonumber
\ee
We have thus arrived at
\be
(
\underbrace{H_0+qV(0)}_H-E_0)U_1(0,t_0;\epsilon)|E_0\rangle
&=&
q\epsilon i\frac{\partial}{\partial q}U_1(0,t_0;\epsilon)|E_0\rangle
+
qU_1(0,t_0;\epsilon)V(t_0)e^{-\epsilon|t_0|}|E_0\rangle
\nonumber\\
\ee
Limits $\epsilon\to 0$ and $t_0\to-\infty$ do not commute. One can show \cite{Brouder} that for some interaction Hamiltonians and some eigenvectors $|E_0\rangle$ the limit (taken in an appropriate way)
\be
\lim_{\epsilon\to 0_+}U_1(0,-\infty;\epsilon)|E_0\rangle
\ee
is an eigenvector of $H=H_0+qV$. Denoting the corresponding eigenvalue by $E$, we find
\be
\Delta E
&=&
E-E_0
\nonumber\\
&=&
\lim_{\epsilon\to 0_+}\frac{\langle E_0|(H-E_0)U_1(0,-\infty;\epsilon)|E_0\rangle}{\langle E_0|U_1(0,-\infty;\epsilon)|E_0\rangle}
\nonumber\\
&=&
q\lim_{\epsilon\to 0_+}\epsilon i\frac{\partial}{\partial q}\ln\langle E_0|U_1(0,-\infty;\epsilon)|E_0\rangle.
\ee
Let us check what kind of $\Delta E$ one obtains in the case of a classical pointlike charge if $|E_0\rangle$ is a vacuum state $|O_0,N\rangle$ typical of a massless $N$-representation of HOLA.

\section{Radiative energy correction for a pointlike classical source in electromagnetic vacuum}

Repeating standard calculations with
\be
q{V}(t)e^{-\epsilon |t|}
&=&
q\int_{\mathbb{R}^3}d^3x\, \delta^{(3)}(\bm x)\phi(t,\bm x)e^{-\epsilon |t|}
\nonumber\\
&=&
q \phi(t,\bm 0)e^{-\epsilon |t|}\\
&=&
\underbrace{q\int dk\, c(\bm k)e^{-i|\bm k|t}e^{-\epsilon |t|}}_{q{V}_-(t)}
+
\underbrace{q\int dk\, c(\bm k)^\dag e^{i|\bm k|t}e^{-\epsilon |t|}}_{q{V}_+(t)}.
\ee
we obtain (for $t_0\leq 0$)
\be
U_1(0,t_0;\epsilon)
&=&
\exp\Big(q\int dk\, c(\bm k)^\dag \frac{1-e^{i|\bm k|t_0}e^{\epsilon t_0}}{-|\bm k|-i\epsilon}\Big)
\exp\Big(q\int dk\, c(\bm k)\frac{1-e^{-i|\bm k|t_0}e^{\epsilon t_0}}{|\bm k|-i\epsilon}\Big)
\nonumber\\
&\pp=&\times\exp\Bigg(-q^2\int dk \,\frac{I_0(\bm k)}{i|\bm k|-\epsilon}
\Big(
\frac{e^{2\epsilon t_0}-1}{2\epsilon}
+
\frac{e^{i|\bm k|t_0}e^{\epsilon t_0}-e^{2\epsilon t_0}}{i|\bm k|+\epsilon}
\Big)
\Bigg)
\ee
The two limits can be explicitly compared
\be
U_1(0,-\infty;\epsilon)
&=&
\exp\Big(q\int dk\, c(\bm k)^\dag \frac{-|\bm k|+i\epsilon}{|\bm k|^2+\epsilon^2}\Big)
\exp\Big(q\int dk\, c(\bm k)\frac{|\bm k|+i\epsilon}{|\bm k|^2+\epsilon^2}\Big)
\nonumber\\
&\pp=&\times\exp\Bigg(-q^2\int dk \,\frac{I_0(\bm k)}{|\bm k|^2+\epsilon^2}
\Big(\frac{i|\bm k|}{2\epsilon}
+
\frac{1}{2}
\Big)
\Bigg),
\\
U_1(0,t_0;0)
&=&
\exp\Big(-iq\int dk\, c(\bm k)^\dag \frac{1-e^{i|\bm k|t_0}}{i|\bm k|}\Big)
\exp\Big(-iq\int dk\, c(\bm k)\frac{1-e^{-i|\bm k|t_0}}{-i|\bm k|}\Big)
\nonumber\\
&\pp=&\times\exp\Bigg(-q^2\int dk \,\frac{I_0(\bm k)}{i|\bm k|}
\Big(
-t_0
+
\frac{e^{i|\bm k|t_0}-1}{i|\bm k|}
\Big)
\Bigg).
\ee
In the $N$-representation of HOLA
\be
U_1(0,-\infty;\epsilon,N)
&=&
\exp\Big(q\int dk\, c(\bm k,N)^\dag \frac{-|\bm k|+i\epsilon}{|\bm k|^2+\epsilon^2}\Big)
\exp\Big(q\int dk\, c(\bm k,N)\frac{|\bm k|+i\epsilon}{|\bm k|^2+\epsilon^2}\Big)
\nonumber\\
&\pp=&\times\exp\Bigg(-q^2\int dk \,\frac{I_0(\bm k,N)}{|\bm k|^2+\epsilon^2}
\Big(\frac{i|\bm k|}{2\epsilon}
+
\frac{1}{2}
\Big)
\Bigg)
\nonumber
\ee
one gets
\be
\langle O_0,N|U_1(0,-\infty;\epsilon,N)|O_0,N\rangle
&=&
\langle O_0,N|\exp\Bigg(-q^2\int dk \,\frac{I_0(\bm k,N)}{|\bm k|^2+\epsilon^2}
\Big(\frac{i|\bm k|}{2\epsilon}
+
\frac{1}{2}
\Big)
\Bigg)|O_0,N\rangle
\nonumber\\
&=&
\langle O_0,1|\exp\Bigg(-\frac{1}{N}q^2\int dk \,\frac{I_0(\bm k,1)}{|\bm k|^2+\epsilon^2}
\Big(\frac{i|\bm k|}{2\epsilon}
+
\frac{1}{2}
\Big)
\Bigg)|O_0,1\rangle^N
\nonumber\\
&=&
\Bigg[\int dk \,|O_0(\bm k)|^2\exp\Bigg(-\frac{1}{N}q^2\frac{1}{|\bm k|^2+\epsilon^2}
\Big(\frac{i|\bm k|}{2\epsilon}
+
\frac{1}{2}
\Big)
\Bigg)\Bigg]^N.
\nonumber\\\label{593}
\ee
With $N\to\infty$ (the Shannon limit)
\be
\langle O_0,N|U_1(0,-\infty;\epsilon,N)|O_0,N\rangle
&\to&
\exp\Bigg(-q^2\int dk \,|O_0(\bm k)|^2\frac{1}{|\bm k|^2+\epsilon^2}
\Big(\frac{i|\bm k|}{2\epsilon}
+
\frac{1}{2}
\Big)
\Bigg)
\nonumber
\ee
so that
\be
&{}&
q\lim_{\epsilon\to 0_+}\epsilon i\frac{\partial}{\partial q}
(-q^2)\int dk \,|O_0(\bm k)|^2\frac{1}{|\bm k|^2+\epsilon^2}
\Big(\frac{i|\bm k|}{2\epsilon}
+
\frac{1}{2}
\Big)
\nonumber\\
&{}&
=
-2iq^2\lim_{\epsilon\to 0_+}
\int dk \,|O_0(\bm k)|^2\frac{1}{|\bm k|^2+\epsilon^2}
\Big(\frac{i|\bm k|}{2}
+
\frac{1}{2}\epsilon
\Big)
\nonumber\\
&{}&
=
q^2 \int \frac{d^3k}{(2\pi)^3 2|\bm k|}\frac{|O_0(\bm k)|^2}{|\bm k|}.
\nonumber
\ee
Comparison with generalized Coulomb field (\ref{gen Coulomb})
\be
\phi_{\rm gC}(\bm x)
&=&
-
q\int \frac{d^3k}{(2\pi)^3 |\bm k|^2} I_0(\bm k)
\cos \bm k\cdot \bm x,\\
\langle O_0,N|\phi_{\rm gC}(\bm x,N)|O_0,N\rangle
&=&
-
q\int \frac{d^3k}{(2\pi)^3 |\bm k|^2} |O_0(\bm k)|^2
\cos \bm k\cdot \bm x
\ee
shows that for large $N$ (in Shannon asymptotics)
\be
q\lim_{\epsilon\to 0_+}\epsilon i\frac{\partial}{\partial q}\ln\langle O_0,N|U_1(0,-\infty;\epsilon,N)|O_0,N\rangle
&\approx&
-\frac{q}{2}\langle O_0,1|\phi_{\rm gC}(\bm 0,1)|O_0,1\rangle
\\
&=&\int dk \,|O_0(\bm k)|^2\Delta E(\bm k) \label{Delta E(k)}
\\
&=&
q_{\rm ph}^2 \int \frac{d^3k}{(2\pi)^3 2|\bm k|^2}\chi(\bm k)
\ee
where
\be
\Delta E(\bm k)
&=&
\frac{q^2}{|\bm k|}
\ee
can be regarded as a radiative correction to vacuum energy typical of a Fourier component of the field with the wave vector $\bm k$. As usual,
$q_{\rm ph}=q\sqrt{Z}$ is the physical charge, and $\chi(\bm k)$ the cutoff function.

The exact result for finite $N$ can be directly derived from (\ref{593}).

It is clear that (\ref{Delta E(k)}) is not a single energy shift but an average. The reason is that the construction from \cite{GL} assumes that $E_0$ is a non-degenerate eigenvalue from discrete part of spectrum, whereas we deal with the whole infinite-dimensional vacuum subspace with continuous spectrum of wave vectors. The case of degenerate $E_0$ is rigorously treated in \cite{Brouder} but only in the case of finite-dimensional subspaces spanned by normalizable eigenvectors of degenerate eigenvalues. None of these assumptions applies to our case. The problem is mathematically complicated and, apparently, not fully understood as yet.

However, it must be stressed that the result does not mean that our radiative correction to the vacuum energy ``depends on cutoff". The correction to the {\it spectrum\/} of the Hamiltonian is given by $\Delta E(\bm k)$ and not by (\ref{Delta E(k)}) which is the average evaluated in $|O_0,N\rangle$.
\medskip

\noindent
{\it Remark\/}: In order to appreciate the conceptual difference between the usual regularization schemes and my ``regularization by quantization" let me recall the case of regularization of $\sum_{n=0}^\infty n$. Within the standard paradigm the sum would mean a divergent {\it eigenvalue corection\/} that arises from zero-point energy. Its standard regularization is based on replacing $\sum_{n=0}^\infty n$ by $\sum_{n=0}^\infty n \chi_n$. Regularization by quantization would replace $\sum_{n=0}^\infty n$ by $\sum_{n=0}^\infty n|n\rangle\langle n|$. Now the ``regularized" correction to the {\it eigenvalue\/} is not $\sum_{n=0}^\infty n \chi_n$, but simply $n$.$\blacktriangle$

In standard quantization, with one-dimensional vacuum space, the result would become
\be
\langle 0|U_1(0,-\infty;\epsilon)|0\rangle
&=&
\exp\Bigg(-q^2\int dk \,\frac{1}{|\bm k|^2+\epsilon^2}
\Big(\frac{i|\bm k|}{2\epsilon}
+
\frac{1}{2}
\Big)
\Bigg)
\ee
and
\be
q\lim_{\epsilon\to 0_+}\epsilon i\frac{\partial}{\partial q}
(-q^2)
\int dk \,\frac{1}{|\bm k|^2+\epsilon^2}
\Big(\frac{i|\bm k|}{2\epsilon}
+
\frac{1}{2}
\Big)
&=&
q^2
\int \frac{d^3k}{(2\pi)^3 2|\bm k|^2}
\ee
would be the divergent contribution from (one half) of the Coulomb energy at location of the charge. This average would indeed coincide with the divergent Coulomb correction to the eigenvalue.

\section{Radiative energy correction for a pointlike classical charge located in front of a plane with Dirichlet boundary condition for both the field and the vacuum}

The example is an interesting prelude to more difficult problems of the Casimir force.
Let us consider the plane $\Sigma=\{\bm x\in \mathbb{R}^3, \,\bm x=(x,y,L)\}$ with Dirichlet boundary conditions
\be
\phi(x_0,\bm x) &=& 0, \quad \bm x\in \Sigma,\\
O_0(x_0,\bm x) &=& 0, \quad \bm x\in \Sigma,
\ee
for solutions of $\Box \phi(x)=0$ and $\Box O_0(x)=0$. Leaving a more detailed analysis for the next section let us give here directly the solutions,
\be
O_0(x)
&=&
C_1\int dk\, O_0(\bm{k})e^{-ikx/2}\Big(1-e^{-ik_z(z-L)}\Big)+{\rm c.c.}
\label{O zC}\\
&=&
C_1\int dk\, \Big(O_0(k_x,k_y,k_z)-O_0(k_x,k_y,-k_z)e^{-ik_zL}\Big)e^{-ikx/2}+{\rm c.c.}
\nonumber\\
&=&
\int dk\, \tilde O_0(\bm k)e^{-ikx/2}+{\rm c.c.}
\nonumber\\
\phi(x)
&=&
C_2\int dk\, c(\bm{k})e^{-ikx}\Big(1-e^{-2ik_z(z-L)}\Big)+{\rm H.c.}
\label{phi zC}\\
&=&
C_2\int dk\, \Big(c(k_x,k_y,k_z)-c(k_x,k_y,-k_z)e^{-2ik_zL}\Big)
e^{-ikx}+{\rm H.c.}
\nonumber\\
&=&
\int dk\, \Big(c'(\bm{k})e^{-ikx}+c'(\bm{k})^\dag e^{ikx}\Big),
\ee
where
\be
\tilde O_0(k_x,k_y,k_z)
&=&
C_1\Big(O_0(k_x,k_y,k_z)-O_0(k_x,k_y,-k_z)e^{-ik_zL}\Big),\label{sym1}\\
c'(k_x,k_y,k_z)
&=&
C_2
\Big(c(k_x,k_y,k_z)-c(k_x,k_y,-k_z)e^{-2ik_zL}\Big),\label{sym2}
\ee
and $C_1$, $C_2$ are normalization constants. Eqs. (\ref{sym1})--(\ref{sym2}) are equivalent to symmetry constraints
\be
c'(k_x,k_y,-k_z)
&=&
C_2
\Big(c(k_x,k_y,-k_z)-c(k_x,k_y,k_z)e^{2ik_zL}\Big)\nonumber\\
&=&
-e^{2ik_zL}
C_2
\Big(-c(k_x,k_y,-k_z)e^{-2ik_zL}+c(k_x,k_y,k_z)\Big)\nonumber\\
&=&
-e^{2ik_zL}c'(k_x,k_y,k_z),\\
\tilde O_0(k_x,k_y,-k_z)
&=&
-e^{ik_zL}\tilde O_0(k_x,k_y,k_z).
\ee

The vacuum state corresponding to $O_0(x)$ is (for $N=1$)
\be
|\tilde O_0,1\rangle
&=&
\int dk \,\tilde O_0(\bm k)|\bm k,0\rangle
\\
&=&
C_1\int dk \,\Big(O_0(k_x,k_y,k_z)-O_0(k_x,k_y,-k_z)e^{-ik_zL}\Big)|k_x,k_y,k_z,0\rangle,
\ee
with the usual conditions imposed on $\tilde O_0(\bm k)$ (square-integrability, vanishing at $\bm k=\bm 0$ faster than any power of $|\bm k|$), but no additional restrictions being assumed about $O_0(\bm k)$.

Since $c'(k_x,k_y,-k_z)=-e^{2ik_zL}c'(k_x,k_y,k_z)$
the degrees of freedom represented by the primed operators are not all independent, as opposed to those represented by $c(\bm{k})$. This is why the $N$ representation of HOLA is assumed to be satisfied by the latter operators.

With these preliminaries we are in position to compute the radiative correction produced by
\be
q{V}(t)e^{-\epsilon |t|}
&=&
q \phi(t,\bm 0)e^{-\epsilon |t|}\\
&=&
\underbrace{C_2q\int dk\, c(\bm k)e^{-i|\bm k|t}\Big(1-e^{2ik_zL}\Big)e^{-\epsilon |t|}}_{q{V}_-(t)}
\nonumber\\
&\pp=&
+
\underbrace{C_2q\int dk\, c(\bm k)^\dag e^{i|\bm k|t}\Big(1-e^{-2ik_zL}\Big)e^{-\epsilon |t|}}_{q{V}_+(t)}.
\ee
In the Shannon limit we get
\be
\langle \tilde O_0,N|U_1(0,-\infty;\epsilon,N)|\tilde O_0,N\rangle
&\to&
\exp\Bigg(-q^2|C_2|^2\int dk \,|\tilde O_0(\bm k)|^2\frac{|1-e^{2ik_zL}|^2}{|\bm k|^2+\epsilon^2}
\Big(\frac{i|\bm k|}{2\epsilon}
+
\frac{1}{2}
\Big)
\Bigg)
\nonumber
\ee
so that for large $N$
\be
q\lim_{\epsilon\to 0_+}\epsilon i\frac{\partial}{\partial q}\ln\langle \tilde O_0,N|U_1(0,-\infty;\epsilon,N)|\tilde O_0,N\rangle
&\approx&
q^2|C_2|^2\int dk \,|\tilde O_0(\bm k)|^2\frac{|1-e^{2ik_zL}|^2}{|\bm k|}
\nonumber\\
&=&
2q^2|C_2|^2\int dk \,|\tilde O_0(\bm k)|^2\frac{1-\cos 2k_zL}{|\bm k|}.
\nonumber\\
\ee
If we move the plane to infinity,
\be
\lim_{L\to \pm \infty}
2q^2|C_2|^2\int dk \,|\tilde O_0(\bm k)|^2\frac{1-\cos 2k_zL}{|\bm k|}
&=&
2q^2|C_2|^2\int \frac{d^3k}{(2\pi)^3 2|\bm k|^2}|\tilde O_0(\bm k)|^2,\label{plane to}
\ee
we should reconstruct the result from the previous section. Note that the symmetry $|\tilde O_0(k_x,k_y,-k_z)|^2=|\tilde O_0(k_x,k_y,k_z)|^2$
is not essential for (\ref{plane to}) since even with no symmetry constraint on $\tilde O_0$
we would get
\be
\int\frac{d^3k}{(2\pi)^3 2|\bm k|^2}|\tilde O_0(k_x,k_y,k_z)|^2=\int\frac{d^3k}{(2\pi)^3 2|\bm k|^2}|\tilde O_0(k_x,k_y,-k_z)|^2
\ee
just by change of variables. We conclude that $|C_2|^2=1/2$ and
\be
\int dk \,|\tilde O_0(\bm k)|^2\Delta E(\bm k)
&=&
q^2\int \frac{d^3k}{(2\pi)^3 2|\bm k|^2}|\tilde O_0(\bm k)|^2\Big(1-\cos 2k_zL\Big).
\ee
In terms of generalized Coulomb field the result
\be
\int dk \,|\tilde O_0(\bm k)|^2\Delta E(\bm k)
&=&
-\frac{q}{2}\langle \tilde O_0,N|\phi_{\rm gC}(\bm 0,N)|\tilde O_0,N\rangle+\frac{q}{2}\langle \tilde O_0,N|\phi_{\rm gC}(0,0,2L,N)|\tilde O_0,N\rangle
\nonumber\\
\ee
is interpretable in terms of the method of images known from classical electrostatics: The presence of the boundary is equivalent to adding a
mirror-reflected opposite charge.

This way of computing interaction of the charge with the boundary, based on the Gell-Mann--Low formula for self-energy, has been inspired by the work
\cite{SH2010} on Casimir-Polder interactions of Bose-Einstein condensates with interfaces. It is interesting that the results (both for the mirror image and Casimir-Polder forces from \cite{SH2010}) are physically correct although the assumptions behind the Gell-Mann--Low construction are not satisfied, neither by my vacuum state nor by the Bose-Einstein condensate discussed in \cite{SH2010}.

\section{Field quantization with less trivial boundary conditions}

In order to discuss the Casimir effect \cite{Casimir} we have to analyze in detail quantization with less trivial boundary conditions. There are many subtleties that go beyond standard treatments of the problem.  Just to give an example, we have to treat both the quantum field $\phi(x)$ and its vacuum classical field $O_0(x)$, but it is by no means evident that they have to fulfill  the same boundary conditions.

Indeed, the vacuum field $O_0(x)$ is a massless neutral scalar {\it classical\/} field, a fact that will remain true also in genuine electrodynamics where our quantum field $\phi(x)$ will be replaced by fields such as $F_{\mu\nu}(x)$, satisfying boundary conditions that cannot be reduced to those for a set of scalar fields. In principle, the walls that are fully reflecting for $\phi(x)$ or $F_{\mu\nu}(x)$ can be fully transparent for $O_0(x)$.

However, in the case discussed in these lecture notes it is convenient to start with the analysis of boundary conditions for the classical field
$O_0(x)$ since the issue is well understood in terms of classical field theory. We can treat it also as an intermediate step that makes an analysis of $\phi(x)$ more natural, just by mimicking the classical derivation.

So, let us consider a cavity consisting of two parallel planes $\Sigma_\pm=\{\bm x\in \mathbb{R}^3, \,\bm x=(x,y,\pm L/2)\}$. The wave equation
\be
\Box O_0(x)
+
\alpha\,\delta(z+L/2)O_0(x)+\beta\,\delta(z-L/2)O_0(x)=0
\ee
represents the vacuum field with appropriate boundary conditions on $\Sigma_\pm$  \cite{Bordag,Munoz}. We will see that the limits $\alpha\to\infty$, $\beta\to\infty$ correspond to fully reflecting planes. For $\alpha=\beta=0$ the field is a superposition
\be
O_0(x)
&=&
\int dk\, \Big(O_0(\bm{k})e^{-ikx/2}+\overline{O_0(\bm{k})}e^{ikx/2}\Big)
\ee
of plane waves
\be
e^{ikx/2}
&=&
e^{i|\bm k|x_0/2}
e^{-i\bm k_\parallel \cdot \bm x/2}
e^{-ik_z z/2},\\
e^{-ikx/2}
&=&
e^{-i|\bm k|x_0/2}
e^{i\bm k_\parallel \cdot \bm x/2}
e^{ik_z z/2},\\
\bm k_\parallel
&=&
(k_x,k_y,0).
\ee
For nonzero $\alpha$ or $\beta$ we expect that the plane-waves decomposition will be replaced according to the recipe
\be
e^{ikx/2}
&\to&
e^{i|\bm k|x_0/2}
e^{-i\bm k_\parallel \cdot \bm x/2}
\overline{f(k_z, z/2)},\\
e^{-ikx/2}
&\to&
e^{-i|\bm k|x_0/2}
e^{i\bm k_\parallel \cdot \bm x/2}
f(k_z, z/2),\\
dk &\to& dk+\textrm{const}\times\sum_n dk_z\,\delta(k_z-k_n),\label{new dk}
\ee
where the new measure takes into account a possibility of additional discrete wave vectors $k_n$ that may appear due to boundary conditions. $f(k_z, z/2)$ can be determined from
\be
\Big(\Box
+
\alpha\,\delta(z+L/2)+\beta\,\delta(z-L/2)
\Big)e^{-i|\bm k|x_0/2}
e^{i\bm k_\parallel \cdot \bm x/2}
f(k_z, z/2)
&=&0,
\ee
or, equivalently, from
\be
\Big(-k_z^2/4
-
\partial_z^2
+
\alpha\,\delta(z+L/2)+\beta\,\delta(z-L/2)
\Big)
f(k_z, z/2)
&=&0,\label{SE f}
\ee
which is a 1-dimensional stationary Schr\"odinger equation with eigenvalue $k_z^2/4$ and Hamiltonian
\be
\hat H=-
\partial_z^2
+
\alpha\,\delta(z+L/2)+\beta\,\delta(z-L/2).
\ee
The first derivative  $\partial_zf(k_z, z/2)$ must have discontinuities at $z=\pm L/2$, but $f(k_z, z/2)$ itself must be continuous at these points (otherwise already the first derivatives would produce deltas, while second derivatives would lead to terms involving $\delta'$, which are absent in the equation). In the regions between $z=\pm L/2$ this is just a free Schr\"odinger equation so the solutions must have a form of superposition of plane waves.

Denoting $g(z)=f(k_z,z/2)$, $g'(z)=f'(k_z,z/2)/2$, integrating (\ref{SE f}) in neighborhoods of $z=\pm L/2$,
\be
\frac{k_z^2}{4}\int_{-L/2-\epsilon}^{-L/2+\epsilon} dz\,g(z)
&=&
\int_{-L/2-\epsilon}^{- L/2+\epsilon} dz\,
\Big(
-g''(z)+\alpha\,\delta(z+L/2)g(z)+\beta\,\delta(z-L/2)g(z)
\Big)
\nonumber\\
&=&
-g'(-L/2+\epsilon)+g'(-L/2-\epsilon)+\alpha\,g(-L/2),\\
\frac{k_z^2}{4}\int_{L/2-\epsilon}^{L/2+\epsilon} dz\,g(z)
&=&
-g'(L/2+\epsilon)+g'(L/2-\epsilon)+\beta\,g(L/2),
\ee
and taking the limit $\epsilon\to 0_+$, we find the boundary conditions
\be
f'(k_z,-L/4+0_+)-f'(k_z,-L/4-0_+)&=&2\alpha\,f(k_z,-L/4),\\
f'(k_z,L/4+0_+)-f'(k_z,L/4-0_+)&=&2\beta\,f(k_z,L/4),\\
f(k_z,-L/4+0_+)-f(k_z,-L/4-0_+)&=&0,\\
f(k_z,L/4+0_+)-f(k_z,L/4-0_+)&=&0.
\ee

Let $k=|k_z|$. Thinking in terms of scattering of a wave packet on a double-delta potential we can distinguish the following three types of elementary solutions.

(i) The wave packet located initially to the left of $z=-L/2$ is a superposition of scattering states of the form
\be
f_1(k_z, z/2)
&=&
\left\{
\begin{array}{cl}
A_1 e^{ikz/2}+B_1e^{-ikz/2} & \textrm{for }z<-L/2,\\
C_1 e^{ikz/2}+D_1e^{-ikz/2} & \textrm{for }-L/2<z<L/2,\\
E_1 e^{ikz/2} & \textrm{for }L/2<z.
\end{array}
\right.
\ee

(ii) The wave packet located initially between $z=-L/2$ and $z=L/2$ is apparently a superposition of scattering states of the form
\be
f_2(k_z, z/2)
&=&
\left\{
\begin{array}{cl}
B_2e^{-ikz/2} & \textrm{for }z<-L/2,\\
C_2 e^{ikz/2}+D_2e^{-ikz/2} & \textrm{for }-L/2<z<L/2,\\
E_2 e^{ikz/2} & \textrm{for }L/2<z.
\end{array}
\right.
\ee

(iii) The wave packet located initially to the right of $z=L/2$ is a superposition of scattering states of the form
\be
f_3(k_z, z/2)
&=&
\left\{
\begin{array}{cl}
B_3e^{-ikz/2} & \textrm{for }z<-L/2,\\
C_3 e^{ikz/2}+D_3e^{-ikz/2} & \textrm{for }-L/2<z<L/2,\\
E_3 e^{ikz/2}+F_3e^{-ikz/2} & \textrm{for }L/2<z.
\end{array}
\right.
\ee
The case (ii) is a special case of both (i) and (iii), so we do not have to treat it separately.

\subsection{Sewing solutions for the case (i)}

Boundary conditions read explicitly
\be
A_1 e^{-ikL/4}+B_1e^{ikL/4} &=& C_1 e^{-ikL/4}+D_1e^{ikL/4},\nonumber\\
C_1 e^{ikL/4}+D_1e^{-ikL/4}  &=& E_1 e^{ikL/4},\nonumber\\
ik(A_1 e^{-ikL/4}-B_1e^{ikL/4}) &=& ik(C_1 e^{-ikL/4}-D_1e^{ikL/4})-2\alpha\Big( C_1 e^{-ikL/4}+D_1e^{ikL/4}\Big),\nonumber\\
ik(C_1 e^{ikL/4}-D_1e^{-ikL/4}) &=& ik E_1 e^{ikL/4}-2\beta E_1 e^{ikL/4}\nonumber.
\ee
If $A_1=0$ then we deal with the case (ii). If $A_1\neq 0$ we can divide all the equations by $A_1$, which is equivalent to starting with $A_1=1$.
So first assume $A_1=1$. Then
\be
1  &=& -B_1e^{ikL/2}+C_1 +D_1e^{ikL/2},\nonumber\\
0  &=&-C_1 e^{ikL/2} -D_1+E_1 e^{ikL/2},\nonumber\\
ik  &=& ik B_1e^{ikL/2}+(ik-2\alpha)C_1 -(ik+2\alpha)D_1e^{ikL/2},\nonumber\\
0 &=& -ikC_1 e^{ikL/2}+ikD_1+(ik-2\beta) E_1 e^{ikL/2}\nonumber.
\ee
hence
\be
B_1 &=& -\frac{i e^{-\frac{1}{2} i k L} \left(k \left(\alpha +e^{i k L} \beta \right)-i \left(-1+e^{i k L}\right) \alpha  \beta \right)}{k^2+i (\alpha +\beta )
   k+\left(-1+e^{i k L}\right) \alpha  \beta },\\
C_1 &=&\frac{k (k+i \beta )}{k^2+i (\alpha +\beta ) k+\left(-1+e^{i k L}\right) \alpha  \beta },\\
D_1 &=&-\frac{i e^{\frac{i k L}{2}} k \beta }{k^2+i (\alpha +\beta ) k+\left(-1+e^{i k L}\right) \alpha  \beta },\\
E_1 &=& \frac{k^2}{k^2+i (\alpha +\beta ) k+\left(-1+e^{i k L}\right) \alpha  \beta }
\ee
For $k\to 0$,
\be
B_1 &\to& -1,\\
C_1 &\to&\frac{\beta }{\alpha +\beta +L\alpha  \beta },\\
D_1 &\to&-\frac{\beta }{\alpha +\beta +L\alpha  \beta },\\
E_1 &\to& 0,
\ee
and
\be
f_1(k_z, z/2)
&\to&
0.
\ee
For $k\to \infty$,
\be
B_1 &\to& 0,\\
C_1 &\to& 1,\\
D_1 &\to& 0,\\
E_1 &\to& 1.
\ee
For large $k$ the cavity becomes transparent, which is much more physical than the Dirichlet conditions that imply full reflectivity no matter how big the $k$s are.

\subsection{Sewing solutions for the case (ii)}

If $A_1=0$ then
\be
0  &=& -B_1e^{ikL/2}+C_1 +D_1e^{ikL/2},\nonumber\\
0  &=&-C_1 e^{ikL/2} -D_1+E_1 e^{ikL/2},\nonumber\\
0  &=& ik B_1e^{ikL/2}+(ik-2\alpha)C_1 -(ik+2\alpha)D_1e^{ikL/2},\nonumber\\
0 &=& -ikC_1 e^{ikL/2}+ikD_1+(ik-2\beta) E_1 e^{ikL/2}\nonumber.
\ee
possess nonzero solutions if
\be
k^2+i (\alpha +\beta ) k+\left(-1+e^{i k L}\right) \alpha  \beta
=0\label{Delta(k)=0}.
\ee
The trivial solution $k=0$ implies $f_2(0,z)=0$.

Now, let us restrict the analysis to $\alpha=\beta$. We rewrite (\ref{Delta(k)=0}) as
\be
(ik-\alpha)^2 &=& \alpha^2e^{i k L},
\ee
hence
\be
ik-\alpha &=& \pm \alpha e^{i k L/2}\nonumber\\
&=&
\pm \alpha e^{(i k-\alpha) L/2}e^{\alpha L/2},\\
e^{(\alpha-ik) L/2}(\alpha-ik)L/2 &=& \pm \alpha e^{\alpha L/2}L/2.
\ee
Denoting $q=(\alpha-i k) L/2$ we reduce the problem to finding $q$ satisfying
\be
qe^q=\pm \alpha e^{\alpha L/2}L/2.\label{qeq}
\ee
Lambert's $W$ function \cite{Boonserm,L Knuth,Dubinov}, defined implicitly by $W(z)e^{W(z)}=z$, has infinite number of branches $W_n(z)$, $n\in \mathbb{Z}$.
By definition of $W$ we find that
\be
W_n(\pm \alpha e^{\alpha L/2}L/2)\exp W_n(\pm \alpha e^{\alpha L/2}L/2)=\pm \alpha e^{\alpha L/2}L/2.
\ee
Accordingly,
\be
(\alpha-i k) L/2=W_n(\pm \alpha e^{\alpha L/2}L/2)
\ee
and, in addition to the trivial solution $k=0$, we find infinitely many discrete values of $k$,
\be
k_{n,\pm}
&=&
-i\frac{L \alpha -2 W_{n}\big(\pm e^{\frac{L \alpha }{2}} L \alpha/2 \big)}{L},\quad n=0,\pm 1,\pm 2\dots
\ee
The nonzero wave vectors are complex. For example, for $\alpha=1$, $L=1$,
\be
k_{0,+}
&=&
0,\nonumber\\
k_{0,-}
&=&
-2.42855 - 1.90448 i,\nonumber\\
k_{1,+}
&=&
-8.66349 - 4.46676 i,\nonumber\\
k_{1,-}
&=&
-15.1274 - 5.51848 i,\nonumber\\
k_{2,+}
&=&
-21.5174 - 6.19436 i,\nonumber\\
k_{2,-}
&=&
-27.8711 - 6.6961 i.\nonumber
\ee
It looks like the imaginary parts are always negative (numerical analysis shows that the imaginary parts tend to 0 with $\alpha\to\infty$, but I don't know yet how to prove it analytically --- probably one should try with the asymptotic expansions given in \cite{L Knuth}). Inserting $k=-k_r-ik_i$ into $f_2(k_z, z/2)$,
\be
f_2(k_z, z/2)
&=&
\left\{
\begin{array}{cl}
B_2e^{-i(-k_r-ik_i)z/2} & \textrm{for }z<-L/2,\\
C_2 e^{i(-k_r-ik_i)z/2}+D_2e^{-i(-k_r-ik_i)z/2} & \textrm{for }-L/2<z<L/2,\\
E_2 e^{i(-k_r-ik_i)z/2} & \textrm{for }L/2<z.
\end{array}
\right.
\nonumber\\
&=&
\left\{
\begin{array}{cl}
B_2e^{(ik_r-k_i)z/2} & \textrm{for }z<-L/2,\\
C_2 e^{(-ik_r+k_i)z/2}+D_2e^{(ik_r-k_i)z/2} & \textrm{for }-L/2<z<L/2,\\
E_2 e^{(-ik_r+k_i)z/2} & \textrm{for }L/2<z.
\end{array}
\right.
\ee
we realize that the wave which is to the left of the cavity not only exponentially grows as $z\to -\infty$ but, surprisingly, propagates to the right (towards the cavity), which contradicts the assumptions that led us to the form of $f_2(k_z, z)$. Moreover, we assumed that $k=|k_z|$, but this is not satisfied by complex $k=-k_r-ik_i$. The same problems occur to the right of the cavity.

So, let us see what will happen if we take the counterintuitive case
\be
f_2(k_z, z/2)
&=&
\left\{
\begin{array}{cl}
A_2e^{ikz/2} & \textrm{for }z<-L/2,\\
C_2 e^{ikz/2}+D_2e^{-ikz/2} & \textrm{for }-L/2<z<L/2,\\
F_2 e^{-ikz/2} & \textrm{for }L/2<z.
\end{array}
\right.
\ee
Mathematically this is the same problem as before but with $k$ replaced by $-k$, so we look for solutions of
\be
k^2-i (\alpha +\beta ) k+\left(-1+e^{-i k L}\right) \alpha  \beta
=0\label{Delta(-k)=0}.
\ee
For $\alpha=\beta$ we find
\be
k_{n,\pm}
&=&
i\frac{L \alpha -2 W_{n}\big(\pm e^{\frac{L \alpha }{2}} L \alpha/2 \big)}{L},\quad n=0,\pm 1,\pm 2\dots
\ee
As we can see $k_{n,\pm}$ has changed sign as well and the difficulties remain.

We conclude that the case (ii) leads to eigenvectors that are as unphysical as negative-energy eigenvectors of harmonic oscillator.
On the other hand, the $A_1\neq 0$ solutions do not impose any constraints on $k$, which are therefore the same as those corresponding to $\alpha=\beta=0$. This observation agrees with the remarks in \cite{Most} (cf. section 1.6 of the book) that for finite $\alpha$, $\beta$ the Hilbert space structure of the model is the same as for fields in free space. It is also a good news from my point of view since we do not have to modify the mode structure of the fields --- modification (\ref{new dk}) is not required.

We will separately consider the limit $\alpha=\beta\to \infty$, but at the end of the calculation.

\subsection{Sewing solutions for the case (iii)}

For completeness let us now switch to the case (iii) which will have properties analogous to those of (i).
We do not have to repeat the calculations. Starting with
\be
f_3(k_z, -z/2)
&=&
\left\{
\begin{array}{cl}
B_3e^{ikz/2} & \textrm{for }z>L/2,\\
C_3 e^{-ikz/2}+D_3e^{ikz/2} & \textrm{for }-L/2<z<L/2,\\
E_3 e^{-ikz/2}+F_3e^{ikz/2} & \textrm{for }-L/2>z,
\end{array}
\right.
\ee
we obtain the same mathematical problem as (i), but the roles of $\alpha$ and $\beta$ have to be interchanged.
Replacing in (i)
$A_1\to F_3=1$, $B_1\to E_3$, $C_1\to D_3$, $D_1\to C_3$, $E_1\to B_3$, $\alpha\leftrightarrow \beta$, we find
\be
E_3 &=& -\frac{i e^{-\frac{1}{2} i k L} \left(k \left(\beta +e^{i k L} \alpha \right)-i \left(-1+e^{i k L}\right) \alpha  \beta \right)}{k^2+i (\alpha +\beta )k+\left(-1+e^{i k L}\right) \alpha  \beta },\\
D_3 &=&\frac{k (k+i \alpha )}{k^2+i (\alpha +\beta ) k+\left(-1+e^{i k L}\right) \alpha  \beta },\\
C_3 &=&-\frac{i e^{\frac{i k L}{2}} k \alpha }{k^2+i (\alpha +\beta ) k+\left(-1+e^{i k L}\right) \alpha  \beta },\\
B_3 &=& \frac{k^2}{k^2+i (\alpha +\beta ) k+\left(-1+e^{i k L}\right) \alpha  \beta }.
\ee
For $k\to 0$
\be
f_3(k_z, z/2)\to 0.
\ee
For $k\to\infty$ the cavity is fully transparent for finite $\alpha$ and $\beta$.

\subsection{Limit of no cavity}

This corresponds to $\alpha=\beta=0$. We find,
\be
B_1 &=& 0,\\
C_1 &=&1,\\
D_1 &=&0,\\
E_1 &=& 1,\\
f_1(k_z, z/2)
&=&
\left\{
\begin{array}{cl}
e^{ikz/2} & \textrm{for }z<-L/2\\
e^{ikz/2} & \textrm{for }-L/2<z<L/2\\
e^{ikz/2} & \textrm{for }L/2<z
\end{array}
\right\}
=e^{i|k_z|z/2},
\ee
$f_1(k_z, z)$ represents plane wave of positive momentum $k_z=|k_z|$. Now,
\be
E_3 &=& 0,\\
D_3 &=&1,\\
C_3 &=&0,\\
B_3 &=& 1,\\
f_3(k_z, z/2)
&=&
\left\{
\begin{array}{cl}
e^{-ikz/2} & \textrm{for }z<-L/2\\
e^{-ikz/2} & \textrm{for }-L/2<z<L/2\\
e^{-ikz/2} & \textrm{for }L/2<z
\end{array}
\right\}
=
e^{-i|k_z|z/2},
\ee
so  $f_3(k_z, z)$ represents plane wave of negative momentum $k_z=-|k_z|$.
Accordingly,
\be
f(k_z, z)
&=&
\left\{
\begin{array}{cl}
f_1(k_z, z) & \textrm{for }k_z\geq 0\\
f_3(k_z, z) & \textrm{for }k_z\leq 0
\end{array}
\right\}
=e^{ik_zz}.
\ee
Let us note that $f(0, z)=1$.

\subsection{Limit of fully reflecting cavity, $\alpha=\beta\to\infty$}

For $\alpha=\beta$,
\be
B_1 &=& -\frac{i e^{-\frac{1}{2} i k L} \left(k \left(1 +e^{i k L} \right)\alpha-i \left(-1+e^{i k L}\right) \alpha^2\right)}
{k^2+2i \alpha k+\left(-1+e^{i k L}\right) \alpha^2}=E_3,\label{BE}\\
C_1 &=&\frac{k (k+i \alpha )}{k^2+2i \alpha k+\left(-1+e^{i k L}\right) \alpha^2}=D_3,\label{CD}\\
D_1 &=&-\frac{i e^{\frac{i k L}{2}} k \alpha }{k^2+2i \alpha k+\left(-1+e^{i k L}\right) \alpha^2}=C_3,\label{DC}\\
E_1 &=& \frac{k^2}{k^2+2i \alpha k+\left(-1+e^{i k L}\right) \alpha^2}=B_3.\label{EB}
\ee
If $-1+e^{i k L}\neq 0$, then in the limit $\alpha\to\infty$
\be
A_1 &=& F_3=1,\\
B_1 &=& E_3=-e^{-\frac{1}{2} i k L},\\
C_1 &=&D_3=0,\\
D_1 &=&C_3=0,\\
E_1 &=& B_3=0.
\ee
and
\be
f_1(k_z, z/2)
&=&
\left\{
\begin{array}{cl}
e^{ikz/2}-e^{-\frac{1}{2} i k L}e^{-ikz/2} & \textrm{for }z<-L/2,\\
0 & \textrm{for }-L/2<z<L/2,\\
0 & \textrm{for }L/2<z.
\end{array}
\right.\nonumber\\
f_3(k_z, z/2)
&=&
\left\{
\begin{array}{cl}
0 & \textrm{for }z<-L/2,\\
0 & \textrm{for }-L/2<z<L/2,\\
-e^{-\frac{1}{2} i k L}e^{ikz/2}+e^{-ikz/2} & \textrm{for }L/2<z.
\end{array}
\right.
\ee
The wave packets arriving from outside of the cavity are reflected from the walls. The fields inside of the cavity vanish if $e^{i k L}\neq 1$.

The situation changes if $-1+e^{i k L}= 0$ (i.e. for $kL=2\pi n$),
\be
A_1 &=& F_3=1,\\
B_1 &=& -\frac{2i  \alpha k e^{-\frac{1}{2} i k L}}
{k^2+2i \alpha k}=E_3 \to -e^{- i k L/2},\\
C_1 &=&\frac{k (k+i \alpha )}{k^2+2i \alpha k}=D_3 \to \frac{1}{2},\\
D_1 &=&-\frac{i e^{\frac{i k L}{2}} k \alpha }{k^2+2i \alpha k}=C_3 \to -\frac{1}{2}e^{i k L/2},\\
E_1 &=& \frac{k^2}{k^2+2i \alpha k}=B_3\to 0.
\ee
As we can see there are two types of solutions since
\be
e^{\pm i k L/2}=e^{i \pi n}
=
\left\{
\begin{array}{cl}
-1 & \textrm{for odd $n$},\\
+1 & \textrm{for even $n$}.
\end{array}
\right.
\ee
Let $n=2m+1$. Then
\be
f_1(k_z, z/2)
&=&
\left\{
\begin{array}{cl}
e^{ikz/2}+e^{-ikz/2} & \textrm{for }z<-L/2,\\
\frac{1}{2}e^{ikz/2}+\frac{1}{2}e^{-ikz/2} & \textrm{for }-L/2<z<L/2,\\
0 & \textrm{for }L/2<z.
\end{array}
\right.
\nonumber\\
f_3(k_z, z/2)
&=&
\left\{
\begin{array}{cl}
0 & \textrm{for }z<-L/2,\\
\frac{1}{2}e^{ikz/2}+\frac{1}{2}e^{-ikz/2} & \textrm{for }-L/2<z<L/2,\\
e^{ikz/2}+e^{-ikz/2} & \textrm{for }L/2<z.
\end{array}
\right.
\ee
Note that for $k_z=|k_z|$
\be
f_1(k_z, z/2)
&=&
\left\{
\begin{array}{cl}
2\cos (k_zz/2) & \textrm{for }z<-L/2,\\
\cos (k_zz/2) & \textrm{for }-L/2<z<L/2,\\
0 & \textrm{for }L/2<z.
\end{array}
\right.
\ee
For $k_z=-|k_z|$
\be
f_3(k_z, z/2)
&=&
\left\{
\begin{array}{cl}
0 & \textrm{for }z<-L/2,\\
\cos (k_zz/2) & \textrm{for }-L/2<z<L/2,\\
2\cos (k_zz/2) & \textrm{for }L/2<z.
\end{array}
\right.
\ee
Let $n=2m$. Then
\be
f_1(k_z, z/2)
&=&
\left\{
\begin{array}{cl}
e^{ikz/2}-e^{-ikz/2} & \textrm{for }z<-L/2,\\
\frac{1}{2}e^{ikz/2}-\frac{1}{2}e^{-ikz/2} & \textrm{for }-L/2<z<L/2,\\
0 & \textrm{for }L/2<z.
\end{array}
\right.
\nonumber\\
f_3(k_z, z/2)
&=&
\left\{
\begin{array}{cl}
0 & \textrm{for }z<-L/2,\\
-\frac{1}{2}e^{ikz/2}+\frac{1}{2}e^{-ikz/2} & \textrm{for }-L/2<z<L/2,\\
-e^{ikz/2}+e^{-ikz/2} & \textrm{for }L/2<z.
\end{array}
\right.
\ee
For $k_z=|k_z|$
\be
f_1(k_z, z/2)
&=&
\left\{
\begin{array}{cl}
2i\sin(k_zz/2) & \textrm{for }z<-L/2,\\
i\sin(k_zz/2) & \textrm{for }-L/2<z<L/2,\\
0 & \textrm{for }L/2<z.
\end{array}
\right.
\ee
For $k_z=-|k_z|$
\be
f_3(k_z, z/2)
&=&
\left\{
\begin{array}{cl}
0 & \textrm{for }z<-L/2,\\
-\frac{1}{2}e^{-ik_zz/2}+\frac{1}{2}e^{ik_zz/2} & \textrm{for }-L/2<z<L/2,\\
-e^{-ik_zz/2}+e^{ik_zz/2} & \textrm{for }L/2<z
\end{array}
\right.
\nonumber\\
&=&
\left\{
\begin{array}{cl}
0 & \textrm{for }z<-L/2,\\
i\sin(k_zz/2) & \textrm{for }-L/2<z<L/2,\\
2i\sin(k_zz/2) & \textrm{for }L/2<z
\end{array}
\right.
\ee

\subsection{Vacuum field for finite $\alpha$ and $\beta$}

The free field (i.e. for $\alpha=\beta=0$) can be written as
\be
O_0(x)
&=&
\int \frac{d^3k}{(2\pi)^32|\bm k|} O_0(\bm{k})e^{-i|\bm k|x_0/2}
e^{i\bm k_\parallel \cdot \bm x/2}
f(k_z,z/2)+{\rm c.c.}
\nonumber\\
&=&
\int_{\mathbb{R}^2} d^2k\int_0^\infty\frac{dk_z}{(2\pi)^32|\bm k|} O_0(\bm{k})e^{-i|\bm k|x_0/2}
e^{i\bm k_\parallel \cdot \bm x/2}
f_1(k_z,z/2)
\nonumber\\
&\pp=&
+
\int_{\mathbb{R}^2} d^2k\int_{-\infty}^0\frac{dk_z}{(2\pi)^32|\bm k|} O_0(\bm{k})e^{-i|\bm k|x_0/2}
e^{i\bm k_\parallel \cdot \bm x/2}
f_3(k_z,z/2)+{\rm c.c.}
\ee
We can extend this definition also for arbitrary finite $\alpha$ and $\beta$.

\subsection{Vacuum field for $\alpha=0$ and $\beta\to\infty$}

This particular case should reconstruct the results of a single infinitely reflecting wall we have discussed in the previous section (the wall is located at $z=L/2$).
Now,
\be
A_1 &=& 1,\\
B_1 &=& -\frac{i e^{\frac{1}{2} i k L}   \beta}{k+i \beta }\to -e^{i k L/2},\\
C_1 &=&1,\\
D_1 &=&-\frac{i e^{\frac{i k L}{2}} \beta }{k+i \beta}\to -e^{i k L/2},\\
E_1 &=& \frac{k}{k+i \beta }\to 0,\\
F_3 &=& 1,\\
E_3 &=& -\frac{i e^{-\frac{1}{2} i k L} \beta}{k+i \beta }\to -e^{-i k L/2},\\
D_3 &=&\frac{k}{k+i \beta }\to 0,\\
C_3 &=&0,\\
B_3 &=& \frac{k}{k+i \beta}\to 0.
\ee
For $k_z=|k_z|=k$
\be
f_1(k_z, z/2)
&=&
\left\{
\begin{array}{cl}
e^{ik_zz/2}-e^{i k_z L/2}e^{-ik_zz/2} & \textrm{for }z<-L/2,\\
e^{ik_zz/2}-e^{i k_z L/2}e^{-ik_zz/2} & \textrm{for }-L/2<z<L/2,\\
0 & \textrm{for }L/2<z.
\end{array}
\right.
\ee
For $k_z=-|k_z|=-k$
\be
f_3(k_z, z/2)
&=&
\left\{
\begin{array}{cl}
0 & \textrm{for }z<-L/2,\\
0 & \textrm{for }-L/2<z<L/2,\\
-e^{-i k L/2}e^{ikz/2}+e^{-ikz/2} & \textrm{for }L/2<z.
\end{array}
\right.
\nonumber\\
&=&
\left\{
\begin{array}{cl}
0 & \textrm{for }z<-L/2,\\
0 & \textrm{for }-L/2<z<L/2,\\
e^{ik_zz/2}-e^{i k_z L/2}e^{-ik_zz/2} & \textrm{for }L/2<z.
\end{array}
\right.\label{f is 0}
\ee
So,
\be
O_0(x)
&=&
\int_{\mathbb{R}^2} d^2k\int_0^\infty\frac{dk_z}{(2\pi)^32|\bm k|} O_0(\bm{k})e^{-i|\bm k|x_0/2}
e^{i\bm k_\parallel \cdot \bm x/2}
f_1(k_z,z/2)
\nonumber\\
&\pp=&
+
\int_{\mathbb{R}^2} d^2k\int_{-\infty}^0\frac{dk_z}{(2\pi)^32|\bm k|} O_0(\bm{k})e^{-i|\bm k|x_0/2}
e^{i\bm k_\parallel \cdot \bm x/2}
f_3(k_z,z/2)+{\rm c.c.}\nonumber\\
&=&
\int\frac{d^3k}{(2\pi)^32|\bm k|} O_0(\bm{k})e^{-i|\bm k|x_0/2}
e^{i\bm k_\parallel \cdot \bm x/2}
\Big(e^{ik_zz/2}-e^{i k_z L/2}e^{-ik_zz/2}\Big)
+{\rm c.c.}
\nonumber\\
&=&
\int dk\, O_0(\bm{k})e^{-ikx/2}
\Big(1-e^{-i k_z (z-L/2)}\Big)
+{\rm c.c.}
\nonumber
\ee
which is exactly of the form we have employed in the previous section in order to discuss interaction of a charge with a plane.

\subsection{Orthonormality of $f(k_z,z/2)$,  $\alpha=\beta>0$}

Let us return to
\be
O_0(x)
&=&
\int_{\mathbb{R}^2} d^2k\int_0^\infty\frac{dk_z}{(2\pi)^32|\bm k|} O_0(\bm{k})e^{-i|\bm k|x_0/2}
e^{i\bm k_\parallel \cdot \bm x/2}
f_1(k_z,z/2)
\nonumber\\
&\pp=&
+
\int_{\mathbb{R}^2} d^2k\int_{-\infty}^0\frac{dk_z}{(2\pi)^32|\bm k|} O_0(\bm{k})e^{-i|\bm k|x_0/2}
e^{i\bm k_\parallel \cdot \bm x/2}
f_3(k_z,z/2)+{\rm c.c.}
\nonumber\\
&=&
\int dk\, O_0(\bm{k})e^{-i|\bm k|x_0/2}
e^{i\bm k_\parallel \cdot \bm x/2}\Big(\underbrace{\theta(k_z)f_1(k_z,z/2)+\theta(-k_z)f_3(k_z,z/2)}_{f(k_z,z/2)}\Big)
+{\rm c.c.}
\ee
with ($k=|k_z|$)
\be
f_1(k_z, z/2)
&=&
\left\{
\begin{array}{cl}
A_1(k) e^{ikz/2}+B_1(k)e^{-ikz/2} & \textrm{for }z<-L/2,\\
C_1(k) e^{ikz/2}+D_1(k)e^{-ikz/2} & \textrm{for }-L/2<z<L/2,\\
E_1(k) e^{ikz/2} & \textrm{for }L/2<z,
\end{array}
\right.
\ee\be
f_3(k_z, z/2)
&=&
\left\{
\begin{array}{cl}
B_3(k)e^{-ikz/2} & \textrm{for }z<-L/2,\\
C_3(k) e^{ikz/2}+D_3(k)e^{-ikz/2} & \textrm{for }-L/2<z<L/2,\\
E_3(k) e^{ikz/2}+F_3(k)e^{-ikz/2} & \textrm{for }L/2<z,
\end{array}
\right.
\ee
\be
f(k_z, z)
&=&
\left\{
\begin{array}{l}
\theta(k_z)\Big(A_1(k_z) e^{ik_zz}+B_1(k_z)e^{-ik_zz} \Big)
+
\theta(-k_z)B_3(-k_z)e^{ik_zz}
,\\
\theta(k_z)\Big(C_1(k_z) e^{ik_zz}+D_1(k_z)e^{-ik_zz}\Big)
+
\theta(-k_z)\Big(C_3(-k_z) e^{-ik_zz}+D_3(-k_z)e^{ik_zz}\Big)
,\\
\theta(k_z)E_1(k_z) e^{ik_zz}
+
\theta(-k_z)\Big(E_3(-k_z) e^{-ik_zz}+F_3(-k_z)e^{ik_zz}\Big)
\end{array}
\right.
\nonumber\\
&=&
\left\{
\begin{array}{l}
\theta(k_z)\Big(A_1(k_z) e^{ik_zz}+B_1(k_z)e^{-ik_zz} \Big)
+
\theta(-k_z)\overline{B_3(k_z)}e^{ik_zz}
,\\
\theta(k_z)\Big(C_1(k_z) e^{ik_zz}+D_1(k_z)e^{-ik_zz}\Big)
+
\theta(-k_z)\Big(\overline{C_3(k_z)} e^{-ik_zz}+\overline{D_3(k_z)}e^{ik_zz}\Big)
,\\
\theta(k_z)E_1(k_z) e^{ik_zz}
+
\theta(-k_z)\Big(\overline{E_3(k_z)} e^{-ik_zz}+\overline{F_3(k_z)}e^{ik_zz}\Big)
\end{array}
\right.
\nonumber\\
&=&
\left\{
\begin{array}{l}
\theta(k_z)\Big(e^{ik_zz}+B_1(k_z)e^{-ik_zz} \Big)
+
\theta(-k_z)\overline{E_1(k_z)}e^{ik_zz}
,\\
\theta(k_z)\Big(C_1(k_z) e^{ik_zz}+D_1(k_z)e^{-ik_zz}\Big)
+
\theta(-k_z)\Big(\overline{D_1(k_z)} e^{-ik_zz}+\overline{C_1(k_z)}e^{ik_zz}\Big)
,\\
\theta(k_z)E_1(k_z) e^{ik_zz}
+
\theta(-k_z)\Big(\overline{B_1(k_z)} e^{-ik_zz}+e^{ik_zz}\Big)
\end{array}
\right.
\nonumber\\
&=&
\left\{
\begin{array}{l}
\theta(k_z)\Big(e^{ik_zz}+B_1(|k_z|)e^{-ik_zz} \Big)
+
\theta(-k_z)E_1(|k_z|)e^{ik_zz}
,\\
\theta(k_z)\Big(C_1(|k_z|) e^{ik_zz}+D_1(|k_z|)e^{-ik_zz}\Big)
+
\theta(-k_z)\Big(D_1(|k_z|) e^{-ik_zz}+C_1(|k_z|)e^{ik_zz}\Big)
,\\
\theta(k_z)E_1(|k_z|) e^{ik_zz}
+
\theta(-k_z)\Big(B_1(|k_z|) e^{-ik_zz}+e^{ik_zz}\Big)
\end{array}
\right.
\nonumber\\
&=&
\left\{
\begin{array}{l}
\Big(\theta(k_z)+\theta(-k_z)E_1(|k_z|)\Big)
e^{ik_zz}
+
\theta(k_z)B_1(|k_z|)e^{-ik_zz}
,\\
C_1(|k_z|) e^{ik_zz}+D_1(|k_z|)e^{-ik_zz},\\
\Big(
\theta(-k_z)
+
\theta(k_z)E_1(|k_z|)\Big)
e^{ik_zz}
+
\theta(-k_z)B_1(|k_z|) e^{-ik_zz}
\end{array}
\right.
\nonumber
\ee
and [cf. (\ref{BE})--(\ref{EB})]
\be
B_1(k) &=& -\frac{ 2ik\alpha \cos \frac{k L}{2}+ 2i\alpha^2 \sin\frac{k L}{2}}
{k^2+2i \alpha k+\left(-1+e^{i k L}\right) \alpha^2}=E_3(k),\\
C_1(k) &=&\frac{k (k+i \alpha )}{k^2+2i \alpha k+\left(-1+e^{i k L}\right) \alpha^2}=D_3(k),\\
D_1(k) &=&-\frac{i e^{\frac{i k L}{2}} k \alpha }{k^2+2i \alpha k+\left(-1+e^{i k L}\right) \alpha^2}=C_3(k),\\
E_1(k) &=& \frac{k^2}{k^2+2i \alpha k+\left(-1+e^{i k L}\right) \alpha^2}=B_3(k).
\ee
Since $\alpha$, $k$, $L$ are real, the coefficients satisfy
$B_j(-k)=\overline{B_j(k)}$,
$C_j(-k)=\overline{C_j(k)}$,
$D_j(-k)=\overline{D_j(k)}$,
$E_j(-k)=\overline{E_j(k)}$, $j=1,3$. The same trivially holds for $A_j$ and $F_j$.

Let us further note that
\be
f(k_z, -z)
&=&
\left\{
\begin{array}{l}
\theta(k_z)E_1(k_z) e^{-ik_zz}
+
\theta(-k_z)\Big(\overline{B_1(k_z)} e^{ik_zz}+e^{-ik_zz}\Big),\\
\theta(k_z)\Big(C_1(k_z) e^{-ik_zz}+D_1(k_z)e^{ik_zz}\Big)
+
\theta(-k_z)\Big(\overline{D_1(k_z)} e^{ik_zz}+\overline{C_1(k_z)}e^{-ik_zz}\Big)
,\\
\theta(k_z)\Big(e^{-ik_zz}+B_1(k_z)e^{ik_zz} \Big)
+
\theta(-k_z)\overline{E_1(k_z)}e^{-ik_zz}
\end{array}
\right.
\nonumber
\ee
and
\be
f(-k_z, -z)
&=&
\left\{
\begin{array}{l}
\theta(-k_z)\overline{E_1(k_z)} e^{ik_zz}
+
\theta(k_z)\Big(B_1(k_z) e^{-ik_zz}+e^{ik_zz}\Big),\\
\theta(-k_z)\Big(\overline{C_1(k_z)} e^{ik_zz}+\overline{D_1(k_z)}e^{-ik_zz}\Big)
+
\theta(k_z)\Big(D_1(k_z) e^{-ik_zz}+C_1(k_z)e^{ik_zz}\Big)
,\\
\theta(-k_z)\Big(e^{ik_zz}+\overline{B_1(k_z)}e^{-ik_zz} \Big)
+
\theta(k_z)E_1(k_z)e^{ik_zz}
\end{array}
\right.
\nonumber\\
&=&
f(k_z, z).
\ee
One can also directly verify that
\be
|A_1|^2=|B_1(k)|^2+|E_1(k)|^2=|B_3(k)|^2+|E_3(k)|^2=|F_3|^2=1,
\ee
a condition typical of scattering of Schr\"odinger particles on a potential wall.

Recall that in the limit of no cavity (i.e. $\alpha=0$) we find no reflection,
$B_1=D_1=0$, $A_1=C_1=E_1=1$,
\be
f(k_z, z)
&=&
\left\{
\begin{array}{l}
\theta(k_z)e^{ik_zz}
+
\theta(-k_z)e^{ik_zz}
\\
\theta(k_z) e^{ik_zz}
+
\theta(-k_z)e^{ik_zz}
\\
\theta(k_z) e^{ik_zz}
+
\theta(-k_z)e^{ik_zz}
\end{array}
\right\}
=
e^{ik_zz}
\nonumber
\ee
Since
\be
\delta(k_z-l_z)
&=&
\int_{-\infty}^\infty \frac{dz}{2\pi}e^{ik_zz}\overline{e^{il_zz}}
\ee
let us check if a similar rule holds for $\alpha>0$.
\medskip

\noindent
{\it Lemma\/}:
\be
\int_{-\infty}^{\infty} \frac{dz}{2\pi}\overline{f(l_z, z)}f(k_z, z)
=
\delta(k_z-l_z).
\ee
\noindent
{\it Proof\/}:
The proof has to be split into several parts. Let us begin with the following observation \cite{Radek}.
Consider two solutions of a one-dimensional Schr\"odinger equation with some potential $U(z)$, satisfying
\be
\overline{f(l_z, z/2)}\Big(-k_z^2/4
-
\partial_z^2
+
U(z)
\Big)
f(k_z, z/2)
&=&0,\label{SE f1}\\
f(k_z, z/2)\Big(-l_z^2/4
-
\partial_z^2
+
U(z)
\Big)
\overline{f(l_z, z/2)}
&=&0.\label{SE f2}
\ee
Subtracting the equations,
\be
(k_z^2/4-l_z^2/4)\overline{f(l_z, z/2)}f(k_z, z/2)
&=&
-\overline{f(l_z, z/2)}\partial_z^2 f(k_z, z/2)
+
f(k_z, z/2)\partial_z^2 \overline{f(l_z, z/2)}
\nonumber\\
&=&
\partial_z\Big(-\overline{f(l_z, z/2)}\partial_z f(k_z, z/2)
+
f(k_z, z/2)\partial_z \overline{f(l_z, z/2)}
\Big),
\nonumber
\ee
and integrating over a finite interval,
\be
\int_{-n}^{n} dz\overline{f(l_z, z)}f(k_z, z)
&=&
\int_{-n}^{n} dz\partial_z\frac{-\overline{f(l_z, z)}\partial_z f(k_z, z)
+
f(k_z, z)\partial_z \overline{f(l_z, z)}
}{k_z^2-l_z^2}
\nonumber\\
&=&
\frac{-\overline{f(l_z,n)}\partial_z f(k_z,n)
+
f(k_z,n)\partial_z \overline{f(l_z,n)}
}{k_z^2-l_z^2}
\nonumber\\
&\pp=&
+
\frac{\overline{f(l_z,-n)}\partial_z f(k_z,-n)
-
f(k_z,-n)\partial_z \overline{f(l_z,-n)}
}{k_z^2-l_z^2}
\nonumber
\ee
we effectively get rid of both the potential and the integral. Moreover, since we are interested in the limit $n\to\infty$, we can take $n$ as large as we want. In our case, taking $n>L/4$, we will deal only with $f$ in regions outside of the two barriers.

Since
\be
\int_{-\infty}^{\infty} dz\overline{f(l_z, z)}f(k_z, z)
&=&
\Big(\theta(l_z)+\theta(-l_z)\Big)\Big(\theta(k_z)+\theta(-k_z)\Big)\int_{-\infty}^{\infty} dz\overline{f(l_z, z)}f(k_z, z)
\nonumber\\
&=&
\theta(l_z)\theta(k_z)\int_{-\infty}^{\infty} dz\overline{f_1(l_z, z)}f_1(k_z, z)
\nonumber\\
&\pp=&
+
\theta(-l_z)\theta(k_z)\int_{-\infty}^{\infty} dz\overline{f_3(l_z, z)}f_1(k_z, z)
\nonumber\\
&\pp=&
+
\theta(l_z)\theta(-k_z)\int_{-\infty}^{\infty} dz\overline{f_1(l_z, z)}f_3(k_z, z)
\nonumber\\
&\pp=&
+
\theta(-l_z)\theta(-k_z)\int_{-\infty}^{\infty} dz\overline{f_3(l_z, z)}f_3(k_z, z)
\nonumber
\ee
we have to separately treat each of the four cases (in fact, the first two are enough).
Moreover, we know that $\lim_{k_z\to 0}f(k_z,z)=0$ for $0<\alpha=\beta<\infty$, so the terms involving $k_z=0$ or $l_z=0$ can be ignored.

\subsubsection{$k_z>0$, $l_z>0$}

In this case $f(k_z,z)=f_1(k_z,z)$, $f(l_z,z)=f_1(l_z,z)$,
\be
\int_{-n}^{n} dz\overline{f(l_z, z)}f(k_z, z)
&=&
\frac{-\overline{f_1(l_z,n)}\partial_z f_1(k_z,n)
+
f_1(k_z,n)\partial_z \overline{f_1(l_z,n)}
}{k_z^2-l_z^2}
\nonumber\\
&\pp=&
+
\frac{\overline{f_1(l_z,-n)}\partial_z f_1(k_z,-n)
-
f_1(k_z,-n)\partial_z \overline{f_1(l_z,-n)}
}{k_z^2-l_z^2}
\label{-a to a}.
\ee
Inserting the explicit forms
\be
f_1(k_z, z)
&=&
\left\{
\begin{array}{ll}
e^{-ik_zn}+B_1(k_z)e^{ik_zn}  & {\rm at}\, z=-n
\\
E_1(k_z) e^{ik_zn} & {\rm at}\, z=n
\end{array}
\right.
\\
\partial_z f_1(k_z, z)
&=&
\left\{
\begin{array}{ll}
ik_ze^{-ik_zn}-ik_zB_1(k_z)e^{ik_zn}  & {\rm at}\, z=-n
\\
ik_zE_1(k_z) e^{ik_zn} & {\rm at}\, z=n
\end{array}
\right.
\ee
into (\ref{-a to a}) we find after some simplifications
\be
(\ref{-a to a})
&=&
-i\frac{1}{k_z-l_z}E_1(k_z) \overline{E_1(l_z)} e^{i(k_z-l_z)n}
\nonumber\\
&\pp=&
+
i\frac{
e^{i(l_z-k_z)n}
-\overline{B_1(l_z)}B_1(k_z)e^{i(k_z-l_z)n}
}{k_z-l_z}
\nonumber\\
&\pp=&
-i
\frac{
B_1(k_z)e^{i(l_z+k_z)n}
-\overline{B_1(l_z)}e^{-i(l_z+k_z)n}
}{k_z+l_z}.
\nonumber
\ee
One recognizes here several delta-sequences of the form (\ref{complex delta s}), and
\be
\theta(l_z)\theta(k_z)\int_{-\infty}^{\infty} dz\overline{f(l_z, z)}f(k_z, z)
&=&
\theta(k_z)^2\pi |E_1(k_z)|^2 \delta(k_z-l_z)
\nonumber\\
&\pp=&
+
\theta(k_z)^2\pi\delta(l_z-k_z)
\nonumber\\
&\pp=&
+\theta(k_z)^2\pi|B_1(k_z)|^2
\delta(k_z-l_z)
\nonumber\\
&\pp=&
+\pi
\theta(-k_z)\theta(k_z)B_1(k_z)\delta(l_z+k_z)
\nonumber\\
&\pp=&
+
\theta(-k_z)\theta(k_z)\pi\overline{B_1(-k_z)}
\delta(l_z+k_z).
\ee
Although $\theta(k_z)+\theta(-k_z)=1$ implies $\theta(0)=1/2$, the part corresponding to $k_z=0$ will not contribute to the final result since $f(0,z)=0$ for nonzero $\alpha$. In the final formula we are thus allowed to set  $\theta(-k_z)\theta(k_z)=0$ (this anyway holds true up to the set
$\{k_z=0\}$ whose measure is zero).
Moreover, we know that $|E_1(k_z)|^2+|B_1(k_z)|^2=1$.
So, finally
\be
\theta(l_z)\theta(k_z)\int_{-\infty}^{\infty} dz\overline{f(l_z, z)}f(k_z, z)
&=&
2\pi \delta(k_z-l_z)\theta(l_z)\theta(k_z).
\ee
In a similar way we show that
\be
\theta(-l_z)\theta(-k_z)\int_{-\infty}^{\infty} dz\overline{f(l_z, z)}f(k_z, z)
&=&
2\pi \delta(k_z-l_z)\theta(-l_z)\theta(-k_z).
\ee
\subsubsection{$k_z>0$, $l_z<0$}

In this case $f(k_z,z)=f_1(k_z,z)$, $f(l_z,z)=f_3(l_z,z)$.
\be
\int_{-n}^{n} dz\overline{f(l_z, z)}f(k_z, z)
&=&
\frac{-\overline{f_3(l_z,n)}\partial_z f_1(k_z,n)
+
f_1(k_z,n)\partial_z \overline{f_3(l_z,n)}
}{k_z^2-l_z^2}
\nonumber\\
&\pp=&
+
\frac{\overline{f_3(l_z,-n)}\partial_z f_1(k_z,-n)
-
f_1(k_z,-n)\partial_z \overline{f_3(l_z,-n)}
}{k_z^2-l_z^2}.\label{755}
\ee
Inserting
\be
\overline{f_3(l_z,z)}
&=&
\left\{
\begin{array}{ll}
E_1(l_z)e^{il_zn}
& {\rm at}\, z=-n \\
B_1(l_z) e^{il_zn}+e^{-il_zn} & {\rm at}\, z=n
\end{array}
\right.
\nonumber\\
\partial_z\overline{f_3(l_z,z)}
&=&
\left\{
\begin{array}{ll}
-il_zE_1(l_z)e^{il_zn}
& {\rm at}\, z=-n \\
il_zB_1(l_z) e^{il_zn}-il_ze^{-il_zn} & {\rm at}\, z=n
\end{array}
\right.
\nonumber
\ee
we get after some transformations
\be
(\ref{755})
&=&
-iB_1(l_z) E_1(k_z) \frac{e^{i(k_z+l_z)n}}{k_z+l_z}
-
iE_1(k_z) \frac{e^{i(k_z-l_z)n}}{k_z-l_z}
\nonumber\\
&\pp=&
-
iB_1(k_z)E_1(l_z)\frac{e^{i(k_z+l_z)n}}{k_z+l_z}
-
iE_1(l_z)\frac{e^{i(l_z-k_z)n}}{l_z-k_z}
\nonumber
\ee
Finally,
\be
\theta(-l_z)\theta(k_z)\int_{-\infty}^{\infty} dz\overline{f(l_z, z)}f(k_z, z)
&=&
\pi B_1(-k_z) E_1(k_z) \delta(k_z+l_z)\theta(-l_z)\theta(k_z)
\nonumber\\
&\pp=&
+
\pi E_1(k_z) \delta(k_z-l_z)\theta(-l_z)\theta(k_z)
\nonumber\\
&\pp=&
+
\pi B_1(k_z)E_1(-k_z)\delta(k_z+l_z)\theta(-l_z)\theta(k_z)
\nonumber\\
&\pp=&
+
\pi E_1(k_z)\delta(l_z-k_z)\theta(-l_z)\theta(k_z)
\nonumber\\
&=&
\pi \Big(\underbrace{\overline{B_1(k_z)} E_1(k_z)+B_1(k_z)\overline{E_1(k_z)}}_0\Big) \delta(k_z+l_z)\theta(k_z)^2
\nonumber\\
&=&
0=2\pi \delta(k_z-l_z)\theta(-l_z)\theta(k_z).\nonumber
\ee
Complex conjugating the above formula we obtain
\be
\theta(l_z)\theta(-k_z)\int_{-\infty}^{\infty} dz\overline{f(l_z, z)}f(k_z, z)
&=&
2\pi \delta(k_z-l_z)\theta(l_z)\theta(-k_z).\nonumber
\ee
This ends the proof. $\blacksquare$

\section{Field operator for finite $\alpha$ and $\beta$}\label{sec tilde f}

For the field operator $\phi(x)$ we have to solve
\be
\Big(\Box
+
\alpha\,\delta(z+L/2)+\beta\,\delta(z-L/2)
\Big)e^{-i|\bm k|x_0}
e^{i\bm k_\parallel \cdot \bm x}
\tilde f(k_z, z)
&=&0,
\ee
or, equivalently,
\be
\Big(-k_z^2
-
\partial_z^2
+
\alpha\,\delta(z+L/2)+\beta\,\delta(z-L/2)
\Big)
\tilde f(k_z, z)
&=&0,
\ee
which is a 1-dimensional stationary Schr\"odinger equation with eigenvalue $k_z^2$ of the same Hamiltonian as before.

Denoting $\tilde g(z)=\tilde f(k_z,z)$, $\tilde g'(z)=\tilde f'(k_z,z)$, integrating in neighborhoods of $z=\pm L/2$,
\be
k_z^2\int_{-L/2-\epsilon}^{-L/2+\epsilon} dz\,\tilde g(z)
&=&
\int_{-L/2-\epsilon}^{- L/2+\epsilon} dz\,
\Big(
-\tilde g''(z)+\alpha\,\delta(z+L/2)\tilde g(z)+\beta\,\delta(z-L/2)\tilde g(z)
\Big)
\nonumber\\
&=&
-\tilde g'(-L/2+\epsilon)+\tilde g'(-L/2-\epsilon)+\alpha\,\tilde g(-L/2),\\
k_z^2\int_{L/2-\epsilon}^{L/2+\epsilon} dz\,\tilde g(z)
&=&
-\tilde g'(L/2+\epsilon)+g'(L/2-\epsilon)+\beta\,\tilde g(L/2),
\ee
and evaluating the limit $\epsilon\to 0_+$, we find the boundary conditions
\be
\tilde f'(k_z,-L/2+0_+)-\tilde f'(k_z,-L/2-0_+)&=&\alpha\,\tilde f(k_z,-L/2),\\
\tilde f'(k_z,L/2+0_+)-\tilde f'(k_z,L/2-0_+)&=&\beta\,\tilde f(k_z,L/2),\\
\tilde f(k_z,-L/2+0_+)-\tilde f(k_z,-L/2-0_+)&=&0,\\
\tilde f(k_z,L/2+0_+)-\tilde f(k_z,L/2-0_+)&=&0.
\ee
Boundary conditions for the analogue of the case (i) from the preceding section read explicitly
\be
\tilde A_1 e^{-ikL/2}+\tilde B_1e^{ikL/2} &=& \tilde C_1 e^{-ikL/2}+\tilde D_1e^{ikL/2},\nonumber\\
\tilde C_1 e^{ikL/2}+\tilde D_1e^{-ikL/2}  &=& \tilde E_1 e^{ikL/2},\nonumber\\
ik(\tilde A_1 e^{-ikL/2}-\tilde B_1e^{ikL/2}) &=& ik(\tilde C_1 e^{-ikL/2}-\tilde D_1e^{ikL/2})-\alpha\Big( \tilde C_1 e^{-ikL/2}+\tilde D_1e^{ikL/2}\Big),\nonumber\\
ik(\tilde C_1 e^{ikL/2}-\tilde D_1e^{-ikL/2}) &=& ik \tilde E_1 e^{ikL/2}-\beta \tilde E_1 e^{ikL/2}\nonumber.
\ee
The rule is clear: We can use the results derived for $O_0(x)$ but we have to replace $L$ by $2L$ in exponents, and $\alpha$ and $\beta$ by $\alpha/2$ and $\beta/2$. Then
\be
\phi(x)
&=&
\int_{\mathbb{R}^2} d^2k\int_0^\infty\frac{dk_z}{(2\pi)^32|\bm k|} c(\bm{k})e^{-i|\bm k|x_0}
e^{i\bm k_\parallel \cdot \bm x}
\tilde f_1(k_z,z)
\nonumber\\
&\pp=&
+
\int_{\mathbb{R}^2} d^2k\int_{-\infty}^0\frac{dk_z}{(2\pi)^32|\bm k|} c(\bm{k})e^{-i|\bm k|x_0}
e^{i\bm k_\parallel \cdot \bm x}
\tilde f_3(k_z,z)+{\rm H.c.}
\nonumber\\
&=&
\int_{\mathbb{R}^3}\frac{d^3k}{(2\pi)^32|\bm k|} c(\bm{k})e^{-i|\bm k|x_0}
e^{i\bm k_\parallel \cdot \bm x}
\tilde f(k_z,z)
+{\rm H.c.}
\ee
\section{Green functions and vacuum averages of field operators}

One has to be aware that for general representations of HOLA the known links between propagators and vacuum averages are no longer true.

Let us denote
\be
\Box_\alpha
&=&
\Box
+
\alpha\,\delta(z+L/2)+\alpha\,\delta(z-L/2)\nonumber\\
&=&
\partial_0^2-\Delta
+
\alpha\,\delta(z+L/2)+\alpha\,\delta(z-L/2)\nonumber\\
&=&
\partial_0^2-\Delta_\alpha.
\ee
I would like to discuss the link between Green functions, that is distributions satisfying
\be
\Box_\alpha G(x,x')
&=&\delta^{(4)}(x-x'),\label{Green}
\ee
and products of field operators quantized with arbitrary representations of the central element $I_0(\bm k)$. For $I_0(\bm k)=1$ we should reconstruct the ordinary Green functions from vacuum averages of products (or commutators) of field operators. Let us begin with the generalization of the Pauli-Jordan function.

\subsection{Generalization of Pauli-Jordan $D(x)$}

Let us define
\be
D^{(+)}(x,x')
&=&
i\int_{\mathbb{R}^3}\frac{d^3k}{(2\pi)^3 2|\bm k|}
e^{-i|\bm k|(x_0-x_0')}
e^{i\bm k_\parallel \cdot (\bm x-\bm x')}
\tilde f(k_z,z)\overline{\tilde f(k_z,z')}
\nonumber\\
D^{(-)}(x,x')
&=&
-i\int_{\mathbb{R}^3}\frac{d^3k}{(2\pi)^3 2|\bm k|}
e^{i|\bm k|(x_0-x_0')}
e^{i\bm k_\parallel \cdot (\bm x-\bm x')}
\tilde f(k_z,z)\overline{\tilde f(k_z,z')},
\nonumber\\
D(x,x')
&=&
D^{(+)}(x,x')
+
D^{(-)}(x,x')
\nonumber\\
&=&
\int_{\mathbb{R}^3}\frac{d^3k}{(2\pi)^3 |\bm k|}
\sin\Big(|\bm k|(x_0-x_0')\Big)
e^{i\bm k_\parallel \cdot (\bm x-\bm x')}
\tilde f(k_z,z)\overline{\tilde f(k_z,z')}
\label{PL alpha},
\nonumber\\
D_1(x,x')
&=&
D^{(+)}(x,x')
-
D^{(-)}(x,x')
\nonumber\\
&=&
i\int_{\mathbb{R}^3}\frac{d^3k}{(2\pi)^3 |\bm k|}
\cos\Big(|\bm k|(x_0-x_0')\Big)
e^{i\bm k_\parallel \cdot (\bm x-\bm x')}
\tilde f(k_z,z)\overline{\tilde f(k_z,z')}
\label{PL1 alpha},
\ee
where $
\bm k_\parallel \cdot (\bm x-\bm x')
=
k_x(x-x')+k_y(y-y')$.
(Note that here $x=x^1$, $y=x^2$, $z=x^3$, whereas $x$ and $x'$ in $D(x,x')$ are space-time points.)
The functions satisfy
\be
\Box_\alpha D^{(\pm)}(x,x')=\Box'_\alpha D^{(\pm)}(x,x')=0.
\ee
Similarly to the usual Pauli-Jordan function one finds
\be
\partial_0 D(x,x')\big|_{x_0=x_0'}
&=&
\int_{\mathbb{R}^3}\frac{d^3k}{(2\pi)^3 }
e^{i\bm k_\parallel \cdot (\bm x-\bm x')}
\tilde f(k_z,z)\overline{\tilde f(k_z,z')}
=
\delta^{(3)}(\bm x-\bm x'),\\
D(x,x')\big|_{x_0=x_0'}
&=&0.
\ee

\subsection{Green functions}

Repeating standard calculations (cf. \cite{Greiner-QFT}, p. 114) we define the Green functions:
\be
G_R(x,x') &=& \theta(x_0-x_0')D(x,x')\quad \textrm{(retarded propagator)},\\
G_A(x,x') &=& -\theta(x_0'-x_0)D(x,x')\quad \textrm{(advanced propagator)},\\
G_F(x,x') &=& \frac{1}{2}\Big(\varepsilon(x_0-x_0')D(x,x')+D_1(x,x')\Big)\quad \textrm{(Feynman propagator)},\\
G_D(x,x') &=& \frac{1}{2}\Big(\varepsilon(x_0-x_0')D(x,x')-D_1(x,x')\Big)\quad \textrm{(Dyson propagator)},\\
\bar G(x,x') &=& \frac{1}{2}\varepsilon(x_0-x_0')D(x,x')\quad \textrm{(principal-part propagator)},\\
\varepsilon(x) &=& 2\theta(x)-1=
\left\{
\begin{array}{cl}
1 & \textrm{for }x>0\\
0 & \textrm{for }x=0\\
-1 & \textrm{for }x<0
\end{array}
\right.
 \quad \textrm{(the sign function)}.
\ee
Let us prove that $\Box_\alpha G_R(x,x')=\delta^{(4)}(x-x')$. All the remaining proofs are essentially identical.
One begins with the general formula
\be
\frac{d}{dx}\int_a^x dy\,f(x,y)
&=&
f(x,x)+\int_a^x dy \frac{\partial f(x,y)}{\partial x}.
\ee
Then
\be
{}&{}&
\Box_\alpha \int_{\mathbb{R}^4}d^4x'\,G_R(x,x')g(x')
\nonumber\\
&{}&\pp=
=
(\partial_0^2-\Delta_\alpha) \int_{-\infty}^{x_0}dx_0'\int_{\mathbb{R}^3}d^3x'\,D(x_0,\bm x,x_0',\bm x')g(x_0',\bm x')
\nonumber\\
&{}&\pp=
=
\partial_0\int_{\mathbb{R}^3}d^3x'\underbrace{D(x_0,\bm x,x_0,\bm x')}_0 g(x_0,\bm x')
+
\partial_0\int_{-\infty}^{x_0}dx_0'\int_{\mathbb{R}^3}d^3x'\,\partial_0 D(x,x')g(x')
\nonumber\\
&{}&\pp{==}
-
 \int_{-\infty}^{x_0}dx_0'\int_{\mathbb{R}^3}d^3x'\,\Delta_\alpha D(x,x')g(x')
\nonumber\\
&{}&\pp=
=
\int_{\mathbb{R}^3}d^3x'\underbrace{\partial_0 D(x_0,\bm x,x_0',\bm x')\big|_{x_0'=x_0}}_{\delta^{(3)}(\bm x-\bm x')}g(x_0,\bm x')
\nonumber\\
&{}&\pp{==}
+
\underbrace{\int_{-\infty}^{x_0}dx_0'\int_{\mathbb{R}^3}d^3x'\,\partial_0^2 D(x,x')g(x')
-
 \int_{-\infty}^{x_0}dx_0'\int_{\mathbb{R}^3}d^3x'\,\Delta_\alpha D(x,x')g(x')}_{\textrm{$0$ since $\Box_\alpha D(x,x')=0$}}
\nonumber\\
&{}&\pp=
=
g(x_0,\bm x)=\int_{\mathbb{R}^4}d^4x'\,\delta^{(4)}(x-x')g(x')\nonumber,
\ee
which was to be demonstrated.

\subsection{Vacuum averages}

Let $\Pi_0$ denote the projector on the vacuum subspace. Then
\be
\Pi_0
\phi(x)\phi(x')
\Pi_0
&=&
\int dk dk'\, \Pi_0c(\bm{k})e^{-i|\bm k|x_0}e^{i\bm k_\parallel \cdot \bm x}\tilde f(k_z,z)
c(\bm{k}')^\dag \Pi_0 e^{i|\bm k'|x_0'}e^{-i\bm k'_\parallel \cdot \bm x'}\overline{\tilde f(k_z',z')}
\nonumber\\
&=&
\int dk \, I_0(\bm{k})e^{-i|\bm k|(x_0-x_0')}e^{i\bm k_\parallel \cdot (\bm x-\bm x')}\tilde f(k_z,z)\overline{\tilde f(k_z',z')}\Pi_0
\nonumber
\ee
For $I_0(\bm k)=1$ one finds
\be
\Pi_0
\phi(x)\phi(x')
\Pi_0
=
-i D^{(+)}(x,x')\Pi_0.
\ee
In the reducible representation of HOLA one finds
\be
\Pi_0
\phi(x,N)\phi(x',N)
\Pi_0
=
-i D^{(+)}(x,x',N)\Pi_0,
\ee
where
\be
D^{(+)}(x,x',N)
&=&
i\int dk \, I_0(\bm{k},N)e^{-i|\bm k|(x_0-x_0')}e^{i\bm k_\parallel \cdot (\bm x-\bm x')}\tilde f(k_z,z)\overline{\tilde f(k_z',z')}
\ee
is the operator whose analogue  we encountered in our analysis of a pointlike charge.

\section{Free-field Hamiltonian with boundary conditions}

Free-field Hamiltonians with nontrivial boundary conditions can be defined in several ways. One is simply to take a free-field Hamiltonian density and integrate over appropriate volumes, taking into account different types of fields in different integration ranges (say, inside and outside of a cavity). Cavities whose boundaries are not fully reflecting (i.e. with finite and non-zero $\alpha$) should be described by Hamiltonian densities that involve the delta-potentials, but integration is then over $\mathbb{R}^3$.

Let us begin with the field operator satisfying
\be
\Box \phi(x)
+
\alpha\,\delta(z+L/2)\phi(x)+\alpha\,\delta(z-L/2)\phi(x)=0.\label{Box phi}
\ee
The potential is nontrivial but time-independent. The Hamiltonian must therefore be independent of time, but the 3-momentum will not be conserved due to the presence of the boundary. Now, let $\phi_1(x)$ and $\phi_2(x)$ be two solutions of (\ref{Box phi}). We know that they can be written as
\be
\phi_j(x)
&=&
\int dk\, c_j(\bm{k})e^{-i|\bm k|x_0}e^{i\bm k_\parallel \cdot \bm x}\tilde f(k_z,z)
+
\int dk\, c_j(\bm{k})^\dag e^{i|\bm k|x_0}e^{-i\bm k_\parallel \cdot \bm x}\overline{\tilde f(k_z,z)}
\ee
with some $c_j(\bm{k})$. $\tilde f(k_z,z)$ is normalized by
\be
\int_{-\infty}^{\infty} \frac{dz}{2\pi}\overline{\tilde f(l_z, z)}\tilde f(k_z, z)
=
\delta(k_z-l_z).
\ee
The Noether current
\be
T{_\mu}{^r}(x)
&=&
\frac{1}{2}\Big(
\partial^r \phi_1(x)\partial_\mu \phi_2(x)
+
\partial_\mu \phi_1(x)\partial^r \phi_2(x)
\Big)
\nonumber\\
&\pp=&
-
\frac{1}{2}\partial_\nu\phi_1(x)\partial^\nu\phi_2(x)g{_\mu}{^r}
+
\frac{1}{2}\alpha\Big(\delta(z+L/2)+\delta(z-L/2)\Big)\phi_1(x)\phi_2(x)
g{_\mu}{^r}
\ee
satisfies
\be
\partial^\mu T{_\mu}{^r}(x)
&=&
\frac{1}{2}\alpha\partial^r\Big(\delta(z+L/2)+\delta(z-L/2)\Big)\phi_1(x)\phi_2(x),
\ee
leading to three conservation laws for $r=0,1,2$, and the non-conserved $z$-component of momentum,
\be
\partial^\mu T{_\mu}{^3}(x)
&=&
-\frac{1}{2}\alpha\Big(\delta'(z+L/2)+\delta'(z-L/2)\Big)\phi_1(x)\phi_2(x)
\nonumber\\
&=&
\frac{1}{2}\alpha\Big(\delta(z+L/2)+\delta(z-L/2)\Big)\frac{\partial}{\partial z}\phi_1(x)\phi_2(x).
\ee
(Note that $\partial^3=\partial/\partial x_3=-\partial/\partial z$.)
The Hamiltonian density ${\cal H}(x)$ is related to $T{_\mu}{^r}(x)$ by
\be
T{_0}{_0}(x)
&=&
\frac{1}{2}
\Big(
\partial_0 \phi_1(x)\partial_0\phi_2(x)
+
\bm\nabla \phi_1(x)\cdot \bm\nabla  \phi_2(x)
\Big)
\nonumber\\
&\pp=&
+
\frac{\alpha}{2}\Big(\delta(z+L/2)+\delta(z-L/2)\Big)\phi_1(x)\phi_2(x)
\\
&=&
\frac{1}{2}
\partial_0 \phi_1(x)\partial_0\phi_2(x)
+
\frac{1}{2}\phi_1(x)\Big(-\Delta+\alpha\delta(z+L/2)+\alpha\delta(z-L/2)\Big)\phi_2(x)
\nonumber\\
&\pp=&
+
\frac{1}{2}\bm\nabla\cdot \Big(\phi_1(x)\bm\nabla  \phi_2(x)\Big)
\\
&=&
\underbrace{\frac{1}{2}
\partial_0 \phi_1(x)\partial_0\phi_2(x)
-
\frac{1}{2}\phi_1(x)\partial_0^2\phi_2(x)}_{{\cal H}(x)}
+
\underbrace{\frac{1}{2}\bm\nabla\cdot \Big(\phi_1(x)\bm\nabla  \phi_2(x)\Big)}_{\textrm{boundary-at-infinity term}}
\ee
In reducible representations of HOLA one can rigorously eliminate the boundary term by an appropriate choice of vacuum. The part involving
$\cal H$ leads to the same Hamiltonian as the one described in Section \ref{sec dig on sf}. We conclude that it is justified to take without any modification the free-field Hamiltonians we have discussed earlier in these lectures, even if $\alpha>0$.

\section{Pressure exerted by cavity walls on vacuum fields}

The $z$-component of the momentum operator,
\be
P^3(x_0)
&=&
\int_{\mathbb{R}^3}d^3x T{_0}^3(x_0,\bm x),
\ee
is not conserved.
The corresponding force operator can be defined as \cite{Barut}
\be
\partial_0 P^k(x_0)
&=&
\int_{\mathbb{R}^3}d^3x\,
\partial_0 T{_0}^k(x_0,\bm x)
+
\sum_{j=1}^3\underbrace{\int_{\mathbb{R}^3}d^3x
\,\partial^j T{_j}^k(x_0,\bm x)}_0
\nonumber\\
&=&
\int_{\mathbb{R}^3}d^3x\, \partial^\mu T{_\mu}^k(x_0,\bm x),
\ee
leading to the force density
\be
{\cal F}^k(x)
&=&
\partial^\mu T{_\mu}^k(x).\label{cal F2}
\ee
In electrodynamics the force density integrated over a volume $V$ and taken with the minus sign yields the Lorentz force acting on charges contained in $V$. In the case of metallic boundary conditions this would be the net mechanical force acting on the metal. In our case, an integral over some $V$ containing a part $S$ of the boundary (and taken with the opposite sign) is expected to coincide with the Casimir force acting on $S$, if the field is in a vacuum state. The change of sign is necessary since the time derivative of field momentum is the force exerted by the boundary on the field, and not the other way around.

The force operator corresponding to the square
\be
S=\{\bm x'=(x',y',L/2);\, x-\epsilon/2\leq x'\leq x+\epsilon/2,\, y-\epsilon/2\leq y'\leq y+\epsilon/2\}
\ee
thus reads
\be
F_\epsilon^3
&=&
\int_{x-\epsilon/2}^{x+\epsilon/2} dx'\int_{y-\epsilon/2}^{y+\epsilon/2} dy'\int_{L/4}^{3L/4} dz'\,\partial^\mu T{_\mu}^3(x_0,\bm x')
\nonumber\\
&=&
\frac{1}{2}\alpha\int_{x-\epsilon/2}^{x+\epsilon/2} dx'\int_{y-\epsilon/2}^{y+\epsilon/2} dy'\int_{-\infty}^{\infty} dz'\,\delta(z'-L/2)\frac{\partial}{\partial z'}\phi_1(x')\phi_2(x').
\ee
The pressure at $\bm x=(x,y,L/2)$ equals $\lim_{\epsilon\to 0}F_\epsilon^3/\epsilon^2$.

Let us now compute the explicit form of $\Pi_0 F_\epsilon^3(x_0,x,y)\Pi_0 $, evaluated  $\phi_1=\phi_2=\phi$. Recall that we first perform the integrals and set $\phi_1=\phi_2=\phi$ afterwards; $\Pi_0$ is the projector on the vacuum subspace.
\be
\Pi_0 F_\epsilon^3\Pi_0
&=&
\frac{1}{2}\alpha\int_{x-\epsilon/2}^{x+\epsilon/2} dx'\int_{y-\epsilon/2}^{y+\epsilon/2} dy'\int_{-\infty}^{\infty} dz'\,\delta(z'-L/2)
\nonumber\\
&\pp=&\times
\frac{\partial}{\partial z'}
\int dk\, \Pi_0 c_2(\bm{k})e^{-i|\bm k|x_0}e^{i\bm k_\parallel \cdot \bm x'}\tilde f(k_z,z')
\int dk'\, c_1(\bm{k}')^\dag \Pi_0 e^{i|\bm k'|x_0}e^{-i\bm k'_\parallel \cdot \bm x'}\overline{\tilde f(k'_z,z')}
\nonumber\\
&=&
\frac{1}{2}\alpha
\int dkdk'\,e^{i(|\bm k'|-|\bm k|)x_0}
\int_{x-\epsilon/2}^{x+\epsilon/2} dx'\,e^{i(k_x-k'_x)x'}\int_{y-\epsilon/2}^{y+\epsilon/2} dy'\,e^{i(k_y-k'_y)y'}
\nonumber\\
&\pp=&\times
\Pi_0 [c_2(\bm{k}),c_1(\bm{k}')^\dag] \Pi_0 \int_{-\infty}^{\infty} dz\,\delta(z-{\textstyle{\frac{L}{2}}})
\Big(\overline{\tilde f'(k'_z,z)}\tilde f(k_z,z)+\overline{\tilde f(k'_z,z)}\tilde f'(k_z,z)\Big)
\nonumber\\
&=&
\frac{1}{2}\alpha\epsilon^2
\int dkdk'\,e^{i(|\bm k'|-|\bm k|)x_0}
\frac{\sin(k_x-k'_x)\epsilon/2}{(k_x-k'_x)\epsilon/2}\frac{\sin(k_y-k'_y)\epsilon/2}{(k_y-k'_y)\epsilon/2}
e^{i(k_x-k'_x)x}e^{i(k_y-k'_y)y}
\nonumber\\
&\pp=&\times
\Pi_0 [c_2(\bm{k}),c_1(\bm{k}')^\dag] \Pi_0
\Big(\overline{\tilde f'(k'_z,L/2)}\tilde f(k_z,L/2)+\overline{\tilde f(k'_z,L/2)}\tilde f'(k_z,L/2)\Big).
\label{F epsilon}
\ee
We have to bear in mind that $\tilde f'(k_z,z)$ is discontinuous at $L/2$. If we assume that Dirac delta is constructed from symmetric delta-sequences, then
\be
\tilde f'(k_z,L/2)
&=&
\frac{1}{2}
\Big(\tilde f'\big(k_z,(L/2)_+\big)+\tilde f'\big(k_z,(L/2)_-\big)\Big).\label{delta z f'}
\ee
However, it is instructive to consider the more general case
\be
\tilde f'(k_z,L/2)
&=&
p_+\tilde f'\big(k_z,(L/2)_+\big)+p_-\tilde f'\big(k_z,(L/2)_-\big),\label{delta z f'p+-}\\
1 &=& p_+ +p_-.\nonumber
\ee
Inserting $\phi_1=\phi_2=\phi$ we obtain
\be
(\ref{F epsilon})
&\to&
\frac{1}{2}\alpha\epsilon^2
\int dkdk'\,e^{i(|\bm k'|-|\bm k|)x_0}
\frac{\sin(k_x-k'_x)\epsilon/2}{(k_x-k'_x)\epsilon/2}\frac{\sin(k_y-k'_y)\epsilon/2}{(k_y-k'_y)\epsilon/2}
e^{i(k_x-k'_x)x}e^{i(k_y-k'_y)y}
\nonumber\\
&\pp=&\times
\Pi_0 I(\bm{k})\delta_0(\bm{k},\bm{k}')
\Big(\overline{\tilde f'(k'_z,L/2)}\tilde f(k_z,L/2)+\overline{\tilde f(k'_z,L/2)}\tilde f'(k_z,L/2)\Big)
\nonumber\\
&=&
\frac{1}{2}\alpha\epsilon^2
\int dk\,
\Pi_0 I(\bm{k})
\Big(\overline{\tilde f'(k_z,L/2)}\tilde f(k_z,L/2)+\overline{\tilde f(k_z,L/2)}\tilde f'(k_z,L/2)\Big).
\ee
In this way we have arrived at the vacuum pressure operator
\be
p_{\rm vac}
&=&
\alpha
\int dk\,
I(\bm{k})
\Re\Big(\overline{\tilde f'(k_z,L/2)}\tilde f(k_z,L/2)\Big)\Pi_0 .
\ee
In the end we have not needed the limit $\epsilon\to 0$. The pressure is the same at all points of the cavity wall.
Knowing that
\be
\tilde f'\big(k_z,(L/2)_+\big)&=&\tilde f'\big(k_z,(L/2)_-\big)+\alpha\,\tilde f(k_z,L/2),
\ee
we can eliminate $\tilde f'\big(k_z,(L/2)_+\big)$ in (\ref{delta z f'p+-}),
\be
\tilde f'(k_z,L/2)
&=&
p_+\Big(\tilde f'\big(k_z,(L/2)_-\big)+\alpha\,\tilde f(k_z,L/2)\Big)
+
p_-\tilde f'\big(k_z,(L/2)_-\big)
\nonumber\\
&=&
\tilde f'\big(k_z,(L/2)_-\big)+\alpha p_+\,\tilde f(k_z,L/2).
\ee
Therefore, alternatively,
\be
p_{\rm vac}
&=&
\alpha
\int dk
I_0(\bm{k})
\Re
\Bigg(
\Big(\overline{\tilde f'\big(k_z,(L/2)_-\big)+\alpha p_+\tilde f(k_z,L/2)}\Big)\tilde f(k_z,L/2)
\Bigg)\Pi_0.
\ee
\section{How to choose $p_+$?}

The dependence on $p_+$ is somewhat disturbing. So, let us reformulate the problem. Consider a region $X=[t,t+\Delta t]\times V$ of the Minkowski space consisting of points $x=(x^0,x^1,x^2,x^3)$ satisfying
$t<x_0<t+\Delta t$, $a_1<x^1<a_2$, $b_1<x^2<b_2$, $c_1<x^3<c_2$, and assume that the energy-momentum tensor $T^{\mu\nu}(x)$ satisfies continuity equation $\partial_\mu T^{\mu\nu}(x)=0$ for $x\in X$ (i.e. $x$ is not at the cavity wall). Then,
\be
0
&=&
\int_Xd^4x\,\partial_\mu T^{\mu\nu}(x)\nonumber\\
&=&
\int_{t}^{t+\Delta t} dx^0\int_{a_1}^{a_2} dx^1\int_{b_1}^{b_2} dx^2\int_{c_1}^{c_2} dx^3\,
\partial_0 T^{0\nu}(x)\nonumber\\
&\pp=&+
\int_{t}^{t+\Delta t} dx^0\int_{a_1}^{a_2} dx\int_{b_1}^{b_2} dy\int_{c_1}^{c_2} dx^3\,\partial_1 T^{1\nu}(x)\nonumber\\
&\pp=&+
\int_{t}^{t+\Delta t} dx^0\int_{a_1}^{a_2} dx\int_{b_1}^{b_2} dy\int_{c_1}^{c_2} dx^3\,\partial_2 T^{2\nu}(x)\nonumber\\
&\pp=&+
\int_{t}^{t+\Delta t} dx^0\int_{a_1}^{a_2} dx\int_{b_1}^{b_2} dy\int_{c_1}^{c_2} dx^3\,\partial_3 T^{3\nu}(x)\nonumber\\
&=&
\int_{a_1}^{a_2} dx\int_{b_1}^{b_2} dy\int_{c_1}^{c_2} dz\,
\Big(T^{0\nu}(t+\Delta t,\bm x)-T^{0\nu}(t,\bm x)\Big)\nonumber\\
&\pp=&+
\int_{t}^{t+\Delta t} dx^0\int_{b_1}^{b_2} dy\int_{c_1}^{c_2} dz\,
\Big(T^{1\nu}(x^0,a_2,y,z)-T^{1\nu}(x^0,a_1,y,z)\Big)\nonumber\\
&\pp=&+
\int_{t}^{t+\Delta t} dx^0\int_{a_1}^{a_2} dx\int_{c_1}^{c_2} dz\,
\Big(T^{2\nu}(x^0,x,b_2,z)-T^{2\nu}(x^0,x,b_1,z)\Big)\nonumber\\
&\pp=&+
\int_{t}^{t+\Delta t} dx^0\int_{a_1}^{a_2} dx\int_{b_1}^{b_2} dy\,
\Big(T^{3\nu}(x^0,x,y,c_2)-T^{3\nu}(x^0,x,y,c_1)\Big).\nonumber
\ee
These formulas lead to the 4-force operator
\be
\frac{d}{dt}
\int_V d^3x\,
T^{0\nu}(t,\bm x)
&=&
-
\int_{b_1}^{b_2} dy\int_{c_1}^{c_2} dz\,
\Big(T^{1\nu}(t,a_2,y,z)-T^{1\nu}(t,a_1,y,z)\Big)\nonumber\\
&\pp=&-
\int_{a_1}^{a_2} dx\int_{c_1}^{c_2} dz\,
\Big(T^{2\nu}(t,x,b_2,z)-T^{2\nu}(t,x,b_1,z)\Big)\nonumber\\
&\pp=&-
\int_{a_1}^{a_2} dx\int_{b_1}^{b_2} dy\,
\Big(T^{3\nu}(t,x,y,c_2)-T^{3\nu}(t,x,y,c_1)\Big)\nonumber\\
&=&
\int_V d^3x\,
{\cal F}^{\nu}(t,\bm x).\label{F density'}
\ee
(\ref{F density'}) consists of six terms concentrated on the six walls of $V$,
\be
{\cal F}^{\nu}(t,\bm x)
&=&
\big(\delta(x-a_1)-\delta(x-a_2)\big)T^{1\nu}(t,\bm x)\nonumber\\
&\pp=&
+
\big(\delta(y-b_1)-\delta(y-b_2)\big)T^{2\nu}(t,\bm x)\nonumber\\
&\pp=&
+
\big(\delta(z-c_1)-\delta(z-c_2)\big)T^{3\nu}(t,\bm x).
\ee
Now consider the $z=L/2$ plane and two volumes: $V_-$ whose $c_2=(L/2)_-$, and $V_+$ with $c_1=(L/2)_+$. The net pressure exerted on the plane at the point $(x,y,L/2)$ equals
\be
p
&=&
\lim_{\epsilon\to 0}\frac{1}{\epsilon^2}\int_{x-\epsilon/2}^{x+\epsilon/2}dx'\int_{y-\epsilon/2}^{y+\epsilon/2}dy'\,
\Big(T^{33}\big(t,x',y',(L/2)_+\big)-T^{33}\big(t,x',y',(L/2)_-\big)\Big).
\label{p_C}
\ee
In principle, (\ref{p_C}) could be further simplified to
\be
p
&=&
T^{33}\big(t,x,y,(L/2)_+\big)-T^{33}\big(t,x,y,(L/2)_-\big),
\ee
but one has to be cautious since in quantum field theory $T^{33}$ involves distributions. It is safer to analyze the more exact form (\ref{p_C}), i.e. first integrate and then take the limit.

Recall that in our construction of $p$ we work with the energy-momentum tensor that does not contain the boundary term,
\be
T{^3}{^3}(x)
&=&
\partial^3 \phi_1(x)\partial^3 \phi_2(x)
-
\frac{1}{2}\partial_\nu\phi_1(x)\partial^\nu\phi_2(x)g^{33}\\
&=&
\frac{1}{2}
\Big(
\partial_0 \phi_1(x)\partial_0 \phi_2(x)
-
\partial_1 \phi_1(x)\partial_1 \phi_2(x)
-
\partial_2 \phi_1(x)\partial_2 \phi_2(x)
+
\partial_3 \phi_1(x)\partial_3 \phi_2(x)
\Big).\nonumber
\ee
However, we could also insert the full $T^{\mu\nu}$ into (\ref{p_C}) --- the terms proportional to $\alpha$ would anyway have canceled out  due to continuity of the fields at $z=L/2$ (the first derivatives over $z$ are  discontinuous, but
\be
\frac{1}{2}\alpha\Big(\delta(z+L/2)+\delta(z-L/2)\Big)\phi_1(x)\phi_2(x)
g{_\mu}{^r}\nonumber
\ee
does not involve derivatives).
Pressure restricted to the vacuum subspace reads
\be
p_{\rm vac}
&=&
\frac{1}{2}\lim_{\epsilon\to 0}\frac{1}{\epsilon^2}\int_{x-\epsilon/2}^{x+\epsilon/2}dx'\int_{y-\epsilon/2}^{y+\epsilon/2}dy'\,
\nonumber\\
&\pp=&\times
\int dk\, \Pi_0 c_2(\bm{k})e^{-i|\bm k|x_0}e^{i\bm k_\parallel \cdot \bm x'}\tilde f'\big(k_z,(L/2)_+\big)
\int dk'\, c_1(\bm{k}')^\dag \Pi_0 e^{i|\bm k'|x_0}e^{-i\bm k'_\parallel \cdot \bm x'}\overline{\tilde f'\big(k'_z,(L/2)_+\big)}
\nonumber\\
&\pp=&
-[(L/2)_+\leftrightarrow (L/2)_-]
\nonumber\\
&=&
\frac{1}{2}\lim_{\epsilon\to 0}\frac{1}{\epsilon^2}
\int dkdk'\int_{x-\epsilon/2}^{x+\epsilon/2}dx'\int_{y-\epsilon/2}^{y+\epsilon/2}dy'\,
e^{i(|\bm k'|-|\bm k|)x_0}e^{i(k_x-k_x') x'}e^{i(k_y-k_y') y'}
\nonumber\\
&\pp=&\times
\Pi_0 c_2(\bm{k}) c_1(\bm{k}')^\dag \Pi_0 \overline{\tilde f'\big(k'_z,(L/2)_+\big)}\tilde f'\big(k_z,(L/2)_+\big)
-[(L/2)_+\leftrightarrow (L/2)_-]\nonumber\\
&=&
\frac{1}{2}\lim_{\epsilon\to 0}
\int dkdk'
e^{i(|\bm k'|-|\bm k|)x_0}e^{i(k_x-k_x') x}e^{i(k_y-k_y') y}
\frac{\sin(k_x-k_x')\epsilon/2}{(k_x-k_x')\epsilon/2}
\frac{\sin(k_y-k_y')\epsilon/2}{(k_y-k_y')\epsilon/2}
\nonumber\\
&\pp=&\times
\Pi_0 c_2(\bm{k}) c_1(\bm{k}')^\dag \Pi_0 \overline{\tilde f'\big(k'_z,(L/2)_+\big)}\tilde f'\big(k_z,(L/2)_+\big)
-[(L/2)_+\leftrightarrow (L/2)_-]\label{drugie to}
\ee
Inserting $\phi_1=\phi_2=\phi$, we obtain
\be
(\ref{drugie to})&\to&
\frac{1}{2}\lim_{\epsilon\to 0}
\int dkdk'
e^{i(|\bm k'|-|\bm k|)x_0}e^{i(k_x-k_x') x}e^{i(k_y-k_y') y}
\frac{\sin(k_x-k_x')\epsilon/2}{(k_x-k_x')\epsilon/2}
\frac{\sin(k_y-k_y')\epsilon/2}{(k_y-k_y')\epsilon/2}
\nonumber\\
&\pp=&\times
I_0(\bm{k})\delta_0(\bm{k},\bm{k}') \Pi_0 \overline{\tilde f'\big(k'_z,(L/2)_+\big)}\tilde f'\big(k_z,(L/2)_+\big)
-[(L/2)_+\leftrightarrow (L/2)_-]\nonumber\\
&=&
\frac{1}{2}
\int dk
I_0(\bm{k})\Pi_0
\Big(
\overline{\tilde f'\big(k_z,(L/2)_+\big)}\tilde f'\big(k_z,(L/2)_+\big)
-
\overline{\tilde f'\big(k_z,(L/2)_-\big)}\tilde f'\big(k_z,(L/2)_-\big)
\Big)\nonumber\\
&=&
\alpha
\int dk
I_0(\bm{k})
\Re
\Bigg(
\Big(\overline{\tilde f'\big(k_z,(L/2)_-\big)+\frac{\alpha}{2}\tilde f(k_z,L/2)}\Big)\tilde f(k_z,L/2)
\Bigg)\Pi_0
\ee
which is the same result as before, but with $p_+=1/2$.

Consistency of the two definitions of pressure leads us back to Dirac deltas constructed from symmetric delta sequences.

\section{Vacuum pressure through Green functions and their reducible-representation generalizations}

In order to phrase my derivation of pressure in a form typically employed in the literature, let me show its link with Green functions [generalized to arbitrary representations of $I_0(\bm k)$].

Let us return to
\be
T{^3}{^3}(x)
&=&
\frac{1}{2}
\Big(
\dots
+
\partial_3 \phi_1(x)\partial_3 \phi_2(x)
\Big).\nonumber
\ee
Its vacuum average can be obtained from
\be
\Pi_0 T{^3}{^3}(x)\Pi_0
&=&
\frac{1}{2}
\Pi_0\Big(
\dots
+
\partial_3 \phi_1(x)\partial_3 \phi_2(x)
\Big)\Pi_0\Big|_{\phi_1=\phi_2=\phi}\nonumber\\
&=&
\dots
+
\frac{1}{2}
\partial_3 \partial_3' \Pi_0\phi(x)\phi(x')\Pi_0\Big|_{x'=x}
\nonumber\\
&=&
\dots
-
\frac{i}{2}
\partial_3 \partial_3' D^{(+)}(x,x')\Pi_0\Big|_{x'=x}.
\label{T33 Pi0}
\ee
Inserting the explicit form
\be
D^{(+)}(x,x')
&=&
i\int dk \, I_0(\bm{k})e^{-i|\bm k|(x_0-x_0')}e^{i\bm k_\parallel \cdot (\bm x-\bm x')}\tilde f(k_z,z)\overline{\tilde f(k_z',z')}
\ee
into (\ref{T33 Pi0}) we find the expressions we have encountered in the preceding two sections,
\be
\Pi_0 T{^3}{^3}(x)\Pi_0
&=&
\dots
+
\frac{1}{2}
\int dk \, I_0(\bm{k})\tilde f'(k_z,z)\overline{\tilde f'(k_z',z)}\Pi_0,
\ee
and
\be
p_{\rm vac}
&=&
\frac{1}{2}
\int dk
I_0(\bm{k})\Pi_0
\Big(
\overline{\tilde f'\big(k_z,(L/2)_+\big)}\tilde f'\big(k_z,(L/2)_+\big)
-
\overline{\tilde f'\big(k_z,(L/2)_-\big)}\tilde f'\big(k_z,(L/2)_-\big)
\Big).
\nonumber
\ee
Let us note that the following rules are also true,
\be
\Pi_0 T{^3}{^3}(x)\Pi_0
&=&
\dots
+
\frac{i}{2}
\partial_3 \partial_3' D^{(-)}(x,x')\Pi_0\Big|_{x'=x}
\nonumber\\
&=&
\dots
+
\frac{i}{4}
\partial_3 \partial_3' \Big(D^{(-)}(x,x')-D^{(+)}(x,x')\Big)\Pi_0\Big|_{x'=x}
\nonumber\\
&=&
\dots
-
\frac{i}{4}
\partial_3 \partial_3' D_1(x,x')\Pi_0\Big|_{x'=x}
\nonumber\\
&=&
\dots
-
\frac{i}{2}
\partial_3 \partial_3' G_F(x,x')\Pi_0\Big|_{x'=x}
\label{G_F(x,x')}\\
&=&
\dots
+
\frac{i}{2}
\partial_3 \partial_3' G_D(x,x')\Pi_0\Big|_{x'=x}.\nonumber
\ee
Of course, all the ``propagators" are here defined in terms of a general $I_0(\bm{k})$, and not necessarily with $I_0(\bm{k})=1$.
Formula (\ref{G_F(x,x')}) is often used in the context of the Casimir effect \cite{Milton}.

\section{Explicit form of pressure operator at $z=L/2$ for parallel plates}

$\tilde f(k_z,z)$ can be obtained from $f(k_z,z)$ if one replaces in the latter $L$ by $2L$ and $\alpha$ by $\alpha/2$:
\be
\tilde f(k_z,z)
&=&
\left\{
\begin{array}{l}
\Big(\theta(k_z)+\theta(-k_z)\tilde E_1(|k_z|)\Big)
e^{ik_zz}
+
\theta(k_z)\tilde B_1(|k_z|)e^{-ik_zz}
,\\
\tilde C_1(|k_z|) e^{ik_zz}+\tilde D_1(|k_z|)e^{-ik_zz},\\
\Big(
\theta(-k_z)
+
\theta(k_z)\tilde E_1(|k_z|)\Big)
e^{ik_zz}
+
\theta(-k_z)\tilde B_1(|k_z|) e^{-ik_zz}
\end{array}
\right.
\nonumber
\ee
where
\be
\tilde B_1(k) &=& -\frac{ ik\alpha \cos k L+ i(\alpha^2/2) \sin k L}
{k^2+i \alpha k+\left(-1+e^{2i k L}\right) \alpha^2/4},\nonumber\\
\tilde C_1(k) &=&\frac{k (k+i \alpha/2 )}{k^2+i \alpha k+\left(-1+e^{2i k L}\right) \alpha^2/4},\nonumber\\
\tilde D_1(k) &=&-\frac{i e^{i k L} k \alpha/2 }{k^2+i \alpha k+\left(-1+e^{2i k L}\right) \alpha^2/4},\nonumber\\
\tilde E_1(k) &=& \frac{k^2}{k^2+i \alpha k+\left(-1+e^{2i k L}\right) \alpha^2/4}.\nonumber
\ee
After somewhat tedious and boring calculations we find
\be
{}&{}&
\Re\Bigg(\overline{\Big(\alpha\tilde f'\big(k_z,(L/2)_-\big)
+
\frac{\alpha^2}{2}\tilde f(k_z,L/2)\Big)}
\tilde f(k_z,L/2)\Bigg)\nonumber\\
&{}&\pp=
=
\left\{
\begin{array}{lcl}
\frac{-|k_z|^4\alpha^2/2}{\big||k_z|^2+i \alpha |k_z|+\left(-1+e^{2i |k_z| L}\right) \alpha^2/4\big|^2}
&\textrm{for} & k_z=|k_z|
\\
\frac{|k_z|^4\alpha^2/2}
{\big||k_z|^2+i \alpha |k_z|+\left(-1+e^{2i |k_z| L}\right) \alpha^2/4\big|^2}
+
2k_z^2
-
\frac{2|k_z|^6+\alpha^2|k_z|^4}
{\big||k_z|^2+i \alpha |k_z|+\left(-1+e^{2i |k_z| L}\right) \alpha^2/4\big|^2}
&\textrm{for} & k_z=-|k_z|
\end{array}
\right.\nonumber,\\
\ee
so that the operator representing the Casimir pressure (exerted by the field on the boundary) can be written more explicitly as
\be
p_{\rm C}
&=&
-p_{\rm vac}
\\
&=&
-2\Pi_0
\int_{\mathbb{R}^2} d^2k_\parallel\int_{-\infty}^0\frac{dk_z}{(2\pi)^3 2\sqrt{\bm k_\parallel^2+k_z^2}}
I_0(\bm{k}_\parallel,k_z) k_z^2
\nonumber\\
&\pp=&+
\Pi_0
\int_{\mathbb{R}^2} d^2k_\parallel\int_{-\infty}^0\frac{dk_z}{(2\pi)^3 2\sqrt{\bm k_\parallel^2+k_z^2}}
I_0(\bm{k}_\parallel,k_z)
\frac{2k_z^6+\alpha^2k_z^4}
{\big|k_z^2+i \alpha k_z+\left(-1+e^{2i k_z L}\right) \alpha^2/4\big|^2}
\nonumber\\
&\pp=&+
\Pi_0
\int_{\mathbb{R}^2} d^2k_\parallel\int_{-\infty}^{\infty}\frac{dk_z}{(2\pi)^3 2\sqrt{\bm k_\parallel^2+k_z^2}}
I_0(\bm{k}_\parallel,k_z)
\frac{\varepsilon(k_z)k_z^4\alpha^2/2}
{\big|k_z^2+i \alpha k_z+\left(-1+e^{2i k_z L}\right) \alpha^2/4\big|^2}.\nonumber\\
\ee

\section{Cross-check: $1+1$ space-time in irreducible representations}

Let us cross-check the formula for $p_{\rm C}$ in $1+1$ dimensions and $I_0(\bm k)=I$ (i.e. $|O_0(\bm{k})|^2=1$). We can ignore the contribution from the part involving $\varepsilon(k_z)$ due to its antisymmetry. After some simplifications we get
\be
p_{\rm C}
&=&
\frac{1}{2\pi}\int_0^{\infty}dk_z\,k_z\Bigg(1
-
\frac{1-|r(k_z)|^2}
{\big|1-r(k_z)e^{2i k_z L}\big|^2}\Bigg)
\label{p meijer}\\
&=&
\frac{1}{2\pi}\int_0^{\infty}dk_z\,k_z
\frac{r(k_z)e^{2i k_z L}}{1-r(k_z)e^{2i k_z L}}
+
{\rm c.c.}\label{P_C scatt}
\ee
where
\be
r(k_z) &=& \frac{1}{(1-2ik_z/\alpha)^2}.
\ee
The result (\ref{P_C scatt}) agrees with the general formula obtained in the $S$-matrix approach \cite{Reynaud}.
For $k_z\neq 0$ the integrand in (\ref{P_C scatt}) can be written in a form of an absolutely convergent series, and
\be
\lim_{k_z\to 0}k_z
\frac{r(k_z)e^{2i k_z L}}{1-r(k_z)e^{2i k_z L}}=0.
\ee
Therefore,
\be
p_{\rm C}
&=&
\frac{1}{2\pi}\int_0^{\infty}dk_z\,k_z
\sum_{n=1}^\infty \frac{e^{2ni k_z L}}{(1-2ik_z/\alpha)^{2n}}
+
{\rm c.c.}\label{P_C scatt01}\\
&=&
\frac{1}{2\pi}\sum_{n=1}^\infty \int_0^{\infty}dk_z\,k_z
\frac{e^{2ni k_z L}}{(1-2ik_z/\alpha)^{2n}}
+
{\rm c.c.}\label{P_C scatt1}
\ee
and one is allowed to integrate term-by-term. Let us note that $1/(1-2ik_z/\alpha)^{2n}$ is close to 1 for $k_z\ll \alpha$ and decays to 0 for $k_z\gg \alpha$. A finite $\alpha$ thus creates an effective cutoff in these integrals.

The standard result
\be
p_{\rm C}=-
\frac{1}{2\pi} \Bigg(\int_0^{\infty}dk_z\,k_z
-\frac{\pi^2}{L^2}\sum_{j=1}^\infty j\Bigg)
,\label{P_C scatt2}
\ee
typical of a fully reflecting cavity wall at $z=L/2$ (Dirichlet boundary conditions or, equivalently, $\alpha\to\infty$)
is obtained in one takes the limit $\alpha\to\infty$ in the denominator of (\ref{P_C scatt1}), and then appropriately computes the difference between the sum and the integral. Had one computed the limit under the integral in (\ref{P_C scatt}), the sum in (\ref{P_C scatt2}) would not appear, and the result would be divergent. Let us take a closer look at the above derivations.

\subsection{Euler-MacLaurin formula}

Consider some function $f(x)$ and its integral $\int_0^af(x)dx$. Let us split the segment $[0,a]$ into $n$ segments of length $\Delta x=a/n$. The integral can be approximated by a sum by means of the ``trapezoidal rule"
\be
{}&{}&\Delta x\Bigg(\frac{f(0)+f(\Delta x)}{2}+\frac{f(\Delta x)+f(2\Delta x)}{2}+\dots
+\frac{f\big((n-1)\Delta x\big)+f(n\Delta x)}{2}\Bigg)
\nonumber\\
&{}&\pp==
\Delta x\Bigg(\frac{f(0)}{2}+f(\Delta x)+\dots
+f\big((n-1)\Delta x\big)+\frac{f(a)}{2}\Bigg)
\nonumber\\
&{}&\pp==
\Delta x\sum_{j=0}^{a/\Delta x}f(j\Delta x)
-\Delta x\frac{f(0)+f(a)}{2}
\nonumber\\
&{}&\pp==
\int_0^af(x)dx+R,
\ee
where $R$ is the correction term.
If $f(x)=x$ one gets $R=0$ and the formula is exact.
Indeed,
\be
\Delta x\sum_{j=0}^{a/\Delta x}f(j\Delta x)
-\Delta x\frac{f(0)+f(a)}{2}
&=&
\Delta x\sum_{j=1}^{a/\Delta x}j\Delta x
-\Delta x\frac{a}{2}\nonumber\\
&=&
\Delta x\frac{(1+a/\Delta x)a/\Delta x}{2}\Delta x
-\Delta x\frac{a}{2}\nonumber\\
&=&
\frac{a^2}{2}=\int_0^a xdx\nonumber
\ee
In the general case we can derive a formula for $R$ by means of mathematical induction. One begins with
\be
\int_0^af(x)dx
&=&
\sum_{j=1}^{a/\Delta x}\int_{(j-1)\Delta x}^{{j\Delta x}}f(x)dx\label{P_1(x)}.
\ee
Now let $B_{j}^{\Delta x}(x)=x+c_{j}^{\Delta x}$, $c_{j}^{\Delta x}=$~const, and let us concentrate on a single element of the sum in (\ref{P_1(x)}),
\be
\int_{(j-1)\Delta x}^{{j\Delta x}}f(x)dx
&=&
\int_{(j-1)\Delta x}^{{j\Delta x}}f(x)\frac{dB_{j}^{\Delta x}(x)}{dx}dx\nonumber\\
&=&
-\int_{(j-1)\Delta x}^{{j\Delta x}}\frac{f(x)}{dx}B_{j}^{\Delta x}(x)dx
+f(x)B_{j}^{\Delta x}(x)\Big|_{(j-1)\Delta x}^{{j\Delta x}}
\nonumber\\
&=&
-\int_{(j-1)\Delta x}^{{j\Delta x}}f'(x)B_{j}^{\Delta x}(x)dx\nonumber\\
&\pp=&\pp=
+f(j\Delta x)B_{j}^{\Delta x}(j\Delta x)-f\big((j-1)\Delta x)\big)B_{j}^{\Delta x}\big((j-1)\Delta x)\big)
\nonumber
\ee
If we choose $c_{j}^{\Delta x}$ in such a way that $B_{j}^{\Delta x}(j\Delta x)=\Delta x/2$ and $B_{j}^{\Delta x}\big((j-1)\Delta x)=-\Delta x/2$, then
\be
\int_{(j-1)\Delta x}^{{j\Delta x}}f(x)dx
&=&
-\int_{(j-1)\Delta x}^{{j\Delta x}}f'(x)B_{j}^{\Delta x}(x)dx+
\Delta x\frac{f(j\Delta x)+f\big((j-1)\Delta x)\big)}{2}.
\label{Euler-MacL j}
\ee
Let us explicitly determine $c_{j}^{\Delta x}$. We get
\be
B_1(j\Delta x) &=& j\Delta x+c_{j}^{\Delta x}=\Delta x/2,
\ee
so $c_{j}^{\Delta x}=\Delta x(1/2-j)$, and
\be
B_{j}^{\Delta x}\big((j-1)\Delta x) &=& (j-1)\Delta x+\Delta x(1/2-j)=-\Delta x/2,
\ee
as required. So,
\be
B_{j}^{\Delta x}(x)=x+\Delta x(1/2-j).
\ee
For $\Delta x=1$ the function
\be
B_{1}^{\Delta x}(x)=x-1/2
\ee
is known as the first Bernoulli polynomial $B_1(x)=x-1/2$. In general, the Bernoulli polynomials satisfy $B_0(x)=1$ and
\be
\frac{dB_n(x)}{dx} &=& nB_{n-1}(x),\\
\int_0^1 B_n(x) dx &=& 0,
\ee
for $n>0$.
The procedure leading to (\ref{Euler-MacL j}) could be applied for all $j=1,\dots,n$,
\be
\int_{0}^{{\Delta x}}f(x)dx
&=&
-\int_{0}^{{\Delta x}}f'(x)\Big(x-\frac{1}{2}\Delta x\Big) dx+
\Delta x\frac{f(\Delta x)+f(0)}{2},\nonumber\\
\int_{\Delta x}^{{2\Delta x}}f(x)dx
&=&
-\int_{\Delta x}^{{2\Delta x}}f'(x)\Big(x-\frac{3}{2}\Delta x\Big)dx+
\Delta x\frac{f(2\Delta x)+f(\Delta x)}{2},\nonumber\\
&\vdots&\nonumber\\
\int_{(n-1)\Delta x}^{{n\Delta x}}f(x)dx
&=&
-\int_{(n-1)\Delta x}^{{n\Delta x}}f'(x)\Big(x-\frac{2n-1}{2}\Delta x\Big)dx+
\Delta x\frac{f(a)+f[(n-1)\Delta x]}{2}.\nonumber
\ee
Adding all these formulas one obtains
\be
\int_{0}^{{a}}f(x)dx
&=&
-\int_{0}^{{a}}f'(x)P_1^{\Delta x}(x)dx
\nonumber\\
&\pp=&\pp=
+
\Delta x\sum_{j=0}^{a/\Delta x}f(j\Delta x)
-\Delta x\frac{f(0)+f(a)}{2}
\ee
where $P_1^{\Delta x}(x)$ is a saw-shaped discontinuous function defined, essentially, by $B_{j}^{\Delta x}(x)$ in each of the intervals. $P_1^{\Delta x}(x)$  is unique in each of the open segments $(j-1)\Delta x <x<j\Delta x$, but ambiguities remain at points $\Delta x$, $2\Delta x,\dots,(n-1)\Delta x$. This non-uniqueness is not really essential and the argument will remain unchanged no matter what values of $P_1^{\Delta x}(j\Delta x)$ one selects. In conclusion
\be
R &=& \int_{0}^{{a}}f'(x)P_1^{\Delta x}(x)dx.
\ee
Let us note that for $f(x)=x$ one gets
\be
\int_{0}^{{a}}P_1^{\Delta x}(x)dx=0,
\ee
which is automatically guaranteed by the saw-shaped form of $P_1^{\Delta x}(x)$.

For a more general $f$ we can iterate the procedure,
\be
R
&=&
\int_{0}^{{a}}f'(x)P_1^{\Delta x}(x)dx\nonumber\\
&=&
\int_{0}^{{a}}f'(x)\frac{dP_2^{\Delta x}(x)}{dx}dx\nonumber\\
&=&
f'(a)P_2^{\Delta x}(a)-f'(0)P_2^{\Delta x}(0)- \int_{0}^{{a}}f''(x)P_2^{\Delta x}(x)dx\nonumber,
\ee
and so on. We have introduced a new function $P_2^{\Delta x}(x)$, satisfying for $(j-1)\Delta x <x<j\Delta x$
\be
P_1^{\Delta x}(x)
&=&
\frac{dP_2^{\Delta x}(x)}{dx},\\
P_2^{\Delta x}(x)
&=&
\frac{1}{2}x^2+\Delta x(1/2-j)x+d_j^{\Delta x}.
\ee
Note that at $a=n\Delta x$, $j=n=a/\Delta x$,
\be
P_2^{\Delta x}(a)
&=&
\frac{1}{2}n^2\Delta x^2+\Delta x(1/2-n)n\Delta x+d_n^{\Delta x}
\nonumber\\
&=&
\frac{1}{2}n^2\Delta x^2+\frac{1}{2}n\Delta x^2-n^2\Delta x^2+d_n^{\Delta x}
\nonumber\\
&=&
\frac{1}{2}n(1-n)\Delta x^2+d_n^{\Delta x}
\nonumber
\ee
We can require that for functions piecewise quadratic in each of the intervals the integral occurring in $R$ vanishes, i.e.
\be
0
&=&\int_{(j-1)\Delta x}^{j\Delta x}P_2^{\Delta x}(x)dx\label{0=R}\\
&=&\int_{(j-1)\Delta x}^{j\Delta x}\Big(\frac{1}{2}x^2+\Delta x(1/2-j)x+d_j^{\Delta x}\Big)dx\nonumber
\ee
which implies
\be
d_j^{\Delta x}=\frac{\Delta x^2}{12}+\frac{1}{2}j(j-1)\Delta x^2
\ee
and
\be
P_2^{\Delta x}(a)=P_2^{\Delta x}(0)=\Delta x^2/12.
\ee
Finally,
\be
R
&=&
\frac{\Delta x^2}{12}\big(f'(a)-f'(0)\big)- \int_{0}^{{a}}f''(x)P_2^{\Delta x}(x)dx.\label{R''}
\ee
For $\Delta x=1$, $j=1$, the formula is related to the second Bernoulli polynomial, $B_2(x)$,
\be
P_2^{\Delta x}(a)=\frac{1}{2}\Big(x^2-x+\frac{1}{6}\Big)=\frac{1}{2}B_2(x).
\ee
The procedure that has led us to (\ref{R''}) can be further iterated an arbitrary number of times. In effect, if $f$ is smooth, one replaces the integrals occurring in $R$ by a series. For $\Delta x=1$ and an integer $N$ one arrives at the Euler-MacLaurin formula
\be
\sum_{n=0}^N f(n)-\frac{f(N)+f(0)}{2} -\int_0^N f(x)dx=\sum_{j=1}^\infty \frac{B_{2j}}{(2j)!}\Big( f^{(2j-1)}(N)-f^{(2j-1)}(0)\Big),\label{E-ML}
\ee
where $B_j$ are the Bernoulli numbers. Of particular interest to us is $B_2=1/6=B_2(0)$, as we shall see in a moment.

Obviously, (\ref{R''}) is simple and elegant but it is hard to find a fundamental justification for (\ref{0=R}). In fact, one could continue the procedure an arbitrary number of times with arbitrary values of $d_j^{\Delta x}$.

\subsection{Application to the Casimir pressure}

Let me now explain how the Euler-MacLaurin formula is typically employed \cite{Reynaud} in evaluation of (\ref{P_C scatt2})
\be
p_{\rm C}=
\frac{1}{2\pi} \Bigg(\frac{\pi^2}{L^2}\sum_{j=1}^\infty j-\int_0^{\infty}dk_z\,k_z
\Bigg).
\ee
In order to directly use  (\ref{E-ML}) let us concentrate on the particular case $L=\pi$. Let $\chi(k_z)$ be a cutoff function which is exactly zero for $k_z\geq N$, and let us first compute the auxiliary problem for $f(k_z)=k_z \chi(k_z)$,
\be
p_{\rm C,N} &=&
\frac{1}{2\pi} \Bigg(\sum_{j=1}^\infty j\chi(j)-\int_0^{\infty}dk_z\,k_z\chi(k_z)
\Bigg)\nonumber\\
&=&
\frac{1}{2\pi} \Bigg(\sum_{j=1}^N j\chi(j)-\int_0^{N}dk_z\,k_z\chi(k_z)
\Bigg)\nonumber\\
&=&
\frac{1}{2\pi}\sum_{j=1}^\infty \frac{B_{2j}}{(2j)!}\Big( f^{(2j-1)}(N)-f^{(2j-1)}(0)\Big) \nonumber\\
&=&
\frac{1}{2\pi}\frac{B_{2}}{2}\Big( f'(N)-f'(0)\Big)
+
\frac{1}{2\pi}\sum_{j=2}^\infty \frac{B_{2j}}{(2j)!}\Big( f^{(2j-1)}(N)-f^{(2j-1)}(0)\Big) \nonumber\\
&=&
\frac{1}{2\pi}\frac{B_{2}}{2}\Big( N\chi'(N)+\chi(N)-\chi(0)\Big)
+
\frac{1}{2\pi}\sum_{j=2}^\infty \frac{B_{2j}}{(2j)!}\Big( f^{(2j-1)}(N)-f^{(2j-1)}(0)\Big) \nonumber
\ee
The result becomes independent of $N$ if one makes the assumption that $f(k_z)$ and all its derivatives identically vanish at $k_z=N$ and, moreover, the same holds at 0, with the exception of $\chi(0)=1$ \cite{IZ}. Then
\be
p_{\rm C,N} &=&
-\frac{1}{2\pi}\frac{B_{2}}{2}
=
-\frac{1}{24\pi}.
\ee
For arbitrary $L$ the result is \cite{Reynaud}
\be
p_{\rm C,N} &=&
-\frac{1}{2\pi}\frac{\pi^2}{L^2}\frac{B_{2}}{2}
=
-\frac{\pi}{24 L^2}=p_{\rm C}.
\ee
Under these general assumptions about $\chi$ one finds the Casimir pressure that depends only on the value of $\chi(k_z)$ at the origin. The result is somewhat disturbing from my point of view since, as the readers might have guessed, the cutoff appears in the place where I would expect my vacuum wave function $|O(k_z)|^2$. However, it is essential that $|O(0)|^2=0$, so my guess is that $\chi(0)=0$ together with all its derivatives. Does it mean that my formalism implies a vanishing Casimir pressure in the 1+1 dimensional space-time?

Before we can answer this question we have to return to the form of $p_{\rm C}$ we have derived in (\ref{P_C scatt01}).

\subsection{Direct calculation of $p_{\rm C}$ for $\alpha\to\infty$}

The function $f(k_z)=k_z$ is so simple that one should be able to compute $p_{\rm C}$ more directly, without any need of abstract assumptions about $\chi(k_z)$. So, let us return to the integral (\ref{P_C scatt01})
\be
p_{\rm C}
&=&
\frac{1}{2\pi}\int_0^{\infty}dk_z\,k_z\Bigg(1
-
\frac{1-|r(k_z)|^2}
{\big|1-r(k_z)e^{2i k_z L}\big|^2}\Bigg)
\label{p meijer'}\\
&=&
\frac{1}{2\pi}\int_0^{\infty}dk_z\,k_z
\frac{r(k_z)e^{2i k_z L}}{1-r(k_z)e^{2i k_z L}}
+
{\rm c.c.}\label{P_C scatt'}
\ee
and inspect more closely the properties of the integrand.
\begin{figure}\label{meijers}
\includegraphics[width=8cm]{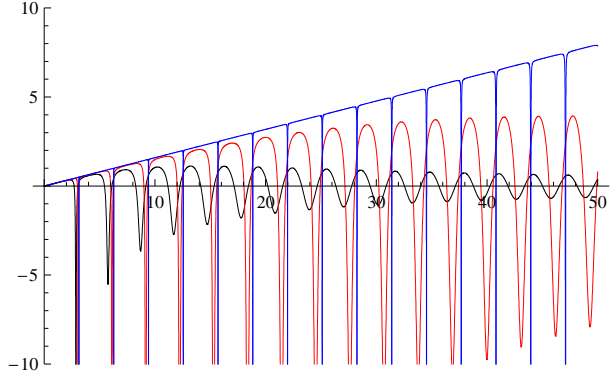}
\caption{The integrand occurring in (\ref{p meijer'}) for $L=1$ and $\alpha=20$ (black), $\alpha=70$ (red), and $\alpha=300$ (blue). With $\alpha\to\infty$ the integrand tends pointwise to $k_z/(2\pi)$, the delta-like peaks disappearing. However, the limit has to be performed in the style of delta-sequences, i.e. first integrate then take the limit.}
\end{figure}
In Fig.~10 the integrand is plotted for three different values of $\alpha$. With $\alpha\to\infty$ it tends pointwise to $k_z/(2\pi)$, the peaks being lost. However, it is evident that we have to perform the limit in a form that preserves the delta-sequence nature of the peaks. A simple procedure that maintains the logic of delta-sequences is to employ the fact that  the integrand of (\ref{P_C scatt'}) is a sum of an absolutely convergent geometric series,
\be
p_{\rm C}
&=&
\frac{1}{2\pi}\int_0^{\infty}dk_z\,k_z
\frac{r(k_z)e^{2i k_z L}}{1-r(k_z)e^{2i k_z L}}
+
{\rm c.c.}
\nonumber\\
&=&
\frac{1}{2\pi}\int_0^{\infty}dk_z\,k_z
\sum_{n=0}^\infty r(k_z)^{n+1}e^{2(n+1)i k_z L}
+
{\rm c.c.}\label{geom s}
\ee
Rewriting (\ref{geom s}) as
\be
p_{\rm C}
&=&
\frac{1}{2\pi}\lim_{N\to\infty}\int_0^{\infty}dk_z\,k_z
\sum_{n=1}^N r(k_z)^{n}e^{2ni k_z L}
+
{\rm c.c.}
\ee
we approximate the integrand by its asymptotic form for $\alpha\to \infty$
\be
p_{\rm C}
&\approx&
\frac{1}{2\pi}\lim_{N\to\infty}\int_0^{\infty}dk_z\,k_z
\sum_{n=1}^N e^{2ni k_z L}
+
{\rm c.c.}\nonumber\\
&=&
\frac{1}{\pi}\lim_{N\to\infty}\int_0^{\infty}dk_z\,k_z
\sum_{n=1}^N \cos(2n k_z L)
\nonumber\\
&=&
\lim_{N\to\infty}\int_0^{\infty}dk_z
\frac{k_z}{2\pi}\Big(\frac{\sin [(2N+1)k_zL]}{\sin(k_zL)}-1\Big).
\label{sin[2N+1]}
\ee
\begin{figure}\label{meijers1}
\includegraphics[width=8cm]{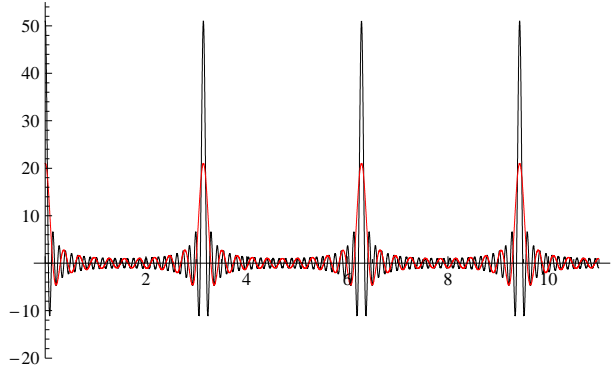}
\caption{Function $\sin [(2N+1)x]/\sin(x)$ for $N=10$ (red) and $N=25$ (black). With $N\to\infty$ the peaks behave like Dirac deltas concentrated at multiples of $\pi$.}
\end{figure}
Fig.~11 shows the function $\sin [(2N+1)x]/\sin(x)$ occurring in the integrand of (\ref{sin[2N+1]}). With $N\to\infty$ the peaks occurring at integer multiples of $\pi$ behave as delta-sequences. In order to see it let us divide the range of integration as follows
\be
\int_0^\infty &=& \int_0^{\frac{\pi}{2L}}+\int_{\frac{\pi}{2L}}^{\frac{3\pi}{2L}}+\dots+\int_{\frac{(2M+1)\pi}{2L}}^{\frac{(2M+3)\pi}{2L}}+\dots
\label{int_0^infty}
\ee
The Riemann-Lebesgue lemma implies that
\be
\lim_{N\to\infty}\int_0^{\frac{\pi}{2L}}dk_z\,k_z
\frac{\sin [(2N+1)k_zL]}{\sin(k_zL)} &=& 0,\\
\lim_{N\to\infty}\int_a^{b}dk_z,k_z
\frac{\sin [(2N+1)k_zL]}{\sin(k_zL)} &=& 0,\label{int_a^b 2N+1}
\ee
if $[a,b]$ does not contain a zero of $\sin(k_zL)$.
For all $L$ and $N$
\be
\frac{1}{2\pi} \int_{\frac{(2j+1)\pi}{2L}}^{\frac{(2j+3)\pi}{2L}}
dk_z\,k_z
\frac{\sin [(2N+1)k_zL]}{\sin(k_zL)}
&=&\frac{(j+1)\pi}{2L^2},\quad \textrm{for }j=0,1,2,\dots \label{int_a^b 2N+1'}
\ee
Taking into account (\ref{int_a^b 2N+1}) and (\ref{int_a^b 2N+1'}) one gets
\be
\lim_{N\to\infty}\frac{1}{2\pi}\int_a^{b}dk_z,k_z
\frac{\sin [(2N+1)k_zL]}{\sin(k_zL)}
&=&\frac{(j+1)\pi}{2L^2},\quad \textrm{for }j=0,1,2,\dots\\
&=&
\frac{1}{2\pi}\int_a^{b}dk_z,k_z\frac{\pi}{L}\delta\Big(k_z-(j+1)\frac{\pi}{L}\Big),
\ee
if $[a,b]$ contains a single point $x=(j+1)\pi/L$.

Therefore, it is justified to write
\be
k_z\frac{\sin [(2N+1)k_zL]}{\sin(k_zL)}
&=&
k_z\sum_{j=-\infty}^\infty \frac{\pi}{L}\delta_N\Big(k_z-j\frac{\pi}{L}\Big)
\ee
where $\delta_N$ denotes a delta-sequence. Returning to the asymptotic (for large $\alpha$) form of the Casimir pressure we find
\be
p_{\rm C}
&\approx&
\lim_{N\to\infty}\int_0^{\infty}dk_z
\frac{k_z}{2\pi}\Big(\frac{\sin [(2N+1)k_zL]}{\sin(k_zL)}-1\Big)\nonumber\\
&=&
\int_0^{\infty}dk_z
\frac{k_z}{2\pi}\Bigg(\sum_{j=-\infty}^\infty \frac{\pi}{L}\delta\Big(k_z-j\frac{\pi}{L}\Big)-1\Bigg)\nonumber
\ee
Now we can proceed in the standard way \cite{IZ,Reynaud}: Introduce a cutoff $K$,
\be
J\pi/L\leq K=J\pi/L+\kappa<(J+1)\pi/L,\nonumber
\ee
for some positive integer $J$, compute the integral from 0 to $K$, and in the end take the limit $K\to\infty$. So,
\be
p_{\rm C}
&\approx&
\lim_{K\to\infty}\int_0^{K}dk_z
\frac{k_z}{2\pi}\Bigg(\sum_{j=-\infty}^\infty \frac{\pi}{L}\delta\Big(k_z-j\frac{\pi}{L}\Big)-1\Bigg)\nonumber\\
&=&
\lim_{K\to\infty}
\Bigg(
\sum_{j=0}^{\lfloor KL/\pi\rfloor} j\frac{\pi}{2L^2}-\frac{K^2}{4\pi}
\Bigg)\nonumber\\
&=&
\lim_{J\to\infty}
\Bigg(
\sum_{j=0}^{J} j\frac{\pi}{2L^2}-\frac{(J\pi/L+\kappa)^2}{4\pi}
\Bigg)\nonumber\\
&=&
\lim_{J\to\infty}
\frac{-L^2 \kappa^2 + J \pi (\pi - 2 L \kappa)}{4 L^2 \pi}
\ee
The result is ambiguous. There is only one case where there is no divergence, and it occurs for $\pi - 2 L \kappa=0$. This is precisely the same splitting of the integral as in (\ref{int_0^infty}), corresponding to $\kappa=\pi/(2L)$ and $M+1=J$. Then,
\be
p_{\rm C}
&\approx&
-\frac{\pi}{16L^2}.\label{sum 0 to M+1}
\ee
The result is finite, but in the denominator we find 16 and not 24, as suggested by the Euler-MacLaurin formula!
\begin{figure}
\includegraphics[width=8cm]{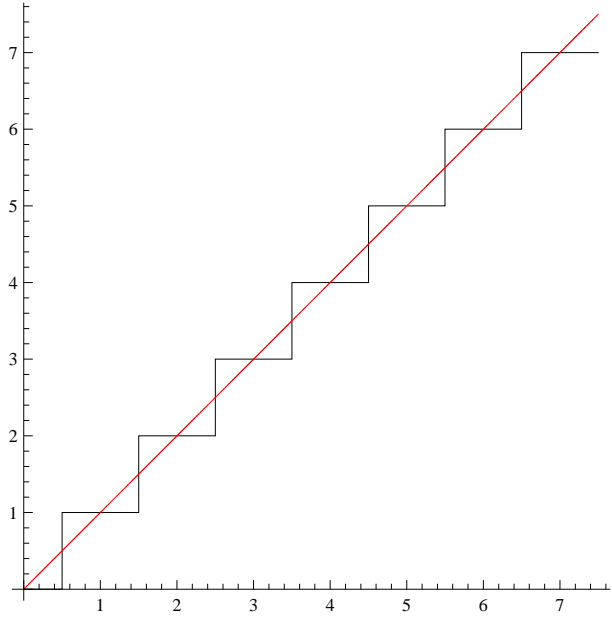}
\caption{Casimir pressure for $L=\pi$ and $J=7$. The area under the red line minus the one under the stairs equals the area of the leftmost triangle, i.e. $1/8$. The pressure is thus $-1/(2\pi)\times 1/8=-1/(16 \pi)$ in agreement with (\ref{sum 0 to M+1}). The area of the triangle is not sensitive to small changes of the red line in a neighborhood of $k_z=0$. The result is thus insensitive to a form of $\chi(k_z)$ at $k_z=0$.}
\end{figure}

Now, what about the putative importance of the value $\chi(0)$ of the cutoff function at $k_z=0$? The plot at Fig.~12 shows that the result {\it is not sensitive\/} to small changes of the red line in the neighborhood of $k_z=0$. One concludes that for small $k_z$ one could take a vanishing cutoff function, in contrast to the argument based on the Euler-MacLaurin formula. The latter has to be used with caution and one has to be more careful with argumentation based on series expansions of $\chi(k_z)$. There exist nontrivial smooth functions that vanish together with all their derivatives at $k_z=0$, and Taylor-like expansions at zero are then meaningless. This happens, in particular, for
the wave function (\ref{O_0-1/k}),
\be
|O_0(\bm k)|^2 &=& \frac{2\pi^2y^2}{\lambda^2K_2(2\lambda)} e^{-\frac{\lambda^2}{y\cdot k}-y\cdot k}.
\ee
The moral following from all these calculations is that, contrary to what one often reads in the literature, the Casimir effect --- even in simplest cases --- depends on cutoffs. For finite $\alpha$ the role of the cutoff function is played by $r(k_z)$, which indeed explicitly depends on $\alpha$. The very fact that the Casimir pressure can depend on cutoffs is in itself not new \cite{Hagen,Horsley}, a fact that is somewhat embarrassing for standard field quantization, but I have never encountered a paper where where one would arrive at the problem at such an elementary calculation.

Still, from my perspective the effect is very interesting as opening possibilities of directly testing the form of the vacuum wave function $O_0(\bm k)$.

\section{Casimir pressure in reducible representations of HOLA}

Let us return to the full 3+1 dimensional calculation. In reducible $N$-representations of HOLA one finds
\be
\langle O_0,N|p_{\rm C}|O_0,N\rangle
&=&
-2\int_{\mathbb{R}^3} \frac{d^3k}{(2\pi)^3 2|\bm k|}|O_0(\bm{k})|^2\theta(-k_z)
k_z^2
\nonumber\\
&\pp=&+
\int_{\mathbb{R}^3} \frac{d^3k}{(2\pi)^3 2|\bm k|}|O_0(\bm{k})|^2\theta(-k_z)
\frac{2k_z^6+\alpha^2k_z^4}
{\big|k_z^2+i \alpha k_z+\left(-1+e^{2i k_z L}\right) \alpha^2/4\big|^2}
\nonumber\\
&\pp=&+
\int_{\mathbb{R}^3} \frac{d^3k}{(2\pi)^3 2|\bm k|}|O_0(\bm{k})|^2
\frac{\varepsilon(k_z)k_z^4\alpha^2/2}
{\big|k_z^2+i \alpha k_z+\left(-1+e^{2i k_z L}\right) \alpha^2/4\big|^2}.\nonumber\\
\ee
a result that does not depend on $N$ (this parameter has, of course, nothing to do with $N$ we have used in the previous section).

We know that vacuum probability density $|O_0(\bm{k})|^2$ by assumption decays to 0 with $|\bm k|\to\infty$ and $|\bm k|\to 0$, but
in-between the two (UV and IR) regimes is assumed to be more or less flat, i.e.
$|O_0(\bm{k})|^2\approx Z=\max_{\bm k}\{|O_0(\bm{k})|^2\}$. In order to understand if we can nevertheless say something more about the structure of vacuum by measuring $\langle O_0,N|p_{\rm C}|O_0,N\rangle$, we have to take a closer look at properties of the above two integrands.

In order to compare with the standard result let us assume that
\be
|O_0(k_x,k_y,k_z)|^2\approx |O_0(k_x,k_y,-k_z)|^2\nonumber,
\ee
skip the integral involving the antisymmetric function $\varepsilon(k_z)$, and concentrate on
\be
\langle O_0,N|p_{\rm C}|O_0,N\rangle
&\approx&
-2\int_{\mathbb{R}^3} \frac{d^3k}{(2\pi)^3 2|\bm k|}|O_0(\bm{k})|^2\theta(-k_z)
k_z^2
\nonumber\\
&\pp=&+
\int_{\mathbb{R}^3} \frac{d^3k}{(2\pi)^3 2|\bm k|}|O_0(\bm{k})|^2\theta(-k_z)
\frac{2k_z^6+\alpha^2k_z^4}
{\big|k_z^2+i \alpha k_z+\left(-1+e^{2i k_z L}\right) \alpha^2/4\big|^2}
\nonumber\\
&\stackrel{\alpha\to \infty}{=}&
\int_{\mathbb{R}^2} d^2k\int_0^{\infty}\frac{dk_z\,k_z^2}{(2\pi)^3 \sqrt{k_x^2+k_y^2+k_z^2}}|O_0(\bm{k})|^2
\Bigg(\sum_{j=-\infty}^\infty \frac{\pi}{L}\delta\Big(k_z-j\frac{\pi}{L}\Big)-1\Bigg)\nonumber\\
&=&
\int_{\mathbb{R}^2} d^2k\int_0^{\infty}\frac{dk_z\,k_z^2}{(2\pi)^3 \sqrt{k_x^2+k_y^2+k_z^2}}|O_0(\bm{k})|^2
\sum_{j=-\infty}^\infty \frac{\pi}{L}\delta\Big(k_z-j\frac{\pi}{L}\Big)\nonumber\\
&\pp=&-
\int_{\mathbb{R}^2} d^2k\int_0^{\infty}\frac{dk_z\,k_z^2}{(2\pi)^3 \sqrt{k_x^2+k_y^2+k_z^2}}|O_0(\bm{k})|^2
\nonumber\\
&=&
\sum_{j=0}^\infty
\int_{\mathbb{R}^2} d^2k\frac{j^2 \pi^3}{(2\pi)^3 L^3\sqrt{k_x^2+k_y^2+j^2\pi^2/L^2}}|O_0(k_x,k_y,j\pi/L)|^2
\nonumber\\
&\pp=&-
\int_{\mathbb{R}^2} d^2k\int_0^{\infty}\frac{dk_z\,k_z^2}{(2\pi)^3 \sqrt{k_x^2+k_y^2+k_z^2}}|O_0(\bm{k})|^2
\nonumber
\ee
Let us now continue the calculation with $|O_0(\bm{k})|^2=Ze^{-\frac{\lambda^2}{y_0\sqrt{k_x^2+k_y^2+k_z^2}}-y_0\sqrt{k_x^2+k_y^2+k_z^2}}$, whose parameters were estimated on the basis of the Coulomb law,
\be
\langle O_0,N|p_{\rm C}|O_0,N\rangle
&\approx&
\sum_{j=0}^\infty
Z\int_{\mathbb{R}^2} d^2k\frac{j^2 \pi^3}{(2\pi)^3 L^3\sqrt{k_x^2+k_y^2+j^2\pi^2/L^2}}
\nonumber\\
&\pp=&\pp=
\times
e^{-\frac{\lambda^2}{y_0\sqrt{k_x^2+k_y^2+j^2\pi^2/L^2}}-y_0\sqrt{k_x^2+k_y^2+j^2\pi^2/L^2}}
\nonumber\\
&\pp=&-
Z\int_{\mathbb{R}^2} d^2k\int_0^{\infty}\frac{dk_z\,k_z^2}{(2\pi)^3 \sqrt{k_x^2+k_y^2+k_z^2}}\nonumber\\
&\pp=&\pp=
\times
e^{-\frac{\lambda^2}{y_0\sqrt{k_x^2+k_y^2+k_z^2}}-y_0\sqrt{k_x^2+k_y^2+k_z^2}}
\label{p z K0}
\ee
Switching to polar coordinates
\be
(\ref{p z K0})
&=&
Z\sum_{j=0}^\infty
4\pi \int_0^\infty dr\,r\frac{j^2 \pi^3}{(2\pi)^3 L^3\sqrt{r^2+j^2\pi^2/L^2}}
e^{-\frac{\lambda^2}{y_0\sqrt{r^2+j^2\pi^2/L^2}}-y_0\sqrt{r^2+j^2\pi^2/L^2}}
\nonumber\\
&\pp=&-
4\pi Z\int_0^\infty dr\,r\int_0^{\infty}\frac{dk_z\,k_z^2}{(2\pi)^3 \sqrt{r^2+k_z^2}}
e^{-\frac{\lambda^2}{y_0\sqrt{r^2+k_z^2}}-y_0\sqrt{r^2+k_z^2}}\nonumber\\
&=&
Z\sum_{j=1}^\infty
2\pi \int_0^\infty ds\frac{j^2 \pi^3}{(2\pi)^3 L^3\sqrt{s+j^2\pi^2/L^2}}
e^{-\frac{\lambda^2}{y_0\sqrt{s+j^2\pi^2/L^2}}-y_0\sqrt{s+j^2\pi^2/L^2}}
\nonumber\\
&\pp=&-
2\pi Z\int_0^\infty ds\int_0^{\infty}\frac{dk_z\,k_z^2}{(2\pi)^3 \sqrt{s+k_z^2}}
e^{-\frac{\lambda^2}{y_0\sqrt{s+k_z^2}}-y_0\sqrt{s+k_z^2}}\nonumber
\ee
Now $t=\sqrt{s+k^2}$, $dt=ds/(2\sqrt{s+k^2})$, $k\leq t<\infty$, so
\be
&=&
Z\sum_{j=1}^\infty
 2\int_0^\infty ds\frac{j^2 \pi^3}{(2\pi)^2 L^3 2\sqrt{s+j^2\pi^2/L^2}}
e^{-\frac{\lambda^2}{y_0\sqrt{s+j^2\pi^2/L^2}}-y_0\sqrt{s+j^2\pi^2/L^2}}
\nonumber\\
&\pp=&-
2Z\int_0^\infty ds\int_0^{\infty}\frac{dk_z\,k_z^2}{(2\pi)^2 2\sqrt{s+k_z^2}}
e^{-\frac{\lambda^2}{y_0\sqrt{s+k_z^2}}-y_0\sqrt{s+k_z^2}}\nonumber\\
&=&
\sum_{j=1}^\infty\frac{2Zj^2 \pi^3}{(2\pi)^2 L^3}
\int_0^\infty \frac{ds}{ 2\sqrt{s+j^2\pi^2/L^2}}
e^{-\frac{\lambda^2}{y_0\sqrt{s+j^2\pi^2/L^2}}-y_0\sqrt{s+j^2\pi^2/L^2}}
\nonumber\\
&\pp=&-
\frac{2Z}{(2\pi)^2}\int_0^\infty dk_z\,k_z^2\int_0^{\infty}\frac{ds}{ 2\sqrt{s+k_z^2}}
e^{-\frac{\lambda^2}{y_0\sqrt{s+k_z^2}}-y_0\sqrt{s+k_z^2}}\nonumber\\
&=&
\sum_{j=1}^\infty\frac{Zj^2 \pi}{2 L^3}
\int_{j\pi/L}^\infty dt\,
e^{-\frac{\lambda^2}{y_0t}-y_0t}
-
\frac{Z}{2\pi^2}\int_0^\infty dk_z\,k_z^2\int_{k_z}^{\infty}dt\,
e^{-\frac{\lambda^2}{y_0t}-y_0t}\nonumber\\
&=&
\sum_{j=1}^\infty\frac{Zj^2 \pi}{2 L^3}
\int_{j\pi/L}^\infty dt\,
e^{-\frac{\lambda^2}{y_0t}-y_0t}
-
\frac{Z}{2\pi^2}\int_0^\infty dt\,\int_{0}^{t}dk_z\,k_z^2
e^{-\frac{\lambda^2}{y_0t}-y_0t}\nonumber\\
&=&
\sum_{j=1}^\infty\frac{Zj^2 \pi}{2 L^3}
\int_{j\pi/L}^\infty dt\,
e^{-\frac{\lambda^2}{y_0t}-y_0t}
-
\frac{Z}{6\pi^2}\int_0^\infty dt\,t^3
e^{-\frac{\lambda^2}{y_0t}-y_0t}\nonumber\\
&=&
\sum_{j=1}^\infty\frac{Zj^2 \pi}{2 L^3}
\frac{1}{y_0}\Gamma(1,y_0j\pi/L,\lambda^2)
-
\frac{2Z\lambda^4}{6\pi^2y_0^4}
K_4(2\lambda)
\nonumber
\ee
The result involves the $\alpha=1$ case of the generalized incomplete gamma function \cite{CZub}
\be
\Gamma(\alpha,x,b)=\int_x^\infty t^{\alpha-1}e^{-t-b/t}dt.
\ee
Since $\lambda^2$ is very small we can approximate $\Gamma(1,y_0j\pi/L,\lambda^2)$ by the first two terms of its Taylor expansion (cf. (2.91) in \cite{CZub}),
\be
\Gamma(\alpha,x,b) &=&\sum_{n=0}^\infty \frac{(-1)^n}{n!}\Gamma(\alpha-n,x)b^n,\\
\Gamma(\alpha,x) &=& \int_x^\infty t^{\alpha-1}e^{-t}dt,\\
\Gamma(1,x) &=& \int_x^\infty e^{-t}dt=e^{-x},\\
\Gamma(0,x) &=& \int_x^\infty t^{-1}e^{-t}dt,\\
\Gamma(1,x,b) &\approx& e^{-x}-\Gamma(0,x)b.
\ee
Combining this with
\be
\lambda^4 K_4(2\lambda)=3-\lambda^2+O(\lambda^4)
\ee
we get, up to terms linear in $\lambda^2$,
\be
\langle O_0,N|p_{\rm C}|O_0,N\rangle
&\approx&
\sum_{j=1}^\infty\frac{Zj^2 \pi}{2 L^3}
\frac{1}{y_0}
\Big(
e^{-y_0j\pi/L}-\Gamma(0,y_0j\pi/L)\lambda^2
\Big)
-
\frac{Z}{3\pi^2y_0^4}
(3-\lambda^2)
\nonumber\\
&=&
\frac{Z\pi}{2 L^3y_0}
\sum_{j=1}^\infty
j^2e^{-y_0j\pi/L}
-
\frac{Z}{\pi^2y_0^4}
-
\frac{Z\pi}{2 L^3y_0}
\sum_{j=1}^\infty j^2\Gamma(0,y_0j\pi/L)\lambda^2
+
\frac{Z}{3\pi^2y_0^4}
\lambda^2
\nonumber\\
&=&
\frac{Z \pi}{2 y_0L^3}\frac{e^{\frac{\pi  y_0}{L}} \big(1+e^{\frac{\pi  y_0}{L}}\big)}{\big(-1+e^{\frac{\pi  y_0}{L}}\big)^3}
-
\frac{Z}{\pi^2y_0^4}
\nonumber\\
&\pp=&
+
\Bigg(\frac{Z}{3\pi^2y_0^4}
-
\frac{Z\pi}{2 L^3y_0}
\sum_{j=1}^\infty j^2\Gamma(0,y_0j\pi/L)
\Bigg)
\lambda^2.
\ee
Taking $y_0\ell\sim 10^{-38}$ (the Planck scale) km and $L\ell\sim 10^{-12}$ km (the nanometer scale),
$y_0/L\sim 10^{-26}$, we expand the first term in powers of $y_0/L$,
\be
\langle O_0,N|p_{\rm C}|O_0,N\rangle
&\approx&
-Z\frac{\pi ^2}{240 L^4}
+\frac{Z}{L^4}\Big(\frac{\pi ^4 y_0^2}{3024
   L^2}-\frac{\pi ^6 y_0^4}{57600 L^4}+\frac{\pi ^8 y_0^6}{1330560 L^{6}}+O\left(y_0^7\right)\Big)
\nonumber\\
&\pp=&
+
\frac{Z}{2\pi^2y_0^4}\Bigg(\frac{2}{3}
-
\sum_{j=1}^\infty \frac{y_0\pi}{L}\Big(\frac{j y_0\pi}{L}\Big)^2\Gamma\Big(0,\frac{j y_0\pi}{L}\Big)
\Bigg)
\lambda^2.
\label{Final CP}
\ee
The first term is, up to constants, the standard prediction. The terms proportional to powers of $y_0/L$ are negligible (smallness of these terms illustrates, in standard Casimir-effect terminology, the ``cutoff independence" of the Casimir pressure).
It remains to estimate the correction proportional to $\lambda^2$.

We first observe that
$\int_0^\infty dx \,x^2\Gamma(0,x)=2/3$,
so the expression in braces is the difference
\be
\int_0^\infty dx \,x^2\Gamma(0,x)-\sum_{j=1}^\infty \Delta x \,(j \Delta x)^2\Gamma(0,j \Delta x)
\ee
with $\Delta x=y_0\pi/L\sim 10^{-26}$. Unfortunately, second derivative of $x^2\Gamma(0,x)$ is infinite at $x=0$, so the Euler--MacLaurin formula is not directly applicable. However, representing the sum by an integral from a step function, analogously to the black curve at Fig.~13, we can crudely estimate the order of the difference by
\be
\int_0^{\Delta x/2} dx \,x^2\Gamma(0,x)\sim 10^{-78}
\ee
so that
\be
\frac{L^4}{y_0^4}\Bigg(\frac{2}{3}
-
\sum_{j=1}^\infty \frac{y_0\pi}{L}\Big(\frac{j y_0\pi}{L}\Big)^2\Gamma\Big(0,\frac{j y_0\pi}{L}\Big)
\Bigg)
\sim
10^{26}.\nonumber
\ee
The data for the Coulomb law led to $\lambda^2\sim 10^{-49}$. 
Since Casimir pressure in parallel-plate configuration is currently tested with precision of some 15\% for $L_0=0.5$-3 $\mu$m \cite{Bressi}, the infrared correction we have derived is thus negligible. 

The standard result for the scalar-field Casimir pressure is \cite{Milton}
\be
p_{\rm C}=-\frac{\hbar c\pi ^2}{240 L_0^4}
\ee
where $L_0$ has a dimension of length. Our $L$ is dimensionless, so has to be multiplied by the unit of length, $L_0=L\ell$.
The overall factor $Z$ in (\ref{Final CP}) can be eliminated if one starts with the energy momentum tensor normalized by $C=1/(2Z)$ [see the remarks after
(\ref{C const in T})]. The dimensionless (\ref{Final CP}) has yet to be multiplied by $\hbar c/\ell^4$ to get the correct units of pressure.

\bigskip
{\bf To be continued...}

\end{document}